\documentclass[preprint,12pt]{elsarticle}

\usepackage{graphics,graphicx}
\usepackage{amssymb}
\journal{Physics Reports}

\newcommand{\gtorder}{\mathrel{\raise.3ex\hbox{$>$}\mkern-14mu
            \lower0.6ex\hbox{$\sim$}}}
\newcommand{\ltorder}{\mathrel{\raise.3ex\hbox{$<$}\mkern-14mu
            \lower0.6ex\hbox{$\sim$}}}

\begin{document}

\begin{frontmatter}

\title{The Masses and Spins of Neutron Stars and Stellar-Mass Black Holes}

\author{M. Coleman Miller$^1$ and Jon M. Miller$^2$}

\address{
$^1$University of Maryland, Department of Astronomy and Joint Space-Science Institute, College Park, MD 20742-2421\\
$^2$University of Michigan, Department of Astronomy,
500 Church Street, Ann Arbor, MI 48109-1042\\
}

\begin{abstract}

Stellar-mass black holes and neutron stars represent extremes in
gravity, density, and magnetic fields.  They therefore serve as key
objects in the study of multiple frontiers of physics.  In addition,
their origin (mainly in core-collapse supernovae) and evolution (via
accretion or, for neutron stars, magnetic spindown and
reconfiguration) touch upon multiple open issues in astrophysics.  

In this review, we discuss current mass and spin measurements and their
reliability for neutron stars and stellar-mass black holes, as well as the overall importance of spins and masses for compact object astrophysics.  Current masses are obtained primarily through electromagnetic observations of binaries, although future microlensing observations promise to enhance our understanding substantially.  The spins of neutron stars are straightforward to measure for pulsars, but the birth spins of neutron stars are more difficult to determine.  In contrast, even the current spins of stellar-mass black holes are challenging to measure.  As we discuss, major inroads have been made in black hole spin estimates via analysis of iron lines and continuum emission, with reasonable agreement when both types of estimate are possible for individual objects, and future X-ray polarization measurements may provide additional independent information.  We conclude by exploring the exciting prospects for mass and spin measurements from future gravitational wave detections, which are expected to revolutionize our understanding of strong gravity and compact objects.

\end{abstract}

\begin{keyword}
accretion disks \sep black holes \sep gravitational waves \sep neutron stars


\end{keyword}

\end{frontmatter}
\tableofcontents

\newpage

\section{Introduction}
\label{section:introduction}

Neutron stars were thought to be objects of mostly academic interest until the discovery of pulsars by Jocelyn Bell and Antony Hewish  \cite{1968Natur.217..709H}.  The acceptance of black holes as real sources was much more gradual.  The decades since the first proposals that quasars and other AGN were powered by supermassive black holes  \cite{1964ApJ...140..796S,1969Natur.223..690L} and the discovery of Cyg X-1 \cite{1965Sci...147..394B,1972Natur.235...37W, 1972Natur.235..271B,1972ApJ...177L...5T} and other stellar-mass black hole candidates have seen progressively more definitive identification of some  objects as black holes.  Now the discussion has turned from the reality of these objects to their properties and what they can tell us about fundamental physics and astrophysics.

In this review we will explore what has been learned about the masses and spins of neutron stars and stellar-mass black holes, and what insights the methods and results have yielded.  We begin in \S1.1 by giving context to these measurements, then present an overview of the general theory in \S1.2.  We discuss how masses are measured in \S2, how spins are measured in \S3, and what we can anticipate from future gravitational wave measurements in \S4.  We present our summary in \S5.  This collection of topics has not previously appeared in a single review, but there are several excellent previous articles that discuss some of the individual subjects.  For more detail about the masses and other properties of binary and millisecond pulsars we recommend the review by Lorimer \cite{2008LRR....11....8L}; for additional information about the masses of stellar-mass black holes we suggest the recent work by Kreidberg et al. \cite{2012ApJ...757...36K}; for a summary of the physics of Fe~K$\alpha$ lines a good reference is the review by Reynolds and Nowak \cite{2003PhR...377..389R}; for a discussion of the continuum fitting method of spin determinations we recommend the review by McClintock et al. \cite{2011CQGra..28k4009M}; and for more technical details about the physics and astrophysics of gravitational waves we suggest the reviews by Blanchet \cite{2014LRR....17....2B} and Sathyaprakash and Schutz \cite{2009LRR....12....2S}.

\subsection{Why do we care?}

{\it Fundamental physics.}---Neutron star and black holes are excellent
probes of strong gravity and dense matter.  For example, although Einstein's
theory of gravity, general relativity, has been confirmed in every
experiment and observation yet performed, all such data are restricted
to weak gravity, in the sense that $GM/rc^2$ (where $G$ and $c$ are
respectively Newton's gravitational constant and the speed of light,
$M$ is the gravitational mass of an object, and $r$ is size of the source,
e.g., the separation between the two neutron stars in a binary) 
is less than $\sim 10^{-5}$ (e.g., Figure~1 in
\cite{2008LRR....11....9P}).  Thus the predictions of
strong-field gravity with $GM/rc^2\sim 1$, such as the existence of
unstable circular orbits, are not easily tested.  Neutron stars, with
$GM/Rc^2\sim 0.1-0.3$, and black holes, with $GM/Rc^2\sim 1$ at their
event horizons, are thus natural laboratories for strong gravity.

Neutron stars are also invaluable objects for the study of cold dense
matter.  Their cores are believed to reach densities that are several times 
nuclear saturation density $\rho_0\sim 2.6\times 10^{14}$~g~cm$^{-3}$
(the density of atomic nuclei on Earth), and their composition could
be anything from mainly neutrons and protons to condensates or quark
hybrid matter (see \cite{2011Ap&SS.336...67L} for a recent 
review).  The boundary in mass
between neutron stars and black holes, and the radii of neutron stars, would provide key hints to the
nature of this high-density matter \cite{2013arXiv1312.0029M}.

{\it Astrophysics.}---As astrophysical sources, stellar-mass 
black holes provide prototypes
for the supermassive black holes in active galactic nuclei (AGN), with the
advantage that we receive much greater flux from stellar-mass black holes and their characteristic timescales are much shorter than in AGN.  Thus the study of state changes
and quasi-periodic variability in stellar-mass black holes can yield
unique insights into AGN.  Moreover, jets from accreting neutron stars
and stellar-mass black holes can serve as probes of magnetohydrodynamics
and plasma physics in relativistic conditions and as prototypes for the
wide range of jets in the universe, from protostellar systems to 
supermassive black holes.  These jets are believed to have a significant
impact on their environment, especially on the evolution of galaxies and
galaxy clusters and the formation of structure \cite{2005Natur.433...45M}, so the
relatively nearby neutron stars and stellar-mass black holes give us an
invaluable close up opportunity to study jet phenomena.

As the remnants of core-collapse supernovae, neutron stars and black holes
are also windows into the collapse and explosion processes.  The further
evolution of these systems in accreting binaries can tell us about the
accretion process, e.g., the total amount of mass that is typically
deposited on the compact object.  As we discuss in \S1.2, black holes
that are currently in a binary cannot have accreted enough mass from 
their companion to change their mass substantially.  It is also difficult
for a stellar-mass black hole to accrete enough mass to change significantly
the magnitude of its angular momentum.  Thus the {\it current} mass and
spin magnitude of a stellar-mass black hole is a relatively faithful record 
of its birth mass and spin magnitude, which in turn tells us about the
conditions in the pre-supernova star just before the collapse.  In contrast,
there is an ongoing debate about whether the spin {\it direction} of black
holes can be altered substantially by accretion.  As we will see, this
has a significant impact on the reliability of some types of spin
measurement.

In distinction to black holes, neutron stars can have their mass changed
substantially by accretion from a companion, and the spin magnitudes and
directions of accreting neutron stars are essentially wholly determined
by the accretion in low-mass X-ray binaries, and probably also in 
high-mass X-ray binaries.  It is not possible at this time to measure the 
masses of isolated neutron stars except perhaps by microlensing (and then the identification of the object as a neutron star is not certain).  The current spins of pulsars are obviously straightforward to measure, but these spins have been affected strongly by magnetic spindown.  However, as we discuss in
\S3, there are indirect ways to guess the birth spin of isolated pulsars
based on the spindown energy injected into the surrounding supernova
remnant, and such studies have somewhat surprisingly pointed towards
relatively long birth periods (a few tenths of a second) being typical.

Neutron stars and black holes in dense stellar systems such as globular
clusters and the Galactic center are also probes of the dynamics in such
systems.  For example, millisecond pulsars are thought to be spun up by
accretion from a companion (and what limits their spin is another issue
we will address).  This can happen in an isolated binary, but millisecond
pulsars are much more common per total stellar mass in dense stellar
systems.  This is thought to be because dynamical interactions, particularly
binary-single encounters, can allow previously isolated neutron stars to
swap into binaries and acquire companions that later donate mass and
spin up the neutron star.  As a result, one would expect that such stars
are systematically more massive than neutron stars that have not accreted
and spun up.  There is, likewise, a debate about whether there should
be a substantial number of stellar-mass black holes in globular clusters.
Work for many years has found that the holes would sink into a dense
subcluster and self-eject \cite{1993Natur.364..421K,
1993Natur.364..423S,2000ApJ...528L..17P,2006ApJ...637..937O,
2010MNRAS.402..371B}, but the recent discovery of black hole
candidates in several globular clusters \cite{2007Natur.445..183M,
2010ApJ...725.1805B,2010ApJ...721..323S,2011ApJ...734...79B,
2011MNRAS.410.1655M,2012Natur.490...71S} 
has led to re-examination
of globular cluster black hole dynamics \cite{2013ApJ...763L..15M,2013MNRAS.432.2779B,2013MNRAS.435.3272T,2013MNRAS.436..584B}.  
It could be that black
holes in globular clusters (where they have billions of years to eat
other stars) or low-metallicity environments (where winds and hence mass
loss are weaker) will tend to have higher mass than other black holes.
If so, this gives us valuable information about the evolution of these
objects.

Thus, overall, the study of neutron stars and stellar-mass black holes,
and especially their masses and spins, connects to
many frontier issues of fundamental physics and astrophysics.  We now
discuss some of the general theory of these objects, from their formation
to their later evolution. 

\subsection{General theory of neutron stars and black holes}

In the current universe, neutron stars and stellar-mass black holes
are most commonly created via core-collapse supernovae.  More rarely, black holes could be formed due to the merger of two neutron stars, and it has been suggested that the accretion-induced collapse of a white dwarf could create neutron stars (see \cite{1980AIPC...63....7C} and subsequent references), although observational confirmation remains elusive.  It has also been
proposed that in the early universe primordial black holes might have
formed, particularly during periods of phase change such as the
quark-hadron transition \cite{1974MNRAS.168..399C,1975ApJ...201....1C,
1978SvA....22..129N,1978ApJ...225..237B,1979A&A....80..104N,
1982PThPh..68.1979K,1998AstL...24..413K,1999PhRvD..59l4014J,
1999PhRvD..60f3509G}, but there is no evidence for such
primordial black holes.  We thus focus on the core-collapse formation
route after we lay out some basic physical scales.

{\it Basic scales.}---The range of known neutron star masses is
$1.25~M_\odot-2.01~M_\odot$ (PSR~J0737--3039B \cite{2004Sci...303.1153L} and PSR~J0348+0432 \cite{2013Sci...340..448A}, respectively; here $M_\odot=1.989\times 10^{33}$~g is the mass of the Sun).  Here, and
throughout this review, we use ``mass" to refer to the gravitational
mass of the object (i.e., the mass that would be inferred using 
Kepler's laws and observations of the orbit of a distant satellite)
instead of the baryonic mass (the mass one would obtain by adding
together the separate masses of every constituent particle in the
star).  The difference between the two is the gravitational binding
mass-energy of the star; for a neutron star, the baryonic mass is
expected to be 20-30\% larger than the gravitational mass.

Neutron star radii are not well known; estimates from nuclear theory suggest radii in
the range of $\sim 10-15$~km \cite{2013ApJ...773...11H}.  Here we mean
the circumferential radius, i.e., the radius obtained by dividing
a local measurement of the equatorial circumference by
$2\pi$, rather than other measures of the radius such the proper
distance from the center of the star (which can be different by
tens of percent).  Precise measurements
of the radius are coveted by nuclear physicists, because a well-known
radius and mass for a neutron star would provide important clues about
the nature of the matter in the core.  Unfortunately, the systematic
errors in current radius estimates are so large that they do not provide
useful inputs to nuclear theories (\cite{2013arXiv1312.0029M}; see \cite{2014arXiv1406.1497H} for a discussion in the context of quiescent low-mass X-ray binaries).  As we will discuss more in \S3,
the known spin frequencies of neutron stars range from millihertz to
hundreds of Hertz, with the current record holder at 716~Hz 
(PSR J1748--2446ad, in the globular cluster Terzan 5 \cite{2006Sci...311.1901H}).
Interestingly, the maximum possible spin frequency of a typical
neutron star, beyond which a gravitationally-bound object would fly
apart due to centrifugal acceleration, is probably $\sim 1500-2000$~Hz
\cite{1994ApJ...424..823C}.
Thus, as we will explore further in \S3, something limits the spin
(probably magnetic torques during accretion).  

Black holes, being objects of pure gravity, have no characteristic
mass.  For example, a sufficiently dedicated and powerful alien
species could presumably create mountain-mass black holes, and as we
said above such holes might have been produced in the early universe.
The masses of known black holes range from a few to tens of solar
masses (the stellar-mass black holes that are our focus) to the
galactic central black holes that are millions to billions of solar
masses.  There is an intriguing gap at hundreds to thousands of
solar masses that might be filled by intermediate-mass black holes
\cite{2004IJMPD..13....1M}, but currently there is no {\it dynamical} 
measurement of such a mass, unlike in the other two mass ranges.  

Isolated black holes have only three
possible parameters: mass, angular momentum, and electric charge.  The
net charge is always negligible for astrophysical black holes, as it
is for any macroscopic object, because so many free charges exist in
astrophysical settings that they would quickly accrete to cancel out
the charge.  The angular momentum $J$ of a black hole is usually 
parameterized by the dimensionless combination 
\begin{equation}
{\hat a}=cJ/GM^2\; .
\end{equation}
For a black hole, $0\leq|{\hat a}|\leq 1$.  This is necessary for the hole
to have a horizon, but note that it is not an overall limit for all objects:
for example ${\hat a}$ for the Earth is $\sim 10^3$.  The value of
${\hat a}$ for a maximally rotating neutron star is $\sim 0.7$
\cite{1994ApJ...424..823C}, assuming that the mass is below the stable
mass for a nonrotating star.  A value
of ${\hat a}=0.998$ is sometimes used as an ``astrophysical upper limit"
for black holes, after work by Kip Thorne \cite{1974ApJ...191..507T} 
showed that matter spiraling into a more
rapidly rotating black hole would be spun down by photons emitted retrograde 
to the spin.  This apparently small change, from 
${\hat a}_{\rm max}=1$ to ${\hat a}_{\rm max}=0.998$, has substantial
consequences for the maximum efficiency of energy generation (lowering
it to $\sim 32$\% from $\sim 40$\%) and other properties.  When magnetic
fields in accretion disks are taken into account, the maximum spin could
be lowered yet further, to perhaps ${\hat a}_{\rm max}\sim 0.9$
\cite{2004ApJ...602..312G,2005ApJ...622.1008K}.  
Any spinning object drags reference frames around with it, and
near a black hole this can produce pronounced effects.

\begin{figure}[!htb]
\begin{center}
\includegraphics[scale=0.7]{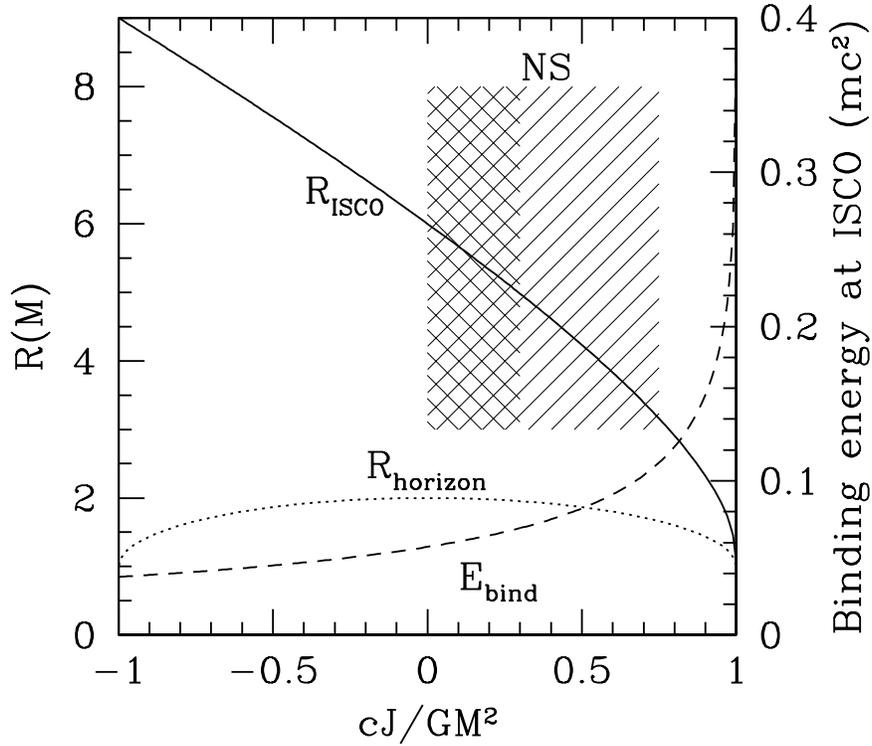}
\vskip -0.5cm
\caption{\footnotesize This figure shows the
dependence of the radius of the innermost stable circular orbit (ISCO, cast in
units of $GM/c^{2}$) on the spin parameter (${\hat a} = cJ/GM^{2}$) of a 
compact object, as well as the radius of the event horizon and the specific binding energy of a test particle in orbit at the ISCO.  Typically, the object of concern is a black hole, but as the hatched region shows the size of neutron stars relative to their ISCO is an open and
interesting question; here the darker hatching shows a plausible range of radii and spin magnitudes for observed stars, and the lighter hatching extends to the possible range of spin magnitudes.  In this plot, positive values of the spin
parameter correspond to prograde orbits, whereas negative values
correspond to retrograde orbits.}
\label{fig:isco}
\end{center}
\end{figure}

In standard Boyer-Lindquist coordinates, where the angles $\theta$ and
$\phi$ are as usual the colatitude and azimuth, the radius $r$ is the
circumferential radius, and intervals in the time $t$ are the intervals
between events as measured by a distant observer in flat spacetime, the
radius of the event horizon is
\begin{equation}
r_H=\left[1+(1-{\hat a}^2)^{1/2}\right]GM/c^2\; .
\end{equation}
Another important radius, which features prominently in estimates of the
spins of black holes, is the radius of the innermost stable circular orbit
(ISCO).  In general relativity, unlike in Newtonian gravity, orbits with
less than a certain radius around a point mass are not stable.  Heuristically,
one can think of this as stemming from the greater gravitational 
accelerations that exist in general relativity; a particle in a circular
orbit must go faster (in some sense) closer to the hole than it would in
Newtonian gravity.  Thus the angular momentum of a circular orbit as a
function of radius has a minimum (unlike in Newtonian gravity, where the
angular momentum decreases monotonically with decreasing radius).  Inside
the ISCO, then, small perturbations tend to move fluid elements in a
rapidly-opening spiral towards the hole.  The circumferential radius
of the ISCO is given by \cite{1972ApJ...178..347B}
\begin{equation}
\begin{array}{rl}
r_{\rm ISCO}&=GM/c^2\left\{3+Z_2\mp\left[(3-Z_1)(3+Z_1+2Z_2)\right]^{1/2}
\right\}\\
{\rm where}\\
Z_1&\equiv 1+(1-{\hat a}^2)^{1/3}\left[(1+{\hat a})^{1/3}+
(1-{\hat a})^{1/3}\right]\\
Z_2&\equiv (3{\hat a}^2+Z_1^2)^{1/2}\; .\\
\end{array}
\end{equation}
Here the -- sign in the expression for $r_{\rm ISCO}$ is for prograde orbits
with respect to the black hole spin, and the + sign is for retrograde orbits.
Thus a maximally spinning black hole with ${\hat a}=1$ has 
$r_{\rm ISCO}=GM/c^2$ for prograde orbits and $r_{\rm ISCO}=
9GM/c^2$ for retrograde orbits, and a nonspinning, or Schwarzschild,
black hole has $r_{\rm ISCO}=6GM/c^2$.  The heuristic explanation for
the role of spin is that frame-dragging gives prograde orbits a free
centrifugal boost, so the angular momentum of a circular orbit
at a given radius is less than it would be around a nonspinning black hole.
Thus the minimum angular momentum orbit is pushed in for prograde orbits,
and is likewise pushed out for retrograde orbits.  See Figure~\ref{fig:isco}
for a plot of the horizon radius, ISCO radius, and specific binding energy
at the ISCO for the range of possible black hole spin parameters.

\medskip

{\it Core-collapse supernovae.}---As
roughly five decades of supernova simulations attest, it is difficult
to include all the relevant physics in such a simulation.  For example,
new phenomena emerge in three dimensions compared to two dimensions
(e.g., the direction in scale of turbulent cascades reverses; see 
\cite{2013RvMP...85..245B} for a recent discussion of this and other issues
in core-collapse theory and simulations).
Neutrino transport, rotation, magnetic fields, and accurate nuclear
networks all play an important role.  One way to understand the delicacy
of this seemingly indelicate phenomenon is to note that the photon energy from core-collapse supernovae is $\sim 10^{48}-10^{49}$~erg, and the kinetic 
energy observed from core-collapse supernovae is commonly
$\sim 10^{51}$~erg, whereas the
total gravitational binding energy released in the formation of a neutron
star is a few times $10^{53}$~erg.  Thus only a small fraction of the
energy, which is released mainly in neutrinos, couples to the matter.
Thus, rather surprisingly for such an energetic event, this is a finely
tuned process.

One consequence of the uncertainty in the simulations is that we do not
have a firm first-principles grasp of the location of the ``mass cut"
(the boundary between matter that falls back and matter that escapes
in a given collapse/explosion), or the expected rotation rate, or the
expected strength of the magnetic field in a neutron star (which affects
subsequent spinup and spindown).  A common argument still found in
textbooks is that the typical birth spin rate and magnetic moment of 
a neutron star can be obtained by imagining that the Sun collapses and
conserves its angular momentum and magnetic flux.  Noting that 
helioseismology suggests a Solar angular momentum of 
$J_\odot\sim 2\times 10^{48}~{\rm g~cm}^2~{\rm s}^{-1}$ 
(e.g., \cite{1998MNRAS.297L..76P}) and that a
reasonable guess for the moment of inertia of a neutron star is
$I\sim MR^2\sim {\rm few}\times 10^{33}~{\rm g}(10^6~{\rm cm})^2
\sim {\rm few}\times 10^{45}~{\rm g~cm}^2$ gives an angular velocity of
$\omega=J/I\sim {\rm few}\times 10^3~{\rm rad~s}^{-1}$, or a rotation
rate of several hundred Hz, consistent with the inferred birth spin
rate of, say, the Crab pulsar.  Similarly, magnetic flux conservation
during the collapse of the Sun to a 10~km radius object would imply
a current surface dipolar field of $B\sim 10^{12}$~G, pleasingly
close to the inferred fields of young pulsars.

Unfortunately, this agreement is spurious.  The Sun will not collapse to
form a neutron star.  Instead, it is the core of 
a massive star that will collapse.  This core is approximately only the size of a white dwarf (i.e., at most a few thousand kilometers), which means that even if magnetic flux has been generated it is likely to be insufficient to explain what we see in neutron stars.  In addition, the likely coupling of the core to the envelope of the pre-supernova giant \cite{1998Natur.393..139S} suggests
that the core might actually rotate quite slowly, on $\sim$year timescales,
meaning that when it contracts from $\sim 10^3$~km to the $\sim 10$~km
neutron star the star would be rotating much more slowly than we see
in practice.  Thus other processes are involved at birth, e.g., off-center
kicks \cite{1998Natur.393..139S} for the spin, and probably some dynamo
action for the magnetic field \cite{1993ApJ...408..194T}, although here unfortunately the
natural field strength is near equipartition, which would imply 
$B\sim 10^{17-18}$~G!  Thus supernova simulators still rely heavily
on observational guidance.

{\it Accretion processes.}---After the initial core collapse, neutron
stars and black holes can acquire additional mass via accretion.  Accretion
from the interstellar medium will be negligible in almost all circumstances.
To see this, note that the standard Bondi-Hoyle rate of accretion from
a medium with a density $\rho$, sound speed $c_s$, and relative bulk speed
$v$ at infinity is
\begin{equation}
{\dot M}=\lambda \pi\rho\left[G^2M^2\over{(c_s^2+v^2)^{3/2}}\right]
\end{equation}
where $\lambda$ is an eigenvalue of order unity that depends somewhat
on the equation of state of the gas.  One might think that this rate
could be substantial; for example, a $10~M_\odot$ black
hole in a dense and cold molecular cloud ($\rho=10^{-20}~{\rm g~cm}^{-3}$
and $c_s\sim v\sim 1~{\rm km~s}^{-1}$) has an accretion rate of 
$\sim 10^{-6}~M_\odot~{\rm yr}^{-1}$, enough to cause e-folding of the
mass in $\sim 10^7$~yr.  However, the accretion will produce a large
amount of luminosity that will strongly heat the accreting matter (thus
increasing $c_s$) and pressure balance will then decrease the density
(see \cite{1994ApJ...436L.127N,1995ApJ...454..370B} and many other 
references).  

Significant accretion therefore requires a companion.  In some cases the
neutron star or black hole might spiral into the companion, producing a
Thorne-\.{Z}ytkow object \cite{1975ApJ...199L..19T,
1977ApJ...212..832T,2014MNRAS.443L..94L}.  Then the accretion rate can be far
above the usual Eddington limit
\begin{equation}
{\dot M}_{\rm Edd}\equiv L_{\rm Edd}/(\eta c^2)=
{4\pi GMc\over{\kappa}}\biggl/(\eta c^2)\sim 10^{18}~{\rm g~s}^{-1}
(M/M_\odot)(\eta/0.1)^{-1}\; ,
\end{equation}
where $\kappa$ is the opacity, $\eta$ is the radiative efficiency, and
for our numerical estimate we have assumed fully ionized matter and
a Thomson opacity;
$\eta\sim 0.1$ is reasonable for moderately rotating
black holes.  A rate of accretion much
larger than ${\dot M}_{\rm Edd}$ can lead to radiation that is trapped
by the advecting flow \cite{1977ApJ...214..840I,1982ApJ...253..873B}
 in black holes, or can lead to the production
of a large number of neutrinos that then escape \cite{1989ApJ...346..847C,
1994ApJ...436..843B} in neutron stars.
A process like this must happen during supernova fallback, and presumably
can change the mass or the spin of a neutron star or black hole on short
timescales.  However, this process does not leave the neutron star or
black hole with a companion unless the compact object was in a triple
system.  We will ignore this possibility henceforth.

{\it Mass and spin evolution due to accretion.}---Stable mass accretion
from a companion to a neutron star or black hole comes in two flavors,
depending on the mass of the companion.  In a low-mass X-ray binary
(LMXB), the companion's mass is typically $M<0.5~M_\odot$, and the
mass is transferred via Roche lobe accretion.  In a high-mass X-ray
binary (HMXB), the companion's mass is typically several solar masses,
and the mass is transferred when the neutron star or black hole captures
part of the wind from the companion.

The distinction is in part because, if all the angular momentum is
retained in the system, transfer of mass from the lighter to the heavier
member of a binary widens the binary, whereas transfer of mass from the
heavier to the lighter tightens the binary.  To see this, note that if
we have a binary with masses $M_1$ and $M_2$, which therefore has total
mass $M=M_1+M_2$ and reduced mass $\mu=M_1M_2/M$, the angular momentum
in a circular orbit of radius $r$ is $L=\mu\sqrt{GMr}$.  Transfer from
the less massive to the more massive object keeps $M$ constant but
reduces $\mu$, which thus requires that $r$ increase to compensate.  Transfer from 
the more massive to the less massive object increases $\mu$, so $r$ must
decrease.  One consequence is that in a binary in which the donor star
is more massive than the neutron star or black hole, if
the mass donation were through the Roche lobe then the system would
shrink, leading to a very large (and thus short lived) mass transfer episode
unless the donor shrank rapidly as it lost mass.
LMXBs, in contrast, are stable accretors and can last for hundreds of
millions of years.  Theoretically a compact object could capture mass
from the wind of a low-mass companion, but other than during the
giant phase these winds are negligible.  

HMXB lifetimes are limited to
the lifetime of the high wind loss portion of the already short lifetime
of a massive star.  Thus HMXBs likely live for only tens to hundreds
of thousands of years, and transfer $<10^{-2}~M_\odot$ 
\cite{1995ApJS..100..217I}.  In
contrast, LMXBs can transfer up to the original mass of the companion,
which might reach $\sim 1~M_\odot$ in some cases.  Another difference is that
Roche lobe accretion is accretion with reasonably high angular momentum,
so the mass flows in an accretion disk.  Winds, in contrast, emerge at
roughly the escape speed of the star, because the radiative acceleration
that drives them will do so until the wind escapes, and the acceleration
is less effective at larger distances so when the speed reaches
approximately the escape speed the wind is already far from the star.
Given that the companion in an HMXB can dominate the total mass of the
system, the escape speed from the companion is at least as large as the
orbital speed, and it can be much larger if the separation between the
companion and the compact object is many times the radius of the companion.
Thus the donated matter need not have enough 
angular momentum to form a disk, and stochastic changes in the wind can alter
the direction of the accretion flow \cite{1976ApJ...204..555S}.  Hence the short lifetimes
of HMXBs means they transfer less total mass than LMXBs do, and the
mass that is accreted may not be as ordered in its flow.

These considerations have led X-ray astronomers to believe for decades
(see \cite{1999MNRAS.305..654K} for a recent discussion) that the masses
and spin magnitudes of stellar-mass black holes are at best moderately
affected by accretion from a binary companion.  The argument is that
even the $\sim 0.5~M_\odot$ maximum that could be accreted in an LMXB
is a small fraction of the $\sim 10~M_\odot$ canonical black hole mass,
and integration of the angular momentum equations at the ISCO indicates
that the spin magnitude also does not change much (although purely
retrograde accretion of $0.5~M_\odot$ onto a maximally spinning $\sim 10~M_\odot$ black
hole could spin it down to ${\hat a}\sim 0.8$).  Thus the masses and
spin magnitudes of stellar-mass black holes are close to what they
were after the supernova.

In contrast, neutron star masses and spin magnitudes could be changed
dramatically by accretion.  A neutron star of initial mass $M\sim 1.5~M_\odot$
that accreted $\sim 0.5-1~M_\odot$ might even be pushed over its limit
and become a black hole.  The spin will change even more substantially.
Unlike uncharged black holes, neutron stars have large magnetic fields,
so the ionized matter that accretes onto them will couple to the fields.
The accretion thus has a large effective lever arm.  The coupling
radius (sometimes called the Alfv\'en radius or magnetospheric radius) is
\begin{equation}
r_{\rm mag}\approx 4\times 10^8~{\rm cm}(B/10^{12}~{\rm G})^{4/7}
({\dot M}/10^{17}~{\rm g~s}^{-1})^{-2/7}
\end{equation}
assuming a dipolar magnetic field.  This is enormously larger than the
$\sim 10^6$~cm radius of the star, so accretion can very rapidly
spin the star up or down, and what we see now is unrelated to the spin
of the neutron star at birth.  Isolated neutron stars also spin down,
by magnetic braking, which in the pure dipolar case causes the spin
period $P$ to evolve as
\begin{equation}
P^2=P_0^2+2P{\dot P}T\; .
\label{eq:spindown}
\end{equation}
Here $T$ is the age of the pulsar, $P_0$ is its initial period, and
$P{\dot P}\propto B^2$ in the magnetic dipole model.  Thus $P{\dot P}$ is
constant if the field strength and geometry do not change.  When
$2P{\dot P}T\gg P_0^2$, the characteristic age is thus
$\tau=P/(2{\dot P})$.
Thus the current spin frequency of an isolated neutron star is only
a lower limit on its birth frequency.  As we discuss further in \S3,
estimates have been made of the birth frequencies of some neutron
stars based on the observation that the spindown energy would have to
go into the surrounding nebula, so initial frequencies that are too
high would lead to nebulae that are too bright.

The direction of the spin of an accreting neutron star is determined
by the accretion, for the reasons indicated above.  The case is not
as clear for accreting black holes.  Another consequence of the frame
dragging around spinning black holes is that matter orbiting in a different
plane than the black hole spin plane will have its orbital axis precess
around the black hole spin axis.  When there is dissipation, this will
lead to a gradual settling of the matter's motion into the spin plane
of the hole (the orbits could be prograde or retrograde, depending on
the conditions; see \cite{2005MNRAS.363...49K}), a process
called the Bardeen-Petterson effect \cite{1975ApJ...195L..65B}.  
The black hole spin direction is 
thus also changed in response.  The relatively small size of the black
hole means that its specific angular momentum is small compared to the
specific angular momentum of the disk; a rough ratio is 
$\ell_{\rm BH}/\ell_{\rm disk}\approx {\hat a}(r_{\rm BH}/r_{\rm disk})^{1/2}$.
Thus for a few$\times 10^{10}$~cm disk around a $10^6$~cm black hole, the
integrated angular momentum of the matter that flows through the disk will be greater than in the black hole even if total accreted mass is only
$\sim 1$\% of the black hole mass (i.e., about $0.1~M_\odot$).

The reason that this does not guarantee alignment is that much of the
disk angular momentum actually flows outwards, where it is ultimately
transferred to the orbital angular momentum of the companion.  The
relevant radius for black hole alignment is where the disk inclination
bends towards the black hole spin plane.  There is considerable
discussion of where this is \cite{1983MNRAS.202.1181P,1992MNRAS.258..811P,1998ApJ...506L..97N,1999ApJ...525..909A,1999MNRAS.304..557O,2000MNRAS.315..570N,2007MNRAS.381.1287L,2010MNRAS.405.1212L,2012ApJ...757L..24N,2013ApJ...768..133S,2013ApJ...777...21S}, with typical estimates being hundreds
to thousands of times the gravitational radius $r_g\equiv GM/c^2$, but
it may also be that alignment is less effective when the accretion
rate approaches Eddington \cite{2007ApJ...668..417F}.  
Thus the current orientation of the
black hole spin might retain some memory of its birth orientation
rather than being fully aligned with the orbit.  As we will discuss in
\S3, this consideration, and the disputed evidence that misalignment
is seen in some systems, has significant implications for spin
estimates using X-ray continua.

We now discuss the methods of estimating the masses of neutron stars
and black holes, and the implications of the results.

\section{Masses}
\label{section:masses}

Ever since the 1767 suggestion by John Michell, the intellectual grandfather of black holes, that some visually close stars are actually physically associated with each other \cite{1767RSPT...57..234M}, binaries have been prime sources for the determination of the masses of their components.  This is because Kepler's laws can be used directly for such systems.  Other techniques are sometimes applied; for example, asteroseismology has been used to find the masses of some white dwarfs, main sequence stars, and giants \cite{1994ARA&A..32...37B,2008ARA&A..46..157W,2012ApJ...749..152M,2013MNRAS.429..423M}.  But neutron stars and black holes have neither observed oscillations nor confirmed surface atomic lines (which provide critical information for ordinary stars).  Future mass measurements using microlensing are promising (see \cite{2012RAA....12..947M} for a recent review, including the discussion of several black hole candidates), particularly if precise astrometric measurements can break degeneracies between the mass, proper motion, and distance to the lens.  However, the current sample of measured masses is dominated by compact objects in binaries.

In this section we begin with a theoretical discussion of the maximum mass of a neutron star.  We then describe the mass function, which places a lower limit on the mass of the {\it companion} to an observed object in a binary, then discuss some of the post-Keplerian observables that can be seen in binary pulsar systems.  We finish by giving references for the masses and estimated uncertainties for a number of neutron stars and black holes.

\subsection{The maximum mass of a neutron star}

The mass boundary between neutron stars and black holes is important for two reasons.  First, it is sensitive to the state of cold matter beyond nuclear density; laboratory experiments cannot access this realm, and nuclear models disagree strongly about the likely properties of such matter.  Second, as we discuss at the end of this section, a mass measurement above the neutron star maximum is the best way to establish that a compact object is a black hole, because other proposed discriminants are unreliable.  Here we focus for simplicity on nonrotating neutron stars, with some comments at the end about the role of uniform and differential rotation.

As our first guess about the maximum mass of a neutron star we can use the elementary Landau \cite{1932PZSow...1..285L} derivation of Chandrasekhar's \cite{1931ApJ....74...81C} maximum mass for white dwarfs.  Suppose that we have a fermion (an electron for white dwarfs, a neutron for neutron stars) whose degeneracy pressure holds up a cold, dense, spherical object.  If the number density of this fermion is $n$ then a given fermion is localized within a region of size $\Delta x\sim n^{-1/3}$ and its Fermi momentum is
\begin{equation}
p_F\sim \hbar/\Delta x\sim \hbar n^{1/3}\; .
\end{equation}
The Fermi energy per fermion is $E_F\approx p_F^2/2m_F$ in the nonrelativistic case (i.e., when $p_F\ll m_Fc$), where $m_F$ is the mass of the fermion, and $E_F\approx p_Fc$ in the relativistic case.  Now suppose that there are $\mu_F$ massive particles (protons and/or neutrons) per fermion; thus $\mu_F\sim 2$ for white dwarfs and $\mu_F\sim 1$ for neutron stars.  In a star of total mass $M$ and radius $R$, this means that the average number density is $n\sim M/(\mu_F m_n R^3)$, where $m_n$ is the mass of a neutron.  Thus $E_F\sim (\hbar^2/2m_F)\left[
M/(\mu_F m_n R^3)\right]^{2/3}\sim 1/R^2$ for nonrelativistic degeneracy, and $E_F\sim \hbar c\left[M/(\mu_F m_n R^3)\right]^{1/3}\sim 1/R$ for relativistic degeneracy.

Ignoring general relativistic corrections and structural details of the star, the gravitational potential energy per degenerate fermion is
\begin{equation}
E_{\rm grav}\sim -{GMm_n\mu_F\over R}\; .
\end{equation}
Thus when we highlight the dependences on the mass, radius, and $\mu_F$, the total energy per degenerate fermion in the nonrelativistically degenerate case is
\begin{equation}
E_{\rm tot}=E_F+E_{\rm grav}={C_1M^{2/3}\over{\mu_F^{2/3}R^2}}-
{C_2M\mu_F\over R}
\end{equation}
where $C_1$ and $C_2$ are dimensional constants.  From this expression we see that minimization of the energy leads to a stable configuration; too large a radius asymptotes the energy to zero (contrasted to the minimum energy, which is negative because the star is bound), and too small a radius yields a positive energy. We also see from energy minimization that at equilibrium, $R\propto M^{-1/3}$, so we recover the result that degenerate objects are smaller when they are more massive.

In the relativistically degenerate case,
\begin{equation}
E_{\rm tot}={C_3M^{1/3}\over{\mu_F^{1/3}R}}-{C_2M\mu_F\over R}\; .
\end{equation}
Both terms have the same $R$ dependence, so if the expression is positive at any $R$ the star will expand to lower its energy until it becomes nonrelativistic, at which point it can settle into a stable equilibrium.  In contrast, if the energy is negative, it can contract indefinitely to reach lower and lower energies, so the star is not stable.  We see that higher masses lead to lower energies at fixed $R$, so the limiting mass is the one that yields $E_{\rm tot}=0$.  This equation shows that $M_{\rm max}\propto \mu_F^{-2}$.  The maximum mass for a white dwarf ($\mu_F\approx 2$) is $M_{\rm WD}\approx 1.4~M_\odot$, so we would expect that the maximum mass for a neutron star ($\mu_F\approx 1$) would be $\sim 4$ times higher, or $\sim 5.6~M_\odot$.

In reality the maximum mass of a neutron star is much smaller than this value.  The main reason is that, unlike white dwarfs, neutron stars are compact enough that mass-energy contributions other than the rest mass play an important role.  Thus the increased pressure that comes with higher density will itself gravitate, so systems of a given mass are more susceptible to collapse then they would be in Newtonian gravity.  A more rigorous approach to the determination of the maximum mass uses the Tolman-Oppenheimer-Volkoff equation \cite{1939PhRv...55..374O}
\begin{equation}
{dP(r)\over{dr}}=-{G\over r^2}\left[\rho(r)+{P(r)\over c^2}\right]
\left[M(r)+4\pi r^3{P(r)\over c^2}\right]\left[1-{2GM(r)\over{c^2 r}}
\right]^{-1}\; .
\end{equation}
This is effectively the equation of hydrostatic equilibrium for a spherically symmetric, nonrotating star in general relativity.  Here $M(r)$ is the gravitational mass contained inside a circumferential radius $r$, and $P(r)$ and $\rho(r)$ are respectively the pressure and density at $r$.  Given a central density and an equation of state that relates $P$ to $\rho$, one can compute the mass and radius of the star.  Note that because the temperature $T<10^{9-10}$~K inside a neutron star with an age of at least hundreds of years is much less than the Fermi temperature $T_F=E_F/k\sim 10^{12}$~K, temperature can be neglected in the equation of state.  Similarly, because the Fermi energy of hundreds of MeV in the core is much greater than the energy differences per particle ($\sim$few MeV) between different compositions, it is usually assumed that the core matter is in its (unknown) equilibrium composition.

Thus given $P(\rho)$, one can calculate the maximum mass of a neutron star.  However, because the central densities are at several times nuclear saturation density (i.e., out of reach of laboratory experiments) and the matter is strongly skewed towards neutrons rather than protons (in contrast to the nearly-symmetric nuclear matter in normal nuclei), $P(\rho)$ is not known from first principles.  One might imagine that the equation of state could be calculated from quantum chromodynamics. However, because fermion wavefunctions switch sign when particles are swapped, estimates of the total energy of a collection of fermions require evaluation of a sum of nearly cancelling terms (this is known as the fermion sign problem, e.g., \cite{1998NuPhS..63..439B,1999NuPhS..73..161A,2002NuPhS.106..142H,2009arXiv0912.4410L}).  Thus the ground state of nuclear matter, especially under significant pressure, is exponentially difficult to compute from first principles.  Thus nuclear physicists rely on astronomers to determine observationally the masses and, ideally, the radii of neutron stars.

The maximum mass of neutron stars can be constrained theoretically using an approach pioneered by Rhoades and Ruffini (\cite{1974PhRvL..32..324R}; see also \cite{1996ApJ...470L..61K}).  They noted that we do have a good understanding of the state of matter from zero density up to nuclear saturation density ($\rho_s\approx 2.6\times 10^{14}~{\rm g~cm}^{-3}$).  They showed that at higher densities the equation of state that gives the largest possible mass is one that is at maximum stiffness, i.e., one for which the speed of sound $c_s=(dP/d\rho)^{1/2}$ is the speed of light.  Applying that assumption to all densities above $\rho_s$ gives a maximum mass of $M_{\rm max}\approx 3.2~M_\odot$.  One can extend this to ask: if we know the equation of state up to some density and then use a $c_s=c$ equation of state, what is the maximum mass?  Kalogera and Baym \cite{1996ApJ...470L..61K} showed that if we trust the (largely converged and calibrated) equation of state up to $2\rho_s$, the maximum mass drops to roughly $M_{\rm max}=2.9~M_\odot$.  Astronomers thus use the rough criterion that if an X-ray emitting object in a binary exceeds $3~M_\odot$, it is most likely a black hole rather than a neutron star.  As a note of caution, even at $\rho_s$ neutron star matter has far more neutrons than protons, so what we can measure in laboratories (see \cite{2014arXiv1401.5839H} for a recent review) does not provide absolutely definitive constraints on the relevant highly asymmetric matter. 

Rotation adds centrifugal support, so rotation can increase the maximum mass of a neutron star, by as much as $\sim$25\% \cite{1987ApJ...314..594F}.  Indeed, there are studies of both supermassive neutron stars (uniformly rotating stars that would be unstable if they stopped rotating) and the yet more massive hypermassive neutron stars (differentially rotating stars that would be unstable if their angular momentum were redistributed to yield uniform rotation).  Such stars could have masses a few tenths of a solar mass in excess of the nonrotating maximum.  The hypermassive case is not relevant for the long-lived neutron stars we see, because differential rotation is damped quickly \cite{1977NYASA.302..528H}.  Supermassive neutron stars are potentially relevant, but in practice the highest rotation rate ever seen from a pulsar (716~Hz; see \cite{2006Sci...311.1901H}) is probably a factor of $\sim$two less than the maximum rotation rate \cite{1994ApJ...424..823C}.  Centrifugal support is a quadratic effect, so rotation is at best a minor contributor to the maximum mass of observed neutron stars.

\subsection{The binary mass function}

Suppose that one observes a star in a binary, which we will call star 1, over at least one full period.  Monitoring of spectral lines (from ordinary stars) or pulses (from a pulsar) can yield both the period $P$ and the maximum line of sight speed $K_1$ of star 1 from our perspective.  It is straightforward to show using Kepler's third law that for an orbit of eccentricity $e$
\begin{equation}
f(M_1,M_2,i)\equiv {K_1^3P(1-e^2)^{3/2}\over{2\pi G}}={(M_2\sin i)^3\over{(M_1+M_2)^2}}
\end{equation}
(e.g., \cite{1984ARA&A..22..537J}).  Here $M_1$ is the mass of star 1, $M_2$ is the mass of its companion, star 2, and $i$ is the inclination of the orbit to our line of sight, defined so that $i=0$ for a face-on orbit and $i=90^\circ$ for an edge-on orbit.

Examination of this equation shows that the {\it mass function} $f$ is a lower limit to $M_2$, the mass of the unobserved star.  Indeed, $f$ is the actual mass of star 2 only if the system is edge-on and $M_1\rightarrow 0$.  If the motion of star 2 can also be measured, e.g., in a double-line spectroscopic binary or a double pulsar, then the mass ratio is also known but the inclination remains a source of uncertainty.  Even if the second star cannot be observed in this way, if $M_1\ll M_2$ then the uncertainty its unknown mass introduces is small.  This is why high-mass X-ray binaries provide more ambiguous evidence for stellar-mass black holes than do low-mass X-ray binaries.  For example, the mass function of Cyg-X--1 is only $\approx 0.25~M_\odot$ (\cite{1973ApJ...179L.129B} and subsequent references), so assumptions must be made about the mass of the companion to make a case that the unseen object exceeds the maximum mass of a neutron star and is thus likely to be a black hole.

The orbital inclination is often difficult to determine for non-eclipsing systems.  A method that has found success for Roche lobe filling binaries is the use of ellipsoidal light variations based on the tidal distortion of the donor \cite{1978pans.proc...43A,1990ApJ...350..386M,2001ApJ...554.1290G}.  The key idea is that a tidally distorted star has a larger cross section as seen from the side than as seen along the line to the compact object, and that the light from when the donor is behind the compact object looks different from when it is in front of the compact object because of edge-darkening effects near the L1 point.  A face-on system will obviously show no variation in the companion's light with orbital phase, whereas an edge-on system will show maximum variation.  Thus the inclination can be estimated even for non-eclipsing systems.  The low luminosity of the companion in a low-mass X-ray binary means that this method is best used for transient systems, where the accretion disk essentially goes away completely and hence does not contaminate the light from the companion (e.g., \cite{2001ApJ...554.1290G}).  Perhaps due to their higher mass ratio, black hole LMXBs are more likely to be transient systems than neutron star LMXBs.  Thus this method has been used successfully for several black hole systems, but few neutron star systems.

\subsection{Binary pulsars and post-Keplerian parameters}

A method unique to neutron stars that can yield high-precision masses is the careful timing of signals from pulsars.  The pulses from pulsars act as superior clocks, with characteristic spindown timescales $P/2{\dot P}$ that can be several billion years for millisecond pulsars, the most stable rotators.  Thus even minute deviations from slow, steady spindown are highly significant and can be used to infer subtle relativistic effects in the binary system.  These considerations led Joseph Taylor and Russell Hulse to search for and discover the first binary neutron star system in 1974 \cite{1975ApJ...195L..51H}.  As we will discuss, the timing of this and similar double neutron star systems has led to precise confirmation of general relativity in weak gravity.  

The extra effects that can currently be measured are:

\begin{itemize}

\item Precession of the pericenter.  To lowest order the frequency of precession of an orbit is
\begin{equation}
{\dot\omega}=3\left(P_b\over{2\pi}\right)^{-5/3}(T_\odot M)^{2/3}(1-e^2)^{-1}\; .
\end{equation}

\item The Einstein delay.  At pericenter, the pulsar is deepest in the gravitational well of the companion, which slows down the received pulse rate.  The magnitude of the effect is
\begin{equation}
\gamma=e\left(P_b\over{2\pi}\right)^{1/3}T_\odot^{2/3}M^{-4/3}m_c(m_p+2m_c)\; .
\end{equation}

\item Binary orbital decay due to gravitational radiation.  The rate of decay of the orbital period is
\begin{equation}
{\dot P}_b=-{192\pi\over 5}\left(P_b\over{2\pi}\right)^{-5/3}f(e)T_\odot^{5/3}
m_pm_c M^{-1/3}\; .
\end{equation}

\item The Shapiro delay.  If the pulses pass close to the companion, the clock of the pulses experiences time dilation.  A face-on orbit will have no dependence of the Shapiro delay on orbital phase, whereas a nearly edge-on orbit will have a sharp increase in the delay at conjunction.  Thus in addition to the magnitude of the effect (the range, or $r$, parameter), there is an effect (the shape, or $s$, parameter) that depends on the binary orientation:
\begin{equation}
\begin{array}{rl}
r&=T_\odot m_c\\
s&=\sin i\; .\\
\end{array}
\end{equation}

\end{itemize}

Here we use the notation of \cite{2009arXiv0907.3219F}.
In this list, $m_p$ is the pulsar mass, $m_c$ is the companion mass, $M=m_p+m_c$ is the total mass (all masses are in solar units), $T_\odot\equiv GM_\odot/c^3=4.925590947\mu{\rm s}$, $P_b$ is the binary orbital period, and $f(e)=(1+73e^2/24+37e^4/96)(1-e^2)^{-7/2}$.  Neutron stars are so small compared to their separations in these systems ($\sim 10^6$~cm versus $\sim 10^{11}$~cm) that they are essentially point sources and there are thus no tidal effects etc. to confuse the analysis of these binaries.  A given system has five associated parameters (two masses and the inclination, semimajor axis, and eccentricity), so when there are more than five independent observables the system is overdetermined and the underlying theory can be tested.  

The neutron stars in double neutron star systems (when at least one appears to us as a pulsar) have masses that can be measured to $\sim 10^{-3}$ relative precision.  The pulsars in these systems rotate with periods of tens of milliseconds, and are not as stable as the true millisecond pulsars (some of which have timing stability one or two orders of magnitude greater; see \cite{1991IEEEP..79.1054T}).  If a millisecond pulsar were to be found in a close binary with another neutron star, the gain in timing precision would eventually allow the measurement of higher-order effects (see \cite{2006Sci...314...97K} for a discussion in the context of the double pulsar PSR~J0737--3039).  

However, because millisecond pulsars are spun up by accretion of at least several hundredths of a solar mass from a companion, and the high-mass stars that are the progenitors of neutron stars do not last long enough to donate such mass \cite{1995ApJS..100..217I}, millisecond pulsars are instead found in binaries with white dwarfs or main sequence stars.  These objects are large enough that tidal effects can be significant.  In addition, the accretion process circularizes the orbit (but not completely; see \cite{1992RSPTA.341...39P} for a remarkable demonstration using the fluctuation-dissipation theorem that residual eccentricities of a few parts per million can be produced, depending on the binary orbital period), so it can be difficult to measure pericenter precession or the Einstein delay, both of which require nonzero eccentricity.  Nonetheless, recently two neutron stars in such systems have been found to have masses $M\approx 2~M_\odot$, which as we will discuss in the next section has important implications for dense matter.

The first of these, PSR~J1614--2230 \cite{2010Natur.467.1081D}, is a pulsar-white dwarf system that had its Shapiro delay measured due to its fortunate nearly edge-on orientation.  For a circular orbit, the two parameters of the Shapiro delay plus the two Keplerian parameters (orbital period and line of sight speed) suffice to determine both masses.  The orientation is also important because for edge-on inclinations the Shapiro delay as a function of orbital phase acquires a cusp near conjunction; more moderate inclinations produce a phase dependence that is partially degenerate with what one would find in an orbit with moderate eccentricity.  The mass inferred for PSR~J1614--2230 was $M=1.97\pm 0.04~M_\odot$ \cite{2010Natur.467.1081D}.

More recently, Antoniadis et al. \cite{2013Sci...340..448A} have found a mass of $M=2.01\pm 0.04~M_\odot$ for PSR~J0348+0432.  This is another pulsar-white dwarf system, but the method of mass estimation was different.  In addition to the pulsar frequency modulation, the authors were able to measure orbital line energy variations from the white dwarf.  This immediately gives a mass ratio between the neutron star and the white dwarf.  They estimated the mass of the white dwarf based on white dwarf models and the measured gravitational redshift of the lines, leading to the final mass estimate for the neutron star.

A convenient updated list of neutron star masses and their uncertainties is maintained at http://www.stellarcollapse.org/nsmasses by Jim Lattimer and Andrew Steiner.  From this it is apparent, and it was noted early, that the masses of neutron stars in double neutron star systems are very tightly clustered, around
$1.35\pm 0.05~M_\odot$.  Indeed, prior to the discovery of more massive neutron stars in binaries with white dwarfs, this clustering was sometimes used to argue that the maximum mass of neutron stars is low, around $1.5~M_\odot$ \cite{1994ApJ...423..659B}.  It now seems clear that if this clustering is significant it stems from a narrowness of evolutionary channels rather than from the fundamental physics of neutron stars.  We also note that millisecond pulsars seem to have systematically larger masses than neutron stars that are not likely to have accreted much \cite{2011A&A...527A..83Z}.  This accords with our expectation that millisecond pulsars could accrete up to several tenths of a solar mass in their lifetimes.

For an early list of estimated black hole mass functions, see \cite{1998ApJ...499..367B}; for more recent updates, see \cite{2010ApJ...725.1918O,2011ApJ...741..103F}, and for an object-by-object discussion see Section~5 of \cite{2012ApJ...757...36K}.  At face value it appears that there is a gap between the masses of neutron stars and black holes \cite{2010ApJ...725.1918O,2011ApJ...741..103F}, although there have been recent concerns raised about the role of selection effects \cite{2012ApJ...757...36K}.  If there is indeed a gap, one possible explanation is that black holes form from failed supernovae and thus have much greater fallback than neutron stars \cite{2014ApJ...785...28K}.  More observations are necessary.  In particular, if there is a large gap in the chirp mass distribution inferred from the gravitational waves emitted in compact binary coalescence (see Section~\ref{section:gw}) this will confirm the mass gap between black holes and neutron stars as well as making it easier to identify which sources contain neutron stars and which contain black holes.

We now turn our attention to the inferred spins of neutron stars and black holes.

\section{Spins}
\label{section:spins}

As we said in Section~\ref{section:introduction}, stellar-mass black holes have spin magnitudes close to what they were at birth.  In contrast, we will show that the spin periods of neutron stars typically bear little relation to their periods at birth.  In addition, even the fastest-spinning neutron stars have far less angular momentum than stellar-mass black holes.  In this section we provide context for these statements by discussing what we know about the spins of neutron stars and black holes.  In Section~\ref{section:conclusions} we will return to the question of spins by discussing the ongoing debate about the degree to which black hole spins affect the powers of their jets.

\subsection{Neutron Star Spins}

The spin periods of more than 2500 neutron stars are known from the periodic modulation of their intensities.  Of these the vast majority --- more than 2300 --- are seen in radio bands, and have periods ranging from 1.4~ms to 12~s (see, e.g., the Australia Telescope National Facility Pulsar Catalog http://www.atnf.csiro.au/research/pulsar/psrcat/ \cite{2006ChJAS...6b.139M} and the catalog http://astro.phys.wvu.edu/GalacticMSPs/GalacticMSPs.txt of Duncan Lorimer; note that {\it Fermi} detections of gamma rays have made a significant contribution to our catalogs of pulsars out of the Galactic plane \cite{2013ApJS..208...17A}).  These pulsars are thought to be rotation-powered, meaning that the spindown energy of the pulsar is the ultimate power source of the radio emission (even though the radio luminosity is a tiny fraction of the spindown power).  There are another $\sim 200$ pulsars that are detected from their bright X-ray emission.  These pulsars are accretion-powered: the source of energy is the gravitational energy released when matter taken from a binary companion is funneled by the neutron star's magnetic field onto a portion of the neutron star's surface.  The spin periods of these pulsars range from 1.7~ms to at least three hours (see \cite{2012arXiv1206.2727P}).  There are also a handful of other neutron stars that have known periods based on phenomena such as thermonuclear burst oscillations \cite{2008ApJS..179..360G} or that are powered by strong magnetic fields rather than rotation (the magnetars; see \cite{2014ApJS..212....6O} and http://www.physics.mcgill.ca/$\sim$pulsar/magnetar/main.html for a catalog).  

\begin{figure}[!htb]
\begin{center}
\includegraphics[scale=0.7]{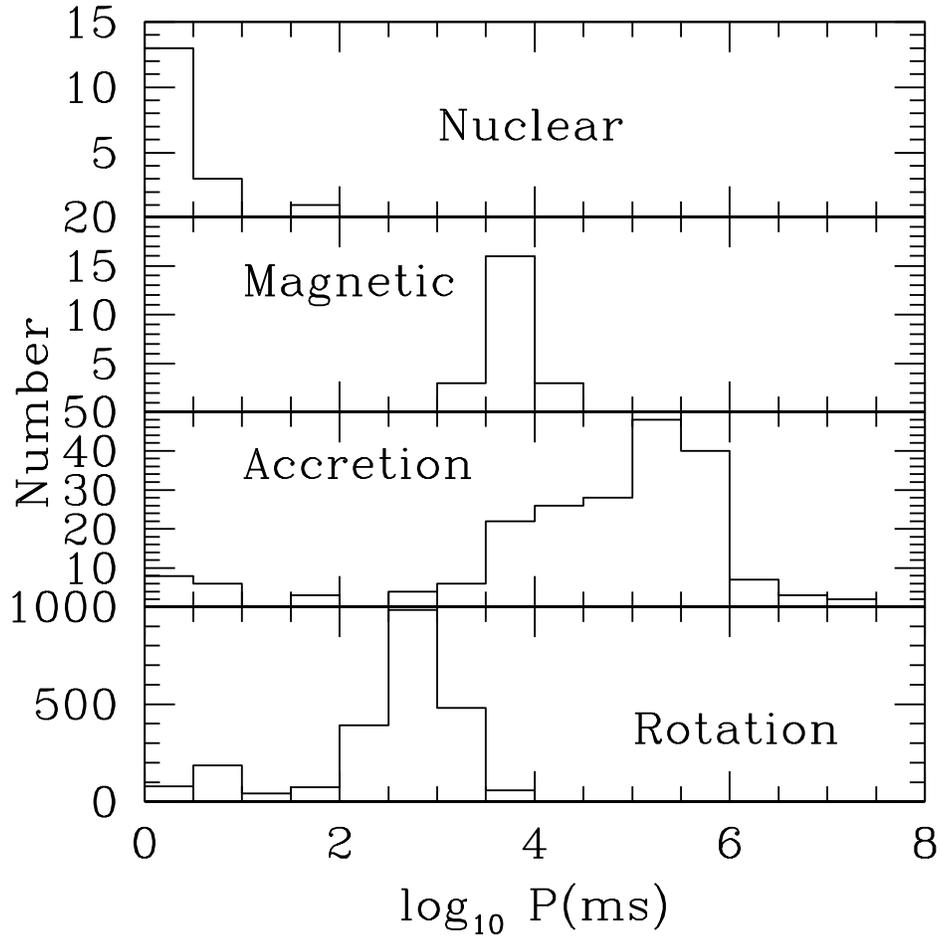}
\caption{\footnotesize Histograms of neutron star spins, categorized by the primary power source.  Some of the nuclear-powered pulsars are also accretion-powered pulsars.  There are numerous selection effects that influence the observed distributions, but the general patterns may be representative of the actual sample.  Data for the rotation-powered pulsars are from http://www.atnf.csiro.au/research/pulsar/psrcat/ \cite{2006ChJAS...6b.139M}, data for accretion-powered pulsars are from http://www.iasfbo.inaf.it/$\sim$mauro/pulsar\_list.html, data for magnetars are from http://www.physics.mcgill.ca/$\sim$pulsar/magnetar/main.html \cite{2014ApJS..212....6O}, and data for nuclear-powered pulsars are from \cite{2012ARA&A..50..609W}.
  }
\end{center}
\end{figure}

For our purposes, the most important question related to neutron star spin is whether a given star is accreting actively.  We now discuss the properties of non-accreting and accreting neutron stars and the spin periods and evolution that have been inferred.

\subsubsection{Non-accreting pulsars}

As we mentioned earlier, the discovery of periodic radio sources in 1967 \cite{1968Natur.217..709H} ushered in the era of neutron star observations.  Even before the discovery of pulsars, Pacini \cite{1967Natur.216..567P} proposed that magnetic dipole radiation from a spinning magnetic neutron star could power supernova remnants.  It was, however, the elegant work of Gold \cite{1968Natur.218..731G} that convinced the community that the periodic pulsations observed by Jocelyn Bell and colleagues were due to rotating neutron stars.  At its core, Gold's argument is that (1)~the shortest periods known in 1968 were tens of milliseconds, which is only accessible by neutron stars or black holes (even white dwarfs are not compact enough for such periods), (2)~black holes have no stable structures that could emit regular beacons, so the objects must be neutron stars, (3)~periodic signals can be produced by rotation, pulsation, or orbits, but (4)~neutron star pulsations have typical periods of at most a few milliseconds, which could not explain the few-second periods that were then known, and (5)~orbital periods would {\it decrease} in time, rapidly, due to emission of gravitational radiation, in contrast to the very nearly stable but increasing periods of known pulsars.  Thus rotating neutron stars are the only possible objects that could produce such short periods that are also nearly stable but increase with time.

The power source of most of these pulsars is likely to be rotation.  As Gold stated:

{\it There are as yet not really enough clues to identify the mechanism of
radio emission.  It could be a process deriving its energy from some
source of internal energy of the star, and thus as difficult to analyse
as solar activity.  But there is another possibility, namely, that the
emission derives its energy from the rotational energy of the star
(very likely the principal remaining energy source), and is a result
of relativistic effects in a co-rotating magnetosphere.}\cite{1968Natur.218..731G}

Note that radio emission carries only a tiny fraction of the spindown power, but higher-energy emission such as gamma rays can sometimes have tens of percent of the spindown luminosity \cite{2013FrPhy...8..679H}.  Indeed, there is a growing category of sources whose integrated X-ray luminosity is considerably greater than the spindown luminosity.  These are known as ``magnetars", and are thought to be powered by the decay and reconfiguration of internal magnetic fields \cite{1995MNRAS.275..255T,1996ApJ...473..322T}.  

Regardless of what powers the observed emission, non-accreting neutron stars are thought to spin down essentially by magnetic dipole radiation (see Section~\ref{section:introduction}).  Putting units to equation~(\ref{eq:spindown}), the period evolves as $P^2=P_0^2+2B_{12}^{2}(T\cdot 10^{-15}~{\rm s})$ for an average surface dipolar field of $10^{12}B_{12}$~G.  Here $T$ is the current age of the pulsar.  This means that memory of the initial spin is lost rapidly if the birth period was much less than a second and the field strengths are in the $\sim 10^{12}$~G range usually considered for young neutron stars.  For example, the Crab pulsar is $\sim 1000$~yr old (the supernova was first seen on 4 July 1054).  Its surface field is inferred to be $B_{12}\approx 5$.  Thus if the initial period was much shorter than the current period of 33~ms then the inferred age of $T\approx 700$~yr is roughly consistent with the known age.  Initial periods anywhere from $\sim 1$~ms to $\sim 10$~ms can be accommodated easily.  For older pulsars an even greater range of initial periods are consistent with the data.  Thus pulsars older than a few thousand years may well have completely lost memory of their initial period.  Indeed, it is evident from Table~1 of \cite{2012Ap&SS.341..457P} that most pulsars in supernova remnants have characteristic ages $T_c=P/2{\dot P}$ that are consistent within the uncertainties with the age of the remnant.  Thus for those pulsars the initial period could have been anything much less than the current period.

We note, incidentally, that the dipole spindown model in its simplest form fails a self-consistency test.  For pure magnetic dipole spindown, the ``braking index" $n_b\equiv \nu{\ddot\nu}/{\dot\nu}^2$, where $\nu$ is the spin frequency, is equal to 3.  Measured values, however, are always less than 3, sometimes by very significant amounts \cite{1977puls.book.....M}.  This suggests that other processes are at work; see \cite{2012NatPh...8..787H,2012A&A...547A...9P} for some recent ideas.

To close this section we mention for completeness that if there is sufficient nonaxisymmetry in a rapidly-rotating neutron star, it could be slowed down by emission of gravitational waves (e.g., via R-mode instabilities on the
surface of a very young, hot neutron star \cite{1998ApJ...502..708A}, \cite{1999ApJ...510..846A}).  However, the
efficiency of this mechanism is expected to have a very steep dependence
on the initial spin period, and is also expected to have a very low
saturation amplitude, so it is not clear if this mechanism will be
important.  Stars with more modest initial spin periods may take
thousands of years to spin down substantially (\cite{2003ApJ...591.1129A}, \cite{2006ApJS..164..130O}).

\subsubsection{Accreting pulsars}

A few years after the discovery of rotation-powered pulsars, pulsating X-ray sources were discovered in several binaries (e.g., Her~X-1; \cite{1972ApJ...174L.143T}).  Initial modeling work did not consider neutron stars because of the assumption that the required supernova would disrupt any binary, but soon several groups proposed that accretion from a companion, guided by the magnetic field of the neutron star, would produce an X-ray emitting hot spot on the surface \cite{1972A&A....21....1P,1974ApJ...189..331D,1973ApJ...184..271L}.  Later work established quantitative relations between the spin frequency of an accreting neutron star and its average accretion rate and dipole magnetic moment, based on the idea that a magnetic spin equilibrium could be established on a timescale much shorter than the X-ray active timescale of the source (e.g., \cite{1977ApJ...217..578G,1978ApJ...223L..83G,1979ApJ...232..259G,1979ApJ...234..296G,1994ApJ...429..781S,1994ApJ...429..797S,1994ApJ...429..808N}).  In spin equilibrium, the rotation frequency of an accreting magnetic neutron star is
\begin{equation}
\nu=0.3~{\rm Hz}~\xi{\dot M}_{17}^{3/7}\mu_{30}^{-6/7}(M/M_\odot)^{5/7}
\label{eq:rotequil}
\end{equation}
for a star with a magnetic moment $10^{30}\mu_{30}$~G~cm$^3$ and an average accretion rate of $10^{17}{\dot M}_{17}$~g~s$^{-1}$.  Here $\xi$ is a factor of order unity that depends on the detailed structure of the flow.  Thus for the neutron stars in high-mass X-ray binaries, which typically have $\mu_{30}\sim{\rm few}$ the equilibrium period is often seconds or longer, but for the neutron stars in low-mass X-ray binaries, which usually have $\mu_{30}\sim 10^{-4}-10^{-3}$, periods of milliseconds can be attained via spinup.

Indeed, accretion-induced spinup was quickly suggested to explain the first millisecond pulsar detected \cite{1982Natur.300..615B,1982CSci...51.1096R,1982Natur.300..728A}.  This explanation has been strengthened by the evidence that millisecond pulsars are usually in binaries ($\sim 70-80$\% versus the $\sim 1$\% of slower pulsars).  The observation of nearly coherent 
millisecond oscillations in thermonuclear X-ray bursts \cite{2008ApJS..179..360G} was taken by most researchers to show that the stars hosting those bursts, which are in LMXBs, are the progenitors of the non-accreting millisecond pulsars.  However, the frequencies of the burst oscillations drift slightly and thus some doubt remained until the discovery of 401~Hz accretion-powered oscillations from SAX~J1808--3658 \cite{1998Natur.394..344W,1998Natur.394..346C}, which were later tied to the frequency and phase of burst oscillations from this source \cite{2003Natur.424...42C}.  Multiple sources now have both accretion-powered and nuclear-powered oscillations at the same frequency.  One source has even been found to transition between accretion-powered pulsations and rotation-powered pulsations when the accretion dies off sufficiently \cite{2013Natur.501..517P}, and there are other sources for which somewhat more indirect evidence of this phenomenon has been suggested \cite{2014ApJ...781L...3P,2014ApJ...789...40B}.

Currently, there are 15 known accretion-powered millisecond X-ray pulsars (see \cite{2012arXiv1206.2727P} for a review).  This includes the X-ray binary Aql X-1, in which $\sim 150$ seconds of coherent oscillations were discovered among many megaseconds of data gathered with the {\it Rossi} X-ray timing explorer (RXTE) (\cite{2008ApJ...674L..41C}).  The observed range of spin frequencies is $\nu = 182-599$~Hz; seven have binary orbital periods of two hours or shorter.

Given that the number of accreting millisecond X-ray pulsars is growing, and especially given that a non-pulsing X-ray binary was found to have momentary coherent pulsations, the question arises: why are coherent pulsations not found in {\it all} neutron star X-ray binaries?  The reasons may be complex, multiple mechanisms may work jointly to hide pulsations, and observations have not yet revealed the answer.  Alignment of the rotation and magnetic axes could serve to hide pulsations (e.g. \cite{2009ApJ...705L..36L}, \cite{2009ApJ...706..417L}).  It has also been suggested that the magnetic field is buried by accretion (e.g., \cite{1986ApJ...305..235T}), only to re-emerge when the accretion rate drops.  This agrees with the observation that accretion-powered millisecond X-ray pulsars are typically under-luminous compared to more standard neutron star X-ray binaries.  However, it is not clear how to reconcile this explanation with the observation that most non-accreting millisecond pulsars obey the spin-up line \cite{1999ApJ...520..696A}, which is the line on a $P-{\dot P}$ diagram that would be obeyed for a source accreting at the Eddington rate or below (for a discussion of some of the complexities involved in defining this line, see Section~4.3 of \cite{2012MNRAS.425.1601T}).  This suggests that most of the accretion occurred at nearly the current magnetic moment, rather than one that was much lower.

Accretion could also slow down the rate of rotation of a neutron star.  As can be seen from Equation~(\ref{eq:rotequil}), if the accretion rate is small enough or the magnetic dipole moment is large enough, the equilibrium period can be hundreds to thousands of seconds.  Periods of this size are indeed seen in some HMXBs (see http://www.iasfbo.inaf.it/$\sim$mauro/pulsar\_list.html) and also in a few LMXBs such as symbiotic X-ray binaries \cite{2012arXiv1206.2727P}.  In a similar vein, if newly-formed neutron stars are surrounded by a fallback accretion disk, early spin down via magnetic coupling may be another means of achieving rapid spin-down (\cite{2006ApJS..164..130O}).  However, it is not clear how common fallback accretion disks may be.  Evidence for such disks is scant.  There is evidence of an infrared excess indicative of a debris disk around one magnetar (\cite{2006Natur.440..772W}), and evidence of cold material being illuminated by a magnetar giant flare (\cite{2000ApJ...537L.111S}).

In sum, for both accreting and non-accreting neutron stars, the spin period can change rapidly enough that the current frequency is not a faithful representation of the initial state of the star.  For accreting neutron stars, it is therefore also the case that the current spin axis of the neutron star should be strongly aligned with the binary orbital axis, as this determines the plane of accretion.  As we now discuss, other more indirect approaches must be tried to determine the natal spins of neutron stars.

\subsection{Natal Neutron Star Spin}

As noted above, numerous mechanisms can drive spin frequency downward
soon after a neutron star is formed.  Over a much longer time
scale, accretion in a binary system can drive spin frequency upward.
This makes it difficult to
estimate natal neutron star spins.  However, it is important to
attempt such estimates because several aspects of massive stellar
evolution and progenitor events can only be resolved observationally;
theoretical treatments leave large uncertainties.  Understanding the
natal spins of neutron stars provides a rare window into the spins of
progenitor cores
(e.g. \cite{1978ApJ...220..279E},\cite{2011ApJ...731L...5M}).
Similarly, the mechanisms that give rise to the (high) observed space
velocities of neutron stars can be constrained by understanding their
natal spin and magnetic fields (e.g. \cite{2002ASPC..271....3K}).

A closer look at Table~1 of \cite{2012Ap&SS.341..457P} reveals that there are a few rotation-powered pulsars for which the characteristic age $T=P/2{\dot P}$ is much greater than the age of the corresponding supernova remnant.  Possible explanations for this discrepancy include misidentification of the correct remnant, a magnetic moment that is reduced on an improbably short time $<10^5$ years, or an initial pulsar period close to the current one.  A different, and likely more reliable, way to constrain the initial period is to note that the spindown energy is ultimately delivered to the surrounding nebula in the form of high-energy charged particles.  Given the reasonable assumption that the synchrotron cooling time is large, this means that the synchrotron luminosity of a supernova remnant gives clues to the initial spin period of the pulsar; if the period was short, the luminosity will be high.

Characterization of the observed distribution of neutron star spins
using Monte Carlo studies that include various physical effects that
alter spin can, in principle, provide an indirect guide to the
distribution of natal spin values.  The difficulty lies in
characterizing the mechanisms that affect spin and stellar evolution.
For instance, some observations suggest that neutron star magnetic fields can
decay on short time scales (e.g. less than 5~Myr;
\cite{1970ApJ...160..979G,1990ApJ...352..222N,2004ApJ...604..775G}), although this is inconsistent with nuclear theory \cite{1969Natur.224..674B,1992ApJ...395..250G} and other statistical studies find decay times $>100$~Myr \cite{1987A&A...178..143S,1992A&A...254..198B,1997MNRAS.289..592L}.  

The issue of ``injection'' also illustrates the difficulty with using
the current neutron star spin distribution to understand the natal
distribution.  In order to explain the flow of pulsars in the period
versus period derivative plane, a population of neutron stars with
slow birth spin periods of $\sim$0.5~s was postulated
\cite{1981JApA....2..315V}, in addition to the population of
Crab-like pulsars with shorter natal spins (100~ms or less).  For many
years, it was uncertain whether or not the injection of slowly-rotating pulsars was really required
by the data, and concerns also arose regarding selection effects in
the surveys from which the data were obtained.  However, the balance of
current opinion seems to be that ``injection'' is not required.  Enforcing a luminosity
cutoff that still provides for significant statistics, Lorimer found
no evidence of injection \cite{1993MNRAS.263..403L}.  A more recent
analysis, using data from the Parkes Multibeam pulsar survey
\cite{2001MNRAS.328...17M}, finds evidence that many pulsars (40\%)
may be born with periods of 0.1--0.5~s, but found no evidence for
distinct sub-populations \cite{2004ApJ...617L.139V}.

Secure estimates of natal neutron star spins require that the magnetic
braking index be measured, and estimates are greatly aided when the
age of an associated supernova remnant is known precisely.  For these
reasons, the Crab pulsar was the first to have its natal period
estimated ($P_0 \simeq 19$~ms, \cite{1977puls.book.....M}).  A
conservative (and likely robust) list of nine natal neutron star
spin estimates is given in Table 7 of Faucher-Giguere \& Kaspi
(\cite{2006ApJ...643..332F}).  The same authors also report Monte Carlo
modeling of the observed pulsar spin distribution, finding that the
distribution is normal, centered at $P_0 = 300$~ms, with a standard
deviation of 150~ms; this is broadly consistent with more recent studies \cite{2012Ap&SS.341..457P,2013MNRAS.430.2281N,2013MNRAS.432..967I}, and evidence for injection is murky at best.  Thus some pulsars could be born with
few-ms periods, but most appear to start their lives rotating moderately.

\subsection{Measuring Black Hole Spin with the Disk}

As we just discussed, the spin of neutron stars can be
measured directly in pulsars.  The fact of a stellar surface also
means that a surface redshift can (in principle) be measured
directly, for instance via redshifted absorption lines from the neutron
star surface (\cite{2002Natur.420...51C}, but see
\cite{2008ApJ...672..504C}).  Size measures are possible via several
means, including thermal continuum emission from the stellar surface,
though continuum emission is always subject to systematic effects
including scattering in the stellar atmosphere, and magnetic
confinement of the observed continuum to a region smaller than the
full surface area.

In contrast, measurements of black hole spin are {\it necessarily}
indirect.  All current methods (and those potentially available in the
near future with the exception of those involving gravitational waves; see Section~\ref{section:gw}) employ the accretion disk to infer the spin of the black hole.  The two most commonly used methods involve fits to the Fe~K$\alpha$ line profile or to the continuum spectrum.  Both methods assume that the disk is geometrically thin and radiatively efficient, and that the fitted emission essentially terminates inside the innermost stable circular orbit (ISCO; see the discussion in Section~\ref{section:introduction}).  The continuum method also assumes that the black hole spin axis is aligned with the orbital axis of the host binary.  We now briefly discuss these assumptions.

\subsubsection{Geometrically thin, radiatively efficient disks with an ISCO}

The classic studies of \cite{1973A&A....24..337S,1973blho.conf..343N} derived the structure of geometrically thin and radiatively efficient disks, with the additional assumption that there is some inner cutoff radius $R_{\rm in}$ to the disk at which there is zero stress.  With these assumptions, one might suppose that energy radiated in an annulus between radii $r$ and $r+dr$ is just the gravitational energy released there: $dE=GMmdr/r^2$, where $M$ is the mass of the central object and $m$ is the mass in the annulus.  In fact, as first realized by Kip Thorne (see the footnote on page 344 of \cite{1973A&A....24..337S}), there is an extra factor of 3 that enters for annuli well outside the inner edge because the transport of angular momentum from inner to outer annuli also transports energy.  With an assumed sharp cutoff at $R_{\rm in}$, the luminosity of an annulus becomes $(3GM{\dot M}dr/r^2)[1-(R_{\rm in}/r)^{1/2}]$ for an accretion rate ${\dot M}$, which integrates to $GM{\dot M}/R_{\rm in}$ over the whole disk.
If $R_{\rm in}$ is the radius of the ISCO, then it has a significant dependence on the spin of the black hole, as shown in Figure~\ref{fig:isco}.  

Observations can partially address the question of whether $R_{\rm in}=R_{\rm ISCO}$ for black hole accretion disks.  Such studies must be
limited to radiatively efficient, geometrically thin accretion disks.  
Recent observations with CCD
spectrometers suggest that radiative efficiency is likely to apply for $L/L_{\rm Edd} \geq
10^{-3}$ (e.g. \cite{2009ApJ...707L..87T,2010MNRAS.402..836R}, \cite{2012ApJ...757...11M},
\cite{2013ApJ...769...16R}); an older, more conservative, and theory-based guideline might be $L/L_{\rm Edd} \geq 10^{-2}$ (\cite{1997ApJ...489..865E}).  The requirement of geometrical thinness imposes an upper limit of $L/L_{\rm Edd}\ltorder 0.3$, depending on the application. Some
studies of disks over a span of relatively high Eddington fractions,
and over long periods of time, strongly suggest that inner disk radii
are stable (\cite{2007ApJ...666.1129R}, \cite{2010ApJ...718L.117S}).  If the
innermost edge of the disk were to fluctuate across the ISCO in
response to variations in the mass accretion rate, the data would
reflect this.  Instead, the emitting area appears to be fairly constant at high
Eddington fractions, and this suggests that disks are truncated at a fixed characteristic radius, which is likely to be at or near the
ISCO.  There is also evidence in some accreting neutron stars for the
predicted effect of the ISCO on kilohertz quasi-periodic brightness
oscillations \cite{1998ApJ...508..791M,2005MNRAS.361..855B,
2005AN....326..808B,2006MNRAS.370.1140B,2007MNRAS.376.1139B}, although the indirectness of the evidence has led to some cautionary notes by other authors \cite{2006MNRAS.371.1925M}.

Theoretical arguments and simulations can also bear on this problem.
As matter within the ISCO plunges inward, it accelerates and becomes more
ionized, and the emission is expected to be optically thin
(\cite{1997ApJ...488..109R}).  This stands in strong contrast to
direct and reflected emission from the accretion disk outside of the
ISCO, which is expected to be distinctive and prominent.  Recently,
several numerical simulations have been attempted with the goal of
addressing this question.  There are essentially two complementary approaches. The first is to model the innermost disk region in the greatest detail possible,
performing general relativistic magnetohydrodynamic (GRMHD) simulations in as many dimensions as is feasible.  Such high-resolution simulations are critical to understand MHD effects near the ISCO \cite{2010ApJ...712.1241S,2011ApJ...738...84H,2012ApJ...749..189S,2013ApJ...772..102H}.  The second is to compromise on
some detail but attempt to simulate for a large number of dynamical
times in order to understand the long-term behavior of the disk in an
equilibrium state.  These simulations are needed to establish inflow equilibrium at larger radii.  The most recent of the latter family of
simulations strongly suggest that accretion disks respect the ISCO,
and thus that direct and reflected emission from the disk can be used to
infer the spin of the black hole within the ISCO \cite{2008ApJ...675.1048R,2010MNRAS.408..752P}, although there remains some concern that these simulations might not have the resolution required to simulate the nonlinear turbulence properly.

There are some cautionary remarks to apply to these disk models.  A natural first assumption is that each annulus emits as a blackbody.  Indeed, such multitemperature models fit the data quite well \cite{1984PASJ...36..741M}.  However, more careful treatment of scattering and absorption processes lead to the theoretical conclusion that although the local spectra are close to a Planck function, the color temperature $T_{\rm col}$ (which is the temperature one gets by fitting the data to a Planck function) is typically larger than the effective temperature $T_{\rm eff}$ (defined such that the local flux is $\sigma_{\rm SB}T_{\rm eff}^4$, where $\sigma_{\rm SB}$ is the Stefan-Boltzmann constant) by a hardening factor $f_{\rm col}=T_{\rm col}/T_{\rm eff}\approx 1.5-1.7$ \cite{2005ApJ...621..372D}.  The more careful models may indeed be correct, but all that can be said from the data is that the the emitting material is optically thick and has a characteristic emitting temperature near $\sim$1 keV for stellar-mass systems.  Whether the emission has the expected hardening factor of $\approx 1.5-1.7$ or whether it is locally blackbody does have implications for the implied spin, and because the observations alone cannot speak to the hardening factor the spins inferred using this method have some potential systematic errors.

A second issue is that recent observations of quasars with foreground microlensing have revealed discrepancies with the standard geometrically
thin disk models \cite{1990ApJ...358L..33W,1991ApJ...381L..39R,2007ApJ...661...19P,2010ApJ...709..278D,2010ApJ...712.1129M,2010ApJ...718.1079B,2012ApJ...751..106J,2014ApJ...783...47J}.  This has led to suggestions that, for example, the temperature might vary along a given annulus \cite{2011ApJ...727L..24D,2012MNRAS.426L..71D,2014ApJ...783..105R}.  If there are similar deviations from the standard model in disks around stellar-mass black holes, this could affect spin determinations.

A third issue is that because radiation-dominated patches of accretion disks in MHD are apparently thermally unstable \cite{2013ApJ...778...65J}, this may limit the reliability of past simulations for determining what happens at high
luminosities where radiation pressure dominates, and therefore for assessing
possible systematic errors in black hole spin measurements.

Finally, most analyses of black hole spins do not take into account that the disks have finite thickness \cite{2008ApJ...675.1048R}, which means that significant emission can persist inside the ISCO by roughly the disk height.  This is especially important for rapidly rotating holes, for which the distance between the ISCO and the horizon need not be much larger than the disk thickness.  

\subsubsection{Black hole spin axis orientation}

As we describe below, spin estimates using iron lines determine the angle between the spin axis and our line of sight as part of their fits.  In
contrast, degeneracies in the continuum models make it difficult to obtain
robust fits for the disk inclination and black hole spin simultaneously \cite{2006ApJ...647..525D}.  This means that practical applications of the method require the disk inclination to be constrained by other data.  A plausible assumption is that when the hole accretes from a companion, its axis is reoriented to align with the orbital axis.  This, however, is not guaranteed.  As we discussed in Section~\ref{section:introduction}, the gas in the disk is torqued by frame-dragging induced by the black hole's spin, and thus in turn torques the black hole.  The relevant radius in the disk is hundreds to thousands of gravitational radii , which means that the hole will be largely aligned by the time that a gas mass that is a fraction of $\sim$few percent of the hole's mass has flowed over.  For a canonical black hole mass of $10~M_\odot$ this means that a few tenths of a solar mass must have accreted.  
This is comparable to the mass of a typical low-mass companion, so it is quite possible that in many systems the required accretion has not happened \cite{2002MNRAS.336.1371M,2008MNRAS.387..188M}.

The data are ambiguous on this issue.  Studies of some black holes conclude that the alignment is good, such as for XTE~J1550--564 \cite{2012ApJ...745..136S}, whereas for other sources such as GRO~J1655--40 (see \cite{1995Natur.375..464H} for the jet orientation and \cite{2001ApJ...554.1290G} for the binary orientation) and V4641 Sgr \cite{2001ApJ...555..489O}, significant misalignment has been inferred.  Interestingly, it has been suggested that a misaligned inner disk could, at high accretion rates, rotate like a solid body and thereby produce the few-Hz quasi-periodic brightness oscillations seen in several accreting stellar-mass black holes \cite{2001ApJ...553..955F,2007ApJ...668..417F,2009MNRAS.397L.101I,2011ApJ...730...36D}.  If this is accurate, it means that spin axes can be somewhat independent of orbital axes.

With this background, we now discuss the two major methods for inferring the spins of stellar-mass black holes.

\subsection{Relativistic Disk Reflection}

Dense gas can scatter or ``reflect'' incident X-ray emission
(\cite{1988ApJ...331..939W}, \cite{1991MNRAS.249..352G}).  This
phenomenon is observed in numerous physical contexts.  For instance,
particular clouds of gas in the Galactic Center are observed to have
strong Fe K emission lines, likely stimulated by ancient X-ray
emission from Sgr A* (e.g. \cite{2004A&A...425L..49R}).  Strong,
narrow Fe K emission lines are also observed in the spectra of
numerous Seyfert-1 and Seyfert-2 AGN, owing to hard X-ray irradiation
of the torus and/or broad line region
(e.g. \cite{2010ApJS..187..581S}, \cite{2011ApJ...738..147S}; also see
\cite{2007MNRAS.382..194N}).  Indeed, in some Seyfert-2 spectra,
emission from the central engine is almost entirely blocked from
direct view, and the observed spectrum is strongly dominated by
reflected light.

The most prominent part of typical reflection spectra is Fe K line
emission, owing to the abundance and fluorescence yield of iron, and
also to the ability of He-like and H-like charge states of iron to retain
electrons at very high temperatures and ionization parameters
(e.g. \cite{1991MNRAS.249..352G}).  The other important features of a
reflection spectrum include photoelectric absorption from the Fe K
shell (which produces an apparent flux decrement), followed by a
recovery of the continuum peaking at about 30~keV (this is where the
albedo of the disk peaks), and a gradual flux decline at higher energy
(the albedo is reduced as incident photons penetrate deeply into the
gas and do not readily re-emerge).  Depending on the specific nature
of the gas and the intensity of the irradiating flux, Fe L lines can
also be quite prominent (\cite{2005MNRAS.358..211R}.  Figure 2 shows
the relativistic line and reflection spectrum in a recent {\it NuSTAR}
spectrum of GRS 1915$+$105 \cite{2013ApJ...775L..45M}.  Careful construction of the reflection models is critical for all spin determinations using this method 
\cite{2005MNRAS.358..211R,2010ApJ...718..695G,2011ApJ...731..131G,2013ApJ...768..146G,2014ApJ...782...76G}.

\begin{figure}[!htb]
\begin{center}
\includegraphics[scale=0.7]{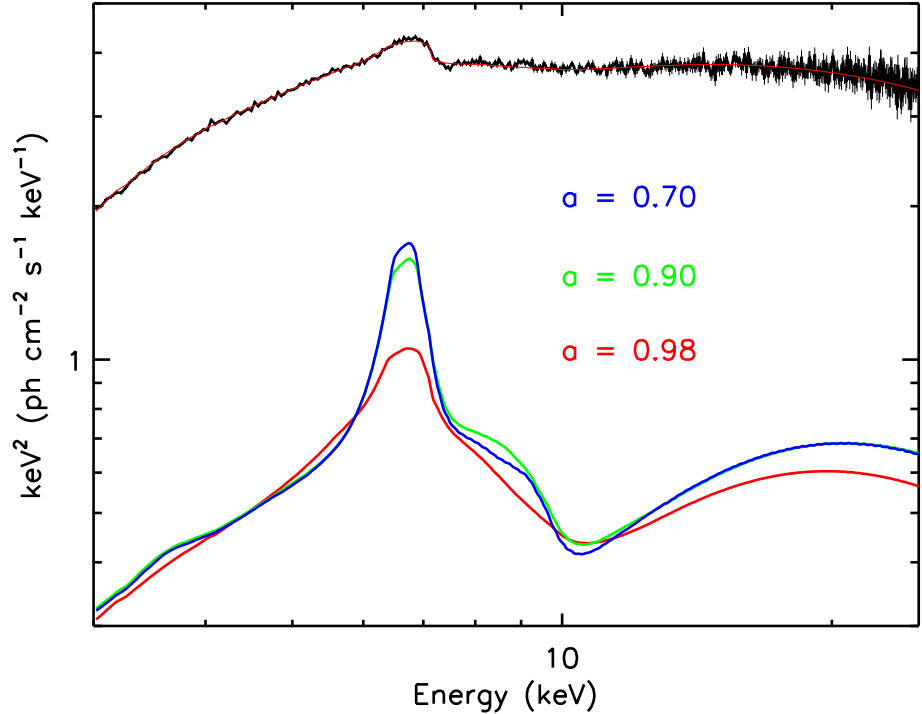}
\caption{\footnotesize This figure shows a {\it NuSTAR}
  spectrum of GRS 1915$+$105 (in black), fit with a disk reflection
  model (in red) that gives a spin of $a = 0.98\pm0.01$
  \cite{2013ApJ...775L..45M}.  The contribution of the best-fit disk
  reflection spectrum to the total spectrum is shown below the data,
  in red.  To illustrate the effect of black hole spin on the
  reflection spectrum, we show models with spins of $a = 0.90$ and $a = 0.70$
  in green and blue, respectively.  The effect of spin is
  clearly to broaden the reflection spectrum, and to give a stronger
  red wing to the Fe K line profile.  Indeed, the power of this
  technique for spin measurements lies partly in the fact that it is a
  relative measurement.  Note that the data must also constrain other
  parameters that are important in shaping the disk reflection
  spectrum, including the inner disk inclination and emissivity, in
  order to determine the spin of the black hole.}
\end{center}
\end{figure}
\medskip

An accretion disk will intercept a large fraction of any hard X-ray
flux generated in the accretion process.  Ignoring potentially
important considerations such as gravitational light bending
\cite{2004MNRAS.349.1435M,2004MNRAS.351..466M,2005MNRAS.360..763R,2013ApJ...763...48R}
and light travel times, the disk should subtend approximately
$2\pi$ steradians as seen from a source of hard X-ray emission.  Utilizing
the disk reflection spectrum requires a change of reference frame.
Reflection spectra are calculated in the fluid frame of the disk --
all of the lines and features are intrinsically narrow.  At infinity,
however, an observer sees a spectrum that has been ``blurred'' by
relativistic Doppler shifts and gravitational red-shifts.  Because the ISCO becomes smaller with higher spin parameters,
the degree of ``blurring'' is set by the spin.  Thus with a simple
relativistic smearing model (usually just a library of rays traced
from close to the black hole to infinity; see
\cite{1989MNRAS.238..729F,1991ApJ...376...90L,2004ApJS..153..205D,2004MNRAS.352..353B,2006ApJ...652.1028B,2013MNRAS.430.1694D}) the spin
can be measured in sensitive X-ray spectra (see, e.g., \cite{2007ARA&A..45..441M}).

In practice, when fitting a blurred disk reflection spectrum to
measure a spin parameter, a few crucial variables must be measured in
addition to the spin.  Parameters central to the blurring include the
{\it emissivity} of the disk reflection spectrum (typically a
power-law in radius, with zero--two breaks; see
\cite{2007ARA&A..45..441M,2009Natur.459..540F,2012MNRAS.424.1284W} ), and the {\it inclination} of the inner
disk.  If the nature of the corona is known a priori, the emissivity
function can be fixed, but this is not a requirement for a spin
measurement.  In the limit of less sensitive spectra, it is often
expedient to assume that a resolved relativistic jet axis traces the
black hole spin axis, and that this vector is normal to the inner
accretion disk (as expected owing to gravito-magneto-hydrodynamic
effects; e.g. \cite{1972ApJ...178..347B,1975ApJ...195L..65B}).
Parameters central to characterizing the reflection include the {\it
  ionization} of the disk, and the {\it flux normalization} of the
reflection spectrum.  In some cases, it may also be necessary to
measure or set the metal abundances of the reflector (see
\cite{2009Natur.459..540F}), though strong abundance anomalies may be
artifacts of an effect similar to ``magnetic levitation'' observed in
the sun.

Several reviews of disk reflection and related topics have been
published (see, e.g., \cite{2003PhR...377..389R,2005CaJPh..83.1177L,2007ARA&A..45..441M}).  The level of
detail in those reviews exceeds what is possible to provide in this
spin--centered treatment, and the reader is referred to the prior work
for additional nuances of disk reflection spectroscopy.  For this
review, we simply proceed to briefly comment on the primary features
of this technique, and to highlight some key results.

This spin measurement technique has some notable strengths and
advantages:\\ 

\noindent$\bullet$ It relies primarily upon lines and atomic physics, and
{\it relative} measurements (e.g. line width) rather than absolute
measurements (e.g. absolute fluxes or absolute magnitudes).
Traditionally, astrophysics is a field that excels at relative
measurement, but faces numerous challenges with absolute measurement
because it is mostly an observational science, not an empirical one.\\

\noindent$\bullet$ Similarly, relativistic distortions to the
reflection spectrum depend on the relative depth of the inner disk
within the potential well -- things scale in a $GM/c^{2}$ sense,
obviating the need to know the mass of the black hole, the distance to the source, or the accretion rate a priori.  Spin
measurements have been obtained from (clear, convincing) black hole
candidates whose mass functions have not been measured
(e.g. \cite{2009ApJ...697..900M,2011MNRAS.411..137H,2011MNRAS.410.2497R}).\\

\noindent$\bullet$ Disk reflection occurs in both stellar-mass black hole
binaries, and in AGN.  This enables tests and explorations of black
hole spin across the mass scale.  This is likely to be very important
for investigations of the role that black hole spin may play in
powering relativistic jets (e.g. \cite{2013ApJ...771...84K}).\\  

\noindent$\bullet$ Reflection spectra are also sensitive to the
inclination of the inner disk, which need not be the same as the
inclination of the outer disk or binary system
(\cite{2002MNRAS.336.1371M}).  Reflection can then serve to obtain a
unique measure of the inner accretion disk inclination.\\

\noindent $\bullet$ In principle, the geometry of the hard X-ray
corona can be inferred through the disk reflection emissivity.  For example, if the observed iron line comes only from a disk illuminated by an isotropic source of hard X-ray photons, the radial dependence of the emissivity will be $J\propto r^{-3}$, but it is not certain that this is the correct emissivity distribution  \cite{2011ApJ...743..115N,2013ApJ...769..156S}.  For instance, if the source of hard photons is above the disk plane on the disk axis, gravitational light bending can produce a significantly stronger radial dependence of the illumination on the disk.  Thus the
emissivity can encode how central or how extended the corona is, what
its height above the disk may be, and whether or not gravitational
light bending is important (\cite{2012MNRAS.424.1284W}, also see
\cite{2013ApJ...769L...7R}).  The emissivity function is often
captured as a power-law in radius, with zero, one, or two breaks.  From the observational standpoint, microlensing data for quasars suggest that the X-ray source size is much smaller than that of the optical or UV \cite{2009ApJ...693..174C,2010ApJ...709..278D}, which might imply that the emissivity index is steeper than $r^{-3}$; this would be compatible with the steep indices required in some disk reflection fits of AGN (e.g., \cite{2001MNRAS.328L..27W} and subsequent papers).  In addition, it is important to note that the decrease in surface density across the ISCO region is likely to create a cutoff in emissivity, but exactly where it happens depends on details.  Thus the shape of this cutoff and exactly how it is positioned with
respect to the ISCO introduce a certain amount of systematic uncertainty in
the spin determination \cite{2008MNRAS.390...21B,2009ApJ...692..411N,2011ApJ...743..115N}.  Recent work \cite{2011MNRAS.414.1183K} suggests that the uncertainty might be $\sim\Delta{\hat a}\sim 0.2$, which is comparable to or less than uncertainties in distance or inclination.\\

Examples of measurements via disk reflection (see Table~\ref{tab:bhspin} for a full list):\\

{\bf GRS 1915$+$105} is one of the best--known stellar-mass black
holes, largely because of its prodigious jet activity.  In some
phases, it produces steady, compact jets; in other states, it has been
observed to launch discrete ``cannonballs'' with apparent speeds in
excess of $c$ (\cite{2000ApJ...543..373D}).  This activity makes GRS
1915$+$105 a particularly interesting source in which to try measure
the black hole spin, since spin may power jets
(\cite{1977MNRAS.179..433B}).  Early efforts at measuring the spin via
reflection were complicated by the detector limitations, but indicated
a high spin consistent with ${\hat a} = 0.9$ (\cite{2006tmgm.meet.1296M};
also see \cite{2002A&A...387..215M}).  Later efforts to measure the
spin using a deep {\it Suzaku} observation were also inconclusive
(\cite{2009ApJ...706...60B}).  A new observation with {\it NuSTAR}
(\cite{2013ApJ...770..103H}), which offers a fertile combination of
spectral resolution, sensitivity, and band pass, has revealed an
excellent blurred disk reflection spectrum and gives ${\hat a} = 0.98\pm
0.01$ (\cite{2013ApJ...775L..45M}).  This spin value is broadly
consistent with one measurement obtained using the disk continuum
(\cite{2006ApJ...652..518M}).  However, for such high spins the finite thickness of the disk will likely soften the limit to $\gtorder 0.9$.  Put more strongly, the assumption of a razor-thin disk that goes into the spectral modeling must break down at the ISCO of a ${\hat a}=0.98$ black hole as close to Eddington as GRS~1915+105.

Some efforts have attempted to move beyond measuring spins in black
holes.  Variability trends between the flux incident upon the disk,
and the reflection spectrum, can potentially serve as an indication of
gravitational light bending (e.g. \cite{2004MNRAS.349.1435M}).
Essentially, in a simple Euclidean scenario, the reflected flux should
track the incident flux in a linear fashion.  In practice, the two are
positively correlated for a range of incident flux levels, but
thereafter the reflection spectrum does not respond.  This is
consistent with a central corona collapsing to small scale heights,
such that most of the hard X-ray emission is focused onto the disk
rather than emitted isotropically (\cite{2004MNRAS.349.1435M,2004MNRAS.351..466M,2005MNRAS.360..763R,2013ApJ...763...48R}).  Light bending is expected if hard X-ray
emission is generated close to spinning black holes, so the potential
detection of light bending serves as a potential confirmation of the
method and measurements.

{\bf XTE J1650$-$500} is a black hole for which spin has been measured
quite well, and its strong disk reflection spectrum supplies the best
evidence of gravitational light bending.  Arguably, the first
reflection-derived spin from a stellar--mass black hole was obtained
in XTE J1650$-$500: a blurred iron line detected with {\it XMM-Newton}
and an inner disk radius consistent with $r = 1.24~GM/c^{2}$ was
measured (\cite{2002ApJ...570L..69M}), indicating a near-maximal spin.
Later modeling using more advanced disk reflection spectra found a
spin of ${\hat a} = 0.79\pm 0.01$ (\cite{2009ApJ...697..900M}).  More
intensive monitoring with {\it BeppoSAX} and RXTE found a non-linear
relationship between the direct and reflected flux, consistent with
the relationship expected if flux is partly modulated by gravitational
light bending close to a spinning black hole
(\cite{2004MNRAS.351..466M,2005MNRAS.360..763R}).  New work
has eliminated the possibility that the flux trend is driven by the
ionization of the disk, and affirmed the signature of light bending
(\cite{2013ApJ...763...48R}).

{\bf GX 339$-$4} is perhaps thee best-known stellar-mass black hole that
is a recurrent transient.  The best measurements of black hole spin in
GX~339$-$4 have been obtained in elevated flux states,
e.g. ``intermediate'' and ``very high'' states using {\it XMM-Newton}
and {\it Suzaku}.  Fits to a very skewed iron line and reflection
spectrum obtained with {\it XMM-Newton} during the 2002-2003 outburst
of GX 339$-$4 measured inner disk radii in the $2-3~GM/c^{2}$ range
(\cite{2004ApJ...606L.131M}), corresponding to spin values of $a
\simeq 0.8-0.9$.  Later observations made using {\it Suzaku} during
the 2006-2007 outburst of GX 339$-$4 measured a spin of $a = 0.89\pm
0.04$ (\cite{2008ApJ...679L.113M}).  Joint fits to the {\it
  XMM-Newton} and {\it Suzaku} data give a spin of $0.93\pm 0.04$
(\cite{2008ApJ...679L.113M}).  Owing to its strong, skewed reflection
spectrum, new reflection models have been tested against data from
GX~339$-$4, enabling one measure of the modeling-based systematic
errors associated with reflection-derived spins.  Related errors
appear to be small, at the level of a few percent or less
(\cite{2008MNRAS.387.1489R}).

{\bf XTE J1752$-$223} is a black hole candidate in which the spin has
been measured well, and like XTE J1650$-$500 it extends the reach of
spin measurements and reflection.  Using {\it XMM-Newton} and {\it
  Suzaku}, disk reflection spectra from this black hole candidate were
measured, and consistency between the ``intermediate'' and
``low/hard'' states was found (\cite{2011MNRAS.410.2497R}).  This is
significant in that the disk may be radially truncated in the low/hard
state, but - at least in these sensitive spectra - the disk instead
appears to remain at the ISCO.  Fitting the disk reflection spectra
jointly, a spin of ${\hat a} = 0.52\pm 0.11$ was measured
(\cite{2011MNRAS.410.2497R}).  This moderate value likely indicates
that reflection--based measures of black hole spin are not
artificially biased toward high values.

The success of disk reflection in measuring black hole spins has
spawned efforts to harness the same technique in accreting neutron
star X-ray binaries (e.g. \cite{2007ApJ...664L.103B,2008ApJ...674..415C,2010ApJ...720..205C}).  In
principle, the extent of relativistic blurring would allow the
dimensionless spin of the neutron star to be measured.  However, this
method is likely to be less practical than timing techniques.  The
more interesting constraint that may be possible via disk reflection
in neutron stars is a constraint on the size of the neutron star, in
cases where the stellar mass is known.  Simply, the stellar radius
must be less than the radius of the inner disk.  Converting an inner
radius measured via disk reflection into physical units from
gravitational radii then provides an upper limit on the neutron star
radius.  Currently, the most convincing relativistic line profile and
best data are from an observation of Serpens X-1 with {\it NuSTAR},
which give $r_{NS} \leq 15$~km (and thus a surface redshift $z_{NS} \geq 0.16$ for $M_{NS} =
1.4~M_{\odot}$) for a broad set of fits, and $r_{NS} \leq 12.6$~km if
the disk extends to the ISCO (\cite{2013ApJ...779L...2M}).  Incidentally, although the spacetime around a rotating object is unique to first order in rotation \cite{1968ApJ...153..807H}, to higher orders the structure depends on the detailed nature of the object.  In particular, the Kerr spacetime is not a good approximation for the external spacetime of a rapidly rotating neutron star \cite{1998ApJ...509..793M}, so for such objects numerical spacetimes must be used.

\subsection{The Accretion Disk Continuum}

It is also possible to measure the spin of a black hole using the
thermal continuum emission from the accretion disk \cite{1997ApJ...482L.155Z,2011CQGra..28k4009M}.  A number of
results suggest that accretion disks in the ``high/soft'' state are
well described in terms of a standard ``thin'' accretion disk
(\cite{1973A&A....24..337S}).  This means that the emission vanishes
at the ISCO as there is zero torque, peaks at a radius just outside of
the ISCO, and follows a $T \propto r^{-3/4}$ profile far from the ISCO.
Essentially, then, the disk can be modeled as a set of
annuli; this is easily encoded into spectral fitting packages (see,
e.g., \cite{1984PASJ...36..741M}).

\begin{figure}[!htb]
\begin{center}
\includegraphics[scale=0.7]{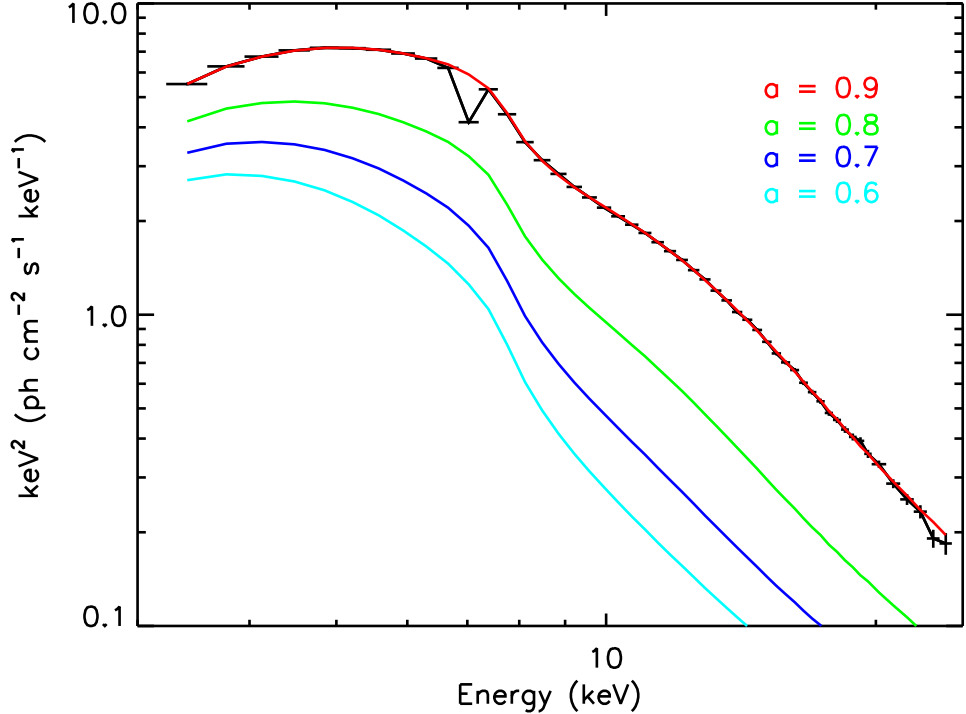}
\caption{\footnotesize This figure shows an {\it RXTE}
  spectrum of GRS 1915$+$105 (in black), fit with a model that
  requires a spin of $a = 0.9$ (in red).  The spectrum is observation
  4 within \cite{2006ApJ...652..518M}; the fitting procedure outlined
  in that work was employed in this fit.  The model also includes a
  hard component, and the choice of the hard component impacts
  spin values.  The notch near to 7 keV is an absorption line included
  in the model.  A disk close to a spinning black hole emits more at
  all wavelengths; to illustrate the effect of spin, then, we show models
  assuming $a = 0.8$, $a = 0.7$, and $a = 0.6$ in green,
  blue, and cyan, respectively.  Continuum-based measurements of spin
  are able to utilize the bulk of the observed X-ray flux, but they
  are an absolute measurement and thus require the mass and distance
  of the black hole to be known precisely.  It is also important to
  know the inner disk inclination and atmospheric scattering
  corrections from other observations or theoretical considerations.  Note that actual fits with different spin
  parameters must allow the hard component to vary and adjust to
  optimize the fit, and that the data must constrain all model
  components in order to deliver robust spin measurements.}
\label{fig:spincontinuum}
\end{center}
\end{figure}

Any blackbody flux nominally gives a ratio of the observed emitting
area over the line-of-sight distance, $r^{2}/d^{2}$, modified by the
inclination angle.  The mass of the black hole must also be known, in
order to translate the physical radii into gravitational radii, and
thus determine the spin.  Thus, to employ continuum fitting, the mass
and distance to the black hole must be known a priori and
independently.  Owing to the fact that thermal emission is a continuum
and does not bear the imprints of Doppler shifts as distinctively as
atomic features, the inclination angle of the inner accretion disk
(which may differ from the binary inclination;
\cite{2002MNRAS.336.1371M}) must also be known a priori.  Last,
radiative transfer through a disk atmosphere will modify the spectrum
emitted in the midplane.  As we discussed above, the observed spectrum
must be corrected by a hardening factor in order to measure a spin.  A
value of $f = 1.7$ is taken to be canonical
(\cite{1995ApJ...445..780S}), though a broader range is possible
depending on the nature of the corona
(e.g. \cite{2000MNRAS.313..193M}).  New models for accretion disks
capture all of these parameters and are able to turn these absolute
photon flux measurements into black hole spins
(e.g. \cite{2005ApJS..157..335L}, \cite{2005ApJ...621..372D}).  Figure~\ref{fig:spincontinuum} depicts the influence of spin on the accretion disk continuum.

\medskip

Distances within the Galaxy are typically difficult to determine, and
errors can be comparable to the distance value itself.  Of course,
carrying forward large errors into spin measurements should result in
large uncertainties.  This situation has recently improved
drastically: VLBA astrometry has made it possible to determine
parallax distances for some black holes (e.g. V404 Cyg and Cyg X-1;
\cite{2009ApJ...706L.230M,2011ApJ...742...83R}), reducing
errors to a very small fraction of the distance value.  Masses are
also difficult to determine, typically giving very large fractional
errors (see, e.g., \cite{2006csxs.book..157M}).  Inner disk
inclinations can be determined via relativistic jets, assuming that
the jet axis is the spin axis and normal to the inner disk \cite{2012ApJ...745..136S}.  Setting
the inner disk inclination to that of the binary system is a less
robust assumption, but provides a path forward when other constraints
are unavailable.

Like the disk reflection method, estimates of spin using the disk continuum have their own strengths:

$\bullet$ In soft spectral states, wherein the disk continuum may be
employed to measure spins, the disk represents the bulk of the photon
flux and the bulk of the energy output.  When the mass, distance,
and inclination are known precisely, this can translate into precise
black hole spin measurements.

$\bullet$ Harnessing the bulk of the observed flux also means that
spectra from stellar-mass sources in the Local Group, and potentially within
galaxies at distances of a few Mpc, can be used to measure spins.  In
such cases, distances are surprisingly known with much better
precision and accuracy than is generally possible within the Milky
Way; the prime difficulty for extra-galactic sources lies in measuring
the black hole mass.

$\bullet$ Monitoring observations obtained with spectrometers that
have low spectral resolution are often sufficient to characterize the
disk continuum well.  Therefore, numerous spectra can be fit together,
and thus jointly determine a spin parameter.  This may reduce the
influence of fluctuations in other parameters in deriving a spin
value.

Examples of measurements via the disk continuum (see Table~\ref{tab:bhspin} for a full list):\\

{\bf GRS 1915$+$105} is measured well using the accretion disk
continuum.  It is not necessarily obvious that this should be the
case, because the line-of-sight column density to GRS 1915$+$105 is
very high and scatters away most of the flux below 2~keV; however, the
high disk temperature compensates and allows robust determinations
based on the Wien tail of the distribution.  Fairly accurate black
hole masses and distances \cite{2013ApJ...768..185S}, and an
inclination defined by extended radio jets
\cite{1999MNRAS.304..865F}, also contribute to good measurements.
Based on fits to numerous spectra selected for a a flux strongly
dominated by the disk and a low Eddington fraction, a spin of ${\hat
  a}\geq 0.98$ was measured \cite{2006ApJ...652..518M}.  It is
noteworthy that the higher spin value agrees with the reflection
measurement \cite{2013ApJ...775L..45M}.  A spin estimate of ${\hat
  a}\simeq 0.7$ was also reported \cite{2006MNRAS.373.1004M}, but
included observations where the source may have been close to
Eddington, potentially inconsistent with the model assumption that the
disk is geometrically thin.

The spin of {\bf Cygnus X-1} was recently measured well using the disk
continuum \cite{2011ApJ...742...85G} and later via reflection
\cite{2012MNRAS.424..217F,2014ApJ...780...78T}, and commensurate
values were again obtained.  Employing a new distance derived using
radio parallax measurements \cite{2011ApJ...742...83R}, and improved
binary system parameters (\cite{2011ApJ...742...84O}; however, see
\cite{2014MNRAS.tmpL..36Z}), a spin of ${\hat a} > 0.95$ is measured
\cite{2011ApJ...742...85G}.  

It is worth noting that reflection-based measurements of the spin of
Cygnus X-1 have improved with better-suited detectors and read-out
modes.  Fits to {\it ASCA} gas spectrometer data obtained in a
high/soft state were consistent with a near-maximal spin, but were not
able to strongly exclude lower values \cite{2006tmgm.meet.1296M}.  In
contrast, spin estimates based on fits to spectra obtained in the
low/hard state using the {\it XMM-Newton}/EPIC-pn CCD camera suggested
a very low spin \cite{2009ApJ...697..900M}.  It is now clear that
photon pile-up (which occurs when the rate of photon arrival is greater than can be processed by the instrumentation) can artificially narrow relativistic skewing
\cite{2010ApJ...724.1441M}, and the value obtained with {\it
  XMM-Newton} might best be regarded as a lower limit.  The most
recent value of ${\hat a} = 0.97^{+0.01}_{-0.02}$ was obtained with {\it
  Suzaku} \cite{2012MNRAS.424..217F}, employing a combination of CCD
spectra at low energy and broad-band coverage up to 400~keV.

{\bf M33 X--7} provides an example of how continuum fitting may be
applied to black holes in nearby galaxies.  The distance to M33 is
known well, and X--7 happens to be an eclipsing source -- immediately
providing a strong inclination constraint and facilitating a mass
constraint.  With these advantages, a spin of ${\hat a}=0.84\pm 0.05$
was measured \cite{2008ApJ...679L..37L,2010ApJ...719L.109L}.

{\bf LMC X-3} is another interesting example of the continuum method.
Following an extensive effort to re-calibrate the flux normalization
of historical X-ray missions relative to the Crab (now known to be
slightly variable), it was possible to show that the inner radius of
the disk in this source may be fairly constant, signaling conditions
that would permit a spin measurement \cite{2010ApJ...718L.117S}.
Binary system parameters can be particularly difficult to derive when
the donor star is massive, however, so despite the compelling X-ray
data, a spin measurement had to await improved system constraints
\cite{2014arXiv1402.0085O}.  A spin of ${\hat a} = 0.21^{+0.18}_{-0.22}$
has now been obtained \cite{2014arXiv1402.0148S}.  It is unclear whether it is significant that this is a smaller spin than has been obtained for other sources.

\subsection{Quasi-Periodic Brightness Oscillations}

In principle, quasi-periodic brightness oscillations (QPOs) provide a simple and precise means of measuring
black hole spin.  The highest frequency QPOs detected in stellar-mass
black holes are broadly consistent with the orbital frequency at the
ISCO, which is $\nu_{ISCO} = 220~{\rm Hz} (10~M_{\odot}/M_{BH})$, for ${\hat a} = 0$ and is larger for prograde spins.  These
features are narrow enough compared to their frequency that they could
in principle provide very specific information.  A hint as to the mechanism is that in a few
systems QPOs are detected in roughly stable 3:2 frequency ratios
(for a review, see \cite{2006csxs.book..157M}), although many models can get approximately this ratio (see the discussion in Section 5.2 of \cite{2009ApJ...692..869R}). Note that there is some evidence that the frequencies can change by at least several percent \cite{2012MNRAS.426.1701B,2013MNRAS.432...19B}, which might be caused by the inner disk becoming marginally optically thin in some circumstances \cite{2014MNRAS.438.3352D}), even in distinct
outbursts separated by several years (XTE J1550$-$564 is one such
case; see \cite{2001ApJS..132..377H,2001ApJ...563..928M}),
suggesting that the frequencies are anchored by gravity and not
fleeting manifestations of processes in an MHD disk that would reveal
nothing of the black hole itself.

The pragmatic difficulties with using QPOs to measure spin are many.
Whereas nearly every stellar-mass black hole shows continuum emission
from the disk, as well as disk reflection, only a minority show high
frequency QPOs, and a smaller minority manifest QPOs in a 3:2
frequency ratio (\cite{2006csxs.book..157M,2006csxs.book...39V}; note that GRS~1915+105 has two such pairs, one pair at 67~Hz and 41~Hz and the other at 164~Hz and 113~Hz).  Moreover, high frequency QPOs are rare,
sometimes only detected in a few observations out of many hundred
sampling an outburst, or in the addition of small segments of hundreds
of observations that satisfy flux and spectral hardness selection
criteria (e.g. \cite{2001ApJS..132..377H},
\cite{2001ApJ...552L..49S}).  Last, spins based on QPO frequencies can
require the mass of the black hole to be known.  These difficulties
stand in marked contrast to the simplicity of disk reflection, for
instance, which can measure a spin using a single sensitive spectrum,
without knowledge of the black hole mass.

The physical difficulties with using QPOs to measure spin are as
severe as the pragmatic hurdles.  The energy spectrum of QPOs is
decidedly not disk--like (e.g. \cite{2001ApJS..132..377H,2001ApJ...552L..49S}).  The rms amplitude of QPOs increases with
increasing energy.  Indeed, the highest frequency yet detected --
450~Hz in power spectra of GRO J1655$-$40 -- was only detected in the
13--27~keV band with RXTE (see \cite{2001ApJ...552L..49S}; a disk
temperature of $kT \simeq 1$~keV is characteristic of a stellar-mass
black hole close to Eddington).  If QPO frequencies are indeed disk
frequencies, the disk must somehow transmit the frequency into a
corona where it is expressed.  Alternatively, it may be the case that
QPO frequencies are not disk frequencies; some models predict QPOs
produced at the coronal base of a jet \cite{2012MNRAS.423.3083M}.

Owing to the degeneracy of different QPO models, examining different
interpretations and spin estimates for a single black hole is likely
more practical than touring the different models in the abstract.
The most interesting case is likely GRO J1655$-$40, because the $450$~Hz
oscillation detected from this source is the highest-frequency yet
detected in a stellar-mass black hole (note that a $300$~Hz
oscillation is detected at the same time, though not in the same
energy range; see \cite{2001ApJ...552L..49S}).

$\bullet$ If the 450~Hz oscillation represents the Keplerian orbital
frequency at the ISCO, constraints on the mass range for GRO
J1655$-$40 then constrain the spin to be $0.15 \leq {\hat a} \leq 0.5$
\cite{2001ApJ...552L..49S}.

$\bullet$ If either the 300~Hz or 450~Hz oscillations are a
Lense-Thirring precession frequency, then the black hole must be
rotating at or near the maximum possible value.

$\bullet$ The oscillations may be resonances between Keplerian,
periastron precession, and nodal precession frequencies
\cite{1999ApJ...524L..63S}.  Such a resonance could account for the
300~Hz and 450~Hz oscillations and would give $0.4 < a < 0.6$, but the
predicted nodal precession could not be associated with the observed
18~Hz QPO also detected in GRO J1655$-$40
\cite{2001ApJ...552L..49S}.  More recently, a spin of ${\hat a} = 0.290\pm
0.003$ has been reported, assuming the QPOs arise via the relativistic
precession mechanism \cite{2014MNRAS.437.2554M}.  Note, however, that this mechanism assumes geodesic motion, but eccentric or out-of-plane motion would be expected to be damped by fluid interactions.

$\bullet$ Normal modes of disk oscillations -- diskoseismic modes --
are another potential model for high-frequency QPOs
\cite{1997ApJ...476..589P,1997ApJ...477L..91N,2001AdSpR..28..505K,2008NewAR..51..828W}, although these have not been seen in magnetohydrodynamic simulations \cite{2009ApJ...692..869R}.  The
fundamental $g$-mode is expected to be the most robust oscillation
mode, and the most readily detectable.  The 450~Hz oscillation of GRO
J1655$-$40 cannot be associated with any $g$-mode.  If the 300~Hz
oscillation is associated with the fundamental $g$-mode, then a spin
of $0.86 < a < 0.98$ is implied \cite{2012ApJ...752L..18W}.

Data from observations of GRO J1655$-$40 and other stellar-mass black
holes do not prefer any specific model for the observed
frequencies.  The same could be said for the few-Hz QPOs seen in many stellar-mass black hole candidates; these are stronger than the high-frequency QPOs, and are often sharper, so although their origin is not known with confidence, if they are linked to the mass and/or spin of the black hole they might ultimately be useful as probes.  Therefore, although it is possible to measure black hole spin via QPOs {\it in principle}, the data do not yet allow for a reliable measurement.  If future data (for instance, from {\it Athena} \cite{2013sf2a.conf..447B} or \textit{LOFT} \cite{2012ExA....34..415F}) are able to identify a particular model as
correct, then true measurements will become possible.  

Current models for high frequency QPOs share some fundamental
difficulties.  The {\it frequencies} are explained within a model, but
not the specific {\it modulation mechanism}.  Nor do models explain
the energy dependence of QPOs, or the nature of other facets such as
phase lags.  Depending on the specific model assumed when interpreting
QPOs, these difficulties may be more or less problematic.

It is additionally unclear why some sources show QPOs in a 3:2
frequency ratio (e.g. GRO J1655$-$40, XTE J1550$-$564, XTE
J1650$-$500, H~1743$-$322; \cite{2001ApJ...552L..49S,2001ApJ...563..928M,2003ApJ...586.1262H,2005ApJ...623..383H}), while others appear to only show a single
QPO (4U 1630$-$472, \cite{2004NuPhS.132..381K}), and many sources show
no high frequency QPOs (e.g. 4U 1543$-$475, GX 339$-$4, XTE
J1817$-$330).

As was discussed by \cite{2006ApJ...642..420S}, it is
remarkable that the sources wherein high frequency QPOs have been
detected are all observed at high inclinations ($i \geq 50^{\circ}$),
at least if the inclination of the inner disk is assumed to be that of
the binary.  In contrast, the sources wherein high-frequency QPOs have
{\it not} been detected are likely viewed at much lower, face-on
inclinations.  This raises intriguing possibilities:


$\bullet$ First, it may be the case that QPOs are more easily detected
at high inclinations owing to special relativistic beaming effects
associated with the orbits and/or resonances that produce the
oscillations.

$\bullet$ Second, if the oscillation is produced via any sort of
precession or oscillation in a vertical direction, an inclination
angle closer to the plane of the disk may be advantageous.  Special
relativistic boosting may also aid detection from high inclination
angles, in such a scenario.

$\bullet$ Third, it may be that sources viewed at low inclinations
produce less Comptonized radiation, and that the scattering process is
somehow important to the production of high frequency QPOs.  The
energy dependence of QPOs may suggest this independently.  This might
be true if scattering through the disk atmosphere is not a simple
constant, as envisioned in the application of simple multiplicative
$f_{\rm col}$ factors.  Naively, the path length through an atmosphere to
an observer at high inclination should be longer.  Potentially, then,
$f_{\rm col}$ may need to be written as $f_{\rm col} (i)$ or $f_{\rm col,0} +
f_{\rm col} (i)$, where $i$ is the inclination angle, and $f_{\rm col,0}$
might be unity or the traditional value of $f_{\rm col} = 1.7$ .  This would have
obvious consequences for spin measurements made using fits to the
accretion disk continuum that have assumed $f_{\rm col} = 1.7$ for all
inclinations.

Last, the recent outburst of the black hole IGR J17091$-$3624 poses
some interesting questions about the use of QPOs to study black hole
spin.  Detections of the the 188~Hz and 276~Hz QPOs in distinct
outbursts of XTE J1550$-$564 strongly signals that the observed
frequencies are anchored by gravity, and not subject to MHD nuances or
$\dot{M}$.  However, 66~Hz QPOs are detected in IGR J17091$-$3624
\cite{2012ApJ...747L...4A}, remarkably similar in frequency and
other properties to the famous 67~Hz QPOs detected in GRS 1915$+$105
\cite{1997ApJ...482..993M}.  It is entirely possible that this is simply coincidental .  However, it raises the prospect that these oscillations
are physically distinct from the higher frequency oscillations
detected in GRO J1655$-$40 and XTE J1550$-$564, in which case some of the QPOs might not be anchored by gravity.

\subsection{Spin from X-ray Polarization}
Currently, it is not possible to measure polarization in the soft X-ray band, where accretion disks dominate the spectrum of stellar-mass black holes.  However, considerable theoretical efforts have been undertaken to develop spin diagnostics based on polarization.  In particular, the closer an accretion disk extends to the black hole owing to black hole spin, the greater the fraction of ``returning radiation'', which is radiation that is gravitationally lensed back to the disk and which is polarized orthogonal to radiation emitted by the disk far from the black hole (\cite{2009ApJ...701.1175S}; see also \cite{2009ApJ...703..569D}).  Thus spin has an energy-dependent effect on both the degree of polarization and the observed polarization angle.  

Because soft X-ray polarization has not yet been measured it is not clear how much of the non-idealities of disk and jet emission that we discussed earlier will cloud inferences about the spin magnitude and direction.  There is, however, plausible optimism that polarization will at least provide complementary measurements, and in the best case it might be free of some of the systematic issues of current spin estimations.  For some references to the possibility of polarization measurements constraining spin and inclination, see \cite{1980ApJ...235..224C,2009ApJ...691..847L}.  There are also some promising suggestions that polarization observations may be able to constrain coronal properties \cite{2010ApJ...712..908S} and that, at least in currently explored models, such measurements in the thermal disk state require no a priori knowledge of the spin, inclination, mass, or distance (\cite{2009ApJ...701.1175S}, although the previous caveats about disk physics complexity must be kept in mind).  Proposed future polarization missions include {\it X-Calibur} \cite{2011APh....34..550K,2011SPIE.8145E.240B} and {\it GEMS} \cite{2010SPIE.7732E..24J,2013arXiv1301.1957S}.

\subsection{Theoretical Expectations for Spins}

Massive black holes can grow in mass and spin through sustained disk
accretion.  However, if accretion is chaotic -- if some accretion
episodes precipitate a retrograde disk -- then accretion
onto massive black holes may not drive them to high spin parameters.
Moreover, massive black holes can grow through mergers with other
black holes; successive additions of random spin vectors will tend to
spin--down the resultant black hole (see, e.g., \cite{2008ApJ...684..822B}).  In Seyfert AGN, at least, the
observed distribution of black hole spins is skewed to high, positive
values, and favors black hole growth through periods of sustained disk
accretion.  Spin measurements are even more sparse among quasars, but
early measurements again point to high spin values and sustained disk
accretion as an important mode of mass growth \cite{2013SSRv..tmp...81R,2013CQGra..30x4004R,2014Natur.507..207R}.

The situation is markedly different for stellar-mass black holes.  As we discussed earlier, the spin of stellar-mass black holes is largely determined at
  birth.  Thus numerous factors may be important in predicting the spin distribution
of stellar-mass black holes, including the mass of the progenitor star, its
spin rate, and its interior structure.  In
particular, different outcomes may result from a very massive star
where magnetic fields have worked to produce nearly solid-body
rotation, versus a scenario where differential rotation is still
possible.  This problem has been examined
in the context of the means by which massive black holes may arise in
galactic centers: the collapse of a single super-massive star.  This situation
may not be directly analogous to the collapse of a normal star.
However, some models suggest that ${\hat a} \sim 0.7$ could typically emerge from core collapse, in contrast to the spin
of ${\hat a} \sim 0.9$ that is predicted over long times due to accretion, as in supermassive black holes \cite{2004ApJ...602..312G}.  

Recently, the distribution of spin parameters in stellar-mass black
holes has been examined with two purposes: to test the predictions of
models, and to compare that distribution to natal
spins inferred for neutron stars as a window into their birth events.
The mean spin measured via reflection was found to be ${\hat a} = 0.66$, and
the mean spin measured via continuum fitting was found to be ${\hat a} =
0.72$ \cite{2011ApJ...731L...5M}.  Not only are these mean values in
good agreement, but so too are the distributions.  Although there are
isolated cases of discrepant measurements, it is likely that both
methods are making solid physical measurements.  Moreover, these spin
values differ markedly from natal spins inferred from neutron stars,
and suggest that black holes may preferentially form in SNe events
with enough angular momentum to drive strong MHD jets, causing them to
appear as GRBs \cite{2011ApJ...731L...5M}.

\begin{figure}[!htb]
\begin{center}
\includegraphics[scale=0.6]{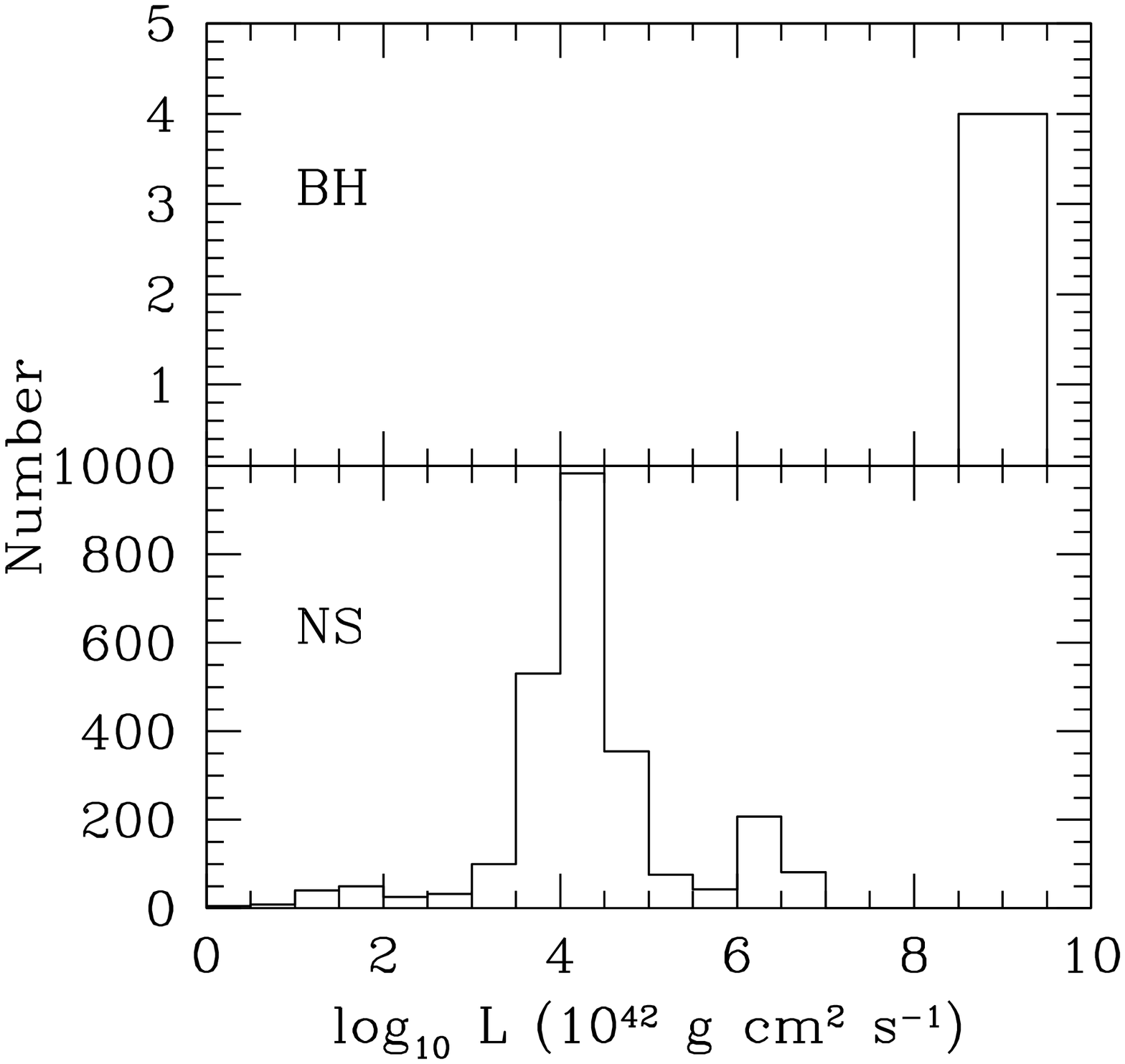}
\caption{\footnotesize Comparison of the spin angular momenta of neutron
stars and black holes.  The key point is that even the fastest-spinning
neutron stars have angular momenta two orders of magnitude less than the
currently estimated angular momenta of black holes.  This implies that
there is a significant difference in the formation process; fallback
from the pre-supernova star for black holes is one candidate.  Here we
assume a moment of inertia of $1.5\times 10^{45}$~g~cm$^2$~s$^{-1}$ for
all the neutron stars (reasonable for a $\sim 1.5~M_\odot$ star; see
\cite{2013PhRvD..88b3009Y}).  For the black holes, we take the spin
parameters for the black holes listed in Table~\ref{tab:bhspin}
(averaging the disk reflection and disk continuum estimates if both
are available) that also have mass estimates.  This leaves us with
eight sources: M33~X--7 (mass from \cite{2007Natur.449..872O}),
LMC~X--1 \cite{2009ApJ...697..573O}, A0620--00 \cite{2010ApJ...710.1127C},
4U~1543--475 \cite{2003IAUS..212..365O}, XTE~J1550--654 \cite{2011ApJ...730...75O}, GRO~J1644--40 \cite{2001ApJ...554.1290G}, GRS~1915+105 \cite{2013MNRAS.430.1832H}, and Cyg~X--1 \cite{2011ApJ...742...84O}.
}
\label{fig:angmom}
\end{center}
\end{figure}

As we see from Figure~\ref{fig:angmom}, the currently known angular momenta of black holes are at least two orders of magnitude larger than the angular momenta of neutron stars, even those spinning within a factor of 2--3 of breakup.  This strongly suggests that black holes acquire vastly larger angular momenta at birth than do neutron stars.  Fallback from the pre-supernova star is one clear possibility for the source of this extra angular momentum.

\begin{table}[htb!]
\title{Stellar-mass Black Hole Spin Measurements}
\begin{footnotesize}
\begin{center}
\begin{tabular}{llll}
Object (by RA) & Disk Reflection & Disk Continuum & References \\
\hline
M33 X-7 &  ~                  & $a=0.84\pm 0.05$ & \cite{2008ApJ...679L..37L,2010ApJ...719L.109L} \\
LMC X-3 & & $0.21^{+0.18}_{-0.22}$ & \cite{2006ApJ...647..525D,2014arXiv1402.0148S} \\
LMC X-1 & $0.97^{+0.02}_{-0.13}$ & $0.92^{+0.05}_{-0.07}$ & \cite{2012MNRAS.427.2552S,2009ApJ...701.1076G} \\
A 0620$-$00 & ~               & $0.12^{+0.19}_{-0.19}$ & \cite{2010ApJ...718L.122G} \\
GS 1124$-$683 & ~             & $-0.24^{+0.05}_{-0.64}$ & \cite{2014ApJ...784L..18M} \\
4U 1543$-$475 & $0.3\pm 0.1$ &  $0.8\pm 0.1$ & \cite{2009ApJ...697..900M}, \cite{2006ApJ...636L.113S} \\
XTE J1550$-$654 & $0.55\pm 0.22$ & $0.34^{+0.37}_{-0.45}$ & \cite{2009ApJ...697..900M,2011MNRAS.416..941S} \\
4U 1630$-$472 & $0.985^{+0.005}_{-0.014}$ & ~             & \cite{2014ApJ...784L...2K} \\
XTE J1650$-$500 & $0.79\pm 0.01$ & ~                  & \cite{2002ApJ...570L..69M,2009ApJ...697..900M} \\
XTE J1652$-$453 & $0.45\pm 0.02$ & ~                  & \cite{2011MNRAS.411..137H} \\
GRO J1655$-$40  & $0.98\pm 0.01$ & $0.7\pm 0.1$       & \cite{2009ApJ...697..900M,2006ApJ...636L.113S} \\
GX 339$-$4 & $0.94\pm 0.02$ & ~                       & \cite{2009ApJ...697..900M} \\
SAX J1711.6$-$3608 & $0.6^{+0.2}_{-0.4}$ & ~             & \cite{2009ApJ...697..900M} \\
XTE J1752$-$223 & $0.52\pm 0.11$ & ~                  & \cite{2011MNRAS.410.2497R} \\
Swift J1753.5$-$0127 & $0.76^{+0.11}_{-0.15}$ & ~        & \cite{2009MNRAS.395.1257R} \\
MAXI J1836$-$194 & $0.88\pm 0.03$ & ~                 & \cite{2012ApJ...751...34R} \\
XTE J1908$+$094 & $0.75\pm 0.09$ & ~                  & \cite{2009MNRAS.395.1257R} \\
Swift J1910.2$-$0546 & $\leq -0.32$ & ~                & \cite{2013ApJ...778..155R} \\
GRS 1915$+$105 & $0.98\pm 0.01$ & $\geq 0.95$         & \cite{2013ApJ...775L..45M,2006ApJ...652..518M} \\
Cygnus X-1 & $0.97^{+0.01}_{-0.02}$ & $\geq 0.95$        & \cite{2012MNRAS.424..217F,2011ApJ...742...85G} \\
\hline
\end{tabular}
\end{center} 
\caption{The errors in the table above generally reflect $1\sigma$
  statistical errors.  Systematic errors are likely to be larger, and
  due to a combination of modeling uncertainties (e.g. the ability of
  the data to constrain the emissivity for reflection--based
  measurements, or the degree and energy at which an underlying hard
  component turns-over at low energy when modeling the disk continuum)
  and theoretical uncertainties (the degree to which a real fluid disk
  respects the test particle ISCO).  The reflection--based spin
  measurement quoted for XTE J1550$-$564 is a composite value based on
  two recent complementary works; other values in this table are based
  on single measurements from the best available current research.
\label{tab:bhspin}
}
\vspace{-1.0\baselineskip}
\end{footnotesize}
\end{table}

\section{Gravitational Waves}
\label{section:gw}

The era of direct gravitational wave detections is expected to commence in a few years, when second-generation ground-based detectors such as Advanced LIGO \cite{2010CQGra..27h4006H}, Advanced Virgo \cite{2011CQGra..28k4002A}, KAGRA \cite{2012CQGra..29l4007S}, and LIGO-India \cite{2013IJMPD..2241010U} reach their full sensitivities.  Depending on the optimism of the modelers, predicted rates of detected compact object coalescence range from a few tenths per year to a few hundred per year \cite{2010CQGra..27q3001A}.  These events, which include the analogs of the known double neutron star binaries as well as currently unknown but likely neutron star -- black hole binaries and double black hole binaries, will open a new window onto compact objects.  For our purposes, such detections will provide neutron star and black hole mass and spin estimates that will at a minimum have different systematics than existing determinations and at best will provide cleaner measurements than we have today.  In this section we discuss how masses and spins can be obtained from gravitational waveforms, and the implications of the expected greatly expanded sample of well-measured compact objects.  We will begin by considering nonspinning systems, then consider the effects of spin.

\subsection{Mass measurements from gravitational waves}

Gravitational radiation is analogous to electromagnetic radiation except that the equivalents of electric dipole and magnetic dipole radiation vanish.  This is because the mass-energy version of electric dipole radiation is eliminated due to linear momentum conservation, and the equivalent of magnetic dipole radiation vanishes due to angular momentum conservation.  Thus the lowest order radiation comes from variation of the mass-energy quadrupole.  Binary systems naturally have time-variable quadrupoles, so they are potentially strong sources of gravitational radiation.  

A classic study of gravitational waves from point-mass binaries was completed by Peters \cite{1964PhRv..136.1224P} to lowest order in the motion.  Effectively, he assumed the adiabatic limit, in which the radiation takes away only a small fraction of the orbital angular momentum and energy in an orbit, so he was able to integrate the energy and angular momentum losses in one orbit of a Kepler ellipse.  He found that the semimajor axis and eccentricity evolve as
\begin{equation}
\begin{array}{rl}
{da\over{dt}}&=-{64\over 5}{G^3m_1m_2(m_1+m_2)\over{c^5a^3(1-e^2)^{7/2}}}
\left(1+73e^2/24+37e^4/96\right)\\
{de\over{dt}}&=-{304\over{15}}e{G^3m_1m_2(m_1+m_2)\over{c^5a^4(1-e^2)^{5/2}}}\left(1+121e^2/304\right)\; .\\
\end{array}
\end{equation}
Here $m_1$ and $m_2$ are the masses of the two objects.  Thus orbits circularize as they shrink.  This can be understood in the impulse approximation: the rate of gravitational radiation is a strong positive function of the relative speed of the objects, which is maximized at the pericenter.  In the limit that energy and angular momentum are only lost at the pericenter, the impulse approximation tells us that the pericenter distance is conserved while the apocenter distance shrinks, so the eccentricity decreases.

When we observe the waves from such a system we see frequencies, so it is convenient to use $\omega=(GM/a^3)^{1/2}$, where $M=m_1+m_2$ is the total mass of the system.  It is also standard to use the symmetric mass ratio $\eta\equiv m_1m_2/M^2$.  Then we find
\begin{equation}
\begin{array}{rl}
{d\omega\over{dt}}={96\over 5}\omega^{11/3}{\eta(GM)^{5/3}\over{c^5(1-e^2)^{7/2}}}\left(1+73e^2/24+37e^4/96\right)\\
{de\over{dt}}={-304\over{15}}e\omega^{8/3}{\eta(GM)^{5/3}\over{c^5(1-e^2)^{5/2}}}\left(1+121e^2/304\right)\; .\\
\end{array}
\end{equation}
We therefore see that the masses only appear in the combination $\eta(GM)^{5/3}$, or $(GM_{\rm ch})^{5/3}$, where the ``chirp mass" is $M_{\rm ch}\equiv \eta^{3/5}M$.  The chirp mass, named thus because it dictates the rate of frequency sweep, can be determined with superior accuracy ($<1$\%; e.g., \cite{2012PhRvD..85j4045V}) by essentially counting cycles.  It will be the most accurately determined parameter for binaries detected with gravitational radiation.

It is much more difficult to determine both masses separately, because different mass combinations only appear in higher-order, and thus much weaker, terms in the post-Newtonian expansion of the motion.  The challenge will be particularly great for double neutron star binaries, or any binary where the components have comparable masses.  The reason is that $\eta$ has a maximum (of 0.25) for equal-mass binaries, and thus has very little variation over a broad range of comparable masses; for example, $\eta=0.24$ for a mass ratio of $1.5:1$.  Thus over almost the entire range of plausible double neutron star mass ratios, $\eta$ is within a few percent of its maximum value.  Given that $\eta$ can be measured at all only due to weak higher-order effects, extremely high signal to noise ratios will be required to disentangle the component masses in a double neutron star binary.

On a somewhat more optimistic note, if one can be sure that a given system has two neutron stars instead of a neutron star and a black hole or two low-mass black holes, then measurement of the chirp mass provides a high-precision measurement of the total mass, and consequently a strong lower limit on the more massive object of the pair.  This is because $M=\eta^{-3/5}M_{\rm ch}$, so if $M_{\rm ch}$ is known then the additional uncertainty on $M$ is just $2-3\%$ larger for plausible neutron star mass ratios.  $M/2$ is then clearly a lower limit on the mass of the more massive neutron star.  Thus if a confirmed double neutron star system has a large enough chirp mass, it could raise the known mass upper limit for neutron stars.

The difficulty, of course, is in being sure that the system has two neutron stars.  If a coincident gamma-ray burst is seen then at least {\it one} of the objects must have been a neutron star.  Such a burst would argue against either object being a black hole with mass $>10~M_\odot$, because the neutron star would just be swallowed \cite{2005ApJ...626L..41M}, but recent fully general relativistic simulations of the inspiral of neutron stars into low-mass black holes, particularly when the hole spins rapidly and the orbit is prograde with respect to the hole, suggest that bursts or other electromagnetic signals might be possible \cite{2006PhRvD..73b4012F,2006PhRvD..73f4027S,2006PhRvD..74l1503S,2007CQGra..24S.125S,2008PhRvD..77h4002E,2008PhRvD..77h4015S,2008PhRvD..78j4015D,2009PhRvD..79d4024E,2010PASJ...62..315C,2010A&A...514A..66R,2010CQGra..27k4106D,2010PhRvL.105k1101C,2011PhRvD..83b4005F,2011ApJ...737L...5S,2011LRR....14....6S,2012PhRvD..85d4015F,2012PhRvD..85f4029E,2012PhRvD..85l4009E,2012PhRvD..86h4026E,2012PhRvD..86l4007F,2013PhRvD..88b1504P,2013PhRvD..88d1503K,2014ApJ...780...31T}.  More research is required.

Black hole -- neutron star binaries have component masses that differ more than in double neutron star binaries, so a given amount of uncertainty in $\eta$ does not translate to such enormous uncertainties in the individual masses.  However, the rate of such coalescences is highly uncertain because we do not know of any black hole -- neutron star examples in our galaxy, and in addition spin-orbit coupling and precession may greatly complicate the analysis of these systems.  We now discuss these effects.

\subsection{The effects of spin}

As we mentioned in Section~\ref{section:introduction}, the spin frequencies of known neutron stars extend up to 716~Hz \cite{2006Sci...311.1901H}, which could correspond to a dimensionless spin ${\hat a}\sim 0.2-0.3$ depending on the star's mass and radius.  However, the highest spin frequency of a neutron star in a binary with another neutron star is only $\sim 44$~Hz.  This is understandable in terms of the evolution of the system: our current understanding of the fastest-rotating neutron stars suggests that their rapid rotation is produced after their birth in supernovae, via long-term accretion from a companion that also somehow reduces their magnetic moments by orders of magnitude (see \cite{2008LRR....11....8L} for a review).  A companion to a neutron star that evolves into another neutron star has a short enough lifetime that it cannot transfer much mass \cite{1995ApJS..100..217I}, and thus such spinup will be difficult.  If the neutron star is paired with a black hole that evolved before the neutron star, then clearly there will be no spinup after the formation of the neutron star, meaning that the neutron stars in NS-BH binaries are likely to have ${\hat a}\ll 1$.  Such spins could affect the long-term phase evolution of the system, because spin-orbit coupling enters at the relatively low 1.5PN order \cite{2007LRR....10....2F}, but these spins will produce only minor effects at high frequencies.  Thus mass and radius inferences for double neutron star systems are unlikely to be affected much by spins.

In contrast, as we described in Section~\ref{section:spins}, current estimates of the spins of stellar-mass black holes suggest that for them ${\hat a}\sim 1$ is possible.  There are no known BH-NS or BH-BH binaries, so extrapolation from BH LMXBs and HMXBs is necessary, but if black hole spins in compact binaries are similarly high then they could have a considerable impact on the gravitational radiation waveforms and the precision with which the properties of the system can be inferred.  This is especially true if black hole spins in compact binaries can be significantly misaligned with the orbital axis (as we discuss in Section~\ref{section:spins}, some observations suggest that such misalignment might exist for LMXBs).  

To see this, consider a neutron star of mass $1.5~M_\odot$ spiraling into a $10~M_\odot$ black hole that has ${\hat a}=0.9$.  For such a system, the spin angular momentum of the black hole exceeds the orbital angular momentum of the neutron star if the orbital radius of the NS is less than $\sim 600$~km, or $\sim 35M$ in gravitational units.  Thus in an extreme case the neutron star orbit could be flipped by the spin of the black hole when the star is close enough.  The effect on the waveform of a comparable-mass BH-BH binary will be more modest, but misaligned spins could still induce considerable precession (see \cite{2008ApJ...688L..61V,2009CQGra..26t4010V,2010CQGra..27k4009R} for references for this and the following discussion).

Such precession can either help or hurt parameter determinations, depending on the details.  For example, compared to a spinless and thus nonprecessing binary, for which the orbital plane is fixed, degeneracies between, e.g., the sky direction and the normal to the orbit can be broken when the binary precesses.  This could help in the localization of an electromagnetic counterpart.  However, rapid spins can also change the overall phase evolution in ways that can be degenerate with a different chirp mass than the actual one.  If, contrary to expectations, there are rapidly-spinning neutron stars in compact binaries, their oblateness could complicate analysis of the high-frequency portion of the coalescence.  For the expected essentially nonspinning neutron stars in such systems, the recently elucidated relations between the moment of inertia, Love number, and quadrupole moment (the so-called I-Love-Q relations; see \cite{2013PhRvD..88b3009Y,2013Sci...341..365Y,2013PhRvD..88b3007M}) will help in the inference of neutron star masses and radii.

The phase space produced when spins are included is large enough that it has not yet been possible to do a comprehensive study.  Some preliminary results \cite{2008ApJ...688L..61V,2009CQGra..26t4010V,2010CQGra..27k4009R} suggest that the precision with which masses can be inferred will not be problematically reduced in systems with spins.  Thus the challenge might not be parameter estimation once a detection is made, but rather the detection itself, given the expanded library of waveforms that will be needed to accommodate spin.

\section{Conclusions}
\label{section:conclusions}

We have reviewed current estimates of the masses and spins of neutron stars and black holes, in isolation, and in binary systems with a variety of possible companions.  Taking a broad view of recent results and progress, it is increasingly clear that stellar-mass compact objects are vital laboratories for the exploration of strong gravity and nuclear physics.

Precise neutron star masses, which until recently were limited to the narrow $1.25~M_\odot-1.44~M_\odot$ range thus far observed from double neutron star systems, have now been extended to $2~M_\odot$ \cite{2010Natur.467.1081D,2013Sci...340..448A}.  This has immediate implications for the equation of state of cold supranuclear matter; in particular, it requires that this matter be relatively hard, whatever its precise composition.  Truly strong constraints on this matter await reliable neutron star radius estimates \cite{2013arXiv1312.0029M}, but the increasing precision of millisecond pulsar timing suggests that measurements of Shapiro delay and other post-Keplerian parameters will in the near future be possible for a broader range of neutron star systems.  If any of those systems have well-determined masses that are significantly higher than the current $2~M_\odot$ maximum then equations of state involving exotic non-nucleonic components will be strongly challenged \cite{2010arXiv1012.3208L}.

The masses of stellar black holes are not as easy to measure with precision, because they require a special set of circumstances (a transient system that fully shuts off its accretion to allow measurements of ellipsoidal light variations).  Sensitivity is not the limiting factor (at least in the Galaxy), nor is timing.  Rather, patience is the key: more sources in the Galaxy means waiting a greater time for currently inactive transients to reveal themselves.  Qualitative advances with current data may be possible if high-frequency QPOs are understood from first-principles simulations, so in this field we may primarily be waiting on theoretical developments.

An intriguing question is whether there is actually a mass gap between neutron stars and black holes.  Some initial analyses imply that such a gap exists (or at least that low-mass black holes are much less common than might be suggested by the initial mass function of main sequence stars), but more recent work has suggested that at least part of this result could be due to selection effects.  Whether or not there is a gap, an understanding of the true black hole and neutron star mass distributions at formation will give us key information about the evolution of massive stars and the nature of core collapse.

The current spin of any pulsar is by definition easy to measure.  It is much more challenging to determine the natal spin of such neutron stars.  Estimates have been made in a few cases based on, for example, the argument that an extremely rapid initial spin would produce a highly luminous nebula.  Such arguments suggest in a handful of cases that the initial spin periods are more like tenths of a second than milliseconds, but given that this is based on $\sim 10$ out of $\sim 2000$ pulsars we must be wary of selection effects.

Unlike neutron star spins (and masses), black hole spins are believed to essentially be natal, with one possible exception being that the spins of black holes in a hyperaccreting Thorne-\.{Z}ytkow phase might be altered significantly.  For the majority of black holes, therefore, if the current spin can be measured it serves as a record of the formation process.  Estimates of black hole spin are necessarily far more indirect than they are for neutron stars.  We have reviewed two different methods, one involving characterization of Fe~K$\alpha$ lines and the other using continuum spectral fits.  Both methods have strengths and weaknesses, but there is reason for cautious optimism because spin estimates for the overlap sources are in fairly good agreement.  Other methods, e.g., those involving QPO modeling or polarization, will require significant theoretical or instrumental development to be reliable, but when this happens it will produce an independent channel of information.  In any case, current estimates of black hole angular momenta indicate that most are far beyond any possible angular momentum for neutron stars, which has implications for the rotation distribution of their progenitor stars.

Measurements of black hole spin also provide an avenue by which
relativistic jet launching mechanisms can be explored.  The
Blandford-Znajek effect is a particularly appealing jet-launching
mechanism that taps the spin of the black hole
\cite{1977MNRAS.179..433B}; if it is correct, then proxies for jet
power should be positively correlated with spin values.  Based on very
few sources, correlations between proxies for jet power and black
hole spin have been claimed (\cite{2012MNRAS.419L..69N};
however, other work suggests that the apparent relation is an
artifact of selection effects (\cite{2013MNRAS.431..405R}).  Deeper
appreciation of the mechanism may not even predict a simple
relationship between spin and a jet proxy like radio flux.  Spin
might set the maximum possible power that a jet can carry; however,
instantaneous jet power is likely determined by how the disk and
ergosphere couple, and this must be a function of several accretion
parameters that affect the magnetic field.  That is, spin may set
the maximum horsepower of the engine, but the accretion flow is a
throttle (see, e.g., \cite{2013ApJ...771...84K}). A large collection
of jet proxy measurements for a collection of sources with known
spins might be expected to fill a large portion of a $P_{jet}$
versus $a$ plane, with evidence of spin driving jets manifesting as
an upper envelope.

An exciting near-future prospect is the entry of gravitational wave observations into the mass and spin constraint game.  For the stellar-mass objects we consider, this will be the domain of ground-based detectors rather than the equally exciting developments in, e.g., pulsar timing arrays.  The initial detections of binary compact object coalescence will measure precisely only the chirp masses of the binaries, rather than the individual masses.  However, as sensitivity (particularly at high frequencies) improves, detection of higher-order effects will allow the measurements of individual masses.  This will clarify whether, for example, the current tight range of neutron star masses in double neutron star systems is maintained (perhaps due to a narrow evolutionary channel for such systems) or whether occasional much higher-mass neutron stars will be seen.  Spin measurements of neutron stars will be challenging if they rotate as slowly as the stars in currently known double neutron star systems.  There is reason to be optimistic that black hole spins will be easier to measure, and it will be especially interesting to determine the distribution of relative spin orientations of binary black holes.  This is because different origins (evolution from a massive binary, versus dynamical capture in a dense stellar environment) imply very different distributions of the relative orientations.

Finally, it is important to emphasize that electromagnetic and gravitational wave measurements of neutron stars and black holes will provide strongly complementary constraints on the nature of strong gravity.  The detection of gravitational radiation, particularly of a double black hole coalescence, will yield direct constraints on the nature of strong gravity, and in particular whether general relativity can pass such clean tests as it has passed the weak gravity tests available from binary pulsars and in our Solar System.  However, such observations cannot by themselves test our theories of how matter and electromagnetic fields interact in strong gravity.  This information can only come from current and future electromagnetic observations of compact objects, especially those that are accreting.  For example, the excellent fits to current iron line and continuum observations both rely on the existence of an innermost stable circular orbit, which is a key prediction of models of strong gravity.  Another example is that if we can detect a pulsar in a binary around a black hole, this will produce new levels of precision in black hole mass and possibly spin measurements as well as yielding new types of tests of the predictions of general relativity in weak gravity.

In general, the theoretical framework for and observations of compact object masses and spins have evolved rapidly.  The extreme nature of the gravity and material properties of these systems mean that they are crucial testbeds for some of our most fundamental physical theories.  We hope for and expect continued qualitative improvements in our understanding, and we look forward to the unanticipated in our study of neutron stars and black holes.

\bigskip


\section{Acknowledgements}
\label{section:acknowledgements}

We thank Charles Bailyn, Didier Barret, Ed Brown, Ed Cackett, Shane Davis, Jason Dexter, Zach Etienne, Julian Krolik, Alessandro Patruno, Chris Reynolds, and Jeremy Schnittman for helpful discussions and comments on an early version of this manuscript.  We give special thanks to our editor, Marc Kamionkowski, for proposing this review and for many helpful suggestions.  MCM was supported in part by NASA ATP grant NNX12AG29G.  This work was also supported in part by National Science Foundation Grant No. PHYS-1066293 and the hospitality of the Aspen Center for Physics. 

\newpage

\bibliography{PhysRep}

\begin{thebibliography}{399}
\expandafter\ifx\csname natexlab\endcsname\relax\def\natexlab#1{#1}\fi
\providecommand{\bibinfo}[2]{#2}
\ifx\xfnm\relax \def\xfnm[#1]{\unskip,\space#1}\fi
\bibitem[{{Hewish} et~al.(1968){Hewish}, {Bell}, {Pilkington}, {Scott}, and
  {Collins}}]{1968Natur.217..709H}
\bibinfo{author}{A.~{Hewish}}, \bibinfo{author}{S.~J. {Bell}},
  \bibinfo{author}{J.~D.~H. {Pilkington}}, \bibinfo{author}{P.~F. {Scott}},
  \bibinfo{author}{R.~A. {Collins}},
\newblock \bibinfo{title}{{Observation of a Rapidly Pulsating Radio Source}},
\newblock \bibinfo{journal}{\nat} \bibinfo{volume}{217} (\bibinfo{year}{1968})
  \bibinfo{pages}{709--713}.
\bibitem[{{Salpeter}(1964)}]{1964ApJ...140..796S}
\bibinfo{author}{E.~E. {Salpeter}},
\newblock \bibinfo{title}{{Accretion of Interstellar Matter by Massive
  Objects.}},
\newblock \bibinfo{journal}{\apj} \bibinfo{volume}{140} (\bibinfo{year}{1964})
  \bibinfo{pages}{796--800}.
\bibitem[{{Lynden-Bell}(1969)}]{1969Natur.223..690L}
\bibinfo{author}{D.~{Lynden-Bell}},
\newblock \bibinfo{title}{{Galactic Nuclei as Collapsed Old Quasars}},
\newblock \bibinfo{journal}{\nat} \bibinfo{volume}{223} (\bibinfo{year}{1969})
  \bibinfo{pages}{690--694}.
\bibitem[{{Bowyer} et~al.(1965){Bowyer}, {Byram}, {Chubb}, and
  {Friedman}}]{1965Sci...147..394B}
\bibinfo{author}{S.~{Bowyer}}, \bibinfo{author}{E.~T. {Byram}},
  \bibinfo{author}{T.~A. {Chubb}}, \bibinfo{author}{H.~{Friedman}},
\newblock \bibinfo{title}{{Cosmic X-ray Sources}},
\newblock \bibinfo{journal}{Science} \bibinfo{volume}{147}
  (\bibinfo{year}{1965}) \bibinfo{pages}{394--398}.
\bibitem[{{Webster} and {Murdin}(1972)}]{1972Natur.235...37W}
\bibinfo{author}{B.~L. {Webster}}, \bibinfo{author}{P.~{Murdin}},
\newblock \bibinfo{title}{{Cygnus X-1-a Spectroscopic Binary with a Heavy
  Companion ?}},
\newblock \bibinfo{journal}{\nat} \bibinfo{volume}{235} (\bibinfo{year}{1972})
  \bibinfo{pages}{37--38}.
\bibitem[{{Bolton}(1972)}]{1972Natur.235..271B}
\bibinfo{author}{C.~T. {Bolton}},
\newblock \bibinfo{title}{{Identification of Cygnus X-1 with HDE 226868}},
\newblock \bibinfo{journal}{\nat} \bibinfo{volume}{235} (\bibinfo{year}{1972})
  \bibinfo{pages}{271--273}.
\bibitem[{{Tananbaum} et~al.(1972){Tananbaum}, {Gursky}, {Kellogg}, {Giacconi},
  and {Jones}}]{1972ApJ...177L...5T}
\bibinfo{author}{H.~{Tananbaum}}, \bibinfo{author}{H.~{Gursky}},
  \bibinfo{author}{E.~{Kellogg}}, \bibinfo{author}{R.~{Giacconi}},
  \bibinfo{author}{C.~{Jones}},
\newblock \bibinfo{title}{{Observation of a Correlated X-Ray Transition in
  Cygnus X-1}},
\newblock \bibinfo{journal}{\apjl} \bibinfo{volume}{177} (\bibinfo{year}{1972})
  \bibinfo{pages}{L5}.
\bibitem[{{Lorimer}(2008)}]{2008LRR....11....8L}
\bibinfo{author}{D.~R. {Lorimer}},
\newblock \bibinfo{title}{{Binary and Millisecond Pulsars}},
\newblock \bibinfo{journal}{Living Reviews in Relativity} \bibinfo{volume}{11}
  (\bibinfo{year}{2008}) \bibinfo{pages}{8}.
\bibitem[{{Kreidberg} et~al.(2012){Kreidberg}, {Bailyn}, {Farr}, and
  {Kalogera}}]{2012ApJ...757...36K}
\bibinfo{author}{L.~{Kreidberg}}, \bibinfo{author}{C.~D. {Bailyn}},
  \bibinfo{author}{W.~M. {Farr}}, \bibinfo{author}{V.~{Kalogera}},
\newblock \bibinfo{title}{{Mass Measurements of Black Holes in X-Ray
  Transients: Is There a Mass Gap?}},
\newblock \bibinfo{journal}{\apj} \bibinfo{volume}{757} (\bibinfo{year}{2012})
  \bibinfo{pages}{36}.
\bibitem[{{Reynolds} and {Nowak}(2003)}]{2003PhR...377..389R}
\bibinfo{author}{C.~S. {Reynolds}}, \bibinfo{author}{M.~A. {Nowak}},
\newblock \bibinfo{title}{{Fluorescent iron lines as a probe of astrophysical
  black hole systems}},
\newblock \bibinfo{journal}{\physrep} \bibinfo{volume}{377}
  (\bibinfo{year}{2003}) \bibinfo{pages}{389--466}.
\bibitem[{{McClintock} et~al.(2011){McClintock}, {Narayan}, {Davis}, {Gou},
  {Kulkarni}, {Orosz}, {Penna}, {Remillard}, and
  {Steiner}}]{2011CQGra..28k4009M}
\bibinfo{author}{J.~E. {McClintock}}, \bibinfo{author}{R.~{Narayan}},
  \bibinfo{author}{S.~W. {Davis}}, \bibinfo{author}{L.~{Gou}},
  \bibinfo{author}{A.~{Kulkarni}}, \bibinfo{author}{J.~A. {Orosz}},
  \bibinfo{author}{R.~F. {Penna}}, \bibinfo{author}{R.~A. {Remillard}},
  \bibinfo{author}{J.~F. {Steiner}},
\newblock \bibinfo{title}{{Measuring the spins of accreting black holes}},
\newblock \bibinfo{journal}{Classical and Quantum Gravity} \bibinfo{volume}{28}
  (\bibinfo{year}{2011}) \bibinfo{pages}{114009}.
\bibitem[{{Blanchet}(2014)}]{2014LRR....17....2B}
\bibinfo{author}{L.~{Blanchet}},
\newblock \bibinfo{title}{{Gravitational Radiation from Post-Newtonian Sources
  and Inspiralling Compact Binaries}},
\newblock \bibinfo{journal}{Living Reviews in Relativity} \bibinfo{volume}{17}
  (\bibinfo{year}{2014}) \bibinfo{pages}{2}.
\bibitem[{{Sathyaprakash} and {Schutz}(2009)}]{2009LRR....12....2S}
\bibinfo{author}{B.~S. {Sathyaprakash}}, \bibinfo{author}{B.~F. {Schutz}},
\newblock \bibinfo{title}{{Physics, Astrophysics and Cosmology with
  Gravitational Waves}},
\newblock \bibinfo{journal}{Living Reviews in Relativity} \bibinfo{volume}{12}
  (\bibinfo{year}{2009}) \bibinfo{pages}{2}.
\bibitem[{{Psaltis}(2008)}]{2008LRR....11....9P}
\bibinfo{author}{D.~{Psaltis}},
\newblock \bibinfo{title}{{Probes and Tests of Strong-Field Gravity with
  Observations in the Electromagnetic Spectrum}},
\newblock \bibinfo{journal}{Living Reviews in Relativity} \bibinfo{volume}{11}
  (\bibinfo{year}{2008}) \bibinfo{pages}{9}.
\bibitem[{{Lattimer}(2011)}]{2011Ap&SS.336...67L}
\bibinfo{author}{J.~M. {Lattimer}},
\newblock \bibinfo{title}{{Neutron stars and the dense matter equation of
  state}},
\newblock \bibinfo{journal}{\apss} \bibinfo{volume}{336} (\bibinfo{year}{2011})
  \bibinfo{pages}{67--74}.
\bibitem[{{Miller}(2013)}]{2013arXiv1312.0029M}
\bibinfo{author}{M.~C. {Miller}},
\newblock \bibinfo{title}{{Astrophysical Constraints on Dense Matter in Neutron
  Stars}},
\newblock \bibinfo{journal}{arXiv:1312.0029}  (\bibinfo{year}{2013}).
\bibitem[{{McNamara} et~al.(2005){McNamara}, {Nulsen}, {Wise}, {Rafferty},
  {Carilli}, {Sarazin}, and {Blanton}}]{2005Natur.433...45M}
\bibinfo{author}{B.~R. {McNamara}}, \bibinfo{author}{P.~E.~J. {Nulsen}},
  \bibinfo{author}{M.~W. {Wise}}, \bibinfo{author}{D.~A. {Rafferty}},
  \bibinfo{author}{C.~{Carilli}}, \bibinfo{author}{C.~L. {Sarazin}},
  \bibinfo{author}{E.~L. {Blanton}},
\newblock \bibinfo{title}{{The heating of gas in a galaxy cluster by X-ray
  cavities and large-scale shock fronts}},
\newblock \bibinfo{journal}{\nat} \bibinfo{volume}{433} (\bibinfo{year}{2005})
  \bibinfo{pages}{45--47}.
\bibitem[{{Kulkarni} et~al.(1993){Kulkarni}, {Hut}, and
  {McMillan}}]{1993Natur.364..421K}
\bibinfo{author}{S.~R. {Kulkarni}}, \bibinfo{author}{P.~{Hut}},
  \bibinfo{author}{S.~{McMillan}},
\newblock \bibinfo{title}{{Stellar black holes in globular clusters}},
\newblock \bibinfo{journal}{\nat} \bibinfo{volume}{364} (\bibinfo{year}{1993})
  \bibinfo{pages}{421--423}.
\bibitem[{{Sigurdsson} and {Hernquist}(1993)}]{1993Natur.364..423S}
\bibinfo{author}{S.~{Sigurdsson}}, \bibinfo{author}{L.~{Hernquist}},
\newblock \bibinfo{title}{{Primordial black holes in globular clusters}},
\newblock \bibinfo{journal}{\nat} \bibinfo{volume}{364} (\bibinfo{year}{1993})
  \bibinfo{pages}{423--425}.
\bibitem[{{Portegies Zwart} and {McMillan}(2000)}]{2000ApJ...528L..17P}
\bibinfo{author}{S.~F. {Portegies Zwart}}, \bibinfo{author}{S.~L.~W.
  {McMillan}},
\newblock \bibinfo{title}{{Black Hole Mergers in the Universe}},
\newblock \bibinfo{journal}{\apjl} \bibinfo{volume}{528} (\bibinfo{year}{2000})
  \bibinfo{pages}{L17--L20}.
\bibitem[{{O'Leary} et~al.(2006){O'Leary}, {Rasio}, {Fregeau}, {Ivanova}, and
  {O'Shaughnessy}}]{2006ApJ...637..937O}
\bibinfo{author}{R.~M. {O'Leary}}, \bibinfo{author}{F.~A. {Rasio}},
  \bibinfo{author}{J.~M. {Fregeau}}, \bibinfo{author}{N.~{Ivanova}},
  \bibinfo{author}{R.~{O'Shaughnessy}},
\newblock \bibinfo{title}{{Binary Mergers and Growth of Black Holes in Dense
  Star Clusters}},
\newblock \bibinfo{journal}{\apj} \bibinfo{volume}{637} (\bibinfo{year}{2006})
  \bibinfo{pages}{937--951}.
\bibitem[{{Banerjee} et~al.(2010){Banerjee}, {Baumgardt}, and
  {Kroupa}}]{2010MNRAS.402..371B}
\bibinfo{author}{S.~{Banerjee}}, \bibinfo{author}{H.~{Baumgardt}},
  \bibinfo{author}{P.~{Kroupa}},
\newblock \bibinfo{title}{{Stellar-mass black holes in star clusters:
  implications for gravitational wave radiation}},
\newblock \bibinfo{journal}{\mnras} \bibinfo{volume}{402}
  (\bibinfo{year}{2010}) \bibinfo{pages}{371--380}.
\bibitem[{{Maccarone} et~al.(2007){Maccarone}, {Kundu}, {Zepf}, and
  {Rhode}}]{2007Natur.445..183M}
\bibinfo{author}{T.~J. {Maccarone}}, \bibinfo{author}{A.~{Kundu}},
  \bibinfo{author}{S.~E. {Zepf}}, \bibinfo{author}{K.~L. {Rhode}},
\newblock \bibinfo{title}{{A black hole in a globular cluster}},
\newblock \bibinfo{journal}{\nat} \bibinfo{volume}{445} (\bibinfo{year}{2007})
  \bibinfo{pages}{183--185}.
\bibitem[{{Brassington} et~al.(2010){Brassington}, {Fabbiano}, {Blake},
  {Zezas}, {Angelini}, {Davies}, {Gallagher}, {Kalogera}, {Kim}, {King},
  {Kundu}, {Trinchieri}, and {Zepf}}]{2010ApJ...725.1805B}
\bibinfo{author}{N.~J. {Brassington}}, \bibinfo{author}{G.~{Fabbiano}},
  \bibinfo{author}{S.~{Blake}}, \bibinfo{author}{A.~{Zezas}},
  \bibinfo{author}{L.~{Angelini}}, \bibinfo{author}{R.~L. {Davies}},
  \bibinfo{author}{J.~{Gallagher}}, \bibinfo{author}{V.~{Kalogera}},
  \bibinfo{author}{D.-W. {Kim}}, \bibinfo{author}{A.~R. {King}},
  \bibinfo{author}{A.~{Kundu}}, \bibinfo{author}{G.~{Trinchieri}},
  \bibinfo{author}{S.~{Zepf}},
\newblock \bibinfo{title}{{The X-ray Spectra of the Luminous LMXBs in NGC 3379:
  Field and Globular Cluster Sources}},
\newblock \bibinfo{journal}{\apj} \bibinfo{volume}{725} (\bibinfo{year}{2010})
  \bibinfo{pages}{1805--1823}.
\bibitem[{{Shih} et~al.(2010){Shih}, {Kundu}, {Maccarone}, {Zepf}, and
  {Joseph}}]{2010ApJ...721..323S}
\bibinfo{author}{I.~C. {Shih}}, \bibinfo{author}{A.~{Kundu}},
  \bibinfo{author}{T.~J. {Maccarone}}, \bibinfo{author}{S.~E. {Zepf}},
  \bibinfo{author}{T.~D. {Joseph}},
\newblock \bibinfo{title}{{A Variable Black Hole X-ray Source in an NGC 1399
  Globular Cluster}},
\newblock \bibinfo{journal}{\apj} \bibinfo{volume}{721} (\bibinfo{year}{2010})
  \bibinfo{pages}{323--328}.
\bibitem[{{Barnard} et~al.(2011){Barnard}, {Garcia}, {Li}, {Primini}, and
  {Murray}}]{2011ApJ...734...79B}
\bibinfo{author}{R.~{Barnard}}, \bibinfo{author}{M.~{Garcia}},
  \bibinfo{author}{Z.~{Li}}, \bibinfo{author}{F.~{Primini}},
  \bibinfo{author}{S.~S. {Murray}},
\newblock \bibinfo{title}{{Four New Black Hole Candidates Identified in M31
  Globular Clusters with Chandra and XMM-Newton}},
\newblock \bibinfo{journal}{\apj} \bibinfo{volume}{734} (\bibinfo{year}{2011})
  \bibinfo{pages}{79}.
\bibitem[{{Maccarone} et~al.(2011){Maccarone}, {Kundu}, {Zepf}, and
  {Rhode}}]{2011MNRAS.410.1655M}
\bibinfo{author}{T.~J. {Maccarone}}, \bibinfo{author}{A.~{Kundu}},
  \bibinfo{author}{S.~E. {Zepf}}, \bibinfo{author}{K.~L. {Rhode}},
\newblock \bibinfo{title}{{A new globular cluster black hole in NGC 4472}},
\newblock \bibinfo{journal}{\mnras} \bibinfo{volume}{410}
  (\bibinfo{year}{2011}) \bibinfo{pages}{1655--1659}.
\bibitem[{{Strader} et~al.(2012){Strader}, {Chomiuk}, {Maccarone},
  {Miller-Jones}, and {Seth}}]{2012Natur.490...71S}
\bibinfo{author}{J.~{Strader}}, \bibinfo{author}{L.~{Chomiuk}},
  \bibinfo{author}{T.~J. {Maccarone}}, \bibinfo{author}{J.~C.~A.
  {Miller-Jones}}, \bibinfo{author}{A.~C. {Seth}},
\newblock \bibinfo{title}{{Two stellar-mass black holes in the globular cluster
  M22}},
\newblock \bibinfo{journal}{\nat} \bibinfo{volume}{490} (\bibinfo{year}{2012})
  \bibinfo{pages}{71--73}.
\bibitem[{{Morscher} et~al.(2013){Morscher}, {Umbreit}, {Farr}, and
  {Rasio}}]{2013ApJ...763L..15M}
\bibinfo{author}{M.~{Morscher}}, \bibinfo{author}{S.~{Umbreit}},
  \bibinfo{author}{W.~M. {Farr}}, \bibinfo{author}{F.~A. {Rasio}},
\newblock \bibinfo{title}{{Retention of Stellar-mass Black Holes in Globular
  Clusters}},
\newblock \bibinfo{journal}{\apjl} \bibinfo{volume}{763} (\bibinfo{year}{2013})
  \bibinfo{pages}{L15}.
\bibitem[{{Breen} and {Heggie}(2013)}]{2013MNRAS.432.2779B}
\bibinfo{author}{P.~G. {Breen}}, \bibinfo{author}{D.~C. {Heggie}},
\newblock \bibinfo{title}{{Dynamical evolution of black hole subsystems in
  idealized star clusters}},
\newblock \bibinfo{journal}{\mnras} \bibinfo{volume}{432}
  (\bibinfo{year}{2013}) \bibinfo{pages}{2779--2797}.
\bibitem[{{Trenti} and {van der Marel}(2013)}]{2013MNRAS.435.3272T}
\bibinfo{author}{M.~{Trenti}}, \bibinfo{author}{R.~{van der Marel}},
\newblock \bibinfo{title}{{No energy equipartition in globular clusters}},
\newblock \bibinfo{journal}{\mnras} \bibinfo{volume}{435}
  (\bibinfo{year}{2013}) \bibinfo{pages}{3272--3282}.
\bibitem[{{Breen} and {Heggie}(2013)}]{2013MNRAS.436..584B}
\bibinfo{author}{P.~G. {Breen}}, \bibinfo{author}{D.~C. {Heggie}},
\newblock \bibinfo{title}{{On black hole subsystems in idealized nuclear star
  clusters}},
\newblock \bibinfo{journal}{\mnras} \bibinfo{volume}{436}
  (\bibinfo{year}{2013}) \bibinfo{pages}{584--589}.
\bibitem[{{Colgate} et~al.(1980){Colgate}, {Petschek}, and
  {Kriese}}]{1980AIPC...63....7C}
\bibinfo{author}{S.~A. {Colgate}}, \bibinfo{author}{A.~G. {Petschek}},
  \bibinfo{author}{J.~T. {Kriese}},
\newblock \bibinfo{title}{{The light curve of type I Supernovae}},
\newblock in: \bibinfo{editor}{R.~{Meyerott}}, \bibinfo{editor}{H.~G.
  {Gillespie}} (Eds.), \bibinfo{booktitle}{Supernovae Spectra},
  volume~\bibinfo{volume}{63} of \textit{\bibinfo{series}{American Institute of
  Physics Conference Series}}, pp. \bibinfo{pages}{7--14}.
\bibitem[{{Carr} and {Hawking}(1974)}]{1974MNRAS.168..399C}
\bibinfo{author}{B.~J. {Carr}}, \bibinfo{author}{S.~W. {Hawking}},
\newblock \bibinfo{title}{{Black holes in the early Universe}},
\newblock \bibinfo{journal}{\mnras} \bibinfo{volume}{168}
  (\bibinfo{year}{1974}) \bibinfo{pages}{399--416}.
\bibitem[{{Carr}(1975)}]{1975ApJ...201....1C}
\bibinfo{author}{B.~J. {Carr}},
\newblock \bibinfo{title}{{The primordial black hole mass spectrum}},
\newblock \bibinfo{journal}{\apj} \bibinfo{volume}{201} (\bibinfo{year}{1975})
  \bibinfo{pages}{1--19}.
\bibitem[{{Nadezhin} et~al.(1978){Nadezhin}, {Novikov}, and
  {Polnarev}}]{1978SvA....22..129N}
\bibinfo{author}{D.~K. {Nadezhin}}, \bibinfo{author}{I.~D. {Novikov}},
  \bibinfo{author}{A.~G. {Polnarev}},
\newblock \bibinfo{title}{{The hydrodynamics of primordial black hole
  formation}},
\newblock \bibinfo{journal}{\sovast} \bibinfo{volume}{22}
  (\bibinfo{year}{1978}) \bibinfo{pages}{129--138}.
\bibitem[{{Bicknell} and {Henriksen}(1978)}]{1978ApJ...225..237B}
\bibinfo{author}{G.~V. {Bicknell}}, \bibinfo{author}{R.~N. {Henriksen}},
\newblock \bibinfo{title}{{Self-similar growth of primordial black holes. II -
  General sound speed}},
\newblock \bibinfo{journal}{\apj} \bibinfo{volume}{225} (\bibinfo{year}{1978})
  \bibinfo{pages}{237--251}.
\bibitem[{{Novikov} et~al.(1979){Novikov}, {Polnarev}, {Starobinskii}, and
  {Zeldovich}}]{1979A&A....80..104N}
\bibinfo{author}{I.~D. {Novikov}}, \bibinfo{author}{A.~G. {Polnarev}},
  \bibinfo{author}{A.~A. {Starobinskii}}, \bibinfo{author}{I.~B. {Zeldovich}},
\newblock \bibinfo{title}{{Primordial black holes}},
\newblock \bibinfo{journal}{\aap} \bibinfo{volume}{80} (\bibinfo{year}{1979})
  \bibinfo{pages}{104--109}.
\bibitem[{{Kodama} et~al.(1982){Kodama}, {Sasaki}, and
  {Sato}}]{1982PThPh..68.1979K}
\bibinfo{author}{H.~{Kodama}}, \bibinfo{author}{M.~{Sasaki}},
  \bibinfo{author}{K.~{Sato}},
\newblock \bibinfo{title}{{Abundance of Primordial Holes Produced by
  Cosmological First-Order Phase Transition}},
\newblock \bibinfo{journal}{Progress of Theoretical Physics}
  \bibinfo{volume}{68} (\bibinfo{year}{1982}) \bibinfo{pages}{1979--1998}.
\bibitem[{{Konoplich} et~al.(1998){Konoplich}, {Rubin}, {Sakharov}, and
  {Khlopov}}]{1998AstL...24..413K}
\bibinfo{author}{R.~V. {Konoplich}}, \bibinfo{author}{S.~G. {Rubin}},
  \bibinfo{author}{A.~S. {Sakharov}}, \bibinfo{author}{M.~Y. {Khlopov}},
\newblock \bibinfo{title}{{Formation of black holes in first-order phase
  transitions in the Universe}},
\newblock \bibinfo{journal}{Astronomy Letters} \bibinfo{volume}{24}
  (\bibinfo{year}{1998}) \bibinfo{pages}{413--417}.
\bibitem[{{Jedamzik} and {Niemeyer}(1999)}]{1999PhRvD..59l4014J}
\bibinfo{author}{K.~{Jedamzik}}, \bibinfo{author}{J.~C. {Niemeyer}},
\newblock \bibinfo{title}{{Primordial black hole formation during first-order
  phase transitions}},
\newblock \bibinfo{journal}{\prd} \bibinfo{volume}{59} (\bibinfo{year}{1999})
  \bibinfo{pages}{124014}.
\bibitem[{{Green} and {Liddle}(1999)}]{1999PhRvD..60f3509G}
\bibinfo{author}{A.~M. {Green}}, \bibinfo{author}{A.~R. {Liddle}},
\newblock \bibinfo{title}{{Critical collapse and the primordial black hole
  initial mass function}},
\newblock \bibinfo{journal}{\prd} \bibinfo{volume}{60} (\bibinfo{year}{1999})
  \bibinfo{pages}{063509}.
\bibitem[{{Lyne} et~al.(2004){Lyne}, {Burgay}, {Kramer}, {Possenti},
  {Manchester}, {Camilo}, {McLaughlin}, {Lorimer}, {D'Amico}, {Joshi},
  {Reynolds}, and {Freire}}]{2004Sci...303.1153L}
\bibinfo{author}{A.~G. {Lyne}}, \bibinfo{author}{M.~{Burgay}},
  \bibinfo{author}{M.~{Kramer}}, \bibinfo{author}{A.~{Possenti}},
  \bibinfo{author}{R.~N. {Manchester}}, \bibinfo{author}{F.~{Camilo}},
  \bibinfo{author}{M.~A. {McLaughlin}}, \bibinfo{author}{D.~R. {Lorimer}},
  \bibinfo{author}{N.~{D'Amico}}, \bibinfo{author}{B.~C. {Joshi}},
  \bibinfo{author}{J.~{Reynolds}}, \bibinfo{author}{P.~C.~C. {Freire}},
\newblock \bibinfo{title}{{A Double-Pulsar System: A Rare Laboratory for
  Relativistic Gravity and Plasma Physics}},
\newblock \bibinfo{journal}{Science} \bibinfo{volume}{303}
  (\bibinfo{year}{2004}) \bibinfo{pages}{1153--1157}.
\bibitem[{{Antoniadis} et~al.(2013){Antoniadis}, {Freire}, {Wex}, {Tauris},
  {Lynch}, {van Kerkwijk}, {Kramer}, {Bassa}, {Dhillon}, {Driebe}, {Hessels},
  {Kaspi}, {Kondratiev}, {Langer}, {Marsh}, {McLaughlin}, {Pennucci}, {Ransom},
  {Stairs}, {van Leeuwen}, {Verbiest}, and {Whelan}}]{2013Sci...340..448A}
\bibinfo{author}{J.~{Antoniadis}}, \bibinfo{author}{P.~C.~C. {Freire}},
  \bibinfo{author}{N.~{Wex}}, \bibinfo{author}{T.~M. {Tauris}},
  \bibinfo{author}{R.~S. {Lynch}}, \bibinfo{author}{M.~H. {van Kerkwijk}},
  \bibinfo{author}{M.~{Kramer}}, \bibinfo{author}{C.~{Bassa}},
  \bibinfo{author}{V.~S. {Dhillon}}, \bibinfo{author}{T.~{Driebe}},
  \bibinfo{author}{J.~W. {Hessels}}, \bibinfo{author}{V.~M. {Kaspi}},
  \bibinfo{author}{V.~I. {Kondratiev}}, \bibinfo{author}{N.~{Langer}},
  \bibinfo{author}{T.~R. {Marsh}}, \bibinfo{author}{M.~A. {McLaughlin}},
  \bibinfo{author}{T.~T. {Pennucci}}, \bibinfo{author}{S.~M. {Ransom}},
  \bibinfo{author}{I.~H. {Stairs}}, \bibinfo{author}{J.~{van Leeuwen}},
  \bibinfo{author}{J.~P.~W. {Verbiest}}, \bibinfo{author}{D.~G. {Whelan}},
\newblock \bibinfo{title}{{A massive pulsar in a compact relativistic
  binary.}},
\newblock \bibinfo{journal}{Science} \bibinfo{volume}{340}
  (\bibinfo{year}{2013}) \bibinfo{pages}{448}.
\bibitem[{{Hebeler} et~al.(2013){Hebeler}, {Lattimer}, {Pethick}, and
  {Schwenk}}]{2013ApJ...773...11H}
\bibinfo{author}{K.~{Hebeler}}, \bibinfo{author}{J.~M. {Lattimer}},
  \bibinfo{author}{C.~J. {Pethick}}, \bibinfo{author}{A.~{Schwenk}},
\newblock \bibinfo{title}{{Equation of State and Neutron Star Properties
  Constrained by Nuclear Physics and Observation}},
\newblock \bibinfo{journal}{\apj} \bibinfo{volume}{773} (\bibinfo{year}{2013})
  \bibinfo{pages}{11}.
\bibitem[{{Heinke} et~al.(2014){Heinke}, {Cohn}, {Lugger}, {Webb}, {Ho},
  {Anderson}, {Campana}, {Bogdanov}, {Haggard}, {Cool}, and
  {Grindlay}}]{2014arXiv1406.1497H}
\bibinfo{author}{C.~O. {Heinke}}, \bibinfo{author}{H.~N. {Cohn}},
  \bibinfo{author}{P.~M. {Lugger}}, \bibinfo{author}{N.~A. {Webb}},
  \bibinfo{author}{W.~C.~G. {Ho}}, \bibinfo{author}{J.~{Anderson}},
  \bibinfo{author}{S.~{Campana}}, \bibinfo{author}{S.~{Bogdanov}},
  \bibinfo{author}{D.~{Haggard}}, \bibinfo{author}{A.~M. {Cool}},
  \bibinfo{author}{J.~E. {Grindlay}},
\newblock \bibinfo{title}{{Improved Mass and Radius Constraints for Quiescent
  Neutron Stars in Omega Cen and NGC 6397}},
\newblock \bibinfo{journal}{ArXiv e-prints}  (\bibinfo{year}{2014}).
\bibitem[{{Hessels} et~al.(2006){Hessels}, {Ransom}, {Stairs}, {Freire},
  {Kaspi}, and {Camilo}}]{2006Sci...311.1901H}
\bibinfo{author}{J.~W.~T. {Hessels}}, \bibinfo{author}{S.~M. {Ransom}},
  \bibinfo{author}{I.~H. {Stairs}}, \bibinfo{author}{P.~C.~C. {Freire}},
  \bibinfo{author}{V.~M. {Kaspi}}, \bibinfo{author}{F.~{Camilo}},
\newblock \bibinfo{title}{{A Radio Pulsar Spinning at 716 Hz}},
\newblock \bibinfo{journal}{Science} \bibinfo{volume}{311}
  (\bibinfo{year}{2006}) \bibinfo{pages}{1901--1904}.
\bibitem[{{Cook} et~al.(1994){Cook}, {Shapiro}, and
  {Teukolsky}}]{1994ApJ...424..823C}
\bibinfo{author}{G.~B. {Cook}}, \bibinfo{author}{S.~L. {Shapiro}},
  \bibinfo{author}{S.~A. {Teukolsky}},
\newblock \bibinfo{title}{{Rapidly rotating neutron stars in general
  relativity: Realistic equations of state}},
\newblock \bibinfo{journal}{\apj} \bibinfo{volume}{424} (\bibinfo{year}{1994})
  \bibinfo{pages}{823--845}.
\bibitem[{{Miller} and {Colbert}(2004)}]{2004IJMPD..13....1M}
\bibinfo{author}{M.~C. {Miller}}, \bibinfo{author}{E.~J.~M. {Colbert}},
\newblock \bibinfo{title}{{Intermediate-Mass Black Holes}},
\newblock \bibinfo{journal}{International Journal of Modern Physics D}
  \bibinfo{volume}{13} (\bibinfo{year}{2004}) \bibinfo{pages}{1--64}.
\bibitem[{{Thorne}(1974)}]{1974ApJ...191..507T}
\bibinfo{author}{K.~S. {Thorne}},
\newblock \bibinfo{title}{{Disk-Accretion onto a Black Hole. II. Evolution of
  the Hole}},
\newblock \bibinfo{journal}{\apj} \bibinfo{volume}{191} (\bibinfo{year}{1974})
  \bibinfo{pages}{507--520}.
\bibitem[{{Gammie} et~al.(2004){Gammie}, {Shapiro}, and
  {McKinney}}]{2004ApJ...602..312G}
\bibinfo{author}{C.~F. {Gammie}}, \bibinfo{author}{S.~L. {Shapiro}},
  \bibinfo{author}{J.~C. {McKinney}},
\newblock \bibinfo{title}{{Black Hole Spin Evolution}},
\newblock \bibinfo{journal}{\apj} \bibinfo{volume}{602} (\bibinfo{year}{2004})
  \bibinfo{pages}{312--319}.
\bibitem[{{Krolik} et~al.(2005){Krolik}, {Hawley}, and
  {Hirose}}]{2005ApJ...622.1008K}
\bibinfo{author}{J.~H. {Krolik}}, \bibinfo{author}{J.~F. {Hawley}},
  \bibinfo{author}{S.~{Hirose}},
\newblock \bibinfo{title}{{Magnetically Driven Accretion Flows in the Kerr
  Metric. IV. Dynamical Properties of the Inner Disk}},
\newblock \bibinfo{journal}{\apj} \bibinfo{volume}{622} (\bibinfo{year}{2005})
  \bibinfo{pages}{1008--1023}.
\bibitem[{{Bardeen} et~al.(1972){Bardeen}, {Press}, and
  {Teukolsky}}]{1972ApJ...178..347B}
\bibinfo{author}{J.~M. {Bardeen}}, \bibinfo{author}{W.~H. {Press}},
  \bibinfo{author}{S.~A. {Teukolsky}},
\newblock \bibinfo{title}{{Rotating Black Holes: Locally Nonrotating Frames,
  Energy Extraction, and Scalar Synchrotron Radiation}},
\newblock \bibinfo{journal}{\apj} \bibinfo{volume}{178} (\bibinfo{year}{1972})
  \bibinfo{pages}{347--370}.
\bibitem[{{Burrows}(2013)}]{2013RvMP...85..245B}
\bibinfo{author}{A.~{Burrows}},
\newblock \bibinfo{title}{{Colloquium: Perspectives on core-collapse supernova
  theory}},
\newblock \bibinfo{journal}{Reviews of Modern Physics} \bibinfo{volume}{85}
  (\bibinfo{year}{2013}) \bibinfo{pages}{245--261}.
\bibitem[{{Pijpers}(1998)}]{1998MNRAS.297L..76P}
\bibinfo{author}{F.~P. {Pijpers}},
\newblock \bibinfo{title}{{Helioseismic determination of the solar
  gravitational quadrupole moment}},
\newblock \bibinfo{journal}{\mnras} \bibinfo{volume}{297}
  (\bibinfo{year}{1998}) \bibinfo{pages}{L76--L80}.
\bibitem[{{Spruit} and {Phinney}(1998)}]{1998Natur.393..139S}
\bibinfo{author}{H.~{Spruit}}, \bibinfo{author}{E.~S. {Phinney}},
\newblock \bibinfo{title}{{Birth kicks as the origin of pulsar rotation}},
\newblock \bibinfo{journal}{\nat} \bibinfo{volume}{393} (\bibinfo{year}{1998})
  \bibinfo{pages}{139--141}.
\bibitem[{{Thompson} and {Duncan}(1993)}]{1993ApJ...408..194T}
\bibinfo{author}{C.~{Thompson}}, \bibinfo{author}{R.~C. {Duncan}},
\newblock \bibinfo{title}{{Neutron star dynamos and the origins of pulsar
  magnetism}},
\newblock \bibinfo{journal}{\apj} \bibinfo{volume}{408} (\bibinfo{year}{1993})
  \bibinfo{pages}{194--217}.
\bibitem[{{Neufeld} et~al.(1994){Neufeld}, {Maloney}, and
  {Conger}}]{1994ApJ...436L.127N}
\bibinfo{author}{D.~A. {Neufeld}}, \bibinfo{author}{P.~R. {Maloney}},
  \bibinfo{author}{S.~{Conger}},
\newblock \bibinfo{title}{{Water maser emission from X-ray-heated circumnuclear
  gas in active galaxies}},
\newblock \bibinfo{journal}{\apjl} \bibinfo{volume}{436} (\bibinfo{year}{1994})
  \bibinfo{pages}{L127}.
\bibitem[{{Blaes} et~al.(1995){Blaes}, {Warren}, and
  {Madau}}]{1995ApJ...454..370B}
\bibinfo{author}{O.~{Blaes}}, \bibinfo{author}{O.~{Warren}},
  \bibinfo{author}{P.~{Madau}},
\newblock \bibinfo{title}{{Accreting, Isolated Neutron Stars. III. Preheating
  of Infalling Gas and Cometary H II Regions}},
\newblock \bibinfo{journal}{\apj} \bibinfo{volume}{454} (\bibinfo{year}{1995})
  \bibinfo{pages}{370}.
\bibitem[{{Thorne} and {Zytkow}(1975)}]{1975ApJ...199L..19T}
\bibinfo{author}{K.~S. {Thorne}}, \bibinfo{author}{A.~N. {Zytkow}},
\newblock \bibinfo{title}{{Red giants and supergiants with degenerate neutron
  cores}},
\newblock \bibinfo{journal}{\apjl} \bibinfo{volume}{199} (\bibinfo{year}{1975})
  \bibinfo{pages}{L19--L24}.
\bibitem[{{Thorne} and {Zytkow}(1977)}]{1977ApJ...212..832T}
\bibinfo{author}{K.~S. {Thorne}}, \bibinfo{author}{A.~N. {Zytkow}},
\newblock \bibinfo{title}{{Stars with degenerate neutron cores. I - Structure
  of equilibrium models}},
\newblock \bibinfo{journal}{\apj} \bibinfo{volume}{212} (\bibinfo{year}{1977})
  \bibinfo{pages}{832--858}.
\bibitem[{{Levesque} et~al.(2014){Levesque}, {Massey}, {{\.Z}ytkow}, and
  {Morrell}}]{2014MNRAS.443L..94L}
\bibinfo{author}{E.~M. {Levesque}}, \bibinfo{author}{P.~{Massey}},
  \bibinfo{author}{A.~N. {{\.Z}ytkow}}, \bibinfo{author}{N.~{Morrell}},
\newblock \bibinfo{title}{{Discovery of a Thorne-{\.Z}ytkow object candidate in
  the Small Magellanic Cloud}},
\newblock \bibinfo{journal}{\mnras} \bibinfo{volume}{443}
  (\bibinfo{year}{2014}) \bibinfo{pages}{L94--L98}.
\bibitem[{{Ichimaru}(1977)}]{1977ApJ...214..840I}
\bibinfo{author}{S.~{Ichimaru}},
\newblock \bibinfo{title}{{Bimodal behavior of accretion disks - Theory and
  application to Cygnus X-1 transitions}},
\newblock \bibinfo{journal}{\apj} \bibinfo{volume}{214} (\bibinfo{year}{1977})
  \bibinfo{pages}{840--855}.
\bibitem[{{Begelman} and {Meier}(1982)}]{1982ApJ...253..873B}
\bibinfo{author}{M.~C. {Begelman}}, \bibinfo{author}{D.~L. {Meier}},
\newblock \bibinfo{title}{{Thick accretion disks - Self-similar, supercritical
  models}},
\newblock \bibinfo{journal}{\apj} \bibinfo{volume}{253} (\bibinfo{year}{1982})
  \bibinfo{pages}{873--896}.
\bibitem[{{Chevalier}(1989)}]{1989ApJ...346..847C}
\bibinfo{author}{R.~A. {Chevalier}},
\newblock \bibinfo{title}{{Neutron star accretion in a supernova}},
\newblock \bibinfo{journal}{\apj} \bibinfo{volume}{346} (\bibinfo{year}{1989})
  \bibinfo{pages}{847--859}.
\bibitem[{{Brown} and {Weingartner}(1994)}]{1994ApJ...436..843B}
\bibinfo{author}{G.~E. {Brown}}, \bibinfo{author}{J.~C. {Weingartner}},
\newblock \bibinfo{title}{{Accretion onto and radiation from the compact object
  formed in SN 1987A}},
\newblock \bibinfo{journal}{\apj} \bibinfo{volume}{436} (\bibinfo{year}{1994})
  \bibinfo{pages}{843--847}.
\bibitem[{{Iben} et~al.(1995){Iben}, {Tutukov}, and
  {Yungelson}}]{1995ApJS..100..217I}
\bibinfo{author}{I.~{Iben}, Jr.}, \bibinfo{author}{A.~V. {Tutukov}},
  \bibinfo{author}{L.~R. {Yungelson}},
\newblock \bibinfo{title}{{A Model of the Galactic X-Ray Binary Population. I.
  High-Mass X-Ray Binaries}},
\newblock \bibinfo{journal}{\apjs} \bibinfo{volume}{100} (\bibinfo{year}{1995})
  \bibinfo{pages}{217}.
\bibitem[{{Shapiro} and {Lightman}(1976)}]{1976ApJ...204..555S}
\bibinfo{author}{S.~L. {Shapiro}}, \bibinfo{author}{A.~P. {Lightman}},
\newblock \bibinfo{title}{{Black holes in X-ray binaries - Marginal existence
  and rotation reversals of accretion disks}},
\newblock \bibinfo{journal}{\apj} \bibinfo{volume}{204} (\bibinfo{year}{1976})
  \bibinfo{pages}{555--560}.
\bibitem[{{King} and {Kolb}(1999)}]{1999MNRAS.305..654K}
\bibinfo{author}{A.~R. {King}}, \bibinfo{author}{U.~{Kolb}},
\newblock \bibinfo{title}{{The evolution of black hole mass and angular
  momentum}},
\newblock \bibinfo{journal}{\mnras} \bibinfo{volume}{305}
  (\bibinfo{year}{1999}) \bibinfo{pages}{654--660}.
\bibitem[{{King} et~al.(2005){King}, {Lubow}, {Ogilvie}, and
  {Pringle}}]{2005MNRAS.363...49K}
\bibinfo{author}{A.~R. {King}}, \bibinfo{author}{S.~H. {Lubow}},
  \bibinfo{author}{G.~I. {Ogilvie}}, \bibinfo{author}{J.~E. {Pringle}},
\newblock \bibinfo{title}{{Aligning spinning black holes and accretion discs}},
\newblock \bibinfo{journal}{\mnras} \bibinfo{volume}{363}
  (\bibinfo{year}{2005}) \bibinfo{pages}{49--56}.
\bibitem[{{Bardeen} and {Petterson}(1975)}]{1975ApJ...195L..65B}
\bibinfo{author}{J.~M. {Bardeen}}, \bibinfo{author}{J.~A. {Petterson}},
\newblock \bibinfo{title}{{The Lense-Thirring Effect and Accretion Disks around
  Kerr Black Holes}},
\newblock \bibinfo{journal}{\apjl} \bibinfo{volume}{195} (\bibinfo{year}{1975})
  \bibinfo{pages}{L65}.
\bibitem[{{Papaloizou} and {Pringle}(1983)}]{1983MNRAS.202.1181P}
\bibinfo{author}{J.~C.~B. {Papaloizou}}, \bibinfo{author}{J.~E. {Pringle}},
\newblock \bibinfo{title}{{The time-dependence of non-planar accretion discs}},
\newblock \bibinfo{journal}{\mnras} \bibinfo{volume}{202}
  (\bibinfo{year}{1983}) \bibinfo{pages}{1181--1194}.
\bibitem[{{Pringle}(1992)}]{1992MNRAS.258..811P}
\bibinfo{author}{J.~E. {Pringle}},
\newblock \bibinfo{title}{{A simple approach to the evolution of twisted
  accretion discs}},
\newblock \bibinfo{journal}{\mnras} \bibinfo{volume}{258}
  (\bibinfo{year}{1992}) \bibinfo{pages}{811--818}.
\bibitem[{{Natarajan} and {Pringle}(1998)}]{1998ApJ...506L..97N}
\bibinfo{author}{P.~{Natarajan}}, \bibinfo{author}{J.~E. {Pringle}},
\newblock \bibinfo{title}{{The Alignment of Disk and Black Hole Spins in Active
  Galactic Nuclei}},
\newblock \bibinfo{journal}{\apjl} \bibinfo{volume}{506} (\bibinfo{year}{1998})
  \bibinfo{pages}{L97--L100}.
\bibitem[{{Armitage} and {Natarajan}(1999)}]{1999ApJ...525..909A}
\bibinfo{author}{P.~J. {Armitage}}, \bibinfo{author}{P.~{Natarajan}},
\newblock \bibinfo{title}{{Lense-Thirring Precession of Accretion Disks around
  Compact Objects}},
\newblock \bibinfo{journal}{\apj} \bibinfo{volume}{525} (\bibinfo{year}{1999})
  \bibinfo{pages}{909--914}.
\bibitem[{{Ogilvie}(1999)}]{1999MNRAS.304..557O}
\bibinfo{author}{G.~I. {Ogilvie}},
\newblock \bibinfo{title}{{The non-linear fluid dynamics of a warped accretion
  disc}},
\newblock \bibinfo{journal}{\mnras} \bibinfo{volume}{304}
  (\bibinfo{year}{1999}) \bibinfo{pages}{557--578}.
\bibitem[{{Nelson} and {Papaloizou}(2000)}]{2000MNRAS.315..570N}
\bibinfo{author}{R.~P. {Nelson}}, \bibinfo{author}{J.~C.~B. {Papaloizou}},
\newblock \bibinfo{title}{{Hydrodynamic simulations of the Bardeen-Petterson
  effect}},
\newblock \bibinfo{journal}{\mnras} \bibinfo{volume}{315}
  (\bibinfo{year}{2000}) \bibinfo{pages}{570--586}.
\bibitem[{{Lodato} and {Pringle}(2007)}]{2007MNRAS.381.1287L}
\bibinfo{author}{G.~{Lodato}}, \bibinfo{author}{J.~E. {Pringle}},
\newblock \bibinfo{title}{{Warp diffusion in accretion discs: a numerical
  investigation}},
\newblock \bibinfo{journal}{\mnras} \bibinfo{volume}{381}
  (\bibinfo{year}{2007}) \bibinfo{pages}{1287--1300}.
\bibitem[{{Lodato} and {Price}(2010)}]{2010MNRAS.405.1212L}
\bibinfo{author}{G.~{Lodato}}, \bibinfo{author}{D.~J. {Price}},
\newblock \bibinfo{title}{{On the diffusive propagation of warps in thin
  accretion discs}},
\newblock \bibinfo{journal}{\mnras} \bibinfo{volume}{405}
  (\bibinfo{year}{2010}) \bibinfo{pages}{1212--1226}.
\bibitem[{{Nixon} et~al.(2012){Nixon}, {King}, {Price}, and
  {Frank}}]{2012ApJ...757L..24N}
\bibinfo{author}{C.~{Nixon}}, \bibinfo{author}{A.~{King}},
  \bibinfo{author}{D.~{Price}}, \bibinfo{author}{J.~{Frank}},
\newblock \bibinfo{title}{{Tearing up the Disk: How Black Holes Accrete}},
\newblock \bibinfo{journal}{\apjl} \bibinfo{volume}{757} (\bibinfo{year}{2012})
  \bibinfo{pages}{L24}.
\bibitem[{{Sorathia} et~al.(2013{\natexlab{a}}){Sorathia}, {Krolik}, and
  {Hawley}}]{2013ApJ...768..133S}
\bibinfo{author}{K.~A. {Sorathia}}, \bibinfo{author}{J.~H. {Krolik}},
  \bibinfo{author}{J.~F. {Hawley}},
\newblock \bibinfo{title}{{Relaxation of Warped Disks: The Case of Pure
  Hydrodynamics}},
\newblock \bibinfo{journal}{\apj} \bibinfo{volume}{768}
  (\bibinfo{year}{2013}{\natexlab{a}}) \bibinfo{pages}{133}.
\bibitem[{{Sorathia} et~al.(2013{\natexlab{b}}){Sorathia}, {Krolik}, and
  {Hawley}}]{2013ApJ...777...21S}
\bibinfo{author}{K.~A. {Sorathia}}, \bibinfo{author}{J.~H. {Krolik}},
  \bibinfo{author}{J.~F. {Hawley}},
\newblock \bibinfo{title}{{Magnetohydrodynamic Simulation of a Disk Subjected
  to Lense-Thirring Precession}},
\newblock \bibinfo{journal}{\apj} \bibinfo{volume}{777}
  (\bibinfo{year}{2013}{\natexlab{b}}) \bibinfo{pages}{21}.
\bibitem[{{Fragile} et~al.(2007){Fragile}, {Blaes}, {Anninos}, and
  {Salmonson}}]{2007ApJ...668..417F}
\bibinfo{author}{P.~C. {Fragile}}, \bibinfo{author}{O.~M. {Blaes}},
  \bibinfo{author}{P.~{Anninos}}, \bibinfo{author}{J.~D. {Salmonson}},
\newblock \bibinfo{title}{{Global General Relativistic Magnetohydrodynamic
  Simulation of a Tilted Black Hole Accretion Disk}},
\newblock \bibinfo{journal}{\apj} \bibinfo{volume}{668} (\bibinfo{year}{2007})
  \bibinfo{pages}{417--429}.
\bibitem[{{Michell}(1767)}]{1767RSPT...57..234M}
\bibinfo{author}{J.~{Michell}},
\newblock \bibinfo{title}{{a Inquiry into the Probable Parallax, and Magnitude
  of the Fixed Stars, from the Quantity of Light Which They Afford us, and the
  Particular Circumstances of Their Situation, by the Rev. John Michell, B. D.
  F. R. S.}},
\newblock \bibinfo{journal}{Royal Society of London Philosophical Transactions
  Series I} \bibinfo{volume}{57} (\bibinfo{year}{1767})
  \bibinfo{pages}{234--264}.
\bibitem[{{Brown} and {Gilliland}(1994)}]{1994ARA&A..32...37B}
\bibinfo{author}{T.~M. {Brown}}, \bibinfo{author}{R.~L. {Gilliland}},
\newblock \bibinfo{title}{{Asteroseismology}},
\newblock \bibinfo{journal}{\araa} \bibinfo{volume}{32} (\bibinfo{year}{1994})
  \bibinfo{pages}{37--82}.
\bibitem[{{Winget} and {Kepler}(2008)}]{2008ARA&A..46..157W}
\bibinfo{author}{D.~E. {Winget}}, \bibinfo{author}{S.~O. {Kepler}},
\newblock \bibinfo{title}{{Pulsating White Dwarf Stars and Precision
  Asteroseismology}},
\newblock \bibinfo{journal}{\araa} \bibinfo{volume}{46} (\bibinfo{year}{2008})
  \bibinfo{pages}{157--199}.
\bibitem[{{Mathur} et~al.(2012){Mathur}, {Metcalfe}, {Woitaszek}, {Bruntt},
  {Verner}, {Christensen-Dalsgaard}, {Creevey}, {Do{\v g}an}, {Basu}, {Karoff},
  {Stello}, {Appourchaux}, {Campante}, {Chaplin}, {Garc{\'{\i}}a}, {Bedding},
  {Benomar}, {Bonanno}, {Deheuvels}, {Elsworth}, {Gaulme}, {Guzik}, {Handberg},
  {Hekker}, {Herzberg}, {Monteiro}, {Piau}, {Quirion}, {R{\'e}gulo}, {Roth},
  {Salabert}, {Serenelli}, {Thompson}, {Trampedach}, {White}, {Ballot},
  {Brand{\~a}o}, {Molenda-{\.Z}akowicz}, {Kjeldsen}, {Twicken}, {Uddin}, and
  {Wohler}}]{2012ApJ...749..152M}
\bibinfo{author}{S.~{Mathur}}, \bibinfo{author}{T.~S. {Metcalfe}},
  \bibinfo{author}{M.~{Woitaszek}}, \bibinfo{author}{H.~{Bruntt}},
  \bibinfo{author}{G.~A. {Verner}},
  \bibinfo{author}{J.~{Christensen-Dalsgaard}}, \bibinfo{author}{O.~L.
  {Creevey}}, \bibinfo{author}{G.~{Do{\v g}an}}, \bibinfo{author}{S.~{Basu}},
  \bibinfo{author}{C.~{Karoff}}, \bibinfo{author}{D.~{Stello}},
  \bibinfo{author}{T.~{Appourchaux}}, \bibinfo{author}{T.~L. {Campante}},
  \bibinfo{author}{W.~J. {Chaplin}}, \bibinfo{author}{R.~A. {Garc{\'{\i}}a}},
  \bibinfo{author}{T.~R. {Bedding}}, \bibinfo{author}{O.~{Benomar}},
  \bibinfo{author}{A.~{Bonanno}}, \bibinfo{author}{S.~{Deheuvels}},
  \bibinfo{author}{Y.~{Elsworth}}, \bibinfo{author}{P.~{Gaulme}},
  \bibinfo{author}{J.~A. {Guzik}}, \bibinfo{author}{R.~{Handberg}},
  \bibinfo{author}{S.~{Hekker}}, \bibinfo{author}{W.~{Herzberg}},
  \bibinfo{author}{M.~J.~P.~F.~G. {Monteiro}}, \bibinfo{author}{L.~{Piau}},
  \bibinfo{author}{P.-O. {Quirion}}, \bibinfo{author}{C.~{R{\'e}gulo}},
  \bibinfo{author}{M.~{Roth}}, \bibinfo{author}{D.~{Salabert}},
  \bibinfo{author}{A.~{Serenelli}}, \bibinfo{author}{M.~J. {Thompson}},
  \bibinfo{author}{R.~{Trampedach}}, \bibinfo{author}{T.~R. {White}},
  \bibinfo{author}{J.~{Ballot}}, \bibinfo{author}{I.~M. {Brand{\~a}o}},
  \bibinfo{author}{J.~{Molenda-{\.Z}akowicz}}, \bibinfo{author}{H.~{Kjeldsen}},
  \bibinfo{author}{J.~D. {Twicken}}, \bibinfo{author}{K.~{Uddin}},
  \bibinfo{author}{B.~{Wohler}},
\newblock \bibinfo{title}{{A Uniform Asteroseismic Analysis of 22 Solar-type
  Stars Observed by Kepler}},
\newblock \bibinfo{journal}{\apj} \bibinfo{volume}{749} (\bibinfo{year}{2012})
  \bibinfo{pages}{152}.
\bibitem[{{Miglio} et~al.(2013){Miglio}, {Chiappini}, {Morel}, {Barbieri},
  {Chaplin}, {Girardi}, {Montalb{\'a}n}, {Valentini}, {Mosser}, {Baudin},
  {Casagrande}, {Fossati}, {Aguirre}, and {Baglin}}]{2013MNRAS.429..423M}
\bibinfo{author}{A.~{Miglio}}, \bibinfo{author}{C.~{Chiappini}},
  \bibinfo{author}{T.~{Morel}}, \bibinfo{author}{M.~{Barbieri}},
  \bibinfo{author}{W.~J. {Chaplin}}, \bibinfo{author}{L.~{Girardi}},
  \bibinfo{author}{J.~{Montalb{\'a}n}}, \bibinfo{author}{M.~{Valentini}},
  \bibinfo{author}{B.~{Mosser}}, \bibinfo{author}{F.~{Baudin}},
  \bibinfo{author}{L.~{Casagrande}}, \bibinfo{author}{L.~{Fossati}},
  \bibinfo{author}{V.~S. {Aguirre}}, \bibinfo{author}{A.~{Baglin}},
\newblock \bibinfo{title}{{Galactic archaeology: mapping and dating stellar
  populations with asteroseismology of red-giant stars}},
\newblock \bibinfo{journal}{\mnras} \bibinfo{volume}{429}
  (\bibinfo{year}{2013}) \bibinfo{pages}{423--428}.
\bibitem[{{Mao}(2012)}]{2012RAA....12..947M}
\bibinfo{author}{S.~{Mao}},
\newblock \bibinfo{title}{{Astrophysical applications of gravitational
  microlensing}},
\newblock \bibinfo{journal}{Research in Astronomy and Astrophysics}
  \bibinfo{volume}{12} (\bibinfo{year}{2012}) \bibinfo{pages}{947--972}.
\bibitem[{{Landau}(1932)}]{1932PZSow...1..285L}
\bibinfo{author}{L.~D. {Landau}},
\newblock \bibinfo{title}{{On the Theory of Stars}},
\newblock \bibinfo{journal}{Phys. Z. Sowjet} \bibinfo{volume}{1}
  (\bibinfo{year}{1932}) \bibinfo{pages}{285}.
\bibitem[{{Chandrasekhar}(1931)}]{1931ApJ....74...81C}
\bibinfo{author}{S.~{Chandrasekhar}},
\newblock \bibinfo{title}{{The Maximum Mass of Ideal White Dwarfs}},
\newblock \bibinfo{journal}{\apj} \bibinfo{volume}{74} (\bibinfo{year}{1931})
  \bibinfo{pages}{81}.
\bibitem[{{Oppenheimer} and {Volkoff}(1939)}]{1939PhRv...55..374O}
\bibinfo{author}{J.~R. {Oppenheimer}}, \bibinfo{author}{G.~M. {Volkoff}},
\newblock \bibinfo{title}{{On Massive Neutron Cores}},
\newblock \bibinfo{journal}{Physical Review} \bibinfo{volume}{55}
  (\bibinfo{year}{1939}) \bibinfo{pages}{374--381}.
\bibitem[{{Barbour} et~al.(1998){Barbour}, {Morrison}, {Klepfish}, {Kogut}, and
  {Lombardo}}]{1998NuPhS..63..439B}
\bibinfo{author}{I.~M. {Barbour}}, \bibinfo{author}{S.~E. {Morrison}},
  \bibinfo{author}{E.~G. {Klepfish}}, \bibinfo{author}{J.~B. {Kogut}},
  \bibinfo{author}{M.-P. {Lombardo}},
\newblock \bibinfo{title}{{The critical points of QCD at non-zero density, T=0,
  g={$\infty$}}},
\newblock \bibinfo{journal}{Nuclear Physics B Proceedings Supplements}
  \bibinfo{volume}{63} (\bibinfo{year}{1998}) \bibinfo{pages}{439--441}.
\bibitem[{{Alford}(1999)}]{1999NuPhS..73..161A}
\bibinfo{author}{M.~{Alford}},
\newblock \bibinfo{title}{{New possibilities for QCD at finite density}},
\newblock \bibinfo{journal}{Nuclear Physics B Proceedings Supplements}
  \bibinfo{volume}{73} (\bibinfo{year}{1999}) \bibinfo{pages}{161--166}.
\bibitem[{{Hands}(2002)}]{2002NuPhS.106..142H}
\bibinfo{author}{S.~{Hands}},
\newblock \bibinfo{title}{{Lattice matter}},
\newblock \bibinfo{journal}{Nuclear Physics B Proceedings Supplements}
  \bibinfo{volume}{106} (\bibinfo{year}{2002}) \bibinfo{pages}{142--150}.
\bibitem[{{Lombardo} et~al.(2009){Lombardo}, {Splittorff}, and
  {Verbaarschot}}]{2009arXiv0912.4410L}
\bibinfo{author}{M.~P. {Lombardo}}, \bibinfo{author}{K.~{Splittorff}},
  \bibinfo{author}{J.~J.~M. {Verbaarschot}},
\newblock \bibinfo{title}{{Lattice QCD and dense quark matter}},
\newblock \bibinfo{journal}{arXiv:0912.4410}  (\bibinfo{year}{2009}).
\bibitem[{{Rhoades} and {Ruffini}(1974)}]{1974PhRvL..32..324R}
\bibinfo{author}{C.~E. {Rhoades}}, \bibinfo{author}{R.~{Ruffini}},
\newblock \bibinfo{title}{{Maximum Mass of a Neutron Star}},
\newblock \bibinfo{journal}{Physical Review Letters} \bibinfo{volume}{32}
  (\bibinfo{year}{1974}) \bibinfo{pages}{324--327}.
\bibitem[{{Kalogera} and {Baym}(1996)}]{1996ApJ...470L..61K}
\bibinfo{author}{V.~{Kalogera}}, \bibinfo{author}{G.~{Baym}},
\newblock \bibinfo{title}{{The Maximum Mass of a Neutron Star}},
\newblock \bibinfo{journal}{\apjl} \bibinfo{volume}{470} (\bibinfo{year}{1996})
  \bibinfo{pages}{L61}.
\bibitem[{{Horowitz} et~al.(2014){Horowitz}, {Brown}, {Kim}, {Lynch},
  {Michaels}, {Ono}, {Piekarewicz}, {Tsang}, and
  {Wolter}}]{2014arXiv1401.5839H}
\bibinfo{author}{C.~J. {Horowitz}}, \bibinfo{author}{E.~F. {Brown}},
  \bibinfo{author}{Y.~{Kim}}, \bibinfo{author}{W.~G. {Lynch}},
  \bibinfo{author}{R.~{Michaels}}, \bibinfo{author}{A.~{Ono}},
  \bibinfo{author}{J.~{Piekarewicz}}, \bibinfo{author}{M.~B. {Tsang}},
  \bibinfo{author}{H.~H. {Wolter}},
\newblock \bibinfo{title}{{A way forward in the study of the symmetry energy:
  experiment, theory, and observation}},
\newblock \bibinfo{journal}{ArXiv e-prints}  (\bibinfo{year}{2014}).
\bibitem[{{Friedman} and {Ipser}(1987)}]{1987ApJ...314..594F}
\bibinfo{author}{J.~L. {Friedman}}, \bibinfo{author}{J.~R. {Ipser}},
\newblock \bibinfo{title}{{On the maximum mass of a uniformly rotating neutron
  star}},
\newblock \bibinfo{journal}{\apj} \bibinfo{volume}{314} (\bibinfo{year}{1987})
  \bibinfo{pages}{594--597}.
\bibitem[{{Hegyi}(1977)}]{1977NYASA.302..528H}
\bibinfo{author}{D.~J. {Hegyi}},
\newblock \bibinfo{title}{{The Damping of Differential Rotation in the Cores of
  Neutron Stars}},
\newblock in: \bibinfo{editor}{M.~D. {Papagiannis}} (Ed.),
  \bibinfo{booktitle}{Eighth Texas Symposium on Relativistic Astrophysics},
  volume \bibinfo{volume}{302} of \textit{\bibinfo{series}{Annals of the New
  York Academy of Sciences}}, p. \bibinfo{pages}{528}.
\bibitem[{{Joss} and {Rappaport}(1984)}]{1984ARA&A..22..537J}
\bibinfo{author}{P.~C. {Joss}}, \bibinfo{author}{S.~A. {Rappaport}},
\newblock \bibinfo{title}{{Neutron Stars in Interacting Binary Systems}},
\newblock \bibinfo{journal}{\araa} \bibinfo{volume}{22} (\bibinfo{year}{1984})
  \bibinfo{pages}{537--592}.
\bibitem[{{Brucato} and {Kristian}(1973)}]{1973ApJ...179L.129B}
\bibinfo{author}{R.~{Brucato}}, \bibinfo{author}{J.~{Kristian}},
\newblock \bibinfo{title}{{Spectroscopic Observations of the Optical Candidate
  for Cygnus X-1}},
\newblock \bibinfo{journal}{\apjl} \bibinfo{volume}{179} (\bibinfo{year}{1973})
  \bibinfo{pages}{L129}.
\bibitem[{{Avni}(1978)}]{1978pans.proc...43A}
\bibinfo{author}{Y.~{Avni}},
\newblock \bibinfo{title}{{Mass estimates from optical-light curves for binary
  X-ray sources}},
\newblock in: \bibinfo{editor}{R.~{Giacconi}}, \bibinfo{editor}{R.~{Ruffini}}
  (Eds.), \bibinfo{booktitle}{Physics and Astrophysics of Neutron Stars and
  Black Holes}, pp. \bibinfo{pages}{43--62}.
\bibitem[{{McClintock} and {Remillard}(1990)}]{1990ApJ...350..386M}
\bibinfo{author}{J.~E. {McClintock}}, \bibinfo{author}{R.~A. {Remillard}},
\newblock \bibinfo{title}{{The X-ray nova Centaurus X-4 - Comparisons with
  A0620 - 00}},
\newblock \bibinfo{journal}{\apj} \bibinfo{volume}{350} (\bibinfo{year}{1990})
  \bibinfo{pages}{386--394}.
\bibitem[{{Greene} et~al.(2001){Greene}, {Bailyn}, and
  {Orosz}}]{2001ApJ...554.1290G}
\bibinfo{author}{J.~{Greene}}, \bibinfo{author}{C.~D. {Bailyn}},
  \bibinfo{author}{J.~A. {Orosz}},
\newblock \bibinfo{title}{{Optical and Infrared Photometry of the Microquasar
  GRO J1655-40 in Quiescence}},
\newblock \bibinfo{journal}{\apj} \bibinfo{volume}{554} (\bibinfo{year}{2001})
  \bibinfo{pages}{1290--1297}.
\bibitem[{{Hulse} and {Taylor}(1975)}]{1975ApJ...195L..51H}
\bibinfo{author}{R.~A. {Hulse}}, \bibinfo{author}{J.~H. {Taylor}},
\newblock \bibinfo{title}{{Discovery of a pulsar in a binary system}},
\newblock \bibinfo{journal}{\apjl} \bibinfo{volume}{195} (\bibinfo{year}{1975})
  \bibinfo{pages}{L51--L53}.
\bibitem[{{Freire}(2009)}]{2009arXiv0907.3219F}
\bibinfo{author}{P.~C.~C. {Freire}},
\newblock \bibinfo{title}{{Eccentric Binary Millisecond Pulsars}},
\newblock \bibinfo{journal}{arXiv:0907.3219}  (\bibinfo{year}{2009}).
\bibitem[{{Taylor}(1991)}]{1991IEEEP..79.1054T}
\bibinfo{author}{J.~H. {Taylor}, Jr.},
\newblock \bibinfo{title}{{Millisecond pulsars - Nature's most stable clocks}},
\newblock \bibinfo{journal}{IEEE Proceedings} \bibinfo{volume}{79}
  (\bibinfo{year}{1991}) \bibinfo{pages}{1054--1062}.
\bibitem[{{Kramer} et~al.(2006){Kramer}, {Stairs}, {Manchester}, {McLaughlin},
  {Lyne}, {Ferdman}, {Burgay}, {Lorimer}, {Possenti}, {D'Amico}, {Sarkissian},
  {Hobbs}, {Reynolds}, {Freire}, and {Camilo}}]{2006Sci...314...97K}
\bibinfo{author}{M.~{Kramer}}, \bibinfo{author}{I.~H. {Stairs}},
  \bibinfo{author}{R.~N. {Manchester}}, \bibinfo{author}{M.~A. {McLaughlin}},
  \bibinfo{author}{A.~G. {Lyne}}, \bibinfo{author}{R.~D. {Ferdman}},
  \bibinfo{author}{M.~{Burgay}}, \bibinfo{author}{D.~R. {Lorimer}},
  \bibinfo{author}{A.~{Possenti}}, \bibinfo{author}{N.~{D'Amico}},
  \bibinfo{author}{J.~M. {Sarkissian}}, \bibinfo{author}{G.~B. {Hobbs}},
  \bibinfo{author}{J.~E. {Reynolds}}, \bibinfo{author}{P.~C.~C. {Freire}},
  \bibinfo{author}{F.~{Camilo}},
\newblock \bibinfo{title}{{Tests of General Relativity from Timing the Double
  Pulsar}},
\newblock \bibinfo{journal}{Science} \bibinfo{volume}{314}
  (\bibinfo{year}{2006}) \bibinfo{pages}{97--102}.
\bibitem[{{Phinney}(1992)}]{1992RSPTA.341...39P}
\bibinfo{author}{E.~S. {Phinney}},
\newblock \bibinfo{title}{{Pulsars as Probes of Newtonian Dynamical Systems}},
\newblock \bibinfo{journal}{Royal Society of London Philosophical Transactions
  Series A} \bibinfo{volume}{341} (\bibinfo{year}{1992})
  \bibinfo{pages}{39--75}.
\bibitem[{{Demorest} et~al.(2010){Demorest}, {Pennucci}, {Ransom}, {Roberts},
  and {Hessels}}]{2010Natur.467.1081D}
\bibinfo{author}{P.~B. {Demorest}}, \bibinfo{author}{T.~{Pennucci}},
  \bibinfo{author}{S.~M. {Ransom}}, \bibinfo{author}{M.~S.~E. {Roberts}},
  \bibinfo{author}{J.~W.~T. {Hessels}},
\newblock \bibinfo{title}{{A two-solar-mass neutron star measured using Shapiro
  delay}},
\newblock \bibinfo{journal}{\nat} \bibinfo{volume}{467} (\bibinfo{year}{2010})
  \bibinfo{pages}{1081--1083}.
\bibitem[{{Brown} and {Bethe}(1994)}]{1994ApJ...423..659B}
\bibinfo{author}{G.~E. {Brown}}, \bibinfo{author}{H.~A. {Bethe}},
\newblock \bibinfo{title}{{A Scenario for a Large Number of Low-Mass Black
  Holes in the Galaxy}},
\newblock \bibinfo{journal}{\apj} \bibinfo{volume}{423} (\bibinfo{year}{1994})
  \bibinfo{pages}{659}.
\bibitem[{{Zhang} et~al.(2011){Zhang}, {Wang}, {Zhao}, {Yin}, {Song},
  {Menezes}, {Wickramasinghe}, {Ferrario}, and
  {Chardonnet}}]{2011A&A...527A..83Z}
\bibinfo{author}{C.~M. {Zhang}}, \bibinfo{author}{J.~{Wang}},
  \bibinfo{author}{Y.~H. {Zhao}}, \bibinfo{author}{H.~X. {Yin}},
  \bibinfo{author}{L.~M. {Song}}, \bibinfo{author}{D.~P. {Menezes}},
  \bibinfo{author}{D.~T. {Wickramasinghe}}, \bibinfo{author}{L.~{Ferrario}},
  \bibinfo{author}{P.~{Chardonnet}},
\newblock \bibinfo{title}{{Study of measured pulsar masses and their possible
  conclusions}},
\newblock \bibinfo{journal}{\aap} \bibinfo{volume}{527} (\bibinfo{year}{2011})
  \bibinfo{pages}{A83}.
\bibitem[{{Bailyn} et~al.(1998){Bailyn}, {Jain}, {Coppi}, and
  {Orosz}}]{1998ApJ...499..367B}
\bibinfo{author}{C.~D. {Bailyn}}, \bibinfo{author}{R.~K. {Jain}},
  \bibinfo{author}{P.~{Coppi}}, \bibinfo{author}{J.~A. {Orosz}},
\newblock \bibinfo{title}{{The Mass Distribution of Stellar Black Holes}},
\newblock \bibinfo{journal}{\apj} \bibinfo{volume}{499} (\bibinfo{year}{1998})
  \bibinfo{pages}{367}.
\bibitem[{{{\"O}zel} et~al.(2010){{\"O}zel}, {Psaltis}, {Narayan}, and
  {McClintock}}]{2010ApJ...725.1918O}
\bibinfo{author}{F.~{{\"O}zel}}, \bibinfo{author}{D.~{Psaltis}},
  \bibinfo{author}{R.~{Narayan}}, \bibinfo{author}{J.~E. {McClintock}},
\newblock \bibinfo{title}{{The Black Hole Mass Distribution in the Galaxy}},
\newblock \bibinfo{journal}{\apj} \bibinfo{volume}{725} (\bibinfo{year}{2010})
  \bibinfo{pages}{1918--1927}.
\bibitem[{{Farr} et~al.(2011){Farr}, {Sravan}, {Cantrell}, {Kreidberg},
  {Bailyn}, {Mandel}, and {Kalogera}}]{2011ApJ...741..103F}
\bibinfo{author}{W.~M. {Farr}}, \bibinfo{author}{N.~{Sravan}},
  \bibinfo{author}{A.~{Cantrell}}, \bibinfo{author}{L.~{Kreidberg}},
  \bibinfo{author}{C.~D. {Bailyn}}, \bibinfo{author}{I.~{Mandel}},
  \bibinfo{author}{V.~{Kalogera}},
\newblock \bibinfo{title}{{The Mass Distribution of Stellar-mass Black Holes}},
\newblock \bibinfo{journal}{\apj} \bibinfo{volume}{741} (\bibinfo{year}{2011})
  \bibinfo{pages}{103}.
\bibitem[{{Kochanek}(2014)}]{2014ApJ...785...28K}
\bibinfo{author}{C.~S. {Kochanek}},
\newblock \bibinfo{title}{{Failed Supernovae Explain the Compact Remnant Mass
  Function}},
\newblock \bibinfo{journal}{\apj} \bibinfo{volume}{785} (\bibinfo{year}{2014})
  \bibinfo{pages}{28}.
\bibitem[{{Manchester}(2006)}]{2006ChJAS...6b.139M}
\bibinfo{author}{R.~N. {Manchester}},
\newblock \bibinfo{title}{{The Parkes Pulsar Timing Array}},
\newblock \bibinfo{journal}{Chinese Journal of Astronomy and Astrophysics
  Supplement} \bibinfo{volume}{6} (\bibinfo{year}{2006})
  \bibinfo{pages}{020000--147}.
\bibitem[{{Abdo} et~al.(2013){Abdo}, {Ajello}, {Allafort}, {Baldini}, {Ballet},
  {Barbiellini}, {Baring}, {Bastieri}, {Belfiore}, {Bellazzini}, and
  et~al.}]{2013ApJS..208...17A}
\bibinfo{author}{A.~A. {Abdo}}, \bibinfo{author}{M.~{Ajello}},
  \bibinfo{author}{A.~{Allafort}}, \bibinfo{author}{L.~{Baldini}},
  \bibinfo{author}{J.~{Ballet}}, \bibinfo{author}{G.~{Barbiellini}},
  \bibinfo{author}{M.~G. {Baring}}, \bibinfo{author}{D.~{Bastieri}},
  \bibinfo{author}{A.~{Belfiore}}, \bibinfo{author}{R.~{Bellazzini}},
  \bibinfo{author}{et~al.},
\newblock \bibinfo{title}{{The Second Fermi Large Area Telescope Catalog of
  Gamma-Ray Pulsars}},
\newblock \bibinfo{journal}{\apjs} \bibinfo{volume}{208} (\bibinfo{year}{2013})
  \bibinfo{pages}{17}.
\bibitem[{{Patruno} and {Watts}(2012)}]{2012arXiv1206.2727P}
\bibinfo{author}{A.~{Patruno}}, \bibinfo{author}{A.~L. {Watts}},
\newblock \bibinfo{title}{{Accreting Millisecond X-Ray Pulsars}},
\newblock \bibinfo{journal}{arXiv:1206.2727}  (\bibinfo{year}{2012}).
\bibitem[{{Galloway} et~al.(2008){Galloway}, {Muno}, {Hartman}, {Psaltis}, and
  {Chakrabarty}}]{2008ApJS..179..360G}
\bibinfo{author}{D.~K. {Galloway}}, \bibinfo{author}{M.~P. {Muno}},
  \bibinfo{author}{J.~M. {Hartman}}, \bibinfo{author}{D.~{Psaltis}},
  \bibinfo{author}{D.~{Chakrabarty}},
\newblock \bibinfo{title}{{Thermonuclear (Type I) X-Ray Bursts Observed by the
  Rossi X-Ray Timing Explorer}},
\newblock \bibinfo{journal}{\apjs} \bibinfo{volume}{179} (\bibinfo{year}{2008})
  \bibinfo{pages}{360--422}.
\bibitem[{{Olausen} and {Kaspi}(2014)}]{2014ApJS..212....6O}
\bibinfo{author}{S.~A. {Olausen}}, \bibinfo{author}{V.~M. {Kaspi}},
\newblock \bibinfo{title}{{The McGill Magnetar Catalog}},
\newblock \bibinfo{journal}{\apjs} \bibinfo{volume}{212} (\bibinfo{year}{2014})
  \bibinfo{pages}{6}.
\bibitem[{{Watts}(2012)}]{2012ARA&A..50..609W}
\bibinfo{author}{A.~L. {Watts}},
\newblock \bibinfo{title}{{Thermonuclear Burst Oscillations}},
\newblock \bibinfo{journal}{\araa} \bibinfo{volume}{50} (\bibinfo{year}{2012})
  \bibinfo{pages}{609--640}.
\bibitem[{{Pacini}(1967)}]{1967Natur.216..567P}
\bibinfo{author}{F.~{Pacini}},
\newblock \bibinfo{title}{{Energy Emission from a Neutron Star}},
\newblock \bibinfo{journal}{\nat} \bibinfo{volume}{216} (\bibinfo{year}{1967})
  \bibinfo{pages}{567--568}.
\bibitem[{{Gold}(1968)}]{1968Natur.218..731G}
\bibinfo{author}{T.~{Gold}},
\newblock \bibinfo{title}{{Rotating Neutron Stars as the Origin of the
  Pulsating Radio Sources}},
\newblock \bibinfo{journal}{\nat} \bibinfo{volume}{218} (\bibinfo{year}{1968})
  \bibinfo{pages}{731--732}.
\bibitem[{{Harding}(2013)}]{2013FrPhy...8..679H}
\bibinfo{author}{A.~K. {Harding}},
\newblock \bibinfo{title}{{The neutron star zoo}},
\newblock \bibinfo{journal}{Frontiers of Physics} \bibinfo{volume}{8}
  (\bibinfo{year}{2013}) \bibinfo{pages}{679--692}.
\bibitem[{{Thompson} and {Duncan}(1995)}]{1995MNRAS.275..255T}
\bibinfo{author}{C.~{Thompson}}, \bibinfo{author}{R.~C. {Duncan}},
\newblock \bibinfo{title}{{The soft gamma repeaters as very strongly magnetized
  neutron stars - I. Radiative mechanism for outbursts}},
\newblock \bibinfo{journal}{\mnras} \bibinfo{volume}{275}
  (\bibinfo{year}{1995}) \bibinfo{pages}{255--300}.
\bibitem[{{Thompson} and {Duncan}(1996)}]{1996ApJ...473..322T}
\bibinfo{author}{C.~{Thompson}}, \bibinfo{author}{R.~C. {Duncan}},
\newblock \bibinfo{title}{{The Soft Gamma Repeaters as Very Strongly Magnetized
  Neutron Stars. II. Quiescent Neutrino, X-Ray, and Alfven Wave Emission}},
\newblock \bibinfo{journal}{\apj} \bibinfo{volume}{473} (\bibinfo{year}{1996})
  \bibinfo{pages}{322}.
\bibitem[{{Popov} and {Turolla}(2012)}]{2012Ap&SS.341..457P}
\bibinfo{author}{S.~B. {Popov}}, \bibinfo{author}{R.~{Turolla}},
\newblock \bibinfo{title}{{Initial spin periods of neutron stars in supernova
  remnants}},
\newblock \bibinfo{journal}{\apss} \bibinfo{volume}{341} (\bibinfo{year}{2012})
  \bibinfo{pages}{457--464}.
\bibitem[{{Manchester} and {Taylor}(1977)}]{1977puls.book.....M}
\bibinfo{author}{R.~N. {Manchester}}, \bibinfo{author}{J.~H. {Taylor}},
  \bibinfo{title}{{Pulsars.}}, \bibinfo{year}{1977}.
\bibitem[{{Ho} and {Andersson}(2012)}]{2012NatPh...8..787H}
\bibinfo{author}{W.~C.~G. {Ho}}, \bibinfo{author}{N.~{Andersson}},
\newblock \bibinfo{title}{{Rotational evolution of young pulsars due to
  superfluid decoupling}},
\newblock \bibinfo{journal}{Nature Physics} \bibinfo{volume}{8}
  (\bibinfo{year}{2012}) \bibinfo{pages}{787--789}.
\bibitem[{{Pons} et~al.(2012){Pons}, {Vigan{\`o}}, and
  {Geppert}}]{2012A&A...547A...9P}
\bibinfo{author}{J.~A. {Pons}}, \bibinfo{author}{D.~{Vigan{\`o}}},
  \bibinfo{author}{U.~{Geppert}},
\newblock \bibinfo{title}{{Pulsar timing irregularities and the imprint of
  magnetic field evolution}},
\newblock \bibinfo{journal}{\aap} \bibinfo{volume}{547} (\bibinfo{year}{2012})
  \bibinfo{pages}{A9}.
\bibitem[{{Andersson}(1998)}]{1998ApJ...502..708A}
\bibinfo{author}{N.~{Andersson}},
\newblock \bibinfo{title}{{A New Class of Unstable Modes of Rotating
  Relativistic Stars}},
\newblock \bibinfo{journal}{\apj} \bibinfo{volume}{502} (\bibinfo{year}{1998})
  \bibinfo{pages}{708}.
\bibitem[{{Andersson} et~al.(1999){Andersson}, {Kokkotas}, and
  {Schutz}}]{1999ApJ...510..846A}
\bibinfo{author}{N.~{Andersson}}, \bibinfo{author}{K.~{Kokkotas}},
  \bibinfo{author}{B.~F. {Schutz}},
\newblock \bibinfo{title}{{Gravitational Radiation Limit on the Spin of Young
  Neutron Stars}},
\newblock \bibinfo{journal}{\apj} \bibinfo{volume}{510} (\bibinfo{year}{1999})
  \bibinfo{pages}{846--853}.
\bibitem[{{Arras} et~al.(2003){Arras}, {Flanagan}, {Morsink}, {Schenk},
  {Teukolsky}, and {Wasserman}}]{2003ApJ...591.1129A}
\bibinfo{author}{P.~{Arras}}, \bibinfo{author}{E.~E. {Flanagan}},
  \bibinfo{author}{S.~M. {Morsink}}, \bibinfo{author}{A.~K. {Schenk}},
  \bibinfo{author}{S.~A. {Teukolsky}}, \bibinfo{author}{I.~{Wasserman}},
\newblock \bibinfo{title}{{Saturation of the r-Mode Instability}},
\newblock \bibinfo{journal}{\apj} \bibinfo{volume}{591} (\bibinfo{year}{2003})
  \bibinfo{pages}{1129--1151}.
\bibitem[{{Ott} et~al.(2006){Ott}, {Burrows}, {Thompson}, {Livne}, and
  {Walder}}]{2006ApJS..164..130O}
\bibinfo{author}{C.~D. {Ott}}, \bibinfo{author}{A.~{Burrows}},
  \bibinfo{author}{T.~A. {Thompson}}, \bibinfo{author}{E.~{Livne}},
  \bibinfo{author}{R.~{Walder}},
\newblock \bibinfo{title}{{The Spin Periods and Rotational Profiles of Neutron
  Stars at Birth}},
\newblock \bibinfo{journal}{\apjs} \bibinfo{volume}{164} (\bibinfo{year}{2006})
  \bibinfo{pages}{130--155}.
\bibitem[{{Tananbaum} et~al.(1972){Tananbaum}, {Gursky}, {Kellogg}, {Levinson},
  {Schreier}, and {Giacconi}}]{1972ApJ...174L.143T}
\bibinfo{author}{H.~{Tananbaum}}, \bibinfo{author}{H.~{Gursky}},
  \bibinfo{author}{E.~M. {Kellogg}}, \bibinfo{author}{R.~{Levinson}},
  \bibinfo{author}{E.~{Schreier}}, \bibinfo{author}{R.~{Giacconi}},
\newblock \bibinfo{title}{{Discovery of a Periodic Pulsating Binary X-Ray
  Source in Hercules from UHURU}},
\newblock \bibinfo{journal}{\apjl} \bibinfo{volume}{174} (\bibinfo{year}{1972})
  \bibinfo{pages}{L143}.
\bibitem[{{Pringle} and {Rees}(1972)}]{1972A&A....21....1P}
\bibinfo{author}{J.~E. {Pringle}}, \bibinfo{author}{M.~J. {Rees}},
\newblock \bibinfo{title}{{Accretion Disc Models for Compact X-Ray Sources}},
\newblock \bibinfo{journal}{\aap} \bibinfo{volume}{21} (\bibinfo{year}{1972})
  \bibinfo{pages}{1}.
\bibitem[{{Davidsen} and {Ostriker}(1974)}]{1974ApJ...189..331D}
\bibinfo{author}{A.~{Davidsen}}, \bibinfo{author}{J.~P. {Ostriker}},
\newblock \bibinfo{title}{{The Nature of Cygnus X-3: a Prototype for
  Old-Population Binary X-Ray Sources}},
\newblock \bibinfo{journal}{\apj} \bibinfo{volume}{189} (\bibinfo{year}{1974})
  \bibinfo{pages}{331--338}.
\bibitem[{{Lamb} et~al.(1973){Lamb}, {Pethick}, and
  {Pines}}]{1973ApJ...184..271L}
\bibinfo{author}{F.~K. {Lamb}}, \bibinfo{author}{C.~J. {Pethick}},
  \bibinfo{author}{D.~{Pines}},
\newblock \bibinfo{title}{{A Model for Compact X-Ray Sources: Accretion by
  Rotating Magnetic Stars}},
\newblock \bibinfo{journal}{\apj} \bibinfo{volume}{184} (\bibinfo{year}{1973})
  \bibinfo{pages}{271--290}.
\bibitem[{{Ghosh} et~al.(1977){Ghosh}, {Pethick}, and
  {Lamb}}]{1977ApJ...217..578G}
\bibinfo{author}{P.~{Ghosh}}, \bibinfo{author}{C.~J. {Pethick}},
  \bibinfo{author}{F.~K. {Lamb}},
\newblock \bibinfo{title}{{Accretion by rotating magnetic neutron stars. I -
  Flow of matter inside the magnetosphere and its implications for spin-up and
  spin-down of the star}},
\newblock \bibinfo{journal}{\apj} \bibinfo{volume}{217} (\bibinfo{year}{1977})
  \bibinfo{pages}{578--596}.
\bibitem[{{Ghosh} and {Lamb}(1978)}]{1978ApJ...223L..83G}
\bibinfo{author}{P.~{Ghosh}}, \bibinfo{author}{F.~K. {Lamb}},
\newblock \bibinfo{title}{{Disk accretion by magnetic neutron stars}},
\newblock \bibinfo{journal}{\apjl} \bibinfo{volume}{223} (\bibinfo{year}{1978})
  \bibinfo{pages}{L83--L87}.
\bibitem[{{Ghosh} and {Lamb}(1979{\natexlab{a}})}]{1979ApJ...232..259G}
\bibinfo{author}{P.~{Ghosh}}, \bibinfo{author}{F.~K. {Lamb}},
\newblock \bibinfo{title}{{Accretion by rotating magnetic neutron stars. II -
  Radial and vertical structure of the transition zone in disk accretion}},
\newblock \bibinfo{journal}{\apj} \bibinfo{volume}{232}
  (\bibinfo{year}{1979}{\natexlab{a}}) \bibinfo{pages}{259--276}.
\bibitem[{{Ghosh} and {Lamb}(1979{\natexlab{b}})}]{1979ApJ...234..296G}
\bibinfo{author}{P.~{Ghosh}}, \bibinfo{author}{F.~K. {Lamb}},
\newblock \bibinfo{title}{{Accretion by rotating magnetic neutron stars. III -
  Accretion torques and period changes in pulsating X-ray sources}},
\newblock \bibinfo{journal}{\apj} \bibinfo{volume}{234}
  (\bibinfo{year}{1979}{\natexlab{b}}) \bibinfo{pages}{296--316}.
\bibitem[{{Shu} et~al.(1994{\natexlab{a}}){Shu}, {Najita}, {Ostriker},
  {Wilkin}, {Ruden}, and {Lizano}}]{1994ApJ...429..781S}
\bibinfo{author}{F.~{Shu}}, \bibinfo{author}{J.~{Najita}},
  \bibinfo{author}{E.~{Ostriker}}, \bibinfo{author}{F.~{Wilkin}},
  \bibinfo{author}{S.~{Ruden}}, \bibinfo{author}{S.~{Lizano}},
\newblock \bibinfo{title}{{Magnetocentrifugally driven flows from young stars
  and disks. 1: A generalized model}},
\newblock \bibinfo{journal}{\apj} \bibinfo{volume}{429}
  (\bibinfo{year}{1994}{\natexlab{a}}) \bibinfo{pages}{781--796}.
\bibitem[{{Shu} et~al.(1994{\natexlab{b}}){Shu}, {Najita}, {Ruden}, and
  {Lizano}}]{1994ApJ...429..797S}
\bibinfo{author}{F.~H. {Shu}}, \bibinfo{author}{J.~{Najita}},
  \bibinfo{author}{S.~P. {Ruden}}, \bibinfo{author}{S.~{Lizano}},
\newblock \bibinfo{title}{{Magnetocentrifugally driven flows from young stars
  and disks. 2: Formulation of the dynamical problem}},
\newblock \bibinfo{journal}{\apj} \bibinfo{volume}{429}
  (\bibinfo{year}{1994}{\natexlab{b}}) \bibinfo{pages}{797--807}.
\bibitem[{{Najita} and {Shu}(1994)}]{1994ApJ...429..808N}
\bibinfo{author}{J.~R. {Najita}}, \bibinfo{author}{F.~H. {Shu}},
\newblock \bibinfo{title}{{Magnetocentrifugally driven flows from young stars
  and disks. 3: Numerical solution of the sub-Alfvenic region}},
\newblock \bibinfo{journal}{\apj} \bibinfo{volume}{429} (\bibinfo{year}{1994})
  \bibinfo{pages}{808--825}.
\bibitem[{{Backer} et~al.(1982){Backer}, {Kulkarni}, {Heiles}, {Davis}, and
  {Goss}}]{1982Natur.300..615B}
\bibinfo{author}{D.~C. {Backer}}, \bibinfo{author}{S.~R. {Kulkarni}},
  \bibinfo{author}{C.~{Heiles}}, \bibinfo{author}{M.~M. {Davis}},
  \bibinfo{author}{W.~M. {Goss}},
\newblock \bibinfo{title}{{A millisecond pulsar}},
\newblock \bibinfo{journal}{\nat} \bibinfo{volume}{300} (\bibinfo{year}{1982})
  \bibinfo{pages}{615--618}.
\bibitem[{{Radhakrishnan} and {Srinivasan}(1982)}]{1982CSci...51.1096R}
\bibinfo{author}{V.~{Radhakrishnan}}, \bibinfo{author}{G.~{Srinivasan}},
\newblock \bibinfo{title}{{On the origin of the recently discovered ultra-rapid
  pulsar}},
\newblock \bibinfo{journal}{Current Science} \bibinfo{volume}{51}
  (\bibinfo{year}{1982}) \bibinfo{pages}{1096--1099}.
\bibitem[{{Alpar} et~al.(1982){Alpar}, {Cheng}, {Ruderman}, and
  {Shaham}}]{1982Natur.300..728A}
\bibinfo{author}{M.~A. {Alpar}}, \bibinfo{author}{A.~F. {Cheng}},
  \bibinfo{author}{M.~A. {Ruderman}}, \bibinfo{author}{J.~{Shaham}},
\newblock \bibinfo{title}{{A new class of radio pulsars}},
\newblock \bibinfo{journal}{\nat} \bibinfo{volume}{300} (\bibinfo{year}{1982})
  \bibinfo{pages}{728--730}.
\bibitem[{{Wijnands} and {van der Klis}(1998)}]{1998Natur.394..344W}
\bibinfo{author}{R.~{Wijnands}}, \bibinfo{author}{M.~{van der Klis}},
\newblock \bibinfo{title}{{A millisecond pulsar in an X-ray binary system}},
\newblock \bibinfo{journal}{\nat} \bibinfo{volume}{394} (\bibinfo{year}{1998})
  \bibinfo{pages}{344--346}.
\bibitem[{{Chakrabarty} and {Morgan}(1998)}]{1998Natur.394..346C}
\bibinfo{author}{D.~{Chakrabarty}}, \bibinfo{author}{E.~H. {Morgan}},
\newblock \bibinfo{title}{{The two-hour orbit of a binary millisecond X-ray
  pulsar}},
\newblock \bibinfo{journal}{\nat} \bibinfo{volume}{394} (\bibinfo{year}{1998})
  \bibinfo{pages}{346--348}.
\bibitem[{{Chakrabarty} et~al.(2003){Chakrabarty}, {Morgan}, {Muno},
  {Galloway}, {Wijnands}, {van der Klis}, and
  {Markwardt}}]{2003Natur.424...42C}
\bibinfo{author}{D.~{Chakrabarty}}, \bibinfo{author}{E.~H. {Morgan}},
  \bibinfo{author}{M.~P. {Muno}}, \bibinfo{author}{D.~K. {Galloway}},
  \bibinfo{author}{R.~{Wijnands}}, \bibinfo{author}{M.~{van der Klis}},
  \bibinfo{author}{C.~B. {Markwardt}},
\newblock \bibinfo{title}{{Nuclear-powered millisecond pulsars and the maximum
  spin frequency of neutron stars}},
\newblock \bibinfo{journal}{\nat} \bibinfo{volume}{424} (\bibinfo{year}{2003})
  \bibinfo{pages}{42--44}.
\bibitem[{{Papitto} et~al.(2013){Papitto}, {Ferrigno}, {Bozzo}, {Rea}, {Pavan},
  {Burderi}, {Burgay}, {Campana}, {di Salvo}, {Falanga}, {Filipovi{\'c}},
  {Freire}, {Hessels}, {Possenti}, {Ransom}, {Riggio}, {Romano}, {Sarkissian},
  {Stairs}, {Stella}, {Torres}, {Wieringa}, and {Wong}}]{2013Natur.501..517P}
\bibinfo{author}{A.~{Papitto}}, \bibinfo{author}{C.~{Ferrigno}},
  \bibinfo{author}{E.~{Bozzo}}, \bibinfo{author}{N.~{Rea}},
  \bibinfo{author}{L.~{Pavan}}, \bibinfo{author}{L.~{Burderi}},
  \bibinfo{author}{M.~{Burgay}}, \bibinfo{author}{S.~{Campana}},
  \bibinfo{author}{T.~{di Salvo}}, \bibinfo{author}{M.~{Falanga}},
  \bibinfo{author}{M.~D. {Filipovi{\'c}}}, \bibinfo{author}{P.~C.~C. {Freire}},
  \bibinfo{author}{J.~W.~T. {Hessels}}, \bibinfo{author}{A.~{Possenti}},
  \bibinfo{author}{S.~M. {Ransom}}, \bibinfo{author}{A.~{Riggio}},
  \bibinfo{author}{P.~{Romano}}, \bibinfo{author}{J.~M. {Sarkissian}},
  \bibinfo{author}{I.~H. {Stairs}}, \bibinfo{author}{L.~{Stella}},
  \bibinfo{author}{D.~F. {Torres}}, \bibinfo{author}{M.~H. {Wieringa}},
  \bibinfo{author}{G.~F. {Wong}},
\newblock \bibinfo{title}{{Swings between rotation and accretion power in a
  binary millisecond pulsar}},
\newblock \bibinfo{journal}{\nat} \bibinfo{volume}{501} (\bibinfo{year}{2013})
  \bibinfo{pages}{517--520}.
\bibitem[{{Patruno} et~al.(2014){Patruno}, {Archibald}, {Hessels}, {Bogdanov},
  {Stappers}, {Bassa}, {Janssen}, {Kaspi}, {Tendulkar}, and
  {Lyne}}]{2014ApJ...781L...3P}
\bibinfo{author}{A.~{Patruno}}, \bibinfo{author}{A.~M. {Archibald}},
  \bibinfo{author}{J.~W.~T. {Hessels}}, \bibinfo{author}{S.~{Bogdanov}},
  \bibinfo{author}{B.~W. {Stappers}}, \bibinfo{author}{C.~G. {Bassa}},
  \bibinfo{author}{G.~H. {Janssen}}, \bibinfo{author}{V.~M. {Kaspi}},
  \bibinfo{author}{S.~{Tendulkar}}, \bibinfo{author}{A.~G. {Lyne}},
\newblock \bibinfo{title}{{A New Accretion Disk around the Missing Link Binary
  System PSR J1023+0038}},
\newblock \bibinfo{journal}{\apjl} \bibinfo{volume}{781} (\bibinfo{year}{2014})
  \bibinfo{pages}{L3}.
\bibitem[{{Bogdanov} et~al.(2014){Bogdanov}, {Patruno}, {Archibald}, {Bassa},
  {Hessels}, {Janssen}, and {Stappers}}]{2014ApJ...789...40B}
\bibinfo{author}{S.~{Bogdanov}}, \bibinfo{author}{A.~{Patruno}},
  \bibinfo{author}{A.~M. {Archibald}}, \bibinfo{author}{C.~{Bassa}},
  \bibinfo{author}{J.~W.~T. {Hessels}}, \bibinfo{author}{G.~H. {Janssen}},
  \bibinfo{author}{B.~W. {Stappers}},
\newblock \bibinfo{title}{{X-Ray Observations of XSS J12270-4859 in a New Low
  State: A Transformation to a Disk-free Rotation-powered Pulsar Binary}},
\newblock \bibinfo{journal}{\apj} \bibinfo{volume}{789} (\bibinfo{year}{2014})
  \bibinfo{pages}{40}.
\bibitem[{{Casella} et~al.(2008){Casella}, {Altamirano}, {Patruno}, {Wijnands},
  and {van der Klis}}]{2008ApJ...674L..41C}
\bibinfo{author}{P.~{Casella}}, \bibinfo{author}{D.~{Altamirano}},
  \bibinfo{author}{A.~{Patruno}}, \bibinfo{author}{R.~{Wijnands}},
  \bibinfo{author}{M.~{van der Klis}},
\newblock \bibinfo{title}{{Discovery of Coherent Millisecond X-Ray Pulsations
  in Aquila X-1}},
\newblock \bibinfo{journal}{\apjl} \bibinfo{volume}{674} (\bibinfo{year}{2008})
  \bibinfo{pages}{L41--L44}.
\bibitem[{{Lamb} et~al.(2009{\natexlab{a}}){Lamb}, {Boutloukos}, {Van
  Wassenhove}, {Chamberlain}, {Lo}, and {Miller}}]{2009ApJ...705L..36L}
\bibinfo{author}{F.~K. {Lamb}}, \bibinfo{author}{S.~{Boutloukos}},
  \bibinfo{author}{S.~{Van Wassenhove}}, \bibinfo{author}{R.~T. {Chamberlain}},
  \bibinfo{author}{K.~H. {Lo}}, \bibinfo{author}{M.~C. {Miller}},
\newblock \bibinfo{title}{{Origin of Intermittent Accretion-Powered X-ray
  Oscillations in Neutron Stars with Millisecond Spin Periods}},
\newblock \bibinfo{journal}{\apjl} \bibinfo{volume}{705}
  (\bibinfo{year}{2009}{\natexlab{a}}) \bibinfo{pages}{L36--L39}.
\bibitem[{{Lamb} et~al.(2009{\natexlab{b}}){Lamb}, {Boutloukos}, {Van
  Wassenhove}, {Chamberlain}, {Lo}, {Clare}, {Yu}, and
  {Miller}}]{2009ApJ...706..417L}
\bibinfo{author}{F.~K. {Lamb}}, \bibinfo{author}{S.~{Boutloukos}},
  \bibinfo{author}{S.~{Van Wassenhove}}, \bibinfo{author}{R.~T. {Chamberlain}},
  \bibinfo{author}{K.~H. {Lo}}, \bibinfo{author}{A.~{Clare}},
  \bibinfo{author}{W.~{Yu}}, \bibinfo{author}{M.~C. {Miller}},
\newblock \bibinfo{title}{{A Model for the Waveform Behavior of Accreting
  Millisecond X-Ray Pulsars: Nearly Aligned Magnetic Fields and Moving Emission
  Regions}},
\newblock \bibinfo{journal}{\apj} \bibinfo{volume}{706}
  (\bibinfo{year}{2009}{\natexlab{b}}) \bibinfo{pages}{417--435}.
\bibitem[{{Taam} and {van den Heuvel}(1986)}]{1986ApJ...305..235T}
\bibinfo{author}{R.~E. {Taam}}, \bibinfo{author}{E.~P.~J. {van den Heuvel}},
\newblock \bibinfo{title}{{Magnetic field decay and the origin of neutron star
  binaries}},
\newblock \bibinfo{journal}{\apj} \bibinfo{volume}{305} (\bibinfo{year}{1986})
  \bibinfo{pages}{235--245}.
\bibitem[{{Arzoumanian} et~al.(1999){Arzoumanian}, {Cordes}, and
  {Wasserman}}]{1999ApJ...520..696A}
\bibinfo{author}{Z.~{Arzoumanian}}, \bibinfo{author}{J.~M. {Cordes}},
  \bibinfo{author}{I.~{Wasserman}},
\newblock \bibinfo{title}{{Pulsar Spin Evolution, Kinematics, and the Birthrate
  of Neutron Star Binaries}},
\newblock \bibinfo{journal}{\apj} \bibinfo{volume}{520} (\bibinfo{year}{1999})
  \bibinfo{pages}{696--705}.
\bibitem[{{Tauris} et~al.(2012){Tauris}, {Langer}, and
  {Kramer}}]{2012MNRAS.425.1601T}
\bibinfo{author}{T.~M. {Tauris}}, \bibinfo{author}{N.~{Langer}},
  \bibinfo{author}{M.~{Kramer}},
\newblock \bibinfo{title}{{Formation of millisecond pulsars with CO white dwarf
  companions - II. Accretion, spin-up, true ages and comparison to MSPs with He
  white dwarf companions}},
\newblock \bibinfo{journal}{\mnras} \bibinfo{volume}{425}
  (\bibinfo{year}{2012}) \bibinfo{pages}{1601--1627}.
\bibitem[{{Wang} et~al.(2006){Wang}, {Chakrabarty}, and
  {Kaplan}}]{2006Natur.440..772W}
\bibinfo{author}{Z.~{Wang}}, \bibinfo{author}{D.~{Chakrabarty}},
  \bibinfo{author}{D.~L. {Kaplan}},
\newblock \bibinfo{title}{{A debris disk around an isolated young neutron
  star}},
\newblock \bibinfo{journal}{\nat} \bibinfo{volume}{440} (\bibinfo{year}{2006})
  \bibinfo{pages}{772--775}.
\bibitem[{{Strohmayer} and {Ibrahim}(2000)}]{2000ApJ...537L.111S}
\bibinfo{author}{T.~E. {Strohmayer}}, \bibinfo{author}{A.~I. {Ibrahim}},
\newblock \bibinfo{title}{{Discovery of a 6.4 KEV Emission Line in a Burst from
  SGR 1900+14}},
\newblock \bibinfo{journal}{\apjl} \bibinfo{volume}{537} (\bibinfo{year}{2000})
  \bibinfo{pages}{L111--L114}.
\bibitem[{{Endal} and {Sofia}(1978)}]{1978ApJ...220..279E}
\bibinfo{author}{A.~S. {Endal}}, \bibinfo{author}{S.~{Sofia}},
\newblock \bibinfo{title}{{The evolution of rotating stars. II - Calculations
  with time-dependent redistribution of angular momentum for 7- and
  10-solar-mass stars}},
\newblock \bibinfo{journal}{\apj} \bibinfo{volume}{220} (\bibinfo{year}{1978})
  \bibinfo{pages}{279--290}.
\bibitem[{{Miller} et~al.(2011){Miller}, {Miller}, and
  {Reynolds}}]{2011ApJ...731L...5M}
\bibinfo{author}{J.~M. {Miller}}, \bibinfo{author}{M.~C. {Miller}},
  \bibinfo{author}{C.~S. {Reynolds}},
\newblock \bibinfo{title}{{The Angular Momenta of Neutron Stars and Black Holes
  as a Window on Supernovae}},
\newblock \bibinfo{journal}{\apjl} \bibinfo{volume}{731} (\bibinfo{year}{2011})
  \bibinfo{pages}{L5}.
\bibitem[{{Kaspi} and {Helfand}(2002)}]{2002ASPC..271....3K}
\bibinfo{author}{V.~M. {Kaspi}}, \bibinfo{author}{D.~J. {Helfand}},
\newblock \bibinfo{title}{{Constraining the Birth Events of Neutron Stars}},
\newblock in: \bibinfo{editor}{P.~O. {Slane}}, \bibinfo{editor}{B.~M.
  {Gaensler}} (Eds.), \bibinfo{booktitle}{Neutron Stars in Supernova Remnants},
  volume \bibinfo{volume}{271} of \textit{\bibinfo{series}{Astronomical Society
  of the Pacific Conference Series}}, p.~\bibinfo{pages}{3}.
\bibitem[{{Gunn} and {Ostriker}(1970)}]{1970ApJ...160..979G}
\bibinfo{author}{J.~E. {Gunn}}, \bibinfo{author}{J.~P. {Ostriker}},
\newblock \bibinfo{title}{{On the Nature of Pulsars. III. Analysis of
  Observations}},
\newblock \bibinfo{journal}{\apj} \bibinfo{volume}{160} (\bibinfo{year}{1970})
  \bibinfo{pages}{979}.
\bibitem[{{Narayan} and {Ostriker}(1990)}]{1990ApJ...352..222N}
\bibinfo{author}{R.~{Narayan}}, \bibinfo{author}{J.~P. {Ostriker}},
\newblock \bibinfo{title}{{Pulsar populations and their evolution}},
\newblock \bibinfo{journal}{\apj} \bibinfo{volume}{352} (\bibinfo{year}{1990})
  \bibinfo{pages}{222--246}.
\bibitem[{{Gonthier} et~al.(2004){Gonthier}, {Van Guilder}, and
  {Harding}}]{2004ApJ...604..775G}
\bibinfo{author}{P.~L. {Gonthier}}, \bibinfo{author}{R.~{Van Guilder}},
  \bibinfo{author}{A.~K. {Harding}},
\newblock \bibinfo{title}{{Role of Beam Geometry in Population Statistics and
  Pulse Profiles of Radio and Gamma-Ray Pulsars}},
\newblock \bibinfo{journal}{\apj} \bibinfo{volume}{604} (\bibinfo{year}{2004})
  \bibinfo{pages}{775--790}.
\bibitem[{{Baym} et~al.(1969){Baym}, {Pethick}, and
  {Pikes}}]{1969Natur.224..674B}
\bibinfo{author}{G.~{Baym}}, \bibinfo{author}{C.~{Pethick}},
  \bibinfo{author}{D.~{Pikes}},
\newblock \bibinfo{title}{{Electrical Conductivity of Neutron Star Matter}},
\newblock \bibinfo{journal}{\nat} \bibinfo{volume}{224} (\bibinfo{year}{1969})
  \bibinfo{pages}{674--675}.
\bibitem[{{Goldreich} and {Reisenegger}(1992)}]{1992ApJ...395..250G}
\bibinfo{author}{P.~{Goldreich}}, \bibinfo{author}{A.~{Reisenegger}},
\newblock \bibinfo{title}{{Magnetic field decay in isolated neutron stars}},
\newblock \bibinfo{journal}{\apj} \bibinfo{volume}{395} (\bibinfo{year}{1992})
  \bibinfo{pages}{250--258}.
\bibitem[{{Stollman}(1987)}]{1987A&A...178..143S}
\bibinfo{author}{G.~M. {Stollman}},
\newblock \bibinfo{title}{{Pulsar statistics}},
\newblock \bibinfo{journal}{\aap} \bibinfo{volume}{178} (\bibinfo{year}{1987})
  \bibinfo{pages}{143--152}.
\bibitem[{{Bhattacharya} et~al.(1992){Bhattacharya}, {Wijers}, {Hartman}, and
  {Verbunt}}]{1992A&A...254..198B}
\bibinfo{author}{D.~{Bhattacharya}}, \bibinfo{author}{R.~A.~M.~J. {Wijers}},
  \bibinfo{author}{J.~W. {Hartman}}, \bibinfo{author}{F.~{Verbunt}},
\newblock \bibinfo{title}{{On the decay of the magnetic fields of single radio
  pulsars}},
\newblock \bibinfo{journal}{\aap} \bibinfo{volume}{254} (\bibinfo{year}{1992})
  \bibinfo{pages}{198--212}.
\bibitem[{{Lorimer} et~al.(1997){Lorimer}, {Bailes}, and
  {Harrison}}]{1997MNRAS.289..592L}
\bibinfo{author}{D.~R. {Lorimer}}, \bibinfo{author}{M.~{Bailes}},
  \bibinfo{author}{P.~A. {Harrison}},
\newblock \bibinfo{title}{{Pulsar statistics - IV. Pulsar velocities}},
\newblock \bibinfo{journal}{\mnras} \bibinfo{volume}{289}
  (\bibinfo{year}{1997}) \bibinfo{pages}{592--604}.
\bibitem[{{Vivekanand} and {Narayan}(1981)}]{1981JApA....2..315V}
\bibinfo{author}{M.~{Vivekanand}}, \bibinfo{author}{R.~{Narayan}},
\newblock \bibinfo{title}{{A new look at pulsar statistics - Birthrate and
  evidence for injection}},
\newblock \bibinfo{journal}{Journal of Astrophysics and Astronomy}
  \bibinfo{volume}{2} (\bibinfo{year}{1981}) \bibinfo{pages}{315--337}.
\bibitem[{{Lorimer} et~al.(1993){Lorimer}, {Bailes}, {Dewey}, and
  {Harrison}}]{1993MNRAS.263..403L}
\bibinfo{author}{D.~R. {Lorimer}}, \bibinfo{author}{M.~{Bailes}},
  \bibinfo{author}{R.~J. {Dewey}}, \bibinfo{author}{P.~A. {Harrison}},
\newblock \bibinfo{title}{{Pulsar Statistics - the Birthrate and Initial Spin
  Periods of Radio Pulsars}},
\newblock \bibinfo{journal}{\mnras} \bibinfo{volume}{263}
  (\bibinfo{year}{1993}) \bibinfo{pages}{403}.
\bibitem[{{Manchester} et~al.(2001){Manchester}, {Lyne}, {Camilo}, {Bell},
  {Kaspi}, {D'Amico}, {McKay}, {Crawford}, {Stairs}, {Possenti}, {Kramer}, and
  {Sheppard}}]{2001MNRAS.328...17M}
\bibinfo{author}{R.~N. {Manchester}}, \bibinfo{author}{A.~G. {Lyne}},
  \bibinfo{author}{F.~{Camilo}}, \bibinfo{author}{J.~F. {Bell}},
  \bibinfo{author}{V.~M. {Kaspi}}, \bibinfo{author}{N.~{D'Amico}},
  \bibinfo{author}{N.~P.~F. {McKay}}, \bibinfo{author}{F.~{Crawford}},
  \bibinfo{author}{I.~H. {Stairs}}, \bibinfo{author}{A.~{Possenti}},
  \bibinfo{author}{M.~{Kramer}}, \bibinfo{author}{D.~C. {Sheppard}},
\newblock \bibinfo{title}{{The Parkes multi-beam pulsar survey - I. Observing
  and data analysis systems, discovery and timing of 100 pulsars}},
\newblock \bibinfo{journal}{\mnras} \bibinfo{volume}{328}
  (\bibinfo{year}{2001}) \bibinfo{pages}{17--35}.
\bibitem[{{Vranesevic} et~al.(2004){Vranesevic}, {Manchester}, {Lorimer},
  {Hobbs}, {Lyne}, {Kramer}, {Camilo}, {Stairs}, {Kaspi}, {D'Amico},
  {Possenti}, {Crawford}, {Faulkner}, and {McLaughlin}}]{2004ApJ...617L.139V}
\bibinfo{author}{N.~{Vranesevic}}, \bibinfo{author}{R.~N. {Manchester}},
  \bibinfo{author}{D.~R. {Lorimer}}, \bibinfo{author}{G.~B. {Hobbs}},
  \bibinfo{author}{A.~G. {Lyne}}, \bibinfo{author}{M.~{Kramer}},
  \bibinfo{author}{F.~{Camilo}}, \bibinfo{author}{I.~H. {Stairs}},
  \bibinfo{author}{V.~M. {Kaspi}}, \bibinfo{author}{N.~{D'Amico}},
  \bibinfo{author}{A.~{Possenti}}, \bibinfo{author}{F.~{Crawford}},
  \bibinfo{author}{A.~J. {Faulkner}}, \bibinfo{author}{M.~A. {McLaughlin}},
\newblock \bibinfo{title}{{Pulsar Birthrates from the Parkes Multibeam
  Survey}},
\newblock \bibinfo{journal}{\apjl} \bibinfo{volume}{617} (\bibinfo{year}{2004})
  \bibinfo{pages}{L139--L142}.
\bibitem[{{Faucher-Gigu{\`e}re} and {Kaspi}(2006)}]{2006ApJ...643..332F}
\bibinfo{author}{C.-A. {Faucher-Gigu{\`e}re}}, \bibinfo{author}{V.~M. {Kaspi}},
\newblock \bibinfo{title}{{Birth and Evolution of Isolated Radio Pulsars}},
\newblock \bibinfo{journal}{\apj} \bibinfo{volume}{643} (\bibinfo{year}{2006})
  \bibinfo{pages}{332--355}.
\bibitem[{{Noutsos} et~al.(2013){Noutsos}, {Schnitzeler}, {Keane}, {Kramer},
  and {Johnston}}]{2013MNRAS.430.2281N}
\bibinfo{author}{A.~{Noutsos}}, \bibinfo{author}{D.~H.~F.~M. {Schnitzeler}},
  \bibinfo{author}{E.~F. {Keane}}, \bibinfo{author}{M.~{Kramer}},
  \bibinfo{author}{S.~{Johnston}},
\newblock \bibinfo{title}{{Pulsar spin-velocity alignment: kinematic ages,
  birth periods and braking indices}},
\newblock \bibinfo{journal}{\mnras} \bibinfo{volume}{430}
  (\bibinfo{year}{2013}) \bibinfo{pages}{2281--2301}.
\bibitem[{{Igoshev} and {Popov}(2013)}]{2013MNRAS.432..967I}
\bibinfo{author}{A.~P. {Igoshev}}, \bibinfo{author}{S.~B. {Popov}},
\newblock \bibinfo{title}{{Neutron star's initial spin period distribution}},
\newblock \bibinfo{journal}{\mnras} \bibinfo{volume}{432}
  (\bibinfo{year}{2013}) \bibinfo{pages}{967--972}.
\bibitem[{{Cottam} et~al.(2002){Cottam}, {Paerels}, and
  {Mendez}}]{2002Natur.420...51C}
\bibinfo{author}{J.~{Cottam}}, \bibinfo{author}{F.~{Paerels}},
  \bibinfo{author}{M.~{Mendez}},
\newblock \bibinfo{title}{{Gravitationally redshifted absorption lines in the
  X-ray burst spectra of a neutron star}},
\newblock \bibinfo{journal}{\nat} \bibinfo{volume}{420} (\bibinfo{year}{2002})
  \bibinfo{pages}{51--54}.
\bibitem[{{Cottam} et~al.(2008){Cottam}, {Paerels}, {M{\'e}ndez}, {Boirin},
  {Lewin}, {Kuulkers}, and {Miller}}]{2008ApJ...672..504C}
\bibinfo{author}{J.~{Cottam}}, \bibinfo{author}{F.~{Paerels}},
  \bibinfo{author}{M.~{M{\'e}ndez}}, \bibinfo{author}{L.~{Boirin}},
  \bibinfo{author}{W.~H.~G. {Lewin}}, \bibinfo{author}{E.~{Kuulkers}},
  \bibinfo{author}{J.~M. {Miller}},
\newblock \bibinfo{title}{{The Burst Spectra of EXO 0748-676 during a Long 2003
  XMM-Newton Observation}},
\newblock \bibinfo{journal}{\apj} \bibinfo{volume}{672} (\bibinfo{year}{2008})
  \bibinfo{pages}{504--509}.
\bibitem[{{Shakura} and {Sunyaev}(1973)}]{1973A&A....24..337S}
\bibinfo{author}{N.~I. {Shakura}}, \bibinfo{author}{R.~A. {Sunyaev}},
\newblock \bibinfo{title}{{Black holes in binary systems. Observational
  appearance.}},
\newblock \bibinfo{journal}{\aap} \bibinfo{volume}{24} (\bibinfo{year}{1973})
  \bibinfo{pages}{337--355}.
\bibitem[{{Novikov} and {Thorne}(1973)}]{1973blho.conf..343N}
\bibinfo{author}{I.~D. {Novikov}}, \bibinfo{author}{K.~S. {Thorne}},
\newblock \bibinfo{title}{{Astrophysics of black holes.}},
\newblock in: \bibinfo{editor}{C.~{Dewitt}}, \bibinfo{editor}{B.~S. {Dewitt}}
  (Eds.), \bibinfo{booktitle}{Black Holes (Les Astres Occlus)}, pp.
  \bibinfo{pages}{343--450}.
\bibitem[{{Tomsick} et~al.(2009){Tomsick}, {Yamaoka}, {Corbel}, {Kaaret},
  {Kalemci}, and {Migliari}}]{2009ApJ...707L..87T}
\bibinfo{author}{J.~A. {Tomsick}}, \bibinfo{author}{K.~{Yamaoka}},
  \bibinfo{author}{S.~{Corbel}}, \bibinfo{author}{P.~{Kaaret}},
  \bibinfo{author}{E.~{Kalemci}}, \bibinfo{author}{S.~{Migliari}},
\newblock \bibinfo{title}{{Truncation of the Inner Accretion Disk Around a
  Black Hole at Low Luminosity}},
\newblock \bibinfo{journal}{\apjl} \bibinfo{volume}{707} (\bibinfo{year}{2009})
  \bibinfo{pages}{L87--L91}.
\bibitem[{{Reis} et~al.(2010){Reis}, {Fabian}, and
  {Miller}}]{2010MNRAS.402..836R}
\bibinfo{author}{R.~C. {Reis}}, \bibinfo{author}{A.~C. {Fabian}},
  \bibinfo{author}{J.~M. {Miller}},
\newblock \bibinfo{title}{{Black hole accretion discs in the canonical low-hard
  state}},
\newblock \bibinfo{journal}{\mnras} \bibinfo{volume}{402}
  (\bibinfo{year}{2010}) \bibinfo{pages}{836--854}.
\bibitem[{{Miller} et~al.(2012){Miller}, {Pooley}, {Fabian}, {Nowak}, {Reis},
  {Cackett}, {Pottschmidt}, and {Wilms}}]{2012ApJ...757...11M}
\bibinfo{author}{J.~M. {Miller}}, \bibinfo{author}{G.~G. {Pooley}},
  \bibinfo{author}{A.~C. {Fabian}}, \bibinfo{author}{M.~A. {Nowak}},
  \bibinfo{author}{R.~C. {Reis}}, \bibinfo{author}{E.~M. {Cackett}},
  \bibinfo{author}{K.~{Pottschmidt}}, \bibinfo{author}{J.~{Wilms}},
\newblock \bibinfo{title}{{On the Role of the Accretion Disk in Black Hole
  Disk-Jet Connections}},
\newblock \bibinfo{journal}{\apj} \bibinfo{volume}{757} (\bibinfo{year}{2012})
  \bibinfo{pages}{11}.
\bibitem[{{Reynolds} and {Miller}(2013)}]{2013ApJ...769...16R}
\bibinfo{author}{M.~T. {Reynolds}}, \bibinfo{author}{J.~M. {Miller}},
\newblock \bibinfo{title}{{A Swift Survey of Accretion onto Stellar-mass Black
  Holes}},
\newblock \bibinfo{journal}{\apj} \bibinfo{volume}{769} (\bibinfo{year}{2013})
  \bibinfo{pages}{16}.
\bibitem[{{Esin} et~al.(1997){Esin}, {McClintock}, and
  {Narayan}}]{1997ApJ...489..865E}
\bibinfo{author}{A.~A. {Esin}}, \bibinfo{author}{J.~E. {McClintock}},
  \bibinfo{author}{R.~{Narayan}},
\newblock \bibinfo{title}{{Advection-dominated Accretion and the Spectral
  States of Black Hole X-Ray Binaries: Application to Nova MUSCAE 1991}},
\newblock \bibinfo{journal}{\apj} \bibinfo{volume}{489} (\bibinfo{year}{1997})
  \bibinfo{pages}{865}.
\bibitem[{{Rykoff} et~al.(2007){Rykoff}, {Miller}, {Steeghs}, and
  {Torres}}]{2007ApJ...666.1129R}
\bibinfo{author}{E.~S. {Rykoff}}, \bibinfo{author}{J.~M. {Miller}},
  \bibinfo{author}{D.~{Steeghs}}, \bibinfo{author}{M.~A.~P. {Torres}},
\newblock \bibinfo{title}{{Swift Observations of the Cooling Accretion Disk of
  XTE J1817-330}},
\newblock \bibinfo{journal}{\apj} \bibinfo{volume}{666} (\bibinfo{year}{2007})
  \bibinfo{pages}{1129--1139}.
\bibitem[{{Steiner} et~al.(2010){Steiner}, {McClintock}, {Remillard}, {Gou},
  {Yamada}, and {Narayan}}]{2010ApJ...718L.117S}
\bibinfo{author}{J.~F. {Steiner}}, \bibinfo{author}{J.~E. {McClintock}},
  \bibinfo{author}{R.~A. {Remillard}}, \bibinfo{author}{L.~{Gou}},
  \bibinfo{author}{S.~{Yamada}}, \bibinfo{author}{R.~{Narayan}},
\newblock \bibinfo{title}{{The Constant Inner-disk Radius of LMC X-3: A Basis
  for Measuring Black Hole Spin}},
\newblock \bibinfo{journal}{\apjl} \bibinfo{volume}{718} (\bibinfo{year}{2010})
  \bibinfo{pages}{L117--L121}.
\bibitem[{{Miller} et~al.(1998){Miller}, {Lamb}, and
  {Psaltis}}]{1998ApJ...508..791M}
\bibinfo{author}{M.~C. {Miller}}, \bibinfo{author}{F.~K. {Lamb}},
  \bibinfo{author}{D.~{Psaltis}},
\newblock \bibinfo{title}{{Sonic-Point Model of Kilohertz Quasi-periodic
  Brightness Oscillations in Low-Mass X-Ray Binaries}},
\newblock \bibinfo{journal}{ApJ} \bibinfo{volume}{508} (\bibinfo{year}{1998})
  \bibinfo{pages}{791--830}.
\bibitem[{{Barret} et~al.(2005{\natexlab{a}}){Barret}, {Olive}, and
  {Miller}}]{2005MNRAS.361..855B}
\bibinfo{author}{D.~{Barret}}, \bibinfo{author}{J.~{Olive}},
  \bibinfo{author}{M.~C. {Miller}},
\newblock \bibinfo{title}{{An abrupt drop in the coherence of the lower kHz
  quasi-periodic oscillations in 4U 1636-536}},
\newblock \bibinfo{journal}{MNRAS} \bibinfo{volume}{361}
  (\bibinfo{year}{2005}{\natexlab{a}}) \bibinfo{pages}{855--860}.
\bibitem[{{Barret} et~al.(2005{\natexlab{b}}){Barret}, {Olive}, and
  {Miller}}]{2005AN....326..808B}
\bibinfo{author}{D.~{Barret}}, \bibinfo{author}{J.~{Olive}},
  \bibinfo{author}{M.~C. {Miller}},
\newblock \bibinfo{title}{{Drop of coherence of the lower kilo-Hz QPO in
  neutron stars: Is there a link with the innermost stable circular orbit?}},
\newblock \bibinfo{journal}{Astronomische Nachrichten} \bibinfo{volume}{326}
  (\bibinfo{year}{2005}{\natexlab{b}}) \bibinfo{pages}{808--811}.
\bibitem[{{Barret} et~al.(2006){Barret}, {Olive}, and
  {Miller}}]{2006MNRAS.370.1140B}
\bibinfo{author}{D.~{Barret}}, \bibinfo{author}{J.~{Olive}},
  \bibinfo{author}{M.~C. {Miller}},
\newblock \bibinfo{title}{{The coherence of kilohertz quasi-periodic
  oscillations in the X-rays from accreting neutron stars}},
\newblock \bibinfo{journal}{MNRAS} \bibinfo{volume}{370} (\bibinfo{year}{2006})
  \bibinfo{pages}{1140--1146}.
\bibitem[{{Barret} et~al.(2007){Barret}, {Olive}, and
  {Miller}}]{2007MNRAS.376.1139B}
\bibinfo{author}{D.~{Barret}}, \bibinfo{author}{J.~{Olive}},
  \bibinfo{author}{M.~C. {Miller}},
\newblock \bibinfo{title}{{Supporting evidence for the signature of the
  innermost stable circular orbit in Rossi X-ray data from 4U 1636-536}},
\newblock \bibinfo{journal}{MNRAS} \bibinfo{volume}{376} (\bibinfo{year}{2007})
  \bibinfo{pages}{1139--1144}.
\bibitem[{{M{\'e}ndez}(2006)}]{2006MNRAS.371.1925M}
\bibinfo{author}{M.~{M{\'e}ndez}},
\newblock \bibinfo{title}{{On the maximum amplitude and coherence of the
  kilohertz quasi-periodic oscillations in low-mass X-ray binaries}},
\newblock \bibinfo{journal}{MNRAS} \bibinfo{volume}{371} (\bibinfo{year}{2006})
  \bibinfo{pages}{1925--1938}.
\bibitem[{{Reynolds} and {Begelman}(1997)}]{1997ApJ...488..109R}
\bibinfo{author}{C.~S. {Reynolds}}, \bibinfo{author}{M.~C. {Begelman}},
\newblock \bibinfo{title}{{Iron Fluorescence from within the Innermost Stable
  Orbit of Black Hole Accretion Disks}},
\newblock \bibinfo{journal}{\apj} \bibinfo{volume}{488} (\bibinfo{year}{1997})
  \bibinfo{pages}{109}.
\bibitem[{{Sorathia} et~al.(2010){Sorathia}, {Reynolds}, and
  {Armitage}}]{2010ApJ...712.1241S}
\bibinfo{author}{K.~A. {Sorathia}}, \bibinfo{author}{C.~S. {Reynolds}},
  \bibinfo{author}{P.~J. {Armitage}},
\newblock \bibinfo{title}{{Connections Between Local and Global Turbulence in
  Accretion Disks}},
\newblock \bibinfo{journal}{\apj} \bibinfo{volume}{712} (\bibinfo{year}{2010})
  \bibinfo{pages}{1241--1247}.
\bibitem[{{Hawley} et~al.(2011){Hawley}, {Guan}, and
  {Krolik}}]{2011ApJ...738...84H}
\bibinfo{author}{J.~F. {Hawley}}, \bibinfo{author}{X.~{Guan}},
  \bibinfo{author}{J.~H. {Krolik}},
\newblock \bibinfo{title}{{Assessing Quantitative Results in Accretion
  Simulations: From Local to Global}},
\newblock \bibinfo{journal}{\apj} \bibinfo{volume}{738} (\bibinfo{year}{2011})
  \bibinfo{pages}{84}.
\bibitem[{{Sorathia} et~al.(2012){Sorathia}, {Reynolds}, {Stone}, and
  {Beckwith}}]{2012ApJ...749..189S}
\bibinfo{author}{K.~A. {Sorathia}}, \bibinfo{author}{C.~S. {Reynolds}},
  \bibinfo{author}{J.~M. {Stone}}, \bibinfo{author}{K.~{Beckwith}},
\newblock \bibinfo{title}{{Global Simulations of Accretion Disks. I.
  Convergence and Comparisons with Local Models}},
\newblock \bibinfo{journal}{\apj} \bibinfo{volume}{749} (\bibinfo{year}{2012})
  \bibinfo{pages}{189}.
\bibitem[{{Hawley} et~al.(2013){Hawley}, {Richers}, {Guan}, and
  {Krolik}}]{2013ApJ...772..102H}
\bibinfo{author}{J.~F. {Hawley}}, \bibinfo{author}{S.~A. {Richers}},
  \bibinfo{author}{X.~{Guan}}, \bibinfo{author}{J.~H. {Krolik}},
\newblock \bibinfo{title}{{Testing Convergence for Global Accretion Disks}},
\newblock \bibinfo{journal}{\apj} \bibinfo{volume}{772} (\bibinfo{year}{2013})
  \bibinfo{pages}{102}.
\bibitem[{{Reynolds} and {Fabian}(2008)}]{2008ApJ...675.1048R}
\bibinfo{author}{C.~S. {Reynolds}}, \bibinfo{author}{A.~C. {Fabian}},
\newblock \bibinfo{title}{{Broad Iron-K{$\alpha$} Emission Lines as a
  Diagnostic of Black Hole Spin}},
\newblock \bibinfo{journal}{\apj} \bibinfo{volume}{675} (\bibinfo{year}{2008})
  \bibinfo{pages}{1048--1056}.
\bibitem[{{Penna} et~al.(2010){Penna}, {McKinney}, {Narayan}, {Tchekhovskoy},
  {Shafee}, and {McClintock}}]{2010MNRAS.408..752P}
\bibinfo{author}{R.~F. {Penna}}, \bibinfo{author}{J.~C. {McKinney}},
  \bibinfo{author}{R.~{Narayan}}, \bibinfo{author}{A.~{Tchekhovskoy}},
  \bibinfo{author}{R.~{Shafee}}, \bibinfo{author}{J.~E. {McClintock}},
\newblock \bibinfo{title}{{Simulations of magnetized discs around black holes:
  effects of black hole spin, disc thickness and magnetic field geometry}},
\newblock \bibinfo{journal}{\mnras} \bibinfo{volume}{408}
  (\bibinfo{year}{2010}) \bibinfo{pages}{752--782}.
\bibitem[{{Mitsuda} et~al.(1984){Mitsuda}, {Inoue}, {Koyama}, {Makishima},
  {Matsuoka}, {Ogawara}, {Suzuki}, {Tanaka}, {Shibazaki}, and
  {Hirano}}]{1984PASJ...36..741M}
\bibinfo{author}{K.~{Mitsuda}}, \bibinfo{author}{H.~{Inoue}},
  \bibinfo{author}{K.~{Koyama}}, \bibinfo{author}{K.~{Makishima}},
  \bibinfo{author}{M.~{Matsuoka}}, \bibinfo{author}{Y.~{Ogawara}},
  \bibinfo{author}{K.~{Suzuki}}, \bibinfo{author}{Y.~{Tanaka}},
  \bibinfo{author}{N.~{Shibazaki}}, \bibinfo{author}{T.~{Hirano}},
\newblock \bibinfo{title}{{Energy spectra of low-mass binary X-ray sources
  observed from TENMA}},
\newblock \bibinfo{journal}{\pasj} \bibinfo{volume}{36} (\bibinfo{year}{1984})
  \bibinfo{pages}{741--759}.
\bibitem[{{Davis} et~al.(2005){Davis}, {Blaes}, {Hubeny}, and
  {Turner}}]{2005ApJ...621..372D}
\bibinfo{author}{S.~W. {Davis}}, \bibinfo{author}{O.~M. {Blaes}},
  \bibinfo{author}{I.~{Hubeny}}, \bibinfo{author}{N.~J. {Turner}},
\newblock \bibinfo{title}{{Relativistic Accretion Disk Models of High-State
  Black Hole X-Ray Binary Spectra}},
\newblock \bibinfo{journal}{\apj} \bibinfo{volume}{621} (\bibinfo{year}{2005})
  \bibinfo{pages}{372--387}.
\bibitem[{{Wambsganss} et~al.(1990){Wambsganss}, {Paczynski}, and
  {Schneider}}]{1990ApJ...358L..33W}
\bibinfo{author}{J.~{Wambsganss}}, \bibinfo{author}{B.~{Paczynski}},
  \bibinfo{author}{P.~{Schneider}},
\newblock \bibinfo{title}{{Interpretation of the microlensing event in QSO 2237
  + 0305}},
\newblock \bibinfo{journal}{\apjl} \bibinfo{volume}{358} (\bibinfo{year}{1990})
  \bibinfo{pages}{L33--L36}.
\bibitem[{{Rauch} and {Blandford}(1991)}]{1991ApJ...381L..39R}
\bibinfo{author}{K.~P. {Rauch}}, \bibinfo{author}{R.~D. {Blandford}},
\newblock \bibinfo{title}{{Microlensing and the structure of active galactic
  nucleus accretion disks}},
\newblock \bibinfo{journal}{\apjl} \bibinfo{volume}{381} (\bibinfo{year}{1991})
  \bibinfo{pages}{L39--L42}.
\bibitem[{{Pooley} et~al.(2007){Pooley}, {Blackburne}, {Rappaport}, and
  {Schechter}}]{2007ApJ...661...19P}
\bibinfo{author}{D.~{Pooley}}, \bibinfo{author}{J.~A. {Blackburne}},
  \bibinfo{author}{S.~{Rappaport}}, \bibinfo{author}{P.~L. {Schechter}},
\newblock \bibinfo{title}{{X-Ray and Optical Flux Ratio Anomalies in Quadruply
  Lensed Quasars. I. Zooming in on Quasar Emission Regions}},
\newblock \bibinfo{journal}{\apj} \bibinfo{volume}{661} (\bibinfo{year}{2007})
  \bibinfo{pages}{19--29}.
\bibitem[{{Dai} et~al.(2010){Dai}, {Kochanek}, {Chartas}, {Koz{\l}owski},
  {Morgan}, {Garmire}, and {Agol}}]{2010ApJ...709..278D}
\bibinfo{author}{X.~{Dai}}, \bibinfo{author}{C.~S. {Kochanek}},
  \bibinfo{author}{G.~{Chartas}}, \bibinfo{author}{S.~{Koz{\l}owski}},
  \bibinfo{author}{C.~W. {Morgan}}, \bibinfo{author}{G.~{Garmire}},
  \bibinfo{author}{E.~{Agol}},
\newblock \bibinfo{title}{{The Sizes of the X-ray and Optical Emission Regions
  of RXJ 1131-1231}},
\newblock \bibinfo{journal}{\apj} \bibinfo{volume}{709} (\bibinfo{year}{2010})
  \bibinfo{pages}{278--285}.
\bibitem[{{Morgan} et~al.(2010){Morgan}, {Kochanek}, {Morgan}, and
  {Falco}}]{2010ApJ...712.1129M}
\bibinfo{author}{C.~W. {Morgan}}, \bibinfo{author}{C.~S. {Kochanek}},
  \bibinfo{author}{N.~D. {Morgan}}, \bibinfo{author}{E.~E. {Falco}},
\newblock \bibinfo{title}{{The Quasar Accretion Disk Size-Black Hole Mass
  Relation}},
\newblock \bibinfo{journal}{\apj} \bibinfo{volume}{712} (\bibinfo{year}{2010})
  \bibinfo{pages}{1129--1136}.
\bibitem[{{Blackburne} and {Kochanek}(2010)}]{2010ApJ...718.1079B}
\bibinfo{author}{J.~A. {Blackburne}}, \bibinfo{author}{C.~S. {Kochanek}},
\newblock \bibinfo{title}{{The Effect of a Time-varying Accretion Disk Size on
  Quasar Microlensing Light Curves}},
\newblock \bibinfo{journal}{\apj} \bibinfo{volume}{718} (\bibinfo{year}{2010})
  \bibinfo{pages}{1079--1084}.
\bibitem[{{Jim{\'e}nez-Vicente} et~al.(2012){Jim{\'e}nez-Vicente},
  {Mediavilla}, {Mu{\~n}oz}, and {Kochanek}}]{2012ApJ...751..106J}
\bibinfo{author}{J.~{Jim{\'e}nez-Vicente}}, \bibinfo{author}{E.~{Mediavilla}},
  \bibinfo{author}{J.~A. {Mu{\~n}oz}}, \bibinfo{author}{C.~S. {Kochanek}},
\newblock \bibinfo{title}{{A Robust Determination of the Size of Quasar
  Accretion Disks Using Gravitational Microlensing}},
\newblock \bibinfo{journal}{\apj} \bibinfo{volume}{751} (\bibinfo{year}{2012})
  \bibinfo{pages}{106}.
\bibitem[{{Jim{\'e}nez-Vicente} et~al.(2014){Jim{\'e}nez-Vicente},
  {Mediavilla}, {Kochanek}, {Mu{\~n}oz}, {Motta}, {Falco}, and
  {Mosquera}}]{2014ApJ...783...47J}
\bibinfo{author}{J.~{Jim{\'e}nez-Vicente}}, \bibinfo{author}{E.~{Mediavilla}},
  \bibinfo{author}{C.~S. {Kochanek}}, \bibinfo{author}{J.~A. {Mu{\~n}oz}},
  \bibinfo{author}{V.~{Motta}}, \bibinfo{author}{E.~{Falco}},
  \bibinfo{author}{A.~M. {Mosquera}},
\newblock \bibinfo{title}{{The Average Size and Temperature Profile of Quasar
  Accretion Disks}},
\newblock \bibinfo{journal}{\apj} \bibinfo{volume}{783} (\bibinfo{year}{2014})
  \bibinfo{pages}{47}.
\bibitem[{{Dexter} and {Agol}(2011)}]{2011ApJ...727L..24D}
\bibinfo{author}{J.~{Dexter}}, \bibinfo{author}{E.~{Agol}},
\newblock \bibinfo{title}{{Quasar Accretion Disks are Strongly Inhomogeneous}},
\newblock \bibinfo{journal}{\apjl} \bibinfo{volume}{727} (\bibinfo{year}{2011})
  \bibinfo{pages}{L24}.
\bibitem[{{Dexter} and {Quataert}(2012)}]{2012MNRAS.426L..71D}
\bibinfo{author}{J.~{Dexter}}, \bibinfo{author}{E.~{Quataert}},
\newblock \bibinfo{title}{{Inhomogeneous accretion discs and the soft states of
  black hole X-ray binaries}},
\newblock \bibinfo{journal}{\mnras} \bibinfo{volume}{426}
  (\bibinfo{year}{2012}) \bibinfo{pages}{L71--L75}.
\bibitem[{{Ruan} et~al.(2014){Ruan}, {Anderson}, {Dexter}, and
  {Agol}}]{2014ApJ...783..105R}
\bibinfo{author}{J.~J. {Ruan}}, \bibinfo{author}{S.~F. {Anderson}},
  \bibinfo{author}{J.~{Dexter}}, \bibinfo{author}{E.~{Agol}},
\newblock \bibinfo{title}{{Evidence for Large Temperature Fluctuations in
  Quasar Accretion Disks from Spectral Variability}},
\newblock \bibinfo{journal}{\apj} \bibinfo{volume}{783} (\bibinfo{year}{2014})
  \bibinfo{pages}{105}.
\bibitem[{{Jiang} et~al.(2013){Jiang}, {Stone}, and
  {Davis}}]{2013ApJ...778...65J}
\bibinfo{author}{Y.-F. {Jiang}}, \bibinfo{author}{J.~M. {Stone}},
  \bibinfo{author}{S.~W. {Davis}},
\newblock \bibinfo{title}{{On the Thermal Stability of Radiation-dominated
  Accretion Disks}},
\newblock \bibinfo{journal}{\apj} \bibinfo{volume}{778} (\bibinfo{year}{2013})
  \bibinfo{pages}{65}.
\bibitem[{{Davis} et~al.(2006){Davis}, {Done}, and
  {Blaes}}]{2006ApJ...647..525D}
\bibinfo{author}{S.~W. {Davis}}, \bibinfo{author}{C.~{Done}},
  \bibinfo{author}{O.~M. {Blaes}},
\newblock \bibinfo{title}{{Testing Accretion Disk Theory in Black Hole X-Ray
  Binaries}},
\newblock \bibinfo{journal}{\apj} \bibinfo{volume}{647} (\bibinfo{year}{2006})
  \bibinfo{pages}{525--538}.
\bibitem[{{Maccarone}(2002)}]{2002MNRAS.336.1371M}
\bibinfo{author}{T.~J. {Maccarone}},
\newblock \bibinfo{title}{{On the misalignment of jets in microquasars}},
\newblock \bibinfo{journal}{\mnras} \bibinfo{volume}{336}
  (\bibinfo{year}{2002}) \bibinfo{pages}{1371--1376}.
\bibitem[{{Martin} et~al.(2008){Martin}, {Tout}, and
  {Pringle}}]{2008MNRAS.387..188M}
\bibinfo{author}{R.~G. {Martin}}, \bibinfo{author}{C.~A. {Tout}},
  \bibinfo{author}{J.~E. {Pringle}},
\newblock \bibinfo{title}{{Alignment time-scale of the microquasar GRO
  J1655-40}},
\newblock \bibinfo{journal}{\mnras} \bibinfo{volume}{387}
  (\bibinfo{year}{2008}) \bibinfo{pages}{188--196}.
\bibitem[{{Steiner} and {McClintock}(2012)}]{2012ApJ...745..136S}
\bibinfo{author}{J.~F. {Steiner}}, \bibinfo{author}{J.~E. {McClintock}},
\newblock \bibinfo{title}{{Modeling the Jet Kinematics of the Black Hole
  Microquasar XTE J1550-564: A Constraint on Spin-Orbit Alignment}},
\newblock \bibinfo{journal}{\apj} \bibinfo{volume}{745} (\bibinfo{year}{2012})
  \bibinfo{pages}{136}.
\bibitem[{{Hjellming} and {Rupen}(1995)}]{1995Natur.375..464H}
\bibinfo{author}{R.~M. {Hjellming}}, \bibinfo{author}{M.~P. {Rupen}},
\newblock \bibinfo{title}{{Episodic ejection of relativistic jets by the X-ray
  transient GRO J1655 - 40}},
\newblock \bibinfo{journal}{\nat} \bibinfo{volume}{375} (\bibinfo{year}{1995})
  \bibinfo{pages}{464--468}.
\bibitem[{{Orosz} et~al.(2001){Orosz}, {Kuulkers}, {van der Klis},
  {McClintock}, {Garcia}, {Callanan}, {Bailyn}, {Jain}, and
  {Remillard}}]{2001ApJ...555..489O}
\bibinfo{author}{J.~A. {Orosz}}, \bibinfo{author}{E.~{Kuulkers}},
  \bibinfo{author}{M.~{van der Klis}}, \bibinfo{author}{J.~E. {McClintock}},
  \bibinfo{author}{M.~R. {Garcia}}, \bibinfo{author}{P.~J. {Callanan}},
  \bibinfo{author}{C.~D. {Bailyn}}, \bibinfo{author}{R.~K. {Jain}},
  \bibinfo{author}{R.~A. {Remillard}},
\newblock \bibinfo{title}{{A Black Hole in the Superluminal Source SAX
  J1819.3-2525 (V4641 Sgr)}},
\newblock \bibinfo{journal}{\apj} \bibinfo{volume}{555} (\bibinfo{year}{2001})
  \bibinfo{pages}{489--503}.
\bibitem[{{Fragile} et~al.(2001){Fragile}, {Mathews}, and
  {Wilson}}]{2001ApJ...553..955F}
\bibinfo{author}{P.~C. {Fragile}}, \bibinfo{author}{G.~J. {Mathews}},
  \bibinfo{author}{J.~R. {Wilson}},
\newblock \bibinfo{title}{{Bardeen-Petterson Effect and Quasi-periodic
  Oscillations in X-Ray Binaries}},
\newblock \bibinfo{journal}{\apj} \bibinfo{volume}{553} (\bibinfo{year}{2001})
  \bibinfo{pages}{955--959}.
\bibitem[{{Ingram} et~al.(2009){Ingram}, {Done}, and
  {Fragile}}]{2009MNRAS.397L.101I}
\bibinfo{author}{A.~{Ingram}}, \bibinfo{author}{C.~{Done}},
  \bibinfo{author}{P.~C. {Fragile}},
\newblock \bibinfo{title}{{Low-frequency quasi-periodic oscillations spectra
  and Lense-Thirring precession}},
\newblock \bibinfo{journal}{\mnras} \bibinfo{volume}{397}
  (\bibinfo{year}{2009}) \bibinfo{pages}{L101--L105}.
\bibitem[{{Dexter} and {Fragile}(2011)}]{2011ApJ...730...36D}
\bibinfo{author}{J.~{Dexter}}, \bibinfo{author}{P.~C. {Fragile}},
\newblock \bibinfo{title}{{Observational Signatures of Tilted Black Hole
  Accretion Disks from Simulations}},
\newblock \bibinfo{journal}{\apj} \bibinfo{volume}{730} (\bibinfo{year}{2011})
  \bibinfo{pages}{36}.
\bibitem[{{White} et~al.(1988){White}, {Lightman}, and
  {Zdziarski}}]{1988ApJ...331..939W}
\bibinfo{author}{T.~R. {White}}, \bibinfo{author}{A.~P. {Lightman}},
  \bibinfo{author}{A.~A. {Zdziarski}},
\newblock \bibinfo{title}{{Compton reflection of gamma rays by cold
  electrons}},
\newblock \bibinfo{journal}{\apj} \bibinfo{volume}{331} (\bibinfo{year}{1988})
  \bibinfo{pages}{939--948}.
\bibitem[{{George} and {Fabian}(1991)}]{1991MNRAS.249..352G}
\bibinfo{author}{I.~M. {George}}, \bibinfo{author}{A.~C. {Fabian}},
\newblock \bibinfo{title}{{X-ray reflection from cold matter in active galactic
  nuclei and X-ray binaries}},
\newblock \bibinfo{journal}{\mnras} \bibinfo{volume}{249}
  (\bibinfo{year}{1991}) \bibinfo{pages}{352--367}.
\bibitem[{{Revnivtsev} et~al.(2004){Revnivtsev}, {Churazov}, {Sazonov},
  {Sunyaev}, {Lutovinov}, {Gilfanov}, {Vikhlinin}, {Shtykovsky}, and
  {Pavlinsky}}]{2004A&A...425L..49R}
\bibinfo{author}{M.~G. {Revnivtsev}}, \bibinfo{author}{E.~M. {Churazov}},
  \bibinfo{author}{S.~Y. {Sazonov}}, \bibinfo{author}{R.~A. {Sunyaev}},
  \bibinfo{author}{A.~A. {Lutovinov}}, \bibinfo{author}{M.~R. {Gilfanov}},
  \bibinfo{author}{A.~A. {Vikhlinin}}, \bibinfo{author}{P.~E. {Shtykovsky}},
  \bibinfo{author}{M.~N. {Pavlinsky}},
\newblock \bibinfo{title}{{Hard X-ray view of the past activity of Sgr A* in a
  natural Compton mirror}},
\newblock \bibinfo{journal}{\aap} \bibinfo{volume}{425} (\bibinfo{year}{2004})
  \bibinfo{pages}{L49--L52}.
\bibitem[{{Shu} et~al.(2010){Shu}, {Yaqoob}, and {Wang}}]{2010ApJS..187..581S}
\bibinfo{author}{X.~W. {Shu}}, \bibinfo{author}{T.~{Yaqoob}},
  \bibinfo{author}{J.~X. {Wang}},
\newblock \bibinfo{title}{{The Cores of the Fe K{$\alpha$} Lines in Active
  Galactic Nuclei: An Extended Chandra High Energy Grating Sample}},
\newblock \bibinfo{journal}{\apjs} \bibinfo{volume}{187} (\bibinfo{year}{2010})
  \bibinfo{pages}{581--606}.
\bibitem[{{Shu} et~al.(2011){Shu}, {Yaqoob}, and {Wang}}]{2011ApJ...738..147S}
\bibinfo{author}{X.~W. {Shu}}, \bibinfo{author}{T.~{Yaqoob}},
  \bibinfo{author}{J.~X. {Wang}},
\newblock \bibinfo{title}{{Chandra High-energy Grating Observations of the Fe
  K{$\alpha$} Line Core in Type II Seyfert Galaxies: A Comparison with Type I
  Nuclei}},
\newblock \bibinfo{journal}{\apj} \bibinfo{volume}{738} (\bibinfo{year}{2011})
  \bibinfo{pages}{147}.
\bibitem[{{Nandra} et~al.(2007){Nandra}, {O'Neill}, {George}, and
  {Reeves}}]{2007MNRAS.382..194N}
\bibinfo{author}{K.~{Nandra}}, \bibinfo{author}{P.~M. {O'Neill}},
  \bibinfo{author}{I.~M. {George}}, \bibinfo{author}{J.~N. {Reeves}},
\newblock \bibinfo{title}{{An XMM-Newton survey of broad iron lines in Seyfert
  galaxies}},
\newblock \bibinfo{journal}{\mnras} \bibinfo{volume}{382}
  (\bibinfo{year}{2007}) \bibinfo{pages}{194--228}.
\bibitem[{{Ross} and {Fabian}(2005)}]{2005MNRAS.358..211R}
\bibinfo{author}{R.~R. {Ross}}, \bibinfo{author}{A.~C. {Fabian}},
\newblock \bibinfo{title}{{A comprehensive range of X-ray ionized-reflection
  models}},
\newblock \bibinfo{journal}{\mnras} \bibinfo{volume}{358}
  (\bibinfo{year}{2005}) \bibinfo{pages}{211--216}.
\bibitem[{{Miller} et~al.(2013){Miller}, {Parker}, {Fuerst}, {Bachetti},
  {Harrison}, {Barret}, {Boggs}, {Chakrabarty}, {Christensen}, {Craig},
  {Fabian}, {Grefenstette}, {Hailey}, {King}, {Stern}, {Tomsick}, {Walton}, and
  {Zhang}}]{2013ApJ...775L..45M}
\bibinfo{author}{J.~M. {Miller}}, \bibinfo{author}{M.~L. {Parker}},
  \bibinfo{author}{F.~{Fuerst}}, \bibinfo{author}{M.~{Bachetti}},
  \bibinfo{author}{F.~A. {Harrison}}, \bibinfo{author}{D.~{Barret}},
  \bibinfo{author}{S.~E. {Boggs}}, \bibinfo{author}{D.~{Chakrabarty}},
  \bibinfo{author}{F.~E. {Christensen}}, \bibinfo{author}{W.~W. {Craig}},
  \bibinfo{author}{A.~C. {Fabian}}, \bibinfo{author}{B.~W. {Grefenstette}},
  \bibinfo{author}{C.~J. {Hailey}}, \bibinfo{author}{A.~L. {King}},
  \bibinfo{author}{D.~K. {Stern}}, \bibinfo{author}{J.~A. {Tomsick}},
  \bibinfo{author}{D.~J. {Walton}}, \bibinfo{author}{W.~W. {Zhang}},
\newblock \bibinfo{title}{{NuSTAR Spectroscopy of GRS 1915+105: Disk
  Reflection, Spin, and Connections to Jets}},
\newblock \bibinfo{journal}{\apjl} \bibinfo{volume}{775} (\bibinfo{year}{2013})
  \bibinfo{pages}{L45}.
\bibitem[{{Garc{\'{\i}}a} and {Kallman}(2010)}]{2010ApJ...718..695G}
\bibinfo{author}{J.~{Garc{\'{\i}}a}}, \bibinfo{author}{T.~R. {Kallman}},
\newblock \bibinfo{title}{{X-ray Reflected Spectra from Accretion Disk Models.
  I. Constant Density Atmospheres}},
\newblock \bibinfo{journal}{\apj} \bibinfo{volume}{718} (\bibinfo{year}{2010})
  \bibinfo{pages}{695--706}.
\bibitem[{{Garc{\'{\i}}a} et~al.(2011){Garc{\'{\i}}a}, {Kallman}, and
  {Mushotzky}}]{2011ApJ...731..131G}
\bibinfo{author}{J.~{Garc{\'{\i}}a}}, \bibinfo{author}{T.~R. {Kallman}},
  \bibinfo{author}{R.~F. {Mushotzky}},
\newblock \bibinfo{title}{{X-ray Reflected Spectra from Accretion Disk Models.
  II. Diagnostic Tools for X-ray Observations}},
\newblock \bibinfo{journal}{\apj} \bibinfo{volume}{731} (\bibinfo{year}{2011})
  \bibinfo{pages}{131}.
\bibitem[{{Garc{\'{\i}}a} et~al.(2013){Garc{\'{\i}}a}, {Dauser}, {Reynolds},
  {Kallman}, {McClintock}, {Wilms}, and {Eikmann}}]{2013ApJ...768..146G}
\bibinfo{author}{J.~{Garc{\'{\i}}a}}, \bibinfo{author}{T.~{Dauser}},
  \bibinfo{author}{C.~S. {Reynolds}}, \bibinfo{author}{T.~R. {Kallman}},
  \bibinfo{author}{J.~E. {McClintock}}, \bibinfo{author}{J.~{Wilms}},
  \bibinfo{author}{W.~{Eikmann}},
\newblock \bibinfo{title}{{X-Ray Reflected Spectra from Accretion Disk Models.
  III. A Complete Grid of Ionized Reflection Calculations}},
\newblock \bibinfo{journal}{\apj} \bibinfo{volume}{768} (\bibinfo{year}{2013})
  \bibinfo{pages}{146}.
\bibitem[{{Garc{\'{\i}}a} et~al.(2014){Garc{\'{\i}}a}, {Dauser}, {Lohfink},
  {Kallman}, {Steiner}, {McClintock}, {Brenneman}, {Wilms}, {Eikmann},
  {Reynolds}, and {Tombesi}}]{2014ApJ...782...76G}
\bibinfo{author}{J.~{Garc{\'{\i}}a}}, \bibinfo{author}{T.~{Dauser}},
  \bibinfo{author}{A.~{Lohfink}}, \bibinfo{author}{T.~R. {Kallman}},
  \bibinfo{author}{J.~F. {Steiner}}, \bibinfo{author}{J.~E. {McClintock}},
  \bibinfo{author}{L.~{Brenneman}}, \bibinfo{author}{J.~{Wilms}},
  \bibinfo{author}{W.~{Eikmann}}, \bibinfo{author}{C.~S. {Reynolds}},
  \bibinfo{author}{F.~{Tombesi}},
\newblock \bibinfo{title}{{Improved Reflection Models of Black Hole Accretion
  Disks: Treating the Angular Distribution of X-Rays}},
\newblock \bibinfo{journal}{\apj} \bibinfo{volume}{782} (\bibinfo{year}{2014})
  \bibinfo{pages}{76}.
\bibitem[{{Miniutti} and {Fabian}(2004)}]{2004MNRAS.349.1435M}
\bibinfo{author}{G.~{Miniutti}}, \bibinfo{author}{A.~C. {Fabian}},
\newblock \bibinfo{title}{{A light bending model for the X-ray temporal and
  spectral properties of accreting black holes}},
\newblock \bibinfo{journal}{\mnras} \bibinfo{volume}{349}
  (\bibinfo{year}{2004}) \bibinfo{pages}{1435--1448}.
\bibitem[{{Miniutti} et~al.(2004){Miniutti}, {Fabian}, and
  {Miller}}]{2004MNRAS.351..466M}
\bibinfo{author}{G.~{Miniutti}}, \bibinfo{author}{A.~C. {Fabian}},
  \bibinfo{author}{J.~M. {Miller}},
\newblock \bibinfo{title}{{The relativistic Fe emission line in XTE J1650-500
  with BeppoSAX: evidence for black hole spin and light-bending effects?}},
\newblock \bibinfo{journal}{\mnras} \bibinfo{volume}{351}
  (\bibinfo{year}{2004}) \bibinfo{pages}{466--472}.
\bibitem[{{Rossi} et~al.(2005){Rossi}, {Homan}, {Miller}, and
  {Belloni}}]{2005MNRAS.360..763R}
\bibinfo{author}{S.~{Rossi}}, \bibinfo{author}{J.~{Homan}},
  \bibinfo{author}{J.~M. {Miller}}, \bibinfo{author}{T.~{Belloni}},
\newblock \bibinfo{title}{{Iron-line and continuum flux variations in the RXTE
  spectra of the black hole candidate XTE J1650-500}},
\newblock \bibinfo{journal}{\mnras} \bibinfo{volume}{360}
  (\bibinfo{year}{2005}) \bibinfo{pages}{763--768}.
\bibitem[{{Reis} et~al.(2013){Reis}, {Miller}, {Reynolds}, {Fabian}, {Walton},
  {Cackett}, and {Steiner}}]{2013ApJ...763...48R}
\bibinfo{author}{R.~C. {Reis}}, \bibinfo{author}{J.~M. {Miller}},
  \bibinfo{author}{M.~T. {Reynolds}}, \bibinfo{author}{A.~C. {Fabian}},
  \bibinfo{author}{D.~J. {Walton}}, \bibinfo{author}{E.~{Cackett}},
  \bibinfo{author}{J.~F. {Steiner}},
\newblock \bibinfo{title}{{Evidence of Light-bending Effects and Its
  Implication for Spectral State Transitions}},
\newblock \bibinfo{journal}{\apj} \bibinfo{volume}{763} (\bibinfo{year}{2013})
  \bibinfo{pages}{48}.
\bibitem[{{Fabian} et~al.(1989){Fabian}, {Rees}, {Stella}, and
  {White}}]{1989MNRAS.238..729F}
\bibinfo{author}{A.~C. {Fabian}}, \bibinfo{author}{M.~J. {Rees}},
  \bibinfo{author}{L.~{Stella}}, \bibinfo{author}{N.~E. {White}},
\newblock \bibinfo{title}{{X-ray fluorescence from the inner disc in Cygnus
  X-1}},
\newblock \bibinfo{journal}{\mnras} \bibinfo{volume}{238}
  (\bibinfo{year}{1989}) \bibinfo{pages}{729--736}.
\bibitem[{{Laor}(1991)}]{1991ApJ...376...90L}
\bibinfo{author}{A.~{Laor}},
\newblock \bibinfo{title}{{Line profiles from a disk around a rotating black
  hole}},
\newblock \bibinfo{journal}{\apj} \bibinfo{volume}{376} (\bibinfo{year}{1991})
  \bibinfo{pages}{90--94}.
\bibitem[{{Dov{\v c}iak} et~al.(2004){Dov{\v c}iak}, {Karas}, and
  {Yaqoob}}]{2004ApJS..153..205D}
\bibinfo{author}{M.~{Dov{\v c}iak}}, \bibinfo{author}{V.~{Karas}},
  \bibinfo{author}{T.~{Yaqoob}},
\newblock \bibinfo{title}{{An Extended Scheme for Fitting X-Ray Data with
  Accretion Disk Spectra in the Strong Gravity Regime}},
\newblock \bibinfo{journal}{\apjs} \bibinfo{volume}{153} (\bibinfo{year}{2004})
  \bibinfo{pages}{205--221}.
\bibitem[{{Beckwith} and {Done}(2004)}]{2004MNRAS.352..353B}
\bibinfo{author}{K.~{Beckwith}}, \bibinfo{author}{C.~{Done}},
\newblock \bibinfo{title}{{Iron line profiles in strong gravity}},
\newblock \bibinfo{journal}{\mnras} \bibinfo{volume}{352}
  (\bibinfo{year}{2004}) \bibinfo{pages}{353--362}.
\bibitem[{{Brenneman} and {Reynolds}(2006)}]{2006ApJ...652.1028B}
\bibinfo{author}{L.~W. {Brenneman}}, \bibinfo{author}{C.~S. {Reynolds}},
\newblock \bibinfo{title}{{Constraining Black Hole Spin via X-Ray
  Spectroscopy}},
\newblock \bibinfo{journal}{\apj} \bibinfo{volume}{652} (\bibinfo{year}{2006})
  \bibinfo{pages}{1028--1043}.
\bibitem[{{Dauser} et~al.(2013){Dauser}, {Garcia}, {Wilms}, {Bock},
  {Brenneman}, {Falanga}, {Fukumura}, and {Reynolds}}]{2013MNRAS.430.1694D}
\bibinfo{author}{T.~{Dauser}}, \bibinfo{author}{J.~{Garcia}},
  \bibinfo{author}{J.~{Wilms}}, \bibinfo{author}{M.~{Bock}},
  \bibinfo{author}{L.~W. {Brenneman}}, \bibinfo{author}{M.~{Falanga}},
  \bibinfo{author}{K.~{Fukumura}}, \bibinfo{author}{C.~S. {Reynolds}},
\newblock \bibinfo{title}{{Irradiation of an accretion disc by a jet: general
  properties and implications for spin measurements of black holes}},
\newblock \bibinfo{journal}{\mnras} \bibinfo{volume}{430}
  (\bibinfo{year}{2013}) \bibinfo{pages}{1694--1708}.
\bibitem[{{Miller}(2007)}]{2007ARA&A..45..441M}
\bibinfo{author}{J.~M. {Miller}},
\newblock \bibinfo{title}{{Relativistic X-Ray Lines from the Inner Accretion
  Disks Around Black Holes}},
\newblock \bibinfo{journal}{\araa} \bibinfo{volume}{45} (\bibinfo{year}{2007})
  \bibinfo{pages}{441--479}.
\bibitem[{{Fabian} et~al.(2009){Fabian}, {Zoghbi}, {Ross}, {Uttley}, {Gallo},
  {Brandt}, {Blustin}, {Boller}, {Caballero-Garcia}, {Larsson}, {Miller},
  {Miniutti}, {Ponti}, {Reis}, {Reynolds}, {Tanaka}, and
  {Young}}]{2009Natur.459..540F}
\bibinfo{author}{A.~C. {Fabian}}, \bibinfo{author}{A.~{Zoghbi}},
  \bibinfo{author}{R.~R. {Ross}}, \bibinfo{author}{P.~{Uttley}},
  \bibinfo{author}{L.~C. {Gallo}}, \bibinfo{author}{W.~N. {Brandt}},
  \bibinfo{author}{A.~J. {Blustin}}, \bibinfo{author}{T.~{Boller}},
  \bibinfo{author}{M.~D. {Caballero-Garcia}}, \bibinfo{author}{J.~{Larsson}},
  \bibinfo{author}{J.~M. {Miller}}, \bibinfo{author}{G.~{Miniutti}},
  \bibinfo{author}{G.~{Ponti}}, \bibinfo{author}{R.~C. {Reis}},
  \bibinfo{author}{C.~S. {Reynolds}}, \bibinfo{author}{Y.~{Tanaka}},
  \bibinfo{author}{A.~J. {Young}},
\newblock \bibinfo{title}{{Broad line emission from iron K- and L-shell
  transitions in the active galaxy 1H0707-495}},
\newblock \bibinfo{journal}{\nat} \bibinfo{volume}{459} (\bibinfo{year}{2009})
  \bibinfo{pages}{540--542}.
\bibitem[{{Wilkins} and {Fabian}(2012)}]{2012MNRAS.424.1284W}
\bibinfo{author}{D.~R. {Wilkins}}, \bibinfo{author}{A.~C. {Fabian}},
\newblock \bibinfo{title}{{Understanding X-ray reflection emissivity profiles
  in AGN: locating the X-ray source}},
\newblock \bibinfo{journal}{\mnras} \bibinfo{volume}{424}
  (\bibinfo{year}{2012}) \bibinfo{pages}{1284--1296}.
\bibitem[{{Liedahl} and {Torres}(2005)}]{2005CaJPh..83.1177L}
\bibinfo{author}{D.~A. {Liedahl}}, \bibinfo{author}{D.~F. {Torres}},
\newblock \bibinfo{title}{{Atomic X-ray spectroscopy of accreting black
  holes}},
\newblock \bibinfo{journal}{Canadian Journal of Physics} \bibinfo{volume}{83}
  (\bibinfo{year}{2005}) \bibinfo{pages}{1177--1240}.
\bibitem[{{Miller} et~al.(2009){Miller}, {Reynolds}, {Fabian}, {Miniutti}, and
  {Gallo}}]{2009ApJ...697..900M}
\bibinfo{author}{J.~M. {Miller}}, \bibinfo{author}{C.~S. {Reynolds}},
  \bibinfo{author}{A.~C. {Fabian}}, \bibinfo{author}{G.~{Miniutti}},
  \bibinfo{author}{L.~C. {Gallo}},
\newblock \bibinfo{title}{{Stellar-Mass Black Hole Spin Constraints from Disk
  Reflection and Continuum Modeling}},
\newblock \bibinfo{journal}{\apj} \bibinfo{volume}{697} (\bibinfo{year}{2009})
  \bibinfo{pages}{900--912}.
\bibitem[{{Hiemstra} et~al.(2011){Hiemstra}, {M{\'e}ndez}, {Done}, {D{\'{\i}}az
  Trigo}, {Altamirano}, and {Casella}}]{2011MNRAS.411..137H}
\bibinfo{author}{B.~{Hiemstra}}, \bibinfo{author}{M.~{M{\'e}ndez}},
  \bibinfo{author}{C.~{Done}}, \bibinfo{author}{M.~{D{\'{\i}}az Trigo}},
  \bibinfo{author}{D.~{Altamirano}}, \bibinfo{author}{P.~{Casella}},
\newblock \bibinfo{title}{{A strong and broad Fe line in the XMM-Newton
  spectrum of the new X-ray transient and black hole candidate XTE J1652-453}},
\newblock \bibinfo{journal}{\mnras} \bibinfo{volume}{411}
  (\bibinfo{year}{2011}) \bibinfo{pages}{137--150}.
\bibitem[{{Reis} et~al.(2011){Reis}, {Miller}, {Fabian}, {Cackett}, {Maitra},
  {Reynolds}, {Rupen}, {Steeghs}, and {Wijnands}}]{2011MNRAS.410.2497R}
\bibinfo{author}{R.~C. {Reis}}, \bibinfo{author}{J.~M. {Miller}},
  \bibinfo{author}{A.~C. {Fabian}}, \bibinfo{author}{E.~M. {Cackett}},
  \bibinfo{author}{D.~{Maitra}}, \bibinfo{author}{C.~S. {Reynolds}},
  \bibinfo{author}{M.~{Rupen}}, \bibinfo{author}{D.~T.~H. {Steeghs}},
  \bibinfo{author}{R.~{Wijnands}},
\newblock \bibinfo{title}{{Multistate observations of the Galactic black hole
  XTE J1752-223: evidence for an intermediate black hole spin}},
\newblock \bibinfo{journal}{\mnras} \bibinfo{volume}{410}
  (\bibinfo{year}{2011}) \bibinfo{pages}{2497--2505}.
\bibitem[{{King} et~al.(2013){King}, {Miller}, {G{\"u}ltekin}, {Walton},
  {Fabian}, {Reynolds}, and {Nandra}}]{2013ApJ...771...84K}
\bibinfo{author}{A.~L. {King}}, \bibinfo{author}{J.~M. {Miller}},
  \bibinfo{author}{K.~{G{\"u}ltekin}}, \bibinfo{author}{D.~J. {Walton}},
  \bibinfo{author}{A.~C. {Fabian}}, \bibinfo{author}{C.~S. {Reynolds}},
  \bibinfo{author}{K.~{Nandra}},
\newblock \bibinfo{title}{{What is on Tap? The Role of Spin in Compact Objects
  and Relativistic Jets}},
\newblock \bibinfo{journal}{\apj} \bibinfo{volume}{771} (\bibinfo{year}{2013})
  \bibinfo{pages}{84}.
\bibitem[{{Noble} et~al.(2011){Noble}, {Krolik}, {Schnittman}, and
  {Hawley}}]{2011ApJ...743..115N}
\bibinfo{author}{S.~C. {Noble}}, \bibinfo{author}{J.~H. {Krolik}},
  \bibinfo{author}{J.~D. {Schnittman}}, \bibinfo{author}{J.~F. {Hawley}},
\newblock \bibinfo{title}{{Radiative Efficiency and Thermal Spectrum of
  Accretion onto Schwarzschild Black Holes}},
\newblock \bibinfo{journal}{\apj} \bibinfo{volume}{743} (\bibinfo{year}{2011})
  \bibinfo{pages}{115}.
\bibitem[{{Schnittman} et~al.(2013){Schnittman}, {Krolik}, and
  {Noble}}]{2013ApJ...769..156S}
\bibinfo{author}{J.~D. {Schnittman}}, \bibinfo{author}{J.~H. {Krolik}},
  \bibinfo{author}{S.~C. {Noble}},
\newblock \bibinfo{title}{{X-Ray Spectra from Magnetohydrodynamic Simulations
  of Accreting Black Holes}},
\newblock \bibinfo{journal}{\apj} \bibinfo{volume}{769} (\bibinfo{year}{2013})
  \bibinfo{pages}{156}.
\bibitem[{{Reis} and {Miller}(2013)}]{2013ApJ...769L...7R}
\bibinfo{author}{R.~C. {Reis}}, \bibinfo{author}{J.~M. {Miller}},
\newblock \bibinfo{title}{{On the Size and Location of the X-Ray Emitting
  Coronae around Black Holes}},
\newblock \bibinfo{journal}{\apjl} \bibinfo{volume}{769} (\bibinfo{year}{2013})
  \bibinfo{pages}{L7}.
\bibitem[{{Chartas} et~al.(2009){Chartas}, {Kochanek}, {Dai}, {Poindexter}, and
  {Garmire}}]{2009ApJ...693..174C}
\bibinfo{author}{G.~{Chartas}}, \bibinfo{author}{C.~S. {Kochanek}},
  \bibinfo{author}{X.~{Dai}}, \bibinfo{author}{S.~{Poindexter}},
  \bibinfo{author}{G.~{Garmire}},
\newblock \bibinfo{title}{{X-Ray Microlensing in RXJ1131-1231 and
  HE1104-1805}},
\newblock \bibinfo{journal}{\apj} \bibinfo{volume}{693} (\bibinfo{year}{2009})
  \bibinfo{pages}{174--185}.
\bibitem[{{Wilms} et~al.(2001){Wilms}, {Reynolds}, {Begelman}, {Reeves},
  {Molendi}, {Staubert}, and {Kendziorra}}]{2001MNRAS.328L..27W}
\bibinfo{author}{J.~{Wilms}}, \bibinfo{author}{C.~S. {Reynolds}},
  \bibinfo{author}{M.~C. {Begelman}}, \bibinfo{author}{J.~{Reeves}},
  \bibinfo{author}{S.~{Molendi}}, \bibinfo{author}{R.~{Staubert}},
  \bibinfo{author}{E.~{Kendziorra}},
\newblock \bibinfo{title}{{XMM-EPIC observation of MCG-6-30-15: direct evidence
  for the extraction of energy from a spinning black hole?}},
\newblock \bibinfo{journal}{\mnras} \bibinfo{volume}{328}
  (\bibinfo{year}{2001}) \bibinfo{pages}{L27--L31}.
\bibitem[{{Beckwith} et~al.(2008){Beckwith}, {Hawley}, and
  {Krolik}}]{2008MNRAS.390...21B}
\bibinfo{author}{K.~{Beckwith}}, \bibinfo{author}{J.~F. {Hawley}},
  \bibinfo{author}{J.~H. {Krolik}},
\newblock \bibinfo{title}{{Where is the radiation edge in magnetized black hole
  accretion discs?}},
\newblock \bibinfo{journal}{\mnras} \bibinfo{volume}{390}
  (\bibinfo{year}{2008}) \bibinfo{pages}{21--38}.
\bibitem[{{Noble} et~al.(2009){Noble}, {Krolik}, and
  {Hawley}}]{2009ApJ...692..411N}
\bibinfo{author}{S.~C. {Noble}}, \bibinfo{author}{J.~H. {Krolik}},
  \bibinfo{author}{J.~F. {Hawley}},
\newblock \bibinfo{title}{{Direct Calculation of the Radiative Efficiency of an
  Accretion Disk Around a Black Hole}},
\newblock \bibinfo{journal}{\apj} \bibinfo{volume}{692} (\bibinfo{year}{2009})
  \bibinfo{pages}{411--421}.
\bibitem[{{Kulkarni} et~al.(2011){Kulkarni}, {Penna}, {Shcherbakov}, {Steiner},
  {Narayan}, {S{\"a} Dowski}, {Zhu}, {McClintock}, {Davis}, and
  {McKinney}}]{2011MNRAS.414.1183K}
\bibinfo{author}{A.~K. {Kulkarni}}, \bibinfo{author}{R.~F. {Penna}},
  \bibinfo{author}{R.~V. {Shcherbakov}}, \bibinfo{author}{J.~F. {Steiner}},
  \bibinfo{author}{R.~{Narayan}}, \bibinfo{author}{A.~{S{\"a} Dowski}},
  \bibinfo{author}{Y.~{Zhu}}, \bibinfo{author}{J.~E. {McClintock}},
  \bibinfo{author}{S.~W. {Davis}}, \bibinfo{author}{J.~C. {McKinney}},
\newblock \bibinfo{title}{{Measuring black hole spin by the continuum-fitting
  method: effect of deviations from the Novikov-Thorne disc model}},
\newblock \bibinfo{journal}{\mnras} \bibinfo{volume}{414}
  (\bibinfo{year}{2011}) \bibinfo{pages}{1183--1194}.
\bibitem[{{Dhawan} et~al.(2000){Dhawan}, {Mirabel}, and
  {Rodr{\'{\i}}guez}}]{2000ApJ...543..373D}
\bibinfo{author}{V.~{Dhawan}}, \bibinfo{author}{I.~F. {Mirabel}},
  \bibinfo{author}{L.~F. {Rodr{\'{\i}}guez}},
\newblock \bibinfo{title}{{AU-Scale Synchrotron Jets and Superluminal Ejecta in
  GRS 1915+105}},
\newblock \bibinfo{journal}{\apj} \bibinfo{volume}{543} (\bibinfo{year}{2000})
  \bibinfo{pages}{373--385}.
\bibitem[{{Blandford} and {Znajek}(1977)}]{1977MNRAS.179..433B}
\bibinfo{author}{R.~D. {Blandford}}, \bibinfo{author}{R.~L. {Znajek}},
\newblock \bibinfo{title}{{Electromagnetic extraction of energy from Kerr black
  holes}},
\newblock \bibinfo{journal}{\mnras} \bibinfo{volume}{179}
  (\bibinfo{year}{1977}) \bibinfo{pages}{433--456}.
\bibitem[{{Miller} et~al.(2006){Miller}, {Fabian}, {Nowak}, and
  {Lewin}}]{2006tmgm.meet.1296M}
\bibinfo{author}{J.~M. {Miller}}, \bibinfo{author}{A.~C. {Fabian}},
  \bibinfo{author}{M.~A. {Nowak}}, \bibinfo{author}{W.~H.~G. {Lewin}},
\newblock \bibinfo{title}{{Relativistic Iron Lines in Galactic Black Holes:.
  Recent Results and Lines in the ASCA Archive}},
\newblock in: \bibinfo{editor}{M.~{Novello}}, \bibinfo{editor}{S.~{Perez
  Bergliaffa}}, \bibinfo{editor}{R.~{Ruffini}} (Eds.), \bibinfo{booktitle}{The
  Tenth Marcel Grossmann Meeting. Proceedings of the MG10 Meeting held at
  Brazilian Center for Research in Physics (CBPF), Rio de Janeiro, Brazil,
  20-26 July 2003, Eds.: M{\'a}rio Novello; Santiago Perez Bergliaffa; Remo
  Ruffini. Singapore: World Scientific Publishing, in 3 volumes, ISBN
  981-256-667-8 (set), ISBN 981-256-980-4 (Part A), ISBN 981-256-979-0 (Part
  B), ISBN 981-256-978-2 (Part C), 2006, XLVIII + 2492 pp.: 2006, p.1296}, p.
  \bibinfo{pages}{1296}.
\bibitem[{{Martocchia} et~al.(2002){Martocchia}, {Matt}, {Karas}, {Belloni},
  and {Feroci}}]{2002A&A...387..215M}
\bibinfo{author}{A.~{Martocchia}}, \bibinfo{author}{G.~{Matt}},
  \bibinfo{author}{V.~{Karas}}, \bibinfo{author}{T.~{Belloni}},
  \bibinfo{author}{M.~{Feroci}},
\newblock \bibinfo{title}{{Evidence for a relativistic iron line in GRS
  1915+105}},
\newblock \bibinfo{journal}{\aap} \bibinfo{volume}{387} (\bibinfo{year}{2002})
  \bibinfo{pages}{215--221}.
\bibitem[{{Blum} et~al.(2009){Blum}, {Miller}, {Fabian}, {Miller}, {Homan},
  {van der Klis}, {Cackett}, and {Reis}}]{2009ApJ...706...60B}
\bibinfo{author}{J.~L. {Blum}}, \bibinfo{author}{J.~M. {Miller}},
  \bibinfo{author}{A.~C. {Fabian}}, \bibinfo{author}{M.~C. {Miller}},
  \bibinfo{author}{J.~{Homan}}, \bibinfo{author}{M.~{van der Klis}},
  \bibinfo{author}{E.~M. {Cackett}}, \bibinfo{author}{R.~C. {Reis}},
\newblock \bibinfo{title}{{Measuring the Spin of GRS 1915+105 with Relativistic
  Disk Reflection}},
\newblock \bibinfo{journal}{\apj} \bibinfo{volume}{706} (\bibinfo{year}{2009})
  \bibinfo{pages}{60--66}.
\bibitem[{{Harrison} et~al.(2013){Harrison}, {Craig}, {Christensen}, {Hailey},
  {Zhang}, {Boggs}, {Stern}, {Cook}, {Forster}, {Giommi}, {Grefenstette},
  {Kim}, {Kitaguchi}, {Koglin}, {Madsen}, {Mao}, {Miyasaka}, {Mori}, {Perri},
  {Pivovaroff}, {Puccetti}, {Rana}, {Westergaard}, {Willis}, {Zoglauer}, {An},
  {Bachetti}, {Barri{\`e}re}, {Bellm}, {Bhalerao}, {Brejnholt}, {Fuerst},
  {Liebe}, {Markwardt}, {Nynka}, {Vogel}, {Walton}, {Wik}, {Alexander},
  {Cominsky}, {Hornschemeier}, {Hornstrup}, {Kaspi}, {Madejski}, {Matt},
  {Molendi}, {Smith}, {Tomsick}, {Ajello}, {Ballantyne}, {Balokovi{\'c}},
  {Barret}, {Bauer}, {Blandford}, {Brandt}, {Brenneman}, {Chiang},
  {Chakrabarty}, {Chenevez}, {Comastri}, {Dufour}, {Elvis}, {Fabian}, {Farrah},
  {Fryer}, {Gotthelf}, {Grindlay}, {Helfand}, {Krivonos}, {Meier}, {Miller},
  {Natalucci}, {Ogle}, {Ofek}, {Ptak}, {Reynolds}, {Rigby}, {Tagliaferri},
  {Thorsett}, {Treister}, and {Urry}}]{2013ApJ...770..103H}
\bibinfo{author}{F.~A. {Harrison}}, \bibinfo{author}{W.~W. {Craig}},
  \bibinfo{author}{F.~E. {Christensen}}, \bibinfo{author}{C.~J. {Hailey}},
  \bibinfo{author}{W.~W. {Zhang}}, \bibinfo{author}{S.~E. {Boggs}},
  \bibinfo{author}{D.~{Stern}}, \bibinfo{author}{W.~R. {Cook}},
  \bibinfo{author}{K.~{Forster}}, \bibinfo{author}{P.~{Giommi}},
  \bibinfo{author}{B.~W. {Grefenstette}}, \bibinfo{author}{Y.~{Kim}},
  \bibinfo{author}{T.~{Kitaguchi}}, \bibinfo{author}{J.~E. {Koglin}},
  \bibinfo{author}{K.~K. {Madsen}}, \bibinfo{author}{P.~H. {Mao}},
  \bibinfo{author}{H.~{Miyasaka}}, \bibinfo{author}{K.~{Mori}},
  \bibinfo{author}{M.~{Perri}}, \bibinfo{author}{M.~J. {Pivovaroff}},
  \bibinfo{author}{S.~{Puccetti}}, \bibinfo{author}{V.~R. {Rana}},
  \bibinfo{author}{N.~J. {Westergaard}}, \bibinfo{author}{J.~{Willis}},
  \bibinfo{author}{A.~{Zoglauer}}, \bibinfo{author}{H.~{An}},
  \bibinfo{author}{M.~{Bachetti}}, \bibinfo{author}{N.~M. {Barri{\`e}re}},
  \bibinfo{author}{E.~C. {Bellm}}, \bibinfo{author}{V.~{Bhalerao}},
  \bibinfo{author}{N.~F. {Brejnholt}}, \bibinfo{author}{F.~{Fuerst}},
  \bibinfo{author}{C.~C. {Liebe}}, \bibinfo{author}{C.~B. {Markwardt}},
  \bibinfo{author}{M.~{Nynka}}, \bibinfo{author}{J.~K. {Vogel}},
  \bibinfo{author}{D.~J. {Walton}}, \bibinfo{author}{D.~R. {Wik}},
  \bibinfo{author}{D.~M. {Alexander}}, \bibinfo{author}{L.~R. {Cominsky}},
  \bibinfo{author}{A.~E. {Hornschemeier}}, \bibinfo{author}{A.~{Hornstrup}},
  \bibinfo{author}{V.~M. {Kaspi}}, \bibinfo{author}{G.~M. {Madejski}},
  \bibinfo{author}{G.~{Matt}}, \bibinfo{author}{S.~{Molendi}},
  \bibinfo{author}{D.~M. {Smith}}, \bibinfo{author}{J.~A. {Tomsick}},
  \bibinfo{author}{M.~{Ajello}}, \bibinfo{author}{D.~R. {Ballantyne}},
  \bibinfo{author}{M.~{Balokovi{\'c}}}, \bibinfo{author}{D.~{Barret}},
  \bibinfo{author}{F.~E. {Bauer}}, \bibinfo{author}{R.~D. {Blandford}},
  \bibinfo{author}{W.~N. {Brandt}}, \bibinfo{author}{L.~W. {Brenneman}},
  \bibinfo{author}{J.~{Chiang}}, \bibinfo{author}{D.~{Chakrabarty}},
  \bibinfo{author}{J.~{Chenevez}}, \bibinfo{author}{A.~{Comastri}},
  \bibinfo{author}{F.~{Dufour}}, \bibinfo{author}{M.~{Elvis}},
  \bibinfo{author}{A.~C. {Fabian}}, \bibinfo{author}{D.~{Farrah}},
  \bibinfo{author}{C.~L. {Fryer}}, \bibinfo{author}{E.~V. {Gotthelf}},
  \bibinfo{author}{J.~E. {Grindlay}}, \bibinfo{author}{D.~J. {Helfand}},
  \bibinfo{author}{R.~{Krivonos}}, \bibinfo{author}{D.~L. {Meier}},
  \bibinfo{author}{J.~M. {Miller}}, \bibinfo{author}{L.~{Natalucci}},
  \bibinfo{author}{P.~{Ogle}}, \bibinfo{author}{E.~O. {Ofek}},
  \bibinfo{author}{A.~{Ptak}}, \bibinfo{author}{S.~P. {Reynolds}},
  \bibinfo{author}{J.~R. {Rigby}}, \bibinfo{author}{G.~{Tagliaferri}},
  \bibinfo{author}{S.~E. {Thorsett}}, \bibinfo{author}{E.~{Treister}},
  \bibinfo{author}{C.~M. {Urry}},
\newblock \bibinfo{title}{{The Nuclear Spectroscopic Telescope Array (NuSTAR)
  High-energy X-Ray Mission}},
\newblock \bibinfo{journal}{\apj} \bibinfo{volume}{770} (\bibinfo{year}{2013})
  \bibinfo{pages}{103}.
\bibitem[{{McClintock} et~al.(2006){McClintock}, {Shafee}, {Narayan},
  {Remillard}, {Davis}, and {Li}}]{2006ApJ...652..518M}
\bibinfo{author}{J.~E. {McClintock}}, \bibinfo{author}{R.~{Shafee}},
  \bibinfo{author}{R.~{Narayan}}, \bibinfo{author}{R.~A. {Remillard}},
  \bibinfo{author}{S.~W. {Davis}}, \bibinfo{author}{L.-X. {Li}},
\newblock \bibinfo{title}{{The Spin of the Near-Extreme Kerr Black Hole GRS
  1915+105}},
\newblock \bibinfo{journal}{\apj} \bibinfo{volume}{652} (\bibinfo{year}{2006})
  \bibinfo{pages}{518--539}.
\bibitem[{{Miller} et~al.(2002){Miller}, {Fabian}, {Wijnands}, {Reynolds},
  {Ehle}, {Freyberg}, {van der Klis}, {Lewin}, {Sanchez-Fernandez}, and
  {Castro-Tirado}}]{2002ApJ...570L..69M}
\bibinfo{author}{J.~M. {Miller}}, \bibinfo{author}{A.~C. {Fabian}},
  \bibinfo{author}{R.~{Wijnands}}, \bibinfo{author}{C.~S. {Reynolds}},
  \bibinfo{author}{M.~{Ehle}}, \bibinfo{author}{M.~J. {Freyberg}},
  \bibinfo{author}{M.~{van der Klis}}, \bibinfo{author}{W.~H.~G. {Lewin}},
  \bibinfo{author}{C.~{Sanchez-Fernandez}}, \bibinfo{author}{A.~J.
  {Castro-Tirado}},
\newblock \bibinfo{title}{{Evidence of Spin and Energy Extraction in a Galactic
  Black Hole Candidate: The XMM-Newton/EPIC-pn Spectrum of XTE J1650-500}},
\newblock \bibinfo{journal}{\apjl} \bibinfo{volume}{570} (\bibinfo{year}{2002})
  \bibinfo{pages}{L69--L73}.
\bibitem[{{Miller} et~al.(2004){Miller}, {Fabian}, {Reynolds}, {Nowak},
  {Homan}, {Freyberg}, {Ehle}, {Belloni}, {Wijnands}, {van der Klis},
  {Charles}, and {Lewin}}]{2004ApJ...606L.131M}
\bibinfo{author}{J.~M. {Miller}}, \bibinfo{author}{A.~C. {Fabian}},
  \bibinfo{author}{C.~S. {Reynolds}}, \bibinfo{author}{M.~A. {Nowak}},
  \bibinfo{author}{J.~{Homan}}, \bibinfo{author}{M.~J. {Freyberg}},
  \bibinfo{author}{M.~{Ehle}}, \bibinfo{author}{T.~{Belloni}},
  \bibinfo{author}{R.~{Wijnands}}, \bibinfo{author}{M.~{van der Klis}},
  \bibinfo{author}{P.~A. {Charles}}, \bibinfo{author}{W.~H.~G. {Lewin}},
\newblock \bibinfo{title}{{Evidence of Black Hole Spin in GX 339-4:
  XMM-Newton/EPIC-pn and RXTE Spectroscopy of the Very High State}},
\newblock \bibinfo{journal}{\apjl} \bibinfo{volume}{606} (\bibinfo{year}{2004})
  \bibinfo{pages}{L131--L134}.
\bibitem[{{Miller} et~al.(2008){Miller}, {Reynolds}, {Fabian}, {Cackett},
  {Miniutti}, {Raymond}, {Steeghs}, {Reis}, and {Homan}}]{2008ApJ...679L.113M}
\bibinfo{author}{J.~M. {Miller}}, \bibinfo{author}{C.~S. {Reynolds}},
  \bibinfo{author}{A.~C. {Fabian}}, \bibinfo{author}{E.~M. {Cackett}},
  \bibinfo{author}{G.~{Miniutti}}, \bibinfo{author}{J.~{Raymond}},
  \bibinfo{author}{D.~{Steeghs}}, \bibinfo{author}{R.~{Reis}},
  \bibinfo{author}{J.~{Homan}},
\newblock \bibinfo{title}{{Initial Measurements of Black Hole Spin in GX 339-4
  from Suzaku Spectroscopy}},
\newblock \bibinfo{journal}{\apjl} \bibinfo{volume}{679} (\bibinfo{year}{2008})
  \bibinfo{pages}{L113--L116}.
\bibitem[{{Reis} et~al.(2008){Reis}, {Fabian}, {Ross}, {Miniutti}, {Miller},
  and {Reynolds}}]{2008MNRAS.387.1489R}
\bibinfo{author}{R.~C. {Reis}}, \bibinfo{author}{A.~C. {Fabian}},
  \bibinfo{author}{R.~R. {Ross}}, \bibinfo{author}{G.~{Miniutti}},
  \bibinfo{author}{J.~M. {Miller}}, \bibinfo{author}{C.~{Reynolds}},
\newblock \bibinfo{title}{{A systematic look at the very high and low/hard
  state of GX339-4: constraining the black hole spin with a new reflection
  model}},
\newblock \bibinfo{journal}{\mnras} \bibinfo{volume}{387}
  (\bibinfo{year}{2008}) \bibinfo{pages}{1489--1498}.
\bibitem[{{Bhattacharyya} and {Strohmayer}(2007)}]{2007ApJ...664L.103B}
\bibinfo{author}{S.~{Bhattacharyya}}, \bibinfo{author}{T.~E. {Strohmayer}},
\newblock \bibinfo{title}{{Evidence of a Broad Relativistic Iron Line from the
  Neutron Star Low-Mass X-Ray Binary Serpens X-1}},
\newblock \bibinfo{journal}{\apjl} \bibinfo{volume}{664} (\bibinfo{year}{2007})
  \bibinfo{pages}{L103--L106}.
\bibitem[{{Cackett} et~al.(2008){Cackett}, {Miller}, {Bhattacharyya},
  {Grindlay}, {Homan}, {van der Klis}, {Miller}, {Strohmayer}, and
  {Wijnands}}]{2008ApJ...674..415C}
\bibinfo{author}{E.~M. {Cackett}}, \bibinfo{author}{J.~M. {Miller}},
  \bibinfo{author}{S.~{Bhattacharyya}}, \bibinfo{author}{J.~E. {Grindlay}},
  \bibinfo{author}{J.~{Homan}}, \bibinfo{author}{M.~{van der Klis}},
  \bibinfo{author}{M.~C. {Miller}}, \bibinfo{author}{T.~E. {Strohmayer}},
  \bibinfo{author}{R.~{Wijnands}},
\newblock \bibinfo{title}{{Relativistic Iron Emission Lines in Neutron Star
  Low-Mass X-Ray Binaries as Probes of Neutron Star Radii}},
\newblock \bibinfo{journal}{\apj} \bibinfo{volume}{674} (\bibinfo{year}{2008})
  \bibinfo{pages}{415--420}.
\bibitem[{{Cackett} et~al.(2010){Cackett}, {Miller}, {Ballantyne}, {Barret},
  {Bhattacharyya}, {Boutelier}, {Miller}, {Strohmayer}, and
  {Wijnands}}]{2010ApJ...720..205C}
\bibinfo{author}{E.~M. {Cackett}}, \bibinfo{author}{J.~M. {Miller}},
  \bibinfo{author}{D.~R. {Ballantyne}}, \bibinfo{author}{D.~{Barret}},
  \bibinfo{author}{S.~{Bhattacharyya}}, \bibinfo{author}{M.~{Boutelier}},
  \bibinfo{author}{M.~C. {Miller}}, \bibinfo{author}{T.~E. {Strohmayer}},
  \bibinfo{author}{R.~{Wijnands}},
\newblock \bibinfo{title}{{Relativistic Lines and Reflection from the Inner
  Accretion Disks Around Neutron Stars}},
\newblock \bibinfo{journal}{\apj} \bibinfo{volume}{720} (\bibinfo{year}{2010})
  \bibinfo{pages}{205--225}.
\bibitem[{{Miller} et~al.(2013){Miller}, {Parker}, {Fuerst}, {Bachetti},
  {Barret}, {Grefenstette}, {Tendulkar}, {Harrison}, {Boggs}, {Chakrabarty},
  {Christensen}, {Craig}, {Fabian}, {Hailey}, {Natalucci}, {Paerels}, {Rana},
  {Stern}, {Tomsick}, and {Zhang}}]{2013ApJ...779L...2M}
\bibinfo{author}{J.~M. {Miller}}, \bibinfo{author}{M.~L. {Parker}},
  \bibinfo{author}{F.~{Fuerst}}, \bibinfo{author}{M.~{Bachetti}},
  \bibinfo{author}{D.~{Barret}}, \bibinfo{author}{B.~W. {Grefenstette}},
  \bibinfo{author}{S.~{Tendulkar}}, \bibinfo{author}{F.~A. {Harrison}},
  \bibinfo{author}{S.~E. {Boggs}}, \bibinfo{author}{D.~{Chakrabarty}},
  \bibinfo{author}{F.~E. {Christensen}}, \bibinfo{author}{W.~W. {Craig}},
  \bibinfo{author}{A.~C. {Fabian}}, \bibinfo{author}{C.~J. {Hailey}},
  \bibinfo{author}{L.~{Natalucci}}, \bibinfo{author}{F.~{Paerels}},
  \bibinfo{author}{V.~{Rana}}, \bibinfo{author}{D.~K. {Stern}},
  \bibinfo{author}{J.~A. {Tomsick}}, \bibinfo{author}{W.~W. {Zhang}},
\newblock \bibinfo{title}{{Constraints on the Neutron Star and Inner Accretion
  Flow in Serpens X-1 Using NuSTAR}},
\newblock \bibinfo{journal}{\apjl} \bibinfo{volume}{779} (\bibinfo{year}{2013})
  \bibinfo{pages}{L2}.
\bibitem[{{Hartle} and {Thorne}(1968)}]{1968ApJ...153..807H}
\bibinfo{author}{J.~B. {Hartle}}, \bibinfo{author}{K.~S. {Thorne}},
\newblock \bibinfo{title}{{Slowly Rotating Relativistic Stars. II. Models for
  Neutron Stars and Supermassive Stars}},
\newblock \bibinfo{journal}{\apj} \bibinfo{volume}{153} (\bibinfo{year}{1968})
  \bibinfo{pages}{807}.
\bibitem[{{Miller} et~al.(1998){Miller}, {Lamb}, and
  {Cook}}]{1998ApJ...509..793M}
\bibinfo{author}{M.~C. {Miller}}, \bibinfo{author}{F.~K. {Lamb}},
  \bibinfo{author}{G.~B. {Cook}},
\newblock \bibinfo{title}{{Effects of Rapid Stellar Rotation on
  Equation-of-State Constraints Derived from Quasi-periodic Brightness
  Oscillations}},
\newblock \bibinfo{journal}{\apj} \bibinfo{volume}{509} (\bibinfo{year}{1998})
  \bibinfo{pages}{793--801}.
\bibitem[{{Zhang} et~al.(1997){Zhang}, {Cui}, and {Chen}}]{1997ApJ...482L.155Z}
\bibinfo{author}{S.~N. {Zhang}}, \bibinfo{author}{W.~{Cui}},
  \bibinfo{author}{W.~{Chen}},
\newblock \bibinfo{title}{{Black Hole Spin in X-Ray Binaries: Observational
  Consequences}},
\newblock \bibinfo{journal}{\apjl} \bibinfo{volume}{482} (\bibinfo{year}{1997})
  \bibinfo{pages}{L155--L158}.
\bibitem[{{Shimura} and {Takahara}(1995)}]{1995ApJ...445..780S}
\bibinfo{author}{T.~{Shimura}}, \bibinfo{author}{F.~{Takahara}},
\newblock \bibinfo{title}{{On the spectral hardening factor of the X-ray
  emission from accretion disks in black hole candidates}},
\newblock \bibinfo{journal}{\apj} \bibinfo{volume}{445} (\bibinfo{year}{1995})
  \bibinfo{pages}{780--788}.
\bibitem[{{Merloni} et~al.(2000){Merloni}, {Fabian}, and
  {Ross}}]{2000MNRAS.313..193M}
\bibinfo{author}{A.~{Merloni}}, \bibinfo{author}{A.~C. {Fabian}},
  \bibinfo{author}{R.~R. {Ross}},
\newblock \bibinfo{title}{{On the interpretation of the multicolour disc model
  for black hole candidates}},
\newblock \bibinfo{journal}{\mnras} \bibinfo{volume}{313}
  (\bibinfo{year}{2000}) \bibinfo{pages}{193--197}.
\bibitem[{{Li} et~al.(2005){Li}, {Zimmerman}, {Narayan}, and
  {McClintock}}]{2005ApJS..157..335L}
\bibinfo{author}{L.-X. {Li}}, \bibinfo{author}{E.~R. {Zimmerman}},
  \bibinfo{author}{R.~{Narayan}}, \bibinfo{author}{J.~E. {McClintock}},
\newblock \bibinfo{title}{{Multitemperature Blackbody Spectrum of a Thin
  Accretion Disk around a Kerr Black Hole: Model Computations and Comparison
  with Observations}},
\newblock \bibinfo{journal}{\apjs} \bibinfo{volume}{157} (\bibinfo{year}{2005})
  \bibinfo{pages}{335--370}.
\bibitem[{{Miller-Jones} et~al.(2009){Miller-Jones}, {Jonker}, {Dhawan},
  {Brisken}, {Rupen}, {Nelemans}, and {Gallo}}]{2009ApJ...706L.230M}
\bibinfo{author}{J.~C.~A. {Miller-Jones}}, \bibinfo{author}{P.~G. {Jonker}},
  \bibinfo{author}{V.~{Dhawan}}, \bibinfo{author}{W.~{Brisken}},
  \bibinfo{author}{M.~P. {Rupen}}, \bibinfo{author}{G.~{Nelemans}},
  \bibinfo{author}{E.~{Gallo}},
\newblock \bibinfo{title}{{The First Accurate Parallax Distance to a Black
  Hole}},
\newblock \bibinfo{journal}{\apjl} \bibinfo{volume}{706} (\bibinfo{year}{2009})
  \bibinfo{pages}{L230--L234}.
\bibitem[{{Reid} et~al.(2011){Reid}, {McClintock}, {Narayan}, {Gou},
  {Remillard}, and {Orosz}}]{2011ApJ...742...83R}
\bibinfo{author}{M.~J. {Reid}}, \bibinfo{author}{J.~E. {McClintock}},
  \bibinfo{author}{R.~{Narayan}}, \bibinfo{author}{L.~{Gou}},
  \bibinfo{author}{R.~A. {Remillard}}, \bibinfo{author}{J.~A. {Orosz}},
\newblock \bibinfo{title}{{The Trigonometric Parallax of Cygnus X-1}},
\newblock \bibinfo{journal}{\apj} \bibinfo{volume}{742} (\bibinfo{year}{2011})
  \bibinfo{pages}{83}.
\bibitem[{{McClintock} and {Remillard}(2006)}]{2006csxs.book..157M}
\bibinfo{author}{J.~E. {McClintock}}, \bibinfo{author}{R.~A. {Remillard}},
  \bibinfo{title}{{Black hole binaries}}, pp. \bibinfo{pages}{157--213}.
\bibitem[{{Steeghs} et~al.(2013){Steeghs}, {McClintock}, {Parsons}, {Reid},
  {Littlefair}, and {Dhillon}}]{2013ApJ...768..185S}
\bibinfo{author}{D.~{Steeghs}}, \bibinfo{author}{J.~E. {McClintock}},
  \bibinfo{author}{S.~G. {Parsons}}, \bibinfo{author}{M.~J. {Reid}},
  \bibinfo{author}{S.~{Littlefair}}, \bibinfo{author}{V.~S. {Dhillon}},
\newblock \bibinfo{title}{{The Not-so-massive Black Hole in the Microquasar
  GRS1915+105}},
\newblock \bibinfo{journal}{\apj} \bibinfo{volume}{768} (\bibinfo{year}{2013})
  \bibinfo{pages}{185}.
\bibitem[{{Fender} et~al.(1999){Fender}, {Garrington}, {McKay}, {Muxlow},
  {Pooley}, {Spencer}, {Stirling}, and {Waltman}}]{1999MNRAS.304..865F}
\bibinfo{author}{R.~P. {Fender}}, \bibinfo{author}{S.~T. {Garrington}},
  \bibinfo{author}{D.~J. {McKay}}, \bibinfo{author}{T.~W.~B. {Muxlow}},
  \bibinfo{author}{G.~G. {Pooley}}, \bibinfo{author}{R.~E. {Spencer}},
  \bibinfo{author}{A.~M. {Stirling}}, \bibinfo{author}{E.~B. {Waltman}},
\newblock \bibinfo{title}{{MERLIN observations of relativistic ejections from
  GRS 1915+105}},
\newblock \bibinfo{journal}{\mnras} \bibinfo{volume}{304}
  (\bibinfo{year}{1999}) \bibinfo{pages}{865--876}.
\bibitem[{{Middleton} et~al.(2006){Middleton}, {Done}, {Gierli{\'n}ski}, and
  {Davis}}]{2006MNRAS.373.1004M}
\bibinfo{author}{M.~{Middleton}}, \bibinfo{author}{C.~{Done}},
  \bibinfo{author}{M.~{Gierli{\'n}ski}}, \bibinfo{author}{S.~W. {Davis}},
\newblock \bibinfo{title}{{Black hole spin in GRS 1915+105}},
\newblock \bibinfo{journal}{\mnras} \bibinfo{volume}{373}
  (\bibinfo{year}{2006}) \bibinfo{pages}{1004--1012}.
\bibitem[{{Gou} et~al.(2011){Gou}, {McClintock}, {Reid}, {Orosz}, {Steiner},
  {Narayan}, {Xiang}, {Remillard}, {Arnaud}, and {Davis}}]{2011ApJ...742...85G}
\bibinfo{author}{L.~{Gou}}, \bibinfo{author}{J.~E. {McClintock}},
  \bibinfo{author}{M.~J. {Reid}}, \bibinfo{author}{J.~A. {Orosz}},
  \bibinfo{author}{J.~F. {Steiner}}, \bibinfo{author}{R.~{Narayan}},
  \bibinfo{author}{J.~{Xiang}}, \bibinfo{author}{R.~A. {Remillard}},
  \bibinfo{author}{K.~A. {Arnaud}}, \bibinfo{author}{S.~W. {Davis}},
\newblock \bibinfo{title}{{The Extreme Spin of the Black Hole in Cygnus X-1}},
\newblock \bibinfo{journal}{\apj} \bibinfo{volume}{742} (\bibinfo{year}{2011})
  \bibinfo{pages}{85}.
\bibitem[{{Fabian} et~al.(2012){Fabian}, {Wilkins}, {Miller}, {Reis},
  {Reynolds}, {Cackett}, {Nowak}, {Pooley}, {Pottschmidt}, {Sanders}, {Ross},
  and {Wilms}}]{2012MNRAS.424..217F}
\bibinfo{author}{A.~C. {Fabian}}, \bibinfo{author}{D.~R. {Wilkins}},
  \bibinfo{author}{J.~M. {Miller}}, \bibinfo{author}{R.~C. {Reis}},
  \bibinfo{author}{C.~S. {Reynolds}}, \bibinfo{author}{E.~M. {Cackett}},
  \bibinfo{author}{M.~A. {Nowak}}, \bibinfo{author}{G.~G. {Pooley}},
  \bibinfo{author}{K.~{Pottschmidt}}, \bibinfo{author}{J.~S. {Sanders}},
  \bibinfo{author}{R.~R. {Ross}}, \bibinfo{author}{J.~{Wilms}},
\newblock \bibinfo{title}{{On the determination of the spin of the black hole
  in Cyg X-1 from X-ray reflection spectra}},
\newblock \bibinfo{journal}{\mnras} \bibinfo{volume}{424}
  (\bibinfo{year}{2012}) \bibinfo{pages}{217--223}.
\bibitem[{{Tomsick} et~al.(2014){Tomsick}, {Nowak}, {Parker}, {Miller},
  {Fabian}, {Harrison}, {Bachetti}, {Barret}, {Boggs}, {Christensen}, {Craig},
  {Forster}, {F{\"u}rst}, {Grefenstette}, {Hailey}, {King}, {Madsen},
  {Natalucci}, {Pottschmidt}, {Ross}, {Stern}, {Walton}, {Wilms}, and
  {Zhang}}]{2014ApJ...780...78T}
\bibinfo{author}{J.~A. {Tomsick}}, \bibinfo{author}{M.~A. {Nowak}},
  \bibinfo{author}{M.~{Parker}}, \bibinfo{author}{J.~M. {Miller}},
  \bibinfo{author}{A.~C. {Fabian}}, \bibinfo{author}{F.~A. {Harrison}},
  \bibinfo{author}{M.~{Bachetti}}, \bibinfo{author}{D.~{Barret}},
  \bibinfo{author}{S.~E. {Boggs}}, \bibinfo{author}{F.~E. {Christensen}},
  \bibinfo{author}{W.~W. {Craig}}, \bibinfo{author}{K.~{Forster}},
  \bibinfo{author}{F.~{F{\"u}rst}}, \bibinfo{author}{B.~W. {Grefenstette}},
  \bibinfo{author}{C.~J. {Hailey}}, \bibinfo{author}{A.~L. {King}},
  \bibinfo{author}{K.~K. {Madsen}}, \bibinfo{author}{L.~{Natalucci}},
  \bibinfo{author}{K.~{Pottschmidt}}, \bibinfo{author}{R.~R. {Ross}},
  \bibinfo{author}{D.~{Stern}}, \bibinfo{author}{D.~J. {Walton}},
  \bibinfo{author}{J.~{Wilms}}, \bibinfo{author}{W.~W. {Zhang}},
\newblock \bibinfo{title}{{The Reflection Component from Cygnus X-1 in the Soft
  State Measured by NuSTAR and Suzaku}},
\newblock \bibinfo{journal}{\apj} \bibinfo{volume}{780} (\bibinfo{year}{2014})
  \bibinfo{pages}{78}.
\bibitem[{{Orosz} et~al.(2011){Orosz}, {McClintock}, {Aufdenberg}, {Remillard},
  {Reid}, {Narayan}, and {Gou}}]{2011ApJ...742...84O}
\bibinfo{author}{J.~A. {Orosz}}, \bibinfo{author}{J.~E. {McClintock}},
  \bibinfo{author}{J.~P. {Aufdenberg}}, \bibinfo{author}{R.~A. {Remillard}},
  \bibinfo{author}{M.~J. {Reid}}, \bibinfo{author}{R.~{Narayan}},
  \bibinfo{author}{L.~{Gou}},
\newblock \bibinfo{title}{{The Mass of the Black Hole in Cygnus X-1}},
\newblock \bibinfo{journal}{\apj} \bibinfo{volume}{742} (\bibinfo{year}{2011})
  \bibinfo{pages}{84}.
\bibitem[{{Zi{\'o}{\l}kowski}(2014)}]{2014MNRAS.tmpL..36Z}
\bibinfo{author}{J.~{Zi{\'o}{\l}kowski}},
\newblock \bibinfo{title}{{Determination of the masses of the components of the
  HDE 226868/Cyg X-1 binary system}},
\newblock \bibinfo{journal}{\mnras}  (\bibinfo{year}{2014}).
\bibitem[{{Miller} et~al.(2010){Miller}, {D'A{\`i}}, {Bautz}, {Bhattacharyya},
  {Burrows}, {Cackett}, {Fabian}, {Freyberg}, {Haberl}, {Kennea}, {Nowak},
  {Reis}, {Strohmayer}, and {Tsujimoto}}]{2010ApJ...724.1441M}
\bibinfo{author}{J.~M. {Miller}}, \bibinfo{author}{A.~{D'A{\`i}}},
  \bibinfo{author}{M.~W. {Bautz}}, \bibinfo{author}{S.~{Bhattacharyya}},
  \bibinfo{author}{D.~N. {Burrows}}, \bibinfo{author}{E.~M. {Cackett}},
  \bibinfo{author}{A.~C. {Fabian}}, \bibinfo{author}{M.~J. {Freyberg}},
  \bibinfo{author}{F.~{Haberl}}, \bibinfo{author}{J.~{Kennea}},
  \bibinfo{author}{M.~A. {Nowak}}, \bibinfo{author}{R.~C. {Reis}},
  \bibinfo{author}{T.~E. {Strohmayer}}, \bibinfo{author}{M.~{Tsujimoto}},
\newblock \bibinfo{title}{{On Relativistic Disk Spectroscopy in Compact Objects
  with X-ray CCD Cameras}},
\newblock \bibinfo{journal}{\apj} \bibinfo{volume}{724} (\bibinfo{year}{2010})
  \bibinfo{pages}{1441--1455}.
\bibitem[{{Liu} et~al.(2008){Liu}, {McClintock}, {Narayan}, {Davis}, and
  {Orosz}}]{2008ApJ...679L..37L}
\bibinfo{author}{J.~{Liu}}, \bibinfo{author}{J.~E. {McClintock}},
  \bibinfo{author}{R.~{Narayan}}, \bibinfo{author}{S.~W. {Davis}},
  \bibinfo{author}{J.~A. {Orosz}},
\newblock \bibinfo{title}{{Precise Measurement of the Spin Parameter of the
  Stellar-Mass Black Hole M33 X-7}},
\newblock \bibinfo{journal}{\apjl} \bibinfo{volume}{679} (\bibinfo{year}{2008})
  \bibinfo{pages}{L37--L40}.
\bibitem[{{Liu} et~al.(2010){Liu}, {McClintock}, {Narayan}, {Davis}, and
  {Orosz}}]{2010ApJ...719L.109L}
\bibinfo{author}{J.~{Liu}}, \bibinfo{author}{J.~E. {McClintock}},
  \bibinfo{author}{R.~{Narayan}}, \bibinfo{author}{S.~W. {Davis}},
  \bibinfo{author}{J.~A. {Orosz}},
\newblock \bibinfo{title}{{Erratum: ''Precise Measurement of the Spin Parameter
  of the Stellar-mass Black Hole M33 X-7''}},
\newblock \bibinfo{journal}{\apjl} \bibinfo{volume}{719} (\bibinfo{year}{2010})
  \bibinfo{pages}{L109}.
\bibitem[{{Orosz} et~al.(2014){Orosz}, {Steiner}, {McClintock}, {Buxton},
  {Bailyn}, {Steeghs}, {Guberman}, and {Torres}}]{2014arXiv1402.0085O}
\bibinfo{author}{J.~A. {Orosz}}, \bibinfo{author}{J.~F. {Steiner}},
  \bibinfo{author}{J.~E. {McClintock}}, \bibinfo{author}{M.~M. {Buxton}},
  \bibinfo{author}{C.~D. {Bailyn}}, \bibinfo{author}{D.~{Steeghs}},
  \bibinfo{author}{A.~{Guberman}}, \bibinfo{author}{M.~A.~P. {Torres}},
\newblock \bibinfo{title}{{The Mass of the Black Hole in LMC X-3}},
\newblock \bibinfo{journal}{ArXiv e-prints}  (\bibinfo{year}{2014}).
\bibitem[{{Steiner} et~al.(2014){Steiner}, {McClintock}, {Orosz}, {Remillard},
  {Bailyn}, {Kolehmainen}, and {Straub}}]{2014arXiv1402.0148S}
\bibinfo{author}{J.~F. {Steiner}}, \bibinfo{author}{J.~E. {McClintock}},
  \bibinfo{author}{J.~A. {Orosz}}, \bibinfo{author}{R.~A. {Remillard}},
  \bibinfo{author}{C.~D. {Bailyn}}, \bibinfo{author}{M.~{Kolehmainen}},
  \bibinfo{author}{O.~{Straub}},
\newblock \bibinfo{title}{{The Low-Spin Black Hole in LMC X-3}},
\newblock \bibinfo{journal}{arXiv:1402.0148}  (\bibinfo{year}{2014}).
\bibitem[{{Reynolds} and {Miller}(2009)}]{2009ApJ...692..869R}
\bibinfo{author}{C.~S. {Reynolds}}, \bibinfo{author}{M.~C. {Miller}},
\newblock \bibinfo{title}{{The Time Variability of Geometrically Thin Black
  Hole Accretion Disks. I. The Search for Modes in Simulated Disks}},
\newblock \bibinfo{journal}{\apj} \bibinfo{volume}{692} (\bibinfo{year}{2009})
  \bibinfo{pages}{869--886}.
\bibitem[{{Belloni} et~al.(2012){Belloni}, {Sanna}, and
  {M{\'e}ndez}}]{2012MNRAS.426.1701B}
\bibinfo{author}{T.~M. {Belloni}}, \bibinfo{author}{A.~{Sanna}},
  \bibinfo{author}{M.~{M{\'e}ndez}},
\newblock \bibinfo{title}{{High-frequency quasi-periodic oscillations in black
  hole binaries}},
\newblock \bibinfo{journal}{\mnras} \bibinfo{volume}{426}
  (\bibinfo{year}{2012}) \bibinfo{pages}{1701--1709}.
\bibitem[{{Belloni} and {Altamirano}(2013)}]{2013MNRAS.432...19B}
\bibinfo{author}{T.~M. {Belloni}}, \bibinfo{author}{D.~{Altamirano}},
\newblock \bibinfo{title}{{Discovery of a 34 Hz quasi-periodic oscillation in
  the X-ray emission of GRS 1915+105}},
\newblock \bibinfo{journal}{\mnras} \bibinfo{volume}{432}
  (\bibinfo{year}{2013}) \bibinfo{pages}{19--22}.
\bibitem[{{Dexter} and {Blaes}(2014)}]{2014MNRAS.438.3352D}
\bibinfo{author}{J.~{Dexter}}, \bibinfo{author}{O.~{Blaes}},
\newblock \bibinfo{title}{{A model of the steep power-law spectra and
  high-frequency quasi-periodic oscillations in luminous black hole X-ray
  binaries}},
\newblock \bibinfo{journal}{\mnras} \bibinfo{volume}{438}
  (\bibinfo{year}{2014}) \bibinfo{pages}{3352--3357}.
\bibitem[{{Homan} et~al.(2001){Homan}, {Wijnands}, {van der Klis}, {Belloni},
  {van Paradijs}, {Klein-Wolt}, {Fender}, and
  {M{\'e}ndez}}]{2001ApJS..132..377H}
\bibinfo{author}{J.~{Homan}}, \bibinfo{author}{R.~{Wijnands}},
  \bibinfo{author}{M.~{van der Klis}}, \bibinfo{author}{T.~{Belloni}},
  \bibinfo{author}{J.~{van Paradijs}}, \bibinfo{author}{M.~{Klein-Wolt}},
  \bibinfo{author}{R.~{Fender}}, \bibinfo{author}{M.~{M{\'e}ndez}},
\newblock \bibinfo{title}{{Correlated X-Ray Spectral and Timing Behavior of the
  Black Hole Candidate XTE J1550-564: A New Interpretation of Black Hole
  States}},
\newblock \bibinfo{journal}{\apjs} \bibinfo{volume}{132} (\bibinfo{year}{2001})
  \bibinfo{pages}{377--402}.
\bibitem[{{Miller} et~al.(2001){Miller}, {Wijnands}, {Homan}, {Belloni},
  {Pooley}, {Corbel}, {Kouveliotou}, {van der Klis}, and
  {Lewin}}]{2001ApJ...563..928M}
\bibinfo{author}{J.~M. {Miller}}, \bibinfo{author}{R.~{Wijnands}},
  \bibinfo{author}{J.~{Homan}}, \bibinfo{author}{T.~{Belloni}},
  \bibinfo{author}{D.~{Pooley}}, \bibinfo{author}{S.~{Corbel}},
  \bibinfo{author}{C.~{Kouveliotou}}, \bibinfo{author}{M.~{van der Klis}},
  \bibinfo{author}{W.~H.~G. {Lewin}},
\newblock \bibinfo{title}{{High-Frequency Quasi-Periodic Oscillations in the
  2000 Outburst of the Galactic Microquasar XTE J1550-564}},
\newblock \bibinfo{journal}{\apj} \bibinfo{volume}{563} (\bibinfo{year}{2001})
  \bibinfo{pages}{928--933}.
\bibitem[{{van der Klis}(2006)}]{2006csxs.book...39V}
\bibinfo{author}{M.~{van der Klis}}, \bibinfo{title}{{Rapid X-ray
  Variability}}, pp. \bibinfo{pages}{39--112}.
\bibitem[{{Strohmayer}(2001)}]{2001ApJ...552L..49S}
\bibinfo{author}{T.~E. {Strohmayer}},
\newblock \bibinfo{title}{{Discovery of a 450 HZ Quasi-periodic Oscillation
  from the Microquasar GRO J1655-40 with the Rossi X-Ray Timing Explorer}},
\newblock \bibinfo{journal}{\apjl} \bibinfo{volume}{552} (\bibinfo{year}{2001})
  \bibinfo{pages}{L49--L53}.
\bibitem[{{McKinney} et~al.(2012){McKinney}, {Tchekhovskoy}, and
  {Blandford}}]{2012MNRAS.423.3083M}
\bibinfo{author}{J.~C. {McKinney}}, \bibinfo{author}{A.~{Tchekhovskoy}},
  \bibinfo{author}{R.~D. {Blandford}},
\newblock \bibinfo{title}{{General relativistic magnetohydrodynamic simulations
  of magnetically choked accretion flows around black holes}},
\newblock \bibinfo{journal}{\mnras} \bibinfo{volume}{423}
  (\bibinfo{year}{2012}) \bibinfo{pages}{3083--3117}.
\bibitem[{{Stella} et~al.(1999){Stella}, {Vietri}, and
  {Morsink}}]{1999ApJ...524L..63S}
\bibinfo{author}{L.~{Stella}}, \bibinfo{author}{M.~{Vietri}},
  \bibinfo{author}{S.~M. {Morsink}},
\newblock \bibinfo{title}{{Correlations in the Quasi-periodic Oscillation
  Frequencies of Low-Mass X-Ray Binaries and the Relativistic Precession
  Model}},
\newblock \bibinfo{journal}{\apjl} \bibinfo{volume}{524} (\bibinfo{year}{1999})
  \bibinfo{pages}{L63--L66}.
\bibitem[{{Motta} et~al.(2014){Motta}, {Belloni}, {Stella}, {Mu{\~n}oz-Darias},
  and {Fender}}]{2014MNRAS.437.2554M}
\bibinfo{author}{S.~E. {Motta}}, \bibinfo{author}{T.~M. {Belloni}},
  \bibinfo{author}{L.~{Stella}}, \bibinfo{author}{T.~{Mu{\~n}oz-Darias}},
  \bibinfo{author}{R.~{Fender}},
\newblock \bibinfo{title}{{Precise mass and spin measurements for a
  stellar-mass black hole through X-ray timing: the case of GRO J1655-40}},
\newblock \bibinfo{journal}{\mnras} \bibinfo{volume}{437}
  (\bibinfo{year}{2014}) \bibinfo{pages}{2554--2565}.
\bibitem[{{Perez} et~al.(1997){Perez}, {Silbergleit}, {Wagoner}, and
  {Lehr}}]{1997ApJ...476..589P}
\bibinfo{author}{C.~A. {Perez}}, \bibinfo{author}{A.~S. {Silbergleit}},
  \bibinfo{author}{R.~V. {Wagoner}}, \bibinfo{author}{D.~E. {Lehr}},
\newblock \bibinfo{title}{{Relativistic Diskoseismology. I. Analytical Results
  for ``Gravity Modes''}},
\newblock \bibinfo{journal}{\apj} \bibinfo{volume}{476} (\bibinfo{year}{1997})
  \bibinfo{pages}{589}.
\bibitem[{{Nowak} et~al.(1997){Nowak}, {Wagoner}, {Begelman}, and
  {Lehr}}]{1997ApJ...477L..91N}
\bibinfo{author}{M.~A. {Nowak}}, \bibinfo{author}{R.~V. {Wagoner}},
  \bibinfo{author}{M.~C. {Begelman}}, \bibinfo{author}{D.~E. {Lehr}},
\newblock \bibinfo{title}{{The 67 HZ Feature in the Black Hole Candidate GRS
  1915+105 as a Possible ``Diskoseismic'' Mode}},
\newblock \bibinfo{journal}{\apjl} \bibinfo{volume}{477} (\bibinfo{year}{1997})
  \bibinfo{pages}{L91}.
\bibitem[{{Kato} et~al.(2001){Kato}, {Hayashi}, {Miyaji}, and
  {Matsumoto}}]{2001AdSpR..28..505K}
\bibinfo{author}{Y.~{Kato}}, \bibinfo{author}{M.~R. {Hayashi}},
  \bibinfo{author}{S.~{Miyaji}}, \bibinfo{author}{R.~{Matsumoto}},
\newblock \bibinfo{title}{{Magnetohydrodynamic simulations of accretion disks
  around a weakly magnetized neutron star in strong gravity}},
\newblock \bibinfo{journal}{Advances in Space Research} \bibinfo{volume}{28}
  (\bibinfo{year}{2001}) \bibinfo{pages}{505--510}.
\bibitem[{{Wagoner}(2008)}]{2008NewAR..51..828W}
\bibinfo{author}{R.~V. {Wagoner}},
\newblock \bibinfo{title}{{Relativistic and Newtonian diskoseismology}},
\newblock \bibinfo{journal}{\nar} \bibinfo{volume}{51} (\bibinfo{year}{2008})
  \bibinfo{pages}{828--834}.
\bibitem[{{Wagoner}(2012)}]{2012ApJ...752L..18W}
\bibinfo{author}{R.~V. {Wagoner}},
\newblock \bibinfo{title}{{Diskoseismology and QPOs Confront Black Hole Spin}},
\newblock \bibinfo{journal}{\apjl} \bibinfo{volume}{752} (\bibinfo{year}{2012})
  \bibinfo{pages}{L18}.
\bibitem[{{Barret} et~al.(2013){Barret}, {Nandra}, {Barcons}, {Fabian}, {den
  Herder}, {Piro}, {Watson}, {Aird}, {Branduardi-Raymont}, {Cappi}, {Carrera},
  {Comastri}, {Costantini}, {Croston}, {Decourchelle}, {Done}, {Dovciak},
  {Ettori}, {Finoguenov}, {Georgakakis}, {Jonker}, {Kaastra}, {Matt}, {Motch},
  {O'Brien}, {Pareschi}, {Pointecouteau}, {Pratt}, {Rauw}, {Reiprich},
  {Sanders}, {Sciortino}, {Willingale}, and {Wilms}}]{2013sf2a.conf..447B}
\bibinfo{author}{D.~{Barret}}, \bibinfo{author}{K.~{Nandra}},
  \bibinfo{author}{X.~{Barcons}}, \bibinfo{author}{A.~{Fabian}},
  \bibinfo{author}{J.~W. {den Herder}}, \bibinfo{author}{L.~{Piro}},
  \bibinfo{author}{M.~{Watson}}, \bibinfo{author}{J.~{Aird}},
  \bibinfo{author}{G.~{Branduardi-Raymont}}, \bibinfo{author}{M.~{Cappi}},
  \bibinfo{author}{F.~{Carrera}}, \bibinfo{author}{A.~{Comastri}},
  \bibinfo{author}{E.~{Costantini}}, \bibinfo{author}{J.~{Croston}},
  \bibinfo{author}{A.~{Decourchelle}}, \bibinfo{author}{C.~{Done}},
  \bibinfo{author}{M.~{Dovciak}}, \bibinfo{author}{S.~{Ettori}},
  \bibinfo{author}{A.~{Finoguenov}}, \bibinfo{author}{A.~{Georgakakis}},
  \bibinfo{author}{P.~{Jonker}}, \bibinfo{author}{J.~{Kaastra}},
  \bibinfo{author}{G.~{Matt}}, \bibinfo{author}{C.~{Motch}},
  \bibinfo{author}{P.~{O'Brien}}, \bibinfo{author}{G.~{Pareschi}},
  \bibinfo{author}{E.~{Pointecouteau}}, \bibinfo{author}{G.~{Pratt}},
  \bibinfo{author}{G.~{Rauw}}, \bibinfo{author}{T.~{Reiprich}},
  \bibinfo{author}{J.~{Sanders}}, \bibinfo{author}{S.~{Sciortino}},
  \bibinfo{author}{R.~{Willingale}}, \bibinfo{author}{J.~{Wilms}},
\newblock \bibinfo{title}{{Athena+: The first Deep Universe X-ray
  Observatory}},
\newblock in: \bibinfo{editor}{L.~{Cambresy}}, \bibinfo{editor}{F.~{Martins}},
  \bibinfo{editor}{E.~{Nuss}}, \bibinfo{editor}{A.~{Palacios}} (Eds.),
  \bibinfo{booktitle}{SF2A-2013: Proceedings of the Annual meeting of the
  French Society of Astronomy and Astrophysics}, pp. \bibinfo{pages}{447--453}.
\bibitem[{{Feroci}(2012)}]{2012ExA....34..415F}
\bibinfo{author}{e.~a. {Feroci}, M.},
\newblock \bibinfo{title}{{The Large Observatory for X-ray Timing (LOFT)}},
\newblock \bibinfo{journal}{Experimental Astronomy} \bibinfo{volume}{34}
  (\bibinfo{year}{2012}) \bibinfo{pages}{415--444}.
\bibitem[{{Homan} et~al.(2003){Homan}, {Klein-Wolt}, {Rossi}, {Miller},
  {Wijnands}, {Belloni}, {van der Klis}, and {Lewin}}]{2003ApJ...586.1262H}
\bibinfo{author}{J.~{Homan}}, \bibinfo{author}{M.~{Klein-Wolt}},
  \bibinfo{author}{S.~{Rossi}}, \bibinfo{author}{J.~M. {Miller}},
  \bibinfo{author}{R.~{Wijnands}}, \bibinfo{author}{T.~{Belloni}},
  \bibinfo{author}{M.~{van der Klis}}, \bibinfo{author}{W.~H.~G. {Lewin}},
\newblock \bibinfo{title}{{High-Frequency Quasi-periodic Oscillations in the
  Black Hole X-Ray Transient XTE J1650-500}},
\newblock \bibinfo{journal}{\apj} \bibinfo{volume}{586} (\bibinfo{year}{2003})
  \bibinfo{pages}{1262--1267}.
\bibitem[{{Homan} et~al.(2005){Homan}, {Miller}, {Wijnands}, {van der Klis},
  {Belloni}, {Steeghs}, and {Lewin}}]{2005ApJ...623..383H}
\bibinfo{author}{J.~{Homan}}, \bibinfo{author}{J.~M. {Miller}},
  \bibinfo{author}{R.~{Wijnands}}, \bibinfo{author}{M.~{van der Klis}},
  \bibinfo{author}{T.~{Belloni}}, \bibinfo{author}{D.~{Steeghs}},
  \bibinfo{author}{W.~H.~G. {Lewin}},
\newblock \bibinfo{title}{{High- and Low-Frequency Quasi-periodic Oscillations
  in the X-Ray Light Curves of the Black Hole Transient H1743-322}},
\newblock \bibinfo{journal}{\apj} \bibinfo{volume}{623} (\bibinfo{year}{2005})
  \bibinfo{pages}{383--391}.
\bibitem[{{Klein-Wolt} et~al.(2004){Klein-Wolt}, {Homan}, and {van der
  Klis}}]{2004NuPhS.132..381K}
\bibinfo{author}{M.~{Klein-Wolt}}, \bibinfo{author}{J.~{Homan}},
  \bibinfo{author}{M.~{van der Klis}},
\newblock \bibinfo{title}{{High frequency features in the 1998 outburst of 4U
  1630-47}},
\newblock \bibinfo{journal}{Nuclear Physics B Proceedings Supplements}
  \bibinfo{volume}{132} (\bibinfo{year}{2004}) \bibinfo{pages}{381--386}.
\bibitem[{{Schnittman} et~al.(2006){Schnittman}, {Homan}, and
  {Miller}}]{2006ApJ...642..420S}
\bibinfo{author}{J.~D. {Schnittman}}, \bibinfo{author}{J.~{Homan}},
  \bibinfo{author}{J.~M. {Miller}},
\newblock \bibinfo{title}{{A Precessing Ring Model for Low-Frequency
  Quasi-periodic Oscillations}},
\newblock \bibinfo{journal}{\apj} \bibinfo{volume}{642} (\bibinfo{year}{2006})
  \bibinfo{pages}{420--426}.
\bibitem[{{Altamirano} and {Belloni}(2012)}]{2012ApJ...747L...4A}
\bibinfo{author}{D.~{Altamirano}}, \bibinfo{author}{T.~{Belloni}},
\newblock \bibinfo{title}{{Discovery of High-frequency Quasi-periodic
  Oscillations in the Black Hole Candidate IGR J17091-3624}},
\newblock \bibinfo{journal}{\apjl} \bibinfo{volume}{747} (\bibinfo{year}{2012})
  \bibinfo{pages}{L4}.
\bibitem[{{Morgan} et~al.(1997){Morgan}, {Remillard}, and
  {Greiner}}]{1997ApJ...482..993M}
\bibinfo{author}{E.~H. {Morgan}}, \bibinfo{author}{R.~A. {Remillard}},
  \bibinfo{author}{J.~{Greiner}},
\newblock \bibinfo{title}{{RXTE Observations of QPOs in the Black Hole
  Candidate GRS 1915+105}},
\newblock \bibinfo{journal}{\apj} \bibinfo{volume}{482} (\bibinfo{year}{1997})
  \bibinfo{pages}{993}.
\bibitem[{{Schnittman} and {Krolik}(2009)}]{2009ApJ...701.1175S}
\bibinfo{author}{J.~D. {Schnittman}}, \bibinfo{author}{J.~H. {Krolik}},
\newblock \bibinfo{title}{{X-ray Polarization from Accreting Black Holes: The
  Thermal State}},
\newblock \bibinfo{journal}{\apj} \bibinfo{volume}{701} (\bibinfo{year}{2009})
  \bibinfo{pages}{1175--1187}.
\bibitem[{{Davis} et~al.(2009){Davis}, {Blaes}, {Hirose}, and
  {Krolik}}]{2009ApJ...703..569D}
\bibinfo{author}{S.~W. {Davis}}, \bibinfo{author}{O.~M. {Blaes}},
  \bibinfo{author}{S.~{Hirose}}, \bibinfo{author}{J.~H. {Krolik}},
\newblock \bibinfo{title}{{The Effects of Magnetic Fields and Inhomogeneities
  on Accretion Disk Spectra and Polarization}},
\newblock \bibinfo{journal}{\apj} \bibinfo{volume}{703} (\bibinfo{year}{2009})
  \bibinfo{pages}{569--584}.
\bibitem[{{Connors} et~al.(1980){Connors}, {Stark}, and
  {Piran}}]{1980ApJ...235..224C}
\bibinfo{author}{P.~A. {Connors}}, \bibinfo{author}{R.~F. {Stark}},
  \bibinfo{author}{T.~{Piran}},
\newblock \bibinfo{title}{{Polarization features of X-ray radiation emitted
  near black holes}},
\newblock \bibinfo{journal}{\apj} \bibinfo{volume}{235} (\bibinfo{year}{1980})
  \bibinfo{pages}{224--244}.
\bibitem[{{Li} et~al.(2009){Li}, {Narayan}, and
  {McClintock}}]{2009ApJ...691..847L}
\bibinfo{author}{L.-X. {Li}}, \bibinfo{author}{R.~{Narayan}},
  \bibinfo{author}{J.~E. {McClintock}},
\newblock \bibinfo{title}{{Inferring the Inclination of a Black Hole Accretion
  Disk from Observations of its Polarized Continuum Radiation}},
\newblock \bibinfo{journal}{\apj} \bibinfo{volume}{691} (\bibinfo{year}{2009})
  \bibinfo{pages}{847--865}.
\bibitem[{{Schnittman} and {Krolik}(2010)}]{2010ApJ...712..908S}
\bibinfo{author}{J.~D. {Schnittman}}, \bibinfo{author}{J.~H. {Krolik}},
\newblock \bibinfo{title}{{X-ray Polarization from Accreting Black Holes:
  Coronal Emission}},
\newblock \bibinfo{journal}{\apj} \bibinfo{volume}{712} (\bibinfo{year}{2010})
  \bibinfo{pages}{908--924}.
\bibitem[{{Krawczynski} et~al.(2011){Krawczynski}, {Garson}, {Guo}, {Baring},
  {Ghosh}, {Beilicke}, and {Lee}}]{2011APh....34..550K}
\bibinfo{author}{H.~{Krawczynski}}, \bibinfo{author}{A.~{Garson}},
  \bibinfo{author}{Q.~{Guo}}, \bibinfo{author}{M.~G. {Baring}},
  \bibinfo{author}{P.~{Ghosh}}, \bibinfo{author}{M.~{Beilicke}},
  \bibinfo{author}{K.~{Lee}},
\newblock \bibinfo{title}{{Scientific prospects for hard X-ray polarimetry}},
\newblock \bibinfo{journal}{Astroparticle Physics} \bibinfo{volume}{34}
  (\bibinfo{year}{2011}) \bibinfo{pages}{550--567}.
\bibitem[{{Beilicke} et~al.(2011){Beilicke}, {Baring}, {Barthelmy}, {Binns},
  {Buckley}, {Cowsik}, {Dowkontt}, {Garson}, {Guo}, {Haba}, {Israel},
  {Kunieda}, {Lee}, {Matsumoto}, {Miyazawa}, {Okajima}, {Schnittman}, {Tamura},
  {Tueller}, and {Krawczynski}}]{2011SPIE.8145E.240B}
\bibinfo{author}{M.~{Beilicke}}, \bibinfo{author}{M.~G. {Baring}},
  \bibinfo{author}{S.~{Barthelmy}}, \bibinfo{author}{W.~R. {Binns}},
  \bibinfo{author}{J.~{Buckley}}, \bibinfo{author}{R.~{Cowsik}},
  \bibinfo{author}{P.~{Dowkontt}}, \bibinfo{author}{A.~{Garson}},
  \bibinfo{author}{Q.~{Guo}}, \bibinfo{author}{Y.~{Haba}},
  \bibinfo{author}{M.~H. {Israel}}, \bibinfo{author}{H.~{Kunieda}},
  \bibinfo{author}{K.~{Lee}}, \bibinfo{author}{H.~{Matsumoto}},
  \bibinfo{author}{T.~{Miyazawa}}, \bibinfo{author}{T.~{Okajima}},
  \bibinfo{author}{J.~{Schnittman}}, \bibinfo{author}{K.~{Tamura}},
  \bibinfo{author}{J.~{Tueller}}, \bibinfo{author}{H.~{Krawczynski}},
\newblock \bibinfo{title}{{Design and tests of the hard x-ray polarimeter
  X-Calibur}},
\newblock in: \bibinfo{booktitle}{Society of Photo-Optical Instrumentation
  Engineers (SPIE) Conference Series}, volume \bibinfo{volume}{8145} of
  \textit{\bibinfo{series}{Society of Photo-Optical Instrumentation Engineers
  (SPIE) Conference Series}}.
\bibitem[{{Jahoda}(2010)}]{2010SPIE.7732E..24J}
\bibinfo{author}{K.~{Jahoda}},
\newblock \bibinfo{title}{{The Gravity and Extreme Magnetism Small Explorer}},
\newblock in: \bibinfo{booktitle}{Society of Photo-Optical Instrumentation
  Engineers (SPIE) Conference Series}, volume \bibinfo{volume}{7732} of
  \textit{\bibinfo{series}{Society of Photo-Optical Instrumentation Engineers
  (SPIE) Conference Series}}.
\bibitem[{{Schnittman} et~al.(2013){Schnittman}, {Angelini}, {Baring},
  {Baumgartner}, {Black}, {Dotson}, {Ghosh}, {Harding}, {Hill}, {Jahoda},
  {Kaaret}, {Kallman}, {Krawczynski}, {Krolik}, {Lai}, {Markwardt}, {Marshall},
  {Martoff}, {Morris}, {Okajima}, {Petre}, {Poutanen}, {Reynolds}, {Scargle},
  {Serlemitsos}, {Soong}, {Strohmayer}, {Swank}, {Tawara}, and
  {Tamagawa}}]{2013arXiv1301.1957S}
\bibinfo{author}{J.~{Schnittman}}, \bibinfo{author}{L.~{Angelini}},
  \bibinfo{author}{M.~{Baring}}, \bibinfo{author}{W.~{Baumgartner}},
  \bibinfo{author}{K.~{Black}}, \bibinfo{author}{J.~{Dotson}},
  \bibinfo{author}{P.~{Ghosh}}, \bibinfo{author}{A.~{Harding}},
  \bibinfo{author}{J.~{Hill}}, \bibinfo{author}{K.~{Jahoda}},
  \bibinfo{author}{P.~{Kaaret}}, \bibinfo{author}{T.~{Kallman}},
  \bibinfo{author}{H.~{Krawczynski}}, \bibinfo{author}{J.~{Krolik}},
  \bibinfo{author}{D.~{Lai}}, \bibinfo{author}{C.~{Markwardt}},
  \bibinfo{author}{H.~{Marshall}}, \bibinfo{author}{J.~{Martoff}},
  \bibinfo{author}{R.~{Morris}}, \bibinfo{author}{T.~{Okajima}},
  \bibinfo{author}{R.~{Petre}}, \bibinfo{author}{J.~{Poutanen}},
  \bibinfo{author}{S.~{Reynolds}}, \bibinfo{author}{J.~{Scargle}},
  \bibinfo{author}{P.~{Serlemitsos}}, \bibinfo{author}{Y.~{Soong}},
  \bibinfo{author}{T.~{Strohmayer}}, \bibinfo{author}{J.~{Swank}},
  \bibinfo{author}{Y.~{Tawara}}, \bibinfo{author}{T.~{Tamagawa}},
\newblock \bibinfo{title}{{X-ray Polarization from Black Holes: GEMS Scientific
  White Paper}},
\newblock \bibinfo{journal}{arXiv:1301.1957}  (\bibinfo{year}{2013}).
\bibitem[{{Berti} and {Volonteri}(2008)}]{2008ApJ...684..822B}
\bibinfo{author}{E.~{Berti}}, \bibinfo{author}{M.~{Volonteri}},
\newblock \bibinfo{title}{{Cosmological Black Hole Spin Evolution by Mergers
  and Accretion}},
\newblock \bibinfo{journal}{\apj} \bibinfo{volume}{684} (\bibinfo{year}{2008})
  \bibinfo{pages}{822--828}.
\bibitem[{{Reynolds}(2013{\natexlab{a}})}]{2013SSRv..tmp...81R}
\bibinfo{author}{C.~S. {Reynolds}},
\newblock \bibinfo{title}{{Measuring Black Hole Spin Using X-Ray Reflection
  Spectroscopy}},
\newblock \bibinfo{journal}{\ssr}  (\bibinfo{year}{2013}{\natexlab{a}}).
\bibitem[{{Reynolds}(2013{\natexlab{b}})}]{2013CQGra..30x4004R}
\bibinfo{author}{C.~S. {Reynolds}},
\newblock \bibinfo{title}{{The spin of supermassive black holes}},
\newblock \bibinfo{journal}{Classical and Quantum Gravity} \bibinfo{volume}{30}
  (\bibinfo{year}{2013}{\natexlab{b}}) \bibinfo{pages}{244004}.
\bibitem[{{Reis} et~al.(2014){Reis}, {Reynolds}, {Miller}, and
  {Walton}}]{2014Natur.507..207R}
\bibinfo{author}{R.~C. {Reis}}, \bibinfo{author}{M.~T. {Reynolds}},
  \bibinfo{author}{J.~M. {Miller}}, \bibinfo{author}{D.~J. {Walton}},
\newblock \bibinfo{title}{{Reflection from the strong gravity regime in a
  lensed quasar at redshift z = 0.658}},
\newblock \bibinfo{journal}{\nat} \bibinfo{volume}{507} (\bibinfo{year}{2014})
  \bibinfo{pages}{207--209}.
\bibitem[{{Steiner} et~al.(2012){Steiner}, {Reis}, {Fabian}, {Remillard},
  {McClintock}, {Gou}, {Cooke}, {Brenneman}, and
  {Sanders}}]{2012MNRAS.427.2552S}
\bibinfo{author}{J.~F. {Steiner}}, \bibinfo{author}{R.~C. {Reis}},
  \bibinfo{author}{A.~C. {Fabian}}, \bibinfo{author}{R.~A. {Remillard}},
  \bibinfo{author}{J.~E. {McClintock}}, \bibinfo{author}{L.~{Gou}},
  \bibinfo{author}{R.~{Cooke}}, \bibinfo{author}{L.~W. {Brenneman}},
  \bibinfo{author}{J.~S. {Sanders}},
\newblock \bibinfo{title}{{A broad iron line in LMC X-1}},
\newblock \bibinfo{journal}{\mnras} \bibinfo{volume}{427}
  (\bibinfo{year}{2012}) \bibinfo{pages}{2552--2561}.
\bibitem[{{Gou} et~al.(2009){Gou}, {McClintock}, {Liu}, {Narayan}, {Steiner},
  {Remillard}, {Orosz}, {Davis}, {Ebisawa}, and
  {Schlegel}}]{2009ApJ...701.1076G}
\bibinfo{author}{L.~{Gou}}, \bibinfo{author}{J.~E. {McClintock}},
  \bibinfo{author}{J.~{Liu}}, \bibinfo{author}{R.~{Narayan}},
  \bibinfo{author}{J.~F. {Steiner}}, \bibinfo{author}{R.~A. {Remillard}},
  \bibinfo{author}{J.~A. {Orosz}}, \bibinfo{author}{S.~W. {Davis}},
  \bibinfo{author}{K.~{Ebisawa}}, \bibinfo{author}{E.~M. {Schlegel}},
\newblock \bibinfo{title}{{A Determination of the Spin of the Black Hole
  Primary in LMC X-1}},
\newblock \bibinfo{journal}{\apj} \bibinfo{volume}{701} (\bibinfo{year}{2009})
  \bibinfo{pages}{1076--1090}.
\bibitem[{{Gou} et~al.(2010){Gou}, {McClintock}, {Steiner}, {Narayan},
  {Cantrell}, {Bailyn}, and {Orosz}}]{2010ApJ...718L.122G}
\bibinfo{author}{L.~{Gou}}, \bibinfo{author}{J.~E. {McClintock}},
  \bibinfo{author}{J.~F. {Steiner}}, \bibinfo{author}{R.~{Narayan}},
  \bibinfo{author}{A.~G. {Cantrell}}, \bibinfo{author}{C.~D. {Bailyn}},
  \bibinfo{author}{J.~A. {Orosz}},
\newblock \bibinfo{title}{{The Spin of the Black Hole in the Soft X-ray
  Transient A0620-00}},
\newblock \bibinfo{journal}{\apjl} \bibinfo{volume}{718} (\bibinfo{year}{2010})
  \bibinfo{pages}{L122--L126}.
\bibitem[{{Morningstar} et~al.(2014){Morningstar}, {Miller}, {Reis}, and
  {Ebisawa}}]{2014ApJ...784L..18M}
\bibinfo{author}{W.~R. {Morningstar}}, \bibinfo{author}{J.~M. {Miller}},
  \bibinfo{author}{R.~C. {Reis}}, \bibinfo{author}{K.~{Ebisawa}},
\newblock \bibinfo{title}{{The Spin of the Black Hole GS 1124-683: Observation
  of a Retrograde Accretion Disk?}},
\newblock \bibinfo{journal}{\apjl} \bibinfo{volume}{784} (\bibinfo{year}{2014})
  \bibinfo{pages}{L18}.
\bibitem[{{Shafee} et~al.(2006){Shafee}, {McClintock}, {Narayan}, {Davis},
  {Li}, and {Remillard}}]{2006ApJ...636L.113S}
\bibinfo{author}{R.~{Shafee}}, \bibinfo{author}{J.~E. {McClintock}},
  \bibinfo{author}{R.~{Narayan}}, \bibinfo{author}{S.~W. {Davis}},
  \bibinfo{author}{L.-X. {Li}}, \bibinfo{author}{R.~A. {Remillard}},
\newblock \bibinfo{title}{{Estimating the Spin of Stellar-Mass Black Holes by
  Spectral Fitting of the X-Ray Continuum}},
\newblock \bibinfo{journal}{\apjl} \bibinfo{volume}{636} (\bibinfo{year}{2006})
  \bibinfo{pages}{L113--L116}.
\bibitem[{{Steiner} et~al.(2011){Steiner}, {Reis}, {McClintock}, {Narayan},
  {Remillard}, {Orosz}, {Gou}, {Fabian}, and {Torres}}]{2011MNRAS.416..941S}
\bibinfo{author}{J.~F. {Steiner}}, \bibinfo{author}{R.~C. {Reis}},
  \bibinfo{author}{J.~E. {McClintock}}, \bibinfo{author}{R.~{Narayan}},
  \bibinfo{author}{R.~A. {Remillard}}, \bibinfo{author}{J.~A. {Orosz}},
  \bibinfo{author}{L.~{Gou}}, \bibinfo{author}{A.~C. {Fabian}},
  \bibinfo{author}{M.~A.~P. {Torres}},
\newblock \bibinfo{title}{{The spin of the black hole microquasar XTE J1550-564
  via the continuum-fitting and Fe-line methods}},
\newblock \bibinfo{journal}{\mnras} \bibinfo{volume}{416}
  (\bibinfo{year}{2011}) \bibinfo{pages}{941--958}.
\bibitem[{{King} et~al.(2014){King}, {Walton}, {Miller}, {Barret}, {Boggs},
  {Christensen}, {Craig}, {Fabian}, {F{\"u}rst}, {Hailey}, {Harrison},
  {Krivonos}, {Mori}, {Natalucci}, {Stern}, {Tomsick}, and
  {Zhang}}]{2014ApJ...784L...2K}
\bibinfo{author}{A.~L. {King}}, \bibinfo{author}{D.~J. {Walton}},
  \bibinfo{author}{J.~M. {Miller}}, \bibinfo{author}{D.~{Barret}},
  \bibinfo{author}{S.~E. {Boggs}}, \bibinfo{author}{F.~E. {Christensen}},
  \bibinfo{author}{W.~W. {Craig}}, \bibinfo{author}{A.~C. {Fabian}},
  \bibinfo{author}{F.~{F{\"u}rst}}, \bibinfo{author}{C.~J. {Hailey}},
  \bibinfo{author}{F.~A. {Harrison}}, \bibinfo{author}{R.~{Krivonos}},
  \bibinfo{author}{K.~{Mori}}, \bibinfo{author}{L.~{Natalucci}},
  \bibinfo{author}{D.~{Stern}}, \bibinfo{author}{J.~A. {Tomsick}},
  \bibinfo{author}{W.~W. {Zhang}},
\newblock \bibinfo{title}{{The Disk Wind in the Rapidly Spinning Stellar-mass
  Black Hole 4U 1630-472 Observed with NuSTAR}},
\newblock \bibinfo{journal}{\apjl} \bibinfo{volume}{784} (\bibinfo{year}{2014})
  \bibinfo{pages}{L2}.
\bibitem[{{Reis} et~al.(2009){Reis}, {Fabian}, {Ross}, and
  {Miller}}]{2009MNRAS.395.1257R}
\bibinfo{author}{R.~C. {Reis}}, \bibinfo{author}{A.~C. {Fabian}},
  \bibinfo{author}{R.~R. {Ross}}, \bibinfo{author}{J.~M. {Miller}},
\newblock \bibinfo{title}{{Determining the spin of two stellar-mass black holes
  from disc reflection signatures}},
\newblock \bibinfo{journal}{\mnras} \bibinfo{volume}{395}
  (\bibinfo{year}{2009}) \bibinfo{pages}{1257--1264}.
\bibitem[{{Reis} et~al.(2012){Reis}, {Miller}, {Reynolds}, {Fabian}, and
  {Walton}}]{2012ApJ...751...34R}
\bibinfo{author}{R.~C. {Reis}}, \bibinfo{author}{J.~M. {Miller}},
  \bibinfo{author}{M.~T. {Reynolds}}, \bibinfo{author}{A.~C. {Fabian}},
  \bibinfo{author}{D.~J. {Walton}},
\newblock \bibinfo{title}{{Suzaku Observation of the Black Hole Candidate Maxi
  J1836-194 in a Hard/Intermediate Spectral State}},
\newblock \bibinfo{journal}{\apj} \bibinfo{volume}{751} (\bibinfo{year}{2012})
  \bibinfo{pages}{34}.
\bibitem[{{Reis} et~al.(2013){Reis}, {Reynolds}, {Miller}, {Walton}, {Maitra},
  {King}, and {Degenaar}}]{2013ApJ...778..155R}
\bibinfo{author}{R.~C. {Reis}}, \bibinfo{author}{M.~T. {Reynolds}},
  \bibinfo{author}{J.~M. {Miller}}, \bibinfo{author}{D.~J. {Walton}},
  \bibinfo{author}{D.~{Maitra}}, \bibinfo{author}{A.~{King}},
  \bibinfo{author}{N.~{Degenaar}},
\newblock \bibinfo{title}{{SWIFT J1910.2-0546: A Possible Black Hole Binary
  with a Retrograde Spin or Truncated Disk}},
\newblock \bibinfo{journal}{\apj} \bibinfo{volume}{778} (\bibinfo{year}{2013})
  \bibinfo{pages}{155}.
\bibitem[{{Yagi} and {Yunes}(2013)}]{2013PhRvD..88b3009Y}
\bibinfo{author}{K.~{Yagi}}, \bibinfo{author}{N.~{Yunes}},
\newblock \bibinfo{title}{{I-Love-Q relations in neutron stars and their
  applications to astrophysics, gravitational waves, and fundamental physics}},
\newblock \bibinfo{journal}{\prd} \bibinfo{volume}{88} (\bibinfo{year}{2013})
  \bibinfo{pages}{023009}.
\bibitem[{{Orosz} et~al.(2007){Orosz}, {McClintock}, {Narayan}, {Bailyn},
  {Hartman}, {Macri}, {Liu}, {Pietsch}, {Remillard}, {Shporer}, and
  {Mazeh}}]{2007Natur.449..872O}
\bibinfo{author}{J.~A. {Orosz}}, \bibinfo{author}{J.~E. {McClintock}},
  \bibinfo{author}{R.~{Narayan}}, \bibinfo{author}{C.~D. {Bailyn}},
  \bibinfo{author}{J.~D. {Hartman}}, \bibinfo{author}{L.~{Macri}},
  \bibinfo{author}{J.~{Liu}}, \bibinfo{author}{W.~{Pietsch}},
  \bibinfo{author}{R.~A. {Remillard}}, \bibinfo{author}{A.~{Shporer}},
  \bibinfo{author}{T.~{Mazeh}},
\newblock \bibinfo{title}{{A 15.65-solar-mass black hole in an eclipsing binary
  in the nearby spiral galaxy M 33}},
\newblock \bibinfo{journal}{\nat} \bibinfo{volume}{449} (\bibinfo{year}{2007})
  \bibinfo{pages}{872--875}.
\bibitem[{{Orosz} et~al.(2009){Orosz}, {Steeghs}, {McClintock}, {Torres},
  {Bochkov}, {Gou}, {Narayan}, {Blaschak}, {Levine}, {Remillard}, {Bailyn},
  {Dwyer}, and {Buxton}}]{2009ApJ...697..573O}
\bibinfo{author}{J.~A. {Orosz}}, \bibinfo{author}{D.~{Steeghs}},
  \bibinfo{author}{J.~E. {McClintock}}, \bibinfo{author}{M.~A.~P. {Torres}},
  \bibinfo{author}{I.~{Bochkov}}, \bibinfo{author}{L.~{Gou}},
  \bibinfo{author}{R.~{Narayan}}, \bibinfo{author}{M.~{Blaschak}},
  \bibinfo{author}{A.~M. {Levine}}, \bibinfo{author}{R.~A. {Remillard}},
  \bibinfo{author}{C.~D. {Bailyn}}, \bibinfo{author}{M.~M. {Dwyer}},
  \bibinfo{author}{M.~{Buxton}},
\newblock \bibinfo{title}{{A New Dynamical Model for the Black Hole Binary LMC
  X-1}},
\newblock \bibinfo{journal}{\apj} \bibinfo{volume}{697} (\bibinfo{year}{2009})
  \bibinfo{pages}{573--591}.
\bibitem[{{Cantrell} et~al.(2010){Cantrell}, {Bailyn}, {Orosz}, {McClintock},
  {Remillard}, {Froning}, {Neilsen}, {Gelino}, and {Gou}}]{2010ApJ...710.1127C}
\bibinfo{author}{A.~G. {Cantrell}}, \bibinfo{author}{C.~D. {Bailyn}},
  \bibinfo{author}{J.~A. {Orosz}}, \bibinfo{author}{J.~E. {McClintock}},
  \bibinfo{author}{R.~A. {Remillard}}, \bibinfo{author}{C.~S. {Froning}},
  \bibinfo{author}{J.~{Neilsen}}, \bibinfo{author}{D.~M. {Gelino}},
  \bibinfo{author}{L.~{Gou}},
\newblock \bibinfo{title}{{The Inclination of the Soft X-Ray Transient A0620-00
  and the Mass of its Black Hole}},
\newblock \bibinfo{journal}{\apj} \bibinfo{volume}{710} (\bibinfo{year}{2010})
  \bibinfo{pages}{1127--1141}.
\bibitem[{{Orosz}(2003)}]{2003IAUS..212..365O}
\bibinfo{author}{J.~A. {Orosz}},
\newblock \bibinfo{title}{{Inventory of black hole binaries}},
\newblock in: \bibinfo{editor}{K.~{van der Hucht}},
  \bibinfo{editor}{A.~{Herrero}}, \bibinfo{editor}{C.~{Esteban}} (Eds.),
  \bibinfo{booktitle}{A Massive Star Odyssey: From Main Sequence to Supernova},
  volume \bibinfo{volume}{212} of \textit{\bibinfo{series}{IAU Symposium}}, p.
  \bibinfo{pages}{365}.
\bibitem[{{Orosz} et~al.(2011){Orosz}, {Steiner}, {McClintock}, {Torres},
  {Remillard}, {Bailyn}, and {Miller}}]{2011ApJ...730...75O}
\bibinfo{author}{J.~A. {Orosz}}, \bibinfo{author}{J.~F. {Steiner}},
  \bibinfo{author}{J.~E. {McClintock}}, \bibinfo{author}{M.~A.~P. {Torres}},
  \bibinfo{author}{R.~A. {Remillard}}, \bibinfo{author}{C.~D. {Bailyn}},
  \bibinfo{author}{J.~M. {Miller}},
\newblock \bibinfo{title}{{An Improved Dynamical Model for the Microquasar XTE
  J1550-564}},
\newblock \bibinfo{journal}{\apj} \bibinfo{volume}{730} (\bibinfo{year}{2011})
  \bibinfo{pages}{75}.
\bibitem[{{Hurley} et~al.(2013){Hurley}, {Callanan}, {Elebert}, and
  {Reynolds}}]{2013MNRAS.430.1832H}
\bibinfo{author}{D.~J. {Hurley}}, \bibinfo{author}{P.~J. {Callanan}},
  \bibinfo{author}{P.~{Elebert}}, \bibinfo{author}{M.~T. {Reynolds}},
\newblock \bibinfo{title}{{The mass of the black hole in GRS 1915+105: new
  constraints from infrared spectroscopy}},
\newblock \bibinfo{journal}{\mnras} \bibinfo{volume}{430}
  (\bibinfo{year}{2013}) \bibinfo{pages}{1832--1838}.
\bibitem[{{Harry} and {LIGO Scientific
  Collaboration}(2010)}]{2010CQGra..27h4006H}
\bibinfo{author}{G.~M. {Harry}}, \bibinfo{author}{{LIGO Scientific
  Collaboration}},
\newblock \bibinfo{title}{{Advanced LIGO: the next generation of gravitational
  wave detectors}},
\newblock \bibinfo{journal}{Classical and Quantum Gravity} \bibinfo{volume}{27}
  (\bibinfo{year}{2010}) \bibinfo{pages}{084006}.
\bibitem[{{Accadia} et~al.(2011){Accadia}, {Acernese}, {Antonucci}, {Astone},
  {Ballardin}, {Barone}, {Barsuglia}, {Basti}, {Bauer}, {Bebronne}, {Beker},
  {Belletoile}, and et~al.}]{2011CQGra..28k4002A}
\bibinfo{author}{T.~{Accadia}}, \bibinfo{author}{F.~{Acernese}},
  \bibinfo{author}{F.~{Antonucci}}, \bibinfo{author}{P.~{Astone}},
  \bibinfo{author}{G.~{Ballardin}}, \bibinfo{author}{F.~{Barone}},
  \bibinfo{author}{M.~{Barsuglia}}, \bibinfo{author}{A.~{Basti}},
  \bibinfo{author}{T.~S. {Bauer}}, \bibinfo{author}{M.~{Bebronne}},
  \bibinfo{author}{M.~G. {Beker}}, \bibinfo{author}{A.~{Belletoile}},
  \bibinfo{author}{et~al.},
\newblock \bibinfo{title}{{Status of the Virgo project}},
\newblock \bibinfo{journal}{Classical and Quantum Gravity} \bibinfo{volume}{28}
  (\bibinfo{year}{2011}) \bibinfo{pages}{114002}.
\bibitem[{{Somiya}(2012)}]{2012CQGra..29l4007S}
\bibinfo{author}{K.~{Somiya}},
\newblock \bibinfo{title}{{Detector configuration of KAGRA-the Japanese
  cryogenic gravitational-wave detector}},
\newblock \bibinfo{journal}{Classical and Quantum Gravity} \bibinfo{volume}{29}
  (\bibinfo{year}{2012}) \bibinfo{pages}{124007}.
\bibitem[{{Unnikrishnan}(2013)}]{2013IJMPD..2241010U}
\bibinfo{author}{C.~S. {Unnikrishnan}},
\newblock \bibinfo{title}{{IndIGO and Ligo-India Scope and Plans for
  Gravitational Wave Research and Precision Metrology in India}},
\newblock \bibinfo{journal}{International Journal of Modern Physics D}
  \bibinfo{volume}{22} (\bibinfo{year}{2013}) \bibinfo{pages}{41010}.
\bibitem[{{Abadie} et~al.(2010){Abadie}, {Abbott}, {Abbott}, {Abernathy},
  {Accadia}, {Acernese}, {Adams}, {Adhikari}, {Ajith}, {Allen}, and
  et~al.}]{2010CQGra..27q3001A}
\bibinfo{author}{J.~{Abadie}}, \bibinfo{author}{B.~P. {Abbott}},
  \bibinfo{author}{R.~{Abbott}}, \bibinfo{author}{M.~{Abernathy}},
  \bibinfo{author}{T.~{Accadia}}, \bibinfo{author}{F.~{Acernese}},
  \bibinfo{author}{C.~{Adams}}, \bibinfo{author}{R.~{Adhikari}},
  \bibinfo{author}{P.~{Ajith}}, \bibinfo{author}{B.~{Allen}},
  \bibinfo{author}{et~al.},
\newblock \bibinfo{title}{{TOPICAL REVIEW: Predictions for the rates of compact
  binary coalescences observable by ground-based gravitational-wave
  detectors}},
\newblock \bibinfo{journal}{Classical and Quantum Gravity} \bibinfo{volume}{27}
  (\bibinfo{year}{2010}) \bibinfo{pages}{173001}.
\bibitem[{{Peters}(1964)}]{1964PhRv..136.1224P}
\bibinfo{author}{P.~C. {Peters}},
\newblock \bibinfo{title}{{Gravitational Radiation and the Motion of Two Point
  Masses}},
\newblock \bibinfo{journal}{Physical Review} \bibinfo{volume}{136}
  (\bibinfo{year}{1964}) \bibinfo{pages}{1224--1232}.
\bibitem[{{Veitch} et~al.(2012){Veitch}, {Mandel}, {Aylott}, {Farr}, {Raymond},
  {Rodriguez}, {van der Sluys}, {Kalogera}, and
  {Vecchio}}]{2012PhRvD..85j4045V}
\bibinfo{author}{J.~{Veitch}}, \bibinfo{author}{I.~{Mandel}},
  \bibinfo{author}{B.~{Aylott}}, \bibinfo{author}{B.~{Farr}},
  \bibinfo{author}{V.~{Raymond}}, \bibinfo{author}{C.~{Rodriguez}},
  \bibinfo{author}{M.~{van der Sluys}}, \bibinfo{author}{V.~{Kalogera}},
  \bibinfo{author}{A.~{Vecchio}},
\newblock \bibinfo{title}{{Estimating parameters of coalescing compact binaries
  with proposed advanced detector networks}},
\newblock \bibinfo{journal}{\prd} \bibinfo{volume}{85} (\bibinfo{year}{2012})
  \bibinfo{pages}{104045}.
\bibitem[{{Miller}(2005)}]{2005ApJ...626L..41M}
\bibinfo{author}{M.~C. {Miller}},
\newblock \bibinfo{title}{{Prompt Mergers of Neutron Stars with Black Holes}},
\newblock \bibinfo{journal}{\apjl} \bibinfo{volume}{626} (\bibinfo{year}{2005})
  \bibinfo{pages}{L41--L44}.
\bibitem[{{Faber} et~al.(2006){Faber}, {Baumgarte}, {Shapiro}, {Taniguchi}, and
  {Rasio}}]{2006PhRvD..73b4012F}
\bibinfo{author}{J.~A. {Faber}}, \bibinfo{author}{T.~W. {Baumgarte}},
  \bibinfo{author}{S.~L. {Shapiro}}, \bibinfo{author}{K.~{Taniguchi}},
  \bibinfo{author}{F.~A. {Rasio}},
\newblock \bibinfo{title}{{Dynamical evolution of black hole-neutron star
  binaries in general relativity: Simulations of tidal disruption}},
\newblock \bibinfo{journal}{\prd} \bibinfo{volume}{73} (\bibinfo{year}{2006})
  \bibinfo{pages}{024012}.
\bibitem[{{Shibata} and {Taniguchi}(2006)}]{2006PhRvD..73f4027S}
\bibinfo{author}{M.~{Shibata}}, \bibinfo{author}{K.~{Taniguchi}},
\newblock \bibinfo{title}{{Merger of binary neutron stars to a black hole: Disk
  mass, short gamma-ray bursts, and quasinormal mode ringing}},
\newblock \bibinfo{journal}{\prd} \bibinfo{volume}{73} (\bibinfo{year}{2006})
  \bibinfo{pages}{064027}.
\bibitem[{{Shibata} and {Uryu}(2006)}]{2006PhRvD..74l1503S}
\bibinfo{author}{M.~{Shibata}}, \bibinfo{author}{K.~{Uryu}},
\newblock \bibinfo{title}{{Merger of black hole-neutron star binaries:
  Nonspinning black hole case}},
\newblock \bibinfo{journal}{\prd} \bibinfo{volume}{74} (\bibinfo{year}{2006})
  \bibinfo{pages}{121503}.
\bibitem[{{Shibata} and {Uryu}(2007)}]{2007CQGra..24S.125S}
\bibinfo{author}{M.~{Shibata}}, \bibinfo{author}{K.~{Uryu}},
\newblock \bibinfo{title}{{Merger of black hole neutron star binaries in full
  general relativity}},
\newblock \bibinfo{journal}{Classical and Quantum Gravity} \bibinfo{volume}{24}
  (\bibinfo{year}{2007}) \bibinfo{pages}{125}.
\bibitem[{{Etienne} et~al.(2008){Etienne}, {Faber}, {Liu}, {Shapiro},
  {Taniguchi}, and {Baumgarte}}]{2008PhRvD..77h4002E}
\bibinfo{author}{Z.~B. {Etienne}}, \bibinfo{author}{J.~A. {Faber}},
  \bibinfo{author}{Y.~T. {Liu}}, \bibinfo{author}{S.~L. {Shapiro}},
  \bibinfo{author}{K.~{Taniguchi}}, \bibinfo{author}{T.~W. {Baumgarte}},
\newblock \bibinfo{title}{{Fully general relativistic simulations of black
  hole-neutron star mergers}},
\newblock \bibinfo{journal}{\prd} \bibinfo{volume}{77} (\bibinfo{year}{2008})
  \bibinfo{pages}{084002}.
\bibitem[{{Shibata} and {Taniguchi}(2008)}]{2008PhRvD..77h4015S}
\bibinfo{author}{M.~{Shibata}}, \bibinfo{author}{K.~{Taniguchi}},
\newblock \bibinfo{title}{{Merger of black hole and neutron star in general
  relativity: Tidal disruption, torus mass, and gravitational waves}},
\newblock \bibinfo{journal}{\prd} \bibinfo{volume}{77} (\bibinfo{year}{2008})
  \bibinfo{pages}{084015}.
\bibitem[{{Duez} et~al.(2008){Duez}, {Foucart}, {Kidder}, {Pfeiffer}, {Scheel},
  and {Teukolsky}}]{2008PhRvD..78j4015D}
\bibinfo{author}{M.~D. {Duez}}, \bibinfo{author}{F.~{Foucart}},
  \bibinfo{author}{L.~E. {Kidder}}, \bibinfo{author}{H.~P. {Pfeiffer}},
  \bibinfo{author}{M.~A. {Scheel}}, \bibinfo{author}{S.~A. {Teukolsky}},
\newblock \bibinfo{title}{{Evolving black hole-neutron star binaries in general
  relativity using pseudospectral and finite difference methods}},
\newblock \bibinfo{journal}{\prd} \bibinfo{volume}{78} (\bibinfo{year}{2008})
  \bibinfo{pages}{104015}.
\bibitem[{{Etienne} et~al.(2009){Etienne}, {Liu}, {Shapiro}, and
  {Baumgarte}}]{2009PhRvD..79d4024E}
\bibinfo{author}{Z.~B. {Etienne}}, \bibinfo{author}{Y.~T. {Liu}},
  \bibinfo{author}{S.~L. {Shapiro}}, \bibinfo{author}{T.~W. {Baumgarte}},
\newblock \bibinfo{title}{{General relativistic simulations of
  black-hole-neutron-star mergers: Effects of black-hole spin}},
\newblock \bibinfo{journal}{\prd} \bibinfo{volume}{79} (\bibinfo{year}{2009})
  \bibinfo{pages}{044024}.
\bibitem[{{Cho} and {Lee}(2010)}]{2010PASJ...62..315C}
\bibinfo{author}{H.-S. {Cho}}, \bibinfo{author}{C.-H. {Lee}},
\newblock \bibinfo{title}{{Analytical Calculation of the Mergers of Black
  Hole-Neutron Star Binaries}},
\newblock \bibinfo{journal}{\pasj} \bibinfo{volume}{62} (\bibinfo{year}{2010})
  \bibinfo{pages}{315--}.
\bibitem[{{Ruffert} and {Janka}(2010)}]{2010A&A...514A..66R}
\bibinfo{author}{M.~{Ruffert}}, \bibinfo{author}{H.-T. {Janka}},
\newblock \bibinfo{title}{{Polytropic neutron star - black hole merger
  simulations with a Paczy{\'n}ski-Wiita potential}},
\newblock \bibinfo{journal}{\aap} \bibinfo{volume}{514} (\bibinfo{year}{2010})
  \bibinfo{pages}{A66}.
\bibitem[{{Duez} et~al.(2010){Duez}, {Foucart}, {Kidder}, {Ott}, and
  {Teukolsky}}]{2010CQGra..27k4106D}
\bibinfo{author}{M.~D. {Duez}}, \bibinfo{author}{F.~{Foucart}},
  \bibinfo{author}{L.~E. {Kidder}}, \bibinfo{author}{C.~D. {Ott}},
  \bibinfo{author}{S.~A. {Teukolsky}},
\newblock \bibinfo{title}{{Equation of state effects in black hole-neutron star
  mergers}},
\newblock \bibinfo{journal}{Classical and Quantum Gravity} \bibinfo{volume}{27}
  (\bibinfo{year}{2010}) \bibinfo{pages}{114106}.
\bibitem[{{Chawla} et~al.(2010){Chawla}, {Anderson}, {Besselman}, {Lehner},
  {Liebling}, {Motl}, and {Neilsen}}]{2010PhRvL.105k1101C}
\bibinfo{author}{S.~{Chawla}}, \bibinfo{author}{M.~{Anderson}},
  \bibinfo{author}{M.~{Besselman}}, \bibinfo{author}{L.~{Lehner}},
  \bibinfo{author}{S.~L. {Liebling}}, \bibinfo{author}{P.~M. {Motl}},
  \bibinfo{author}{D.~{Neilsen}},
\newblock \bibinfo{title}{{Mergers of Magnetized Neutron Stars with Spinning
  Black Holes: Disruption, Accretion, and Fallback}},
\newblock \bibinfo{journal}{Physical Review Letters} \bibinfo{volume}{105}
  (\bibinfo{year}{2010}) \bibinfo{pages}{111101}.
\bibitem[{{Foucart} et~al.(2011){Foucart}, {Duez}, {Kidder}, and
  {Teukolsky}}]{2011PhRvD..83b4005F}
\bibinfo{author}{F.~{Foucart}}, \bibinfo{author}{M.~D. {Duez}},
  \bibinfo{author}{L.~E. {Kidder}}, \bibinfo{author}{S.~A. {Teukolsky}},
\newblock \bibinfo{title}{{Black hole-neutron star mergers: Effects of the
  orientation of the black hole spin}},
\newblock \bibinfo{journal}{\prd} \bibinfo{volume}{83} (\bibinfo{year}{2011})
  \bibinfo{pages}{024005}.
\bibitem[{{Stephens} et~al.(2011){Stephens}, {East}, and
  {Pretorius}}]{2011ApJ...737L...5S}
\bibinfo{author}{B.~C. {Stephens}}, \bibinfo{author}{W.~E. {East}},
  \bibinfo{author}{F.~{Pretorius}},
\newblock \bibinfo{title}{{Eccentric Black-hole-Neutron-star Mergers}},
\newblock \bibinfo{journal}{\apjl} \bibinfo{volume}{737} (\bibinfo{year}{2011})
  \bibinfo{pages}{L5}.
\bibitem[{{Shibata} and {Taniguchi}(2011)}]{2011LRR....14....6S}
\bibinfo{author}{M.~{Shibata}}, \bibinfo{author}{K.~{Taniguchi}},
\newblock \bibinfo{title}{{Coalescence of Black Hole-Neutron Star Binaries}},
\newblock \bibinfo{journal}{Living Reviews in Relativity} \bibinfo{volume}{14}
  (\bibinfo{year}{2011}) \bibinfo{pages}{6}.
\bibitem[{{Foucart} et~al.(2012){Foucart}, {Duez}, {Kidder}, {Scheel},
  {Szilagyi}, and {Teukolsky}}]{2012PhRvD..85d4015F}
\bibinfo{author}{F.~{Foucart}}, \bibinfo{author}{M.~D. {Duez}},
  \bibinfo{author}{L.~E. {Kidder}}, \bibinfo{author}{M.~A. {Scheel}},
  \bibinfo{author}{B.~{Szilagyi}}, \bibinfo{author}{S.~A. {Teukolsky}},
\newblock \bibinfo{title}{{Black hole-neutron star mergers for 10M$_\odot$
  black holes}},
\newblock \bibinfo{journal}{\prd} \bibinfo{volume}{85} (\bibinfo{year}{2012})
  \bibinfo{pages}{044015}.
\bibitem[{{Etienne} et~al.(2012){Etienne}, {Liu}, {Paschalidis}, and
  {Shapiro}}]{2012PhRvD..85f4029E}
\bibinfo{author}{Z.~B. {Etienne}}, \bibinfo{author}{Y.~T. {Liu}},
  \bibinfo{author}{V.~{Paschalidis}}, \bibinfo{author}{S.~L. {Shapiro}},
\newblock \bibinfo{title}{{General relativistic simulations of
  black-hole-neutron-star mergers: Effects of magnetic fields}},
\newblock \bibinfo{journal}{\prd} \bibinfo{volume}{85} (\bibinfo{year}{2012})
  \bibinfo{pages}{064029}.
\bibitem[{{East} et~al.(2012){East}, {Pretorius}, and
  {Stephens}}]{2012PhRvD..85l4009E}
\bibinfo{author}{W.~E. {East}}, \bibinfo{author}{F.~{Pretorius}},
  \bibinfo{author}{B.~C. {Stephens}},
\newblock \bibinfo{title}{{Eccentric black hole-neutron star mergers: Effects
  of black hole spin and equation of state}},
\newblock \bibinfo{journal}{\prd} \bibinfo{volume}{85} (\bibinfo{year}{2012})
  \bibinfo{pages}{124009}.
\bibitem[{{Etienne} et~al.(2012){Etienne}, {Paschalidis}, and
  {Shapiro}}]{2012PhRvD..86h4026E}
\bibinfo{author}{Z.~B. {Etienne}}, \bibinfo{author}{V.~{Paschalidis}},
  \bibinfo{author}{S.~L. {Shapiro}},
\newblock \bibinfo{title}{{General-relativistic simulations of
  black-hole-neutron-star mergers: Effects of tilted magnetic fields}},
\newblock \bibinfo{journal}{\prd} \bibinfo{volume}{86} (\bibinfo{year}{2012})
  \bibinfo{pages}{084026}.
\bibitem[{{Foucart}(2012)}]{2012PhRvD..86l4007F}
\bibinfo{author}{F.~{Foucart}},
\newblock \bibinfo{title}{{Black-hole-neutron-star mergers: Disk mass
  predictions}},
\newblock \bibinfo{journal}{\prd} \bibinfo{volume}{86} (\bibinfo{year}{2012})
  \bibinfo{pages}{124007}.
\bibitem[{{Paschalidis} et~al.(2013){Paschalidis}, {Etienne}, and
  {Shapiro}}]{2013PhRvD..88b1504P}
\bibinfo{author}{V.~{Paschalidis}}, \bibinfo{author}{Z.~B. {Etienne}},
  \bibinfo{author}{S.~L. {Shapiro}},
\newblock \bibinfo{title}{{General-relativistic simulations of binary black
  hole-neutron stars: Precursor electromagnetic signals}},
\newblock \bibinfo{journal}{\prd} \bibinfo{volume}{88} (\bibinfo{year}{2013})
  \bibinfo{pages}{021504}.
\bibitem[{{Kyutoku} et~al.(2013){Kyutoku}, {Ioka}, and
  {Shibata}}]{2013PhRvD..88d1503K}
\bibinfo{author}{K.~{Kyutoku}}, \bibinfo{author}{K.~{Ioka}},
  \bibinfo{author}{M.~{Shibata}},
\newblock \bibinfo{title}{{Anisotropic mass ejection from black hole-neutron
  star binaries: Diversity of electromagnetic counterparts}},
\newblock \bibinfo{journal}{\prd} \bibinfo{volume}{88} (\bibinfo{year}{2013})
  \bibinfo{pages}{041503}.
\bibitem[{{Tanaka} et~al.(2014){Tanaka}, {Hotokezaka}, {Kyutoku}, {Wanajo},
  {Kiuchi}, {Sekiguchi}, and {Shibata}}]{2014ApJ...780...31T}
\bibinfo{author}{M.~{Tanaka}}, \bibinfo{author}{K.~{Hotokezaka}},
  \bibinfo{author}{K.~{Kyutoku}}, \bibinfo{author}{S.~{Wanajo}},
  \bibinfo{author}{K.~{Kiuchi}}, \bibinfo{author}{Y.~{Sekiguchi}},
  \bibinfo{author}{M.~{Shibata}},
\newblock \bibinfo{title}{{Radioactively Powered Emission from Black
  Hole-Neutron Star Mergers}},
\newblock \bibinfo{journal}{\apj} \bibinfo{volume}{780} (\bibinfo{year}{2014})
  \bibinfo{pages}{31}.
\bibitem[{{Futamase} and {Itoh}(2007)}]{2007LRR....10....2F}
\bibinfo{author}{T.~{Futamase}}, \bibinfo{author}{Y.~{Itoh}},
\newblock \bibinfo{title}{{The Post-Newtonian Approximation for Relativistic
  Compact Binaries}},
\newblock \bibinfo{journal}{Living Reviews in Relativity} \bibinfo{volume}{10}
  (\bibinfo{year}{2007}) \bibinfo{pages}{2}.
\bibitem[{{van der Sluys} et~al.(2008){van der Sluys}, {R{\"o}ver}, {Stroeer},
  {Raymond}, {Mandel}, {Christensen}, {Kalogera}, {Meyer}, and
  {Vecchio}}]{2008ApJ...688L..61V}
\bibinfo{author}{M.~V. {van der Sluys}}, \bibinfo{author}{C.~{R{\"o}ver}},
  \bibinfo{author}{A.~{Stroeer}}, \bibinfo{author}{V.~{Raymond}},
  \bibinfo{author}{I.~{Mandel}}, \bibinfo{author}{N.~{Christensen}},
  \bibinfo{author}{V.~{Kalogera}}, \bibinfo{author}{R.~{Meyer}},
  \bibinfo{author}{A.~{Vecchio}},
\newblock \bibinfo{title}{{Gravitational-Wave Astronomy with Inspiral Signals
  of Spinning Compact-Object Binaries}},
\newblock \bibinfo{journal}{\apjl} \bibinfo{volume}{688} (\bibinfo{year}{2008})
  \bibinfo{pages}{L61--L64}.
\bibitem[{{van der Sluys} et~al.(2009){van der Sluys}, {Mandel}, {Raymond},
  {Kalogera}, {R{\"o}ver}, and {Christensen}}]{2009CQGra..26t4010V}
\bibinfo{author}{M.~{van der Sluys}}, \bibinfo{author}{I.~{Mandel}},
  \bibinfo{author}{V.~{Raymond}}, \bibinfo{author}{V.~{Kalogera}},
  \bibinfo{author}{C.~{R{\"o}ver}}, \bibinfo{author}{N.~{Christensen}},
\newblock \bibinfo{title}{{Parameter estimation for signals from compact binary
  inspirals injected into LIGO data}},
\newblock \bibinfo{journal}{Classical and Quantum Gravity} \bibinfo{volume}{26}
  (\bibinfo{year}{2009}) \bibinfo{pages}{204010}.
\bibitem[{{Raymond} et~al.(2010){Raymond}, {van der Sluys}, {Mandel},
  {Kalogera}, {R{\"o}ver}, and {Christensen}}]{2010CQGra..27k4009R}
\bibinfo{author}{V.~{Raymond}}, \bibinfo{author}{M.~V. {van der Sluys}},
  \bibinfo{author}{I.~{Mandel}}, \bibinfo{author}{V.~{Kalogera}},
  \bibinfo{author}{C.~{R{\"o}ver}}, \bibinfo{author}{N.~{Christensen}},
\newblock \bibinfo{title}{{The effects of LIGO detector noise on a
  15-dimensional Markov-chain Monte Carlo analysis of gravitational-wave
  signals}},
\newblock \bibinfo{journal}{Classical and Quantum Gravity} \bibinfo{volume}{27}
  (\bibinfo{year}{2010}) \bibinfo{pages}{114009}.
\bibitem[{{Yagi} and {Yunes}(2013)}]{2013Sci...341..365Y}
\bibinfo{author}{K.~{Yagi}}, \bibinfo{author}{N.~{Yunes}},
\newblock \bibinfo{title}{{I-Love-Q: Unexpected Universal Relations for Neutron
  Stars and Quark Stars}},
\newblock \bibinfo{journal}{Science} \bibinfo{volume}{341}
  (\bibinfo{year}{2013}) \bibinfo{pages}{365--368}.
\bibitem[{{Maselli} et~al.(2013){Maselli}, {Cardoso}, {Ferrari}, {Gualtieri},
  and {Pani}}]{2013PhRvD..88b3007M}
\bibinfo{author}{A.~{Maselli}}, \bibinfo{author}{V.~{Cardoso}},
  \bibinfo{author}{V.~{Ferrari}}, \bibinfo{author}{L.~{Gualtieri}},
  \bibinfo{author}{P.~{Pani}},
\newblock \bibinfo{title}{{Equation-of-state-independent relations in neutron
  stars}},
\newblock \bibinfo{journal}{\prd} \bibinfo{volume}{88} (\bibinfo{year}{2013})
  \bibinfo{pages}{023007}.
\bibitem[{{Lattimer} and {Prakash}(2010)}]{2010arXiv1012.3208L}
\bibinfo{author}{J.~M. {Lattimer}}, \bibinfo{author}{M.~{Prakash}},
\newblock \bibinfo{title}{{What a Two Solar Mass Neutron Star Really Means}},
\newblock \bibinfo{journal}{arXiv:1012.3208}  (\bibinfo{year}{2010}).
\bibitem[{{Narayan} and {McClintock}(2012)}]{2012MNRAS.419L..69N}
\bibinfo{author}{R.~{Narayan}}, \bibinfo{author}{J.~E. {McClintock}},
\newblock \bibinfo{title}{{Observational evidence for a correlation between jet
  power and black hole spin}},
\newblock \bibinfo{journal}{\mnras} \bibinfo{volume}{419}
  (\bibinfo{year}{2012}) \bibinfo{pages}{L69--L73}.
\bibitem[{{Russell} et~al.(2013){Russell}, {Gallo}, and
  {Fender}}]{2013MNRAS.431..405R}
\bibinfo{author}{D.~M. {Russell}}, \bibinfo{author}{E.~{Gallo}},
  \bibinfo{author}{R.~P. {Fender}},
\newblock \bibinfo{title}{{Observational constraints on the powering mechanism
  of transient relativistic jets}},
\newblock \bibinfo{journal}{\mnras} \bibinfo{volume}{431}
  (\bibinfo{year}{2013}) \bibinfo{pages}{405--414}.

\end{thebibliography}

\end{document}